\documentclass[12pt]{article}
\usepackage[utf8]{inputenc}

\usepackage[authoryear]{natbib}
\usepackage{amssymb}
\usepackage{amsmath}
\usepackage{graphicx}
\usepackage{txfonts}
\usepackage{xspace}
\usepackage{subfigure}
\usepackage{makeidx}
\usepackage{emptypage}
\usepackage{hyperref}
\usepackage{rotating}
\usepackage{mwe}

\usepackage{tikz}
	\usetikzlibrary{trees}
	\usetikzlibrary{decorations.pathmorphing}
	\usetikzlibrary{shapes,arrows}
\usepackage{wrapfig}

\makeindex
\newcommand{\DFT}{\textbf{DFT}\xspace}
\newcommand{\FFT}{\textbf{FFT}\xspace}
\newcommand{\mFFT}{\textbf{mFFT}\xspace}
\newcommand{\FFTs}{\textbf{FFTs}\xspace}
\newcommand{\EPG}{\textbf{EPG}\xspace}
\newcommand{\EPGs}{\textbf{EPG}s\xspace}
\newcommand{\PCR}{\textbf{PCR}\xspace}
\newcommand{\PGF}{\textbf{PGF}\xspace}
\newcommand{\PGFs}{\textbf{PGFs}\xspace}
\newcommand{\STR}{\textbf{STR}\xspace}
\newcommand{\LTD}{\textbf{LTDNA}\xspace}
\newcommand{\mcl}[1]{${#1}{\mu}$\textsl{L}}

\newcommand{\RFU}{\textbf{RFU}\xspace}

\newcommand{\pref}[1]{\mbox{page~\pageref{#1}}}
\renewcommand{\eqref}[1]{\mbox{(\ref{#1})}}

\newcommand{\appref}[1]{\mbox{Appendix~\ref{#1}}}
\newcommand{\secref}[1]{\mbox{Section~\ref{#1}}}
\newcommand{\figref}[1]{\mbox{Figure~\ref{#1}}}

\newcommand{\tabref}[1]{\mbox{Table~\ref{#1}}}

\newcommand{\Theoremref}[1]{\mbox{Theorem~\ref{#1}}}

\newcommand{\Figref}[1]{\mbox{Figure~\ref{#1}}}

\newcommand{\partref}[1]{\mbox{Part~\ref{#1}}}

\newcommand{\algref}[1]{\mbox{Algorithm~\ref{#1}}}
\newtheorem{expl}{Example}[section]
\newtheorem{definer}[expl]{Definition}

\newtheorem{theorem}{Theorem}[section]
\newtheorem{algor}[expl]{Algorithm}

\newcommand{\halm}{\hspace*{\fill} $\Box$\par}

\newenvironment{alg}{\begin{algor}\rm}{\halm\end{algor}}

\newcommand{\cd}{\,|\,}

\newcommand{\E}{{\varmathbb{E}}}
\newcommand{\V}{{\varmathbb{V}}}
\newcommand{\cov}{{\rm Cov}}
\newcommand{\cor}{{\rm Cor}}
\newcommand{\hyp}{{\cal H}}

\newcommand\sbackup{{\vspace{-18pt}}}

\newcommand{\etc}{{\rm etc.}\xspace}

\begin{document}
\pagenumbering{roman}
\author{Robert G. Cowell}
\title{A unifying
  framework for the modelling and analysis of STR DNA samples arising
  in forensic casework}
\date{\today}

\maketitle
\begin{abstract}

  This paper presents a new framework for analysing forensic DNA
  samples using probabilistic genotyping. Specifically it presents a
  mathematical framework for specifying and combining the steps in
  producing forensic casework electropherograms of short tandem repeat
  loci from DNA samples. It is applicable to both high and low
  template DNA samples, that is, samples containing either high or low
  amounts DNA. A specific model is developed within the framework, by
  way of particular modelling assumptions and approximations, and its
  interpretive power presented on examples using simulated data and
  data from a publicly available dataset.

  The framework relies heavily on the use of univariate and
  multivariate probability generating functions.  It is shown that
  these provide a succinct and elegant mathematical scaffolding to
  model the key steps in the process. A significant development in
  this paper is that of new numerical methods for accurately and
  efficiently evaluating the probability distribution of amplicons
  arising from the polymerase chain reaction process, which is
  modelled as a discrete multi-type branching process.  Source code in
  the scripting languages Python, R and Julia is provided for
  illustration of these methods.  These new developments will be of
  general interest to persons working outside the province of forensic
  DNA interpretation that this paper focuses on. \newline

  \textbf{Keywords}: {DNA profiles; forensic statistics; polymerase
    chain reaction; branching process; probability generating
    functions; Fast Fourier Transform.}

\end{abstract}

\clearpage

\tableofcontents
\newpage
\pagenumbering{arabic}

\setcounter{page}{1}

\section*{Introduction}
\label{sec:dnaintro}

Genetic fingerprinting, also called DNA profiling, has grown to be an
indispensable tool for identification of individuals in the
investigative and judicial process associated with criminal cases.
Judiciaries throughout the world have built up large databases of DNA
profiles of convicted, and additionally in some cases non-convicted,
individuals. These databases are based, in the main, upon
short-tandem-repeat (\STR) loci. Although newer DNA identification
techniques based upon genomic sequencing (next generation sequencing)
are being actively developed, the size of the currently existing
databases, and the comparative low cost and reliability of the
laboratory process for dealing with \STR loci means that \STR DNA
profiling will remain in use for several years to some.

Ever since the pioneering work instigated by \cite{jeffreys:85},
advances of techniques in the collection and processing of crime scene
DNA samples have led to forensic laboratories routinely processing
small amounts of DNA, for example as may be contained in just a few
cells extracted from a fingerprint. Indeed, current technology allows
for the amplification of even as little DNA as is contained within one
cell \citep{Findlay1997}.  Such \textit{low template DNA} (\LTD)
samples can present challenges to the forensic scientist in their
interpretation. Such examples typically show signs of containing DNA
from two or more persons.  These \LTD mixtures are also subject to
various artefacts such as degradation, drop-in, drop-out and stutter
which further exacerbate the problems of their interpretation. Over
the years a variety of methods have been developed and applied to
provide interpretation in court.

It is convenient to distinguish two phases of DNA processing. The
first is the laboratory processing of the physical DNA. This part is
described in detail in \citep{butler2011advancedmethod}. The end
result of this is one or more \textit{electropherograms} (\EPGs),
described further in \secref{sec:pcr} below. Each \EPG is then subject
to interpretation by the forensic statistician; interpretation issues
are described in detail in \citep{butler2014advancedinterp}.

In recent years \textit{probabilistic genotyping} models have been
gaining acceptance within the forensic science community and in court
for interpreting challenging DNA mixtures. These models typically
describe themselves as \textit{fully continuous}, by which is meant
(at least as understood by the author) that they model the peak
heights seen in the \EPG as realisations of continuous random
variables in statistical models for evaluating the \textit{weight of
  evidence} of competing hypotheses presented to a court.

It is perhaps fair to say that that such models begin their
interpretation with the \EPG, or \EPGs in case of the analysis of
multiple replicates from a sample, and that much of the information
contained in the processing of the DNA sample that lead to the \EPG is
not used.  It is the contention of this paper that, particularly for
low template analyses, these steps need to be explicitly or implicitly
incorporated into any statistical model for analysing \STR DNA single
source samples and mixtures. To not do so is to lose important
information that could lead to biases or incorrect inferences. This
paper presents a framework in which such information can readily be
incorporated.

An early model for the whole process from sample to \EPG was given by
\citet{gill:etal:2005}. Their model was used only to simulate \EPG
peaks, in order to relate some of the parameters of their model to the
procedures in their laboratory and observed values from \PCR runs. The
model was not concerned interpreting given \EPGs. An \textbf{R}
package \citep{Rmanual} called \texttt{forensim} \citep{forensim},
implements the simulation model. Recently, another model for
simulating \EPGs has been proposed \citep{duffy:etal:2016}.

The framework presented in this book can be thought of as an extensive
elaboration of the simulation model of \citet{gill:etal:2005}. We say
\textit{framework} rather than \textit{model} as many models may be
derived from the framework; in addition the framework may be extended
to cover other aspects of the processing not covered in this
paper. One key elaboration, described in more detail below, is that
the \citet{gill:etal:2005} model starts the branching process
simulation using amplicons; it ignores the fact that the \PCR
processing of real DNA samples starts with genomic strands. A second
elaboration is that it does not take account of the tagging of
amplicons with dyes so that they can be observed for the \EPG. We
incorporate explicitly such the tagging in our framework. We also
extend the \cite{gill:etal:2005} model to include forward and double
backward stutters. Higher order stutters could also be included if
desired, but because of their smallness relative to other peaks in the
\EPG they are not considered important (as they would be
indistinguishable from noise).

However, not only can our framework be used to make simulated \EPGs,
but it can also be used for statistical interpretation.  A key part of
this is the use of \textit{multivariate probability generating
  functions } to model the steps in the laboratory process leading to
the \EPGs. This provides a compact and elegant probability model, or
rather set of models based on approximations that are assumed. Using
standard techniques, means and variances of peak height distributions
may be found from these generating functions, and these may be used to
model peak height distributions using standard distributions, such as
normal, lognormal or gamma distributions via moment matching.  Such
distributions have been used in fully continuous models as simple and
convenient distributions: in part this has been because there is no
analytic form the for full distribution of amplicons arising in the
\PCR branching process.  However, this paper shows that such full
branching process distributions may be obtained efficiently from the
probability generating functions; there is therefore no need for
probabilistic genotyping software to assume simple standard
distributions for peak height distributions.  We show in examples that they do not
capture the intricate \textit{multi-modal} distributional behaviour of
the full branching process that can occur for \LTD samples.

This paper consists of three main parts. The first introduces
background information to the objectives and problems of interpreting
forensic DNA samples.  The second part introduces elements of the
mathematical approach used in the paper, and shows how full
distributions can be efficiently computed for the branching processes
by combining a specification in terms of probability generating
functions with their evaluation using discrete Fourier transforms. The
third part presents a detailed comprehensive framework for modelling
forensic DNA problems, from which specific models may be formed by
specialisation and approximations. It describes a particular model
specialisation based on the general framework. The efficacy of the
model is exhibited with simulated and real data. 

\clearpage
\part{Background}
In this part of the paper we introduce background information to the
problems we are addressing with the framework developed here.

\section{DNA background}

In this section some background information on DNA is introduced,
sufficient for the remainder of the paper; readers unfamiliar with
this background are recommended to consult
\citep{butler2011advancedmethod,butler2014advancedinterp} for more
details.

\subsection{Short Tandem Repeat (STR) markers}

Forensic scientists encode an individual's genetic profile using the
composition of DNA at various positions on the chromosomes.  A
specific position on a chromosome is called a \textit{locus}, or
\textit{marker}.  Human DNA has twenty three pairs of chromosomes:
twenty two autosomal chromosome pairs and a sex-linked pair, the $X$
and $Y$ chromosomes.

The information at each (autosomal) locus consists of an unordered
pair of \textit{alleles}\footnote{This is also true for the Amelogenin
  locus which occurs on both X and Y chromosomes.  However for other
  sex-linked loci the alleles might also possibly occur singly or not
  at all. For example a female does not have a Y chromosome, and so
  will not have an allele in any Y-linked locus.}  which forms the
\textit{genotype} at that locus; a pair because chromosomes come in
pairs, one inherited from the father and one from the mother, and
unordered because it is not recorded from which chromosome of the pair
each allele originates.

The loci used for forensic identification have been chosen for various
reasons. Among these, we point out the two.

The first is that at each locus there is a wide variability between
individuals in the alleles that may be observed. This variability can
therefore be exploited to differentiate people.

The second reason is that, at least until recently, each (autosomal)
locus is either on a distinct chromosome, or if a pair of loci are on
the same chromosome then they are widely separated. When this occurs,
the alleles at the various loci may be treated as mutually
independent, thus simplifying the statistical analysis.  However in
recent years the numbers of loci used has increased, so that there are
now some pairs loci that are close together on the same chromosome. We
return to this issue of \textit{genetic linkage} later.

The alleles of a marker are sequences of the four amino acid
nucleotides \emph{adenine, cytosine, guanine} and \emph{thymine},
which we represent by the letters A, C, G and T. Each amino acid is
also called a \textit{base}, and because the DNA molecule has a double
helix structure, each amino acid on one strand is linked to a
complementary amino acid on the other strand; a complementary pair of
amino acids is called a \textit{base pair}.

An allele is typically named by its \textit{repeat number}, usually an
integer. For example, consider the allele with repeat number 5
(commonly also referred to as allele 5 for brevity) of the marker
TH01.  This allele includes the sequence of four nucleotides AATG
repeated consecutively five times. It can be designated by the formula
$[AATG]_5$. Likewise allele 8 of TH01 has eight consecutive
repetitions of the AATG sequence, which may be denoted by
$[AATG]_8$. Repeat numbers are not always integers. For example,
allele 8.3 of TH01 has the chemical sequence $[AATG]_5ATG[AATG]_3$, in
which `8' refers to the eight complete four-word bases $[AATG]$ and
the `.3' refers to the three base-long word sequence $ATG$ in the
middle. Repeat numbers with decimal `.1' and `.2' endings are also
possible, indicating the presence of a word of one or two bases. Note
that the integer part of the repeat number counts how many complete
words of four bases make up the allele sequence but the words need not
be all identical and may vary even within loci. For example, allele 11
of the marker vWA has the base sequence $TCTA[TCTG]_3[TCTA]_7$. Also,
some markers are based on tri- or pentanucleotide motifs rather than
tetranucleotides as above. The base-letter sequences for many alleles
may be found in \cite{book:Butler}.

When the repeat numbers of the two alleles of an individual at a
marker are the same, then the genotype for that marker is said to be
\emph{homozygous}; when the repeat numbers differ, the genotype for
that marker is said to be \emph{heterozygous}.

The repetitive structure in the alleles gives rise to the term
\emph{short tandem repeat} (\STR) marker to describe these loci; they
also go by the name of \textit{microsatellites}.  Note that for other
purposes of genetic analysis it is common to use single-nucleotide
polymorphisms (SNPs) which are defined as DNA sequence variations that
occur when a single nucleotide (A, T, C, or G) in the genome sequence
is altered.

Forensic identification using {\STR}s is based upon the size of the
allele. However, for some loci, different people might not have the
same nucleotide sequence for one or more alleles of a specific repeat
number. This information is not used by the framework developed in
this paper, but is the basis of finer genetic discrimination available
to the next-generation sequencing methods currently being developed
that were mentioned in the introduction.

Within a population the various alleles of \STR markers do not occur
equally often, some can be quite common and some quite rare.  When
carrying out probability calculations based on DNA, forensic
scientists use estimates of probabilities based on allele frequencies
in profiles of a sample of individuals. The sample sizes typically
range from a few hundred to thousands of individuals.  For example,
\cite{butler:etal:03} presents tables of US-population \STR-allele
frequencies for Caucasians, African-Americans, and Hispanics based on
sample sizes of 302, 258 and 140 individuals. These are used to
estimate the genetic profile probabilities of an individual: for the
autosomal loci and the locus Amelogenin, independence of loci
expressed through the lack of genetic linkage means that they may be
found by multiplying probabilities across the loci.  Although the \PCR
process described below applies to other sex-linked loci, the linkage
of loci on the sex-linked loci presents special problems in estimating
genetic profile probabilities. This issue will be addressed in another
paper, although some of the discussion below will not be specific to
autosomal loci.

\subsection{The PCR process}
\label{sec:pcr}

The DNA collected from a crime scene for forensic analysis consists of
a number of human cells from one or more individuals. Note that each
cell of an individual will contain two alleles (diploid cells) for
each autosomal marker, whereas sperm cells have only one allele
(haploid cells).  This means that in a mixture, a particular
individual will contribute the same number of alleles for each
marker. In order to identify the alleles that are present, a DNA
sample is first subjected to chemical reagents that break down the
cell walls so that the individual chromosomes are released into a
solution. A small amount of this solution is used to quantify the
concentration of DNA; the typical unit of measurement is picograms per
microlitre, the DNA in a single human cell having a mass of between
six and seven picograms. Having determined the density of DNA in the
sample, a volume is extracted that is estimated to contain a certain
quantity of DNA, typically around 0.5 nanograms, equivalent to around
75 human cells.  The DNA in the extract is then amplified using the
\emph{polymerase chain reaction} (PCR) process. This involves adding
\textit{primers} and other biochemicals to the extract, and then
subjecting it to a number of rapid heating and cooling cycles. Heating
the extract has the effect of splitting apart the two complementary
strands of DNA, the cooling phase then allows free floating primers
and amino acids to bind with these individual strands in such a way
that the DNA is copied. By the action of repeated heating and cooling
cycles, typically around 28 altogether, an initially small amount of
DNA is amplified to an amount large enough for
quantification. Mathematically, the amplification process may be
modelled as a branching process
\citep{article:SunPCR,article:StolovitzkyCecchi}. The amplification
process is not 100\% efficient, that is, not every allele gets copied
in each cycle. This means that if two distinct alleles in a marker are
present in the extract in the same amount prior to amplification, they
will typically occur in different amounts at the end of the PCR
process.

Note that after breaking down the cell walls to release the genomes,
there could be sufficient DNA in the sample to carry out several
independent PCR amplifications with sub-extracts. When this is done it
is called a \textit{replicate run}.

To understand the quantification stage of the post PCR amplified DNA,
it is important to know that the amplification process does not copy
only the repeated DNA word segment of a marker, it also copies DNA at
either end. These are called \textit{flanking regions}, and their
presence is important in performing the PCR process. Regions at either
end of the flanking region, called \textit{primer binding regions},
are where the primers bind to the DNA to initiate copying. These
regions are labelled as 5' and 3'; there is one for each of the two
strands making up the double helix of the DNA. During the heating
cycle the double helix separates into two strands, which for reasons
that will become apparent later we denote by the two genomic strands
by $g$ and $g_d$. During the cooling phase one of the strands primers
will bind at the 5' end, and on the other at the 3' end. After the
primers have attached the Taq polymerase adds individual bases to
complete the copying. This leads to two strands that we call $h$ and
$h_d$, (with $h$ for $h$alf-strand).

The process is illustrated in the following sequence of figures. In
the first figure we have a fragment of DNA consisting of two
complimentary joined strands $g$ and $g_d$ which are long enough to
contain the repeat structure of an allele of interest and primer
binding regions on either side of the allele. The `teeth' in the
figure indicates the individual bases making up each strand.

\begin{center}
  \includegraphics[scale=0.6]{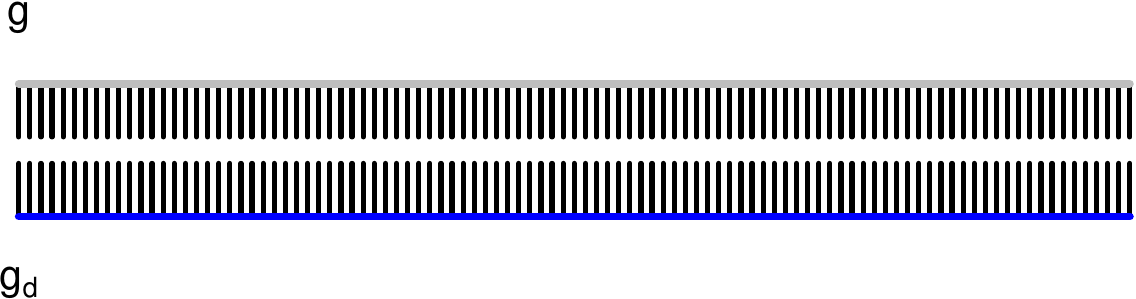}
\end{center}

Heating breaks the bonds of the base-pairs of the DNA molecule, so
separating the strands.  During cooling primers,indicated by the
smaller sets of teeth. bind to each of the 5' ends of the strands.
Note that the sequences of bases in the primers attaching to the two
ends are in general different, because of the differences in the base
sequence in the primer-binding regions of the two strands.

\begin{center}
  \includegraphics[scale=0.6]{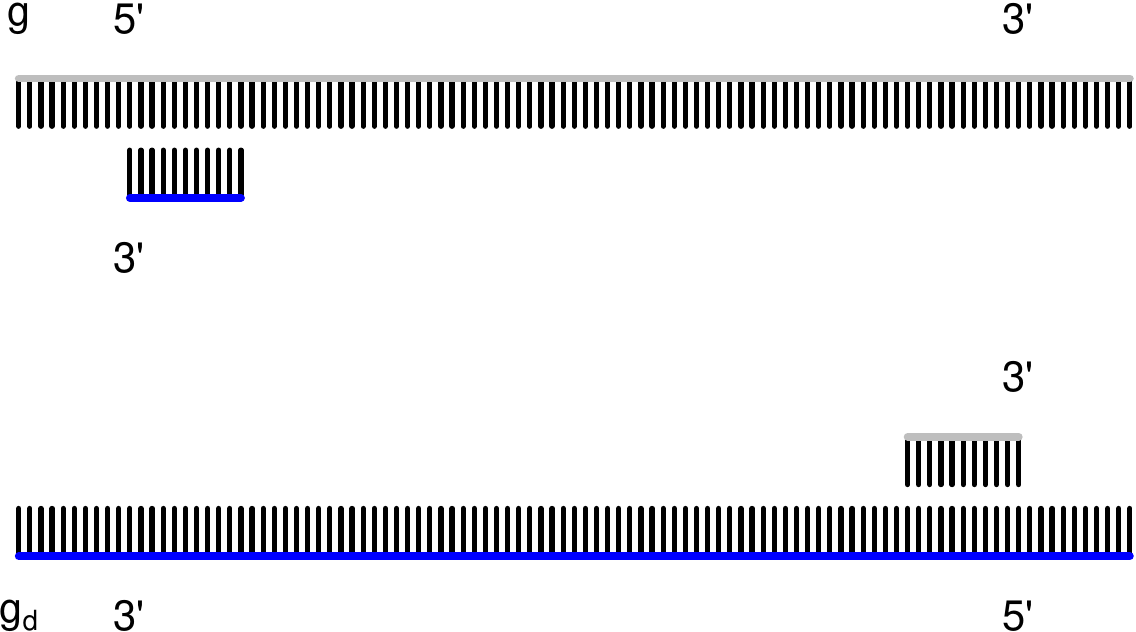}
\end{center}

After the primer binds at one end the Taq polymerase extends to copy
to beyond the primer region at the other end.  The $g$ strand
generates the $h_d$ strand for its copy, and the $g_d$ strand the $h$
strand for its copy.

\begin{center}
  \includegraphics[scale=0.6]{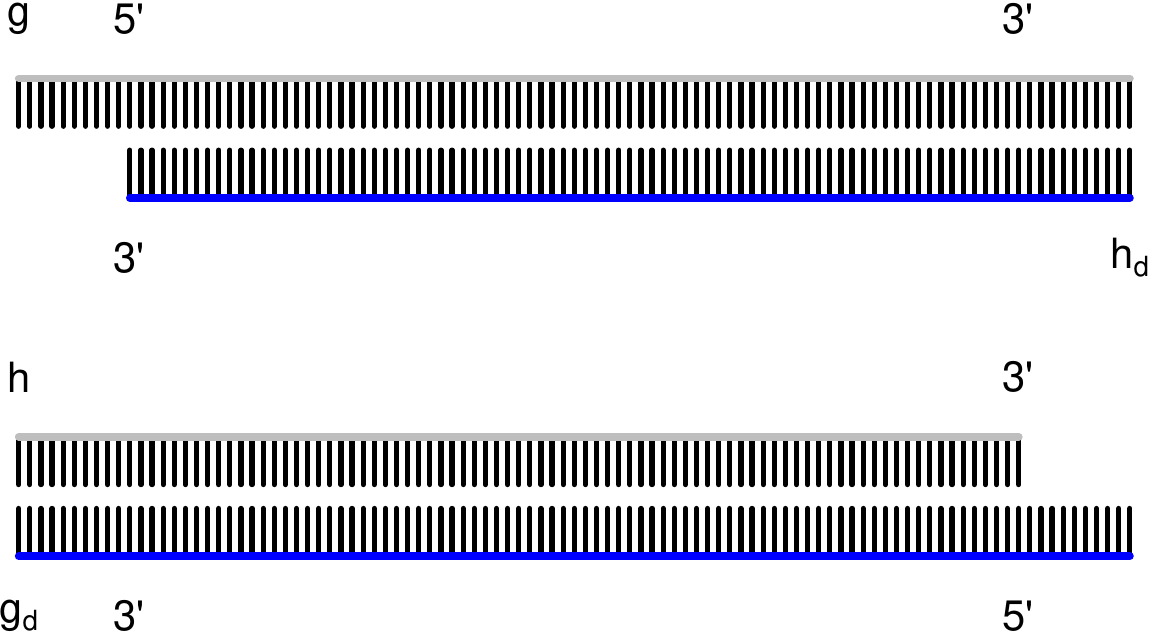}
\end{center}

During the next heating cycle the $g/h_d$ strand pair separate, and
the $g_d/h$ pair separate. The $g$ and $g_d$ strands repeat the
process as above. During the cooling phase, primers bind to the 5'
ends of the $h$ and $h_d$ strands,

\begin{center}
  \includegraphics[scale=0.6]{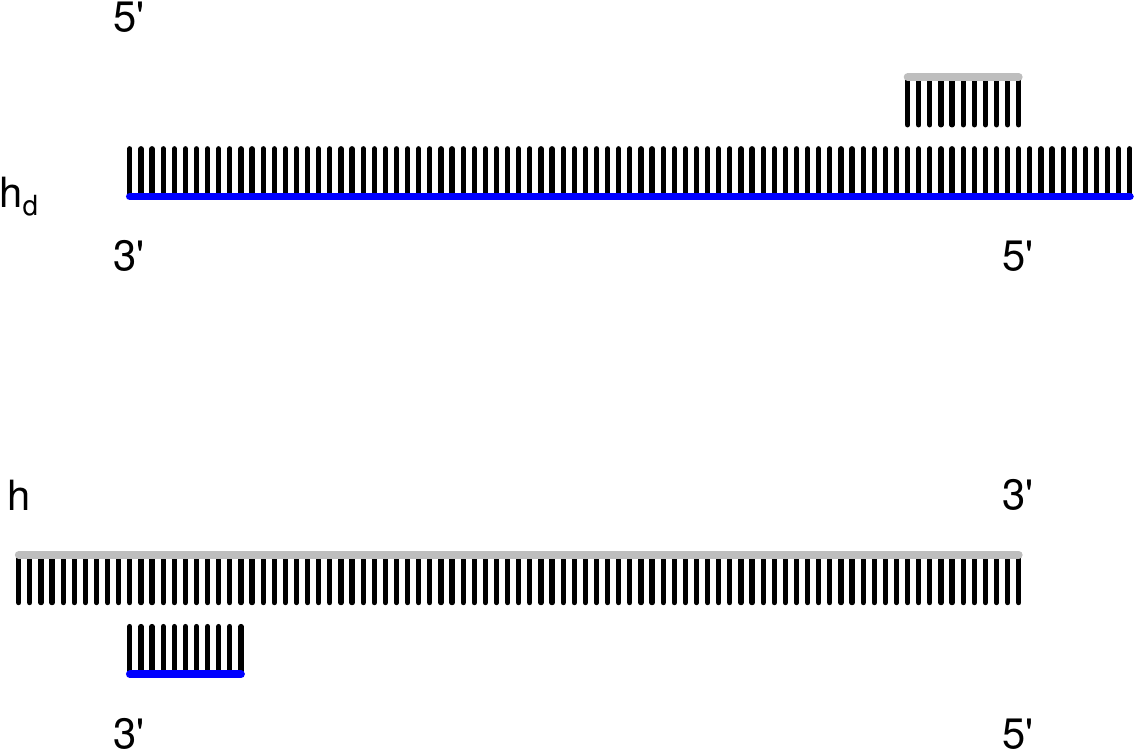}
\end{center}

and the Taq polymerase adds bases to the ends, thus making with the
$h$ strand a complementary $a_d$ strand, and with the $h_d$ strand, a
complementary $a$ strand.
\begin{center}
  \includegraphics[scale=0.6]{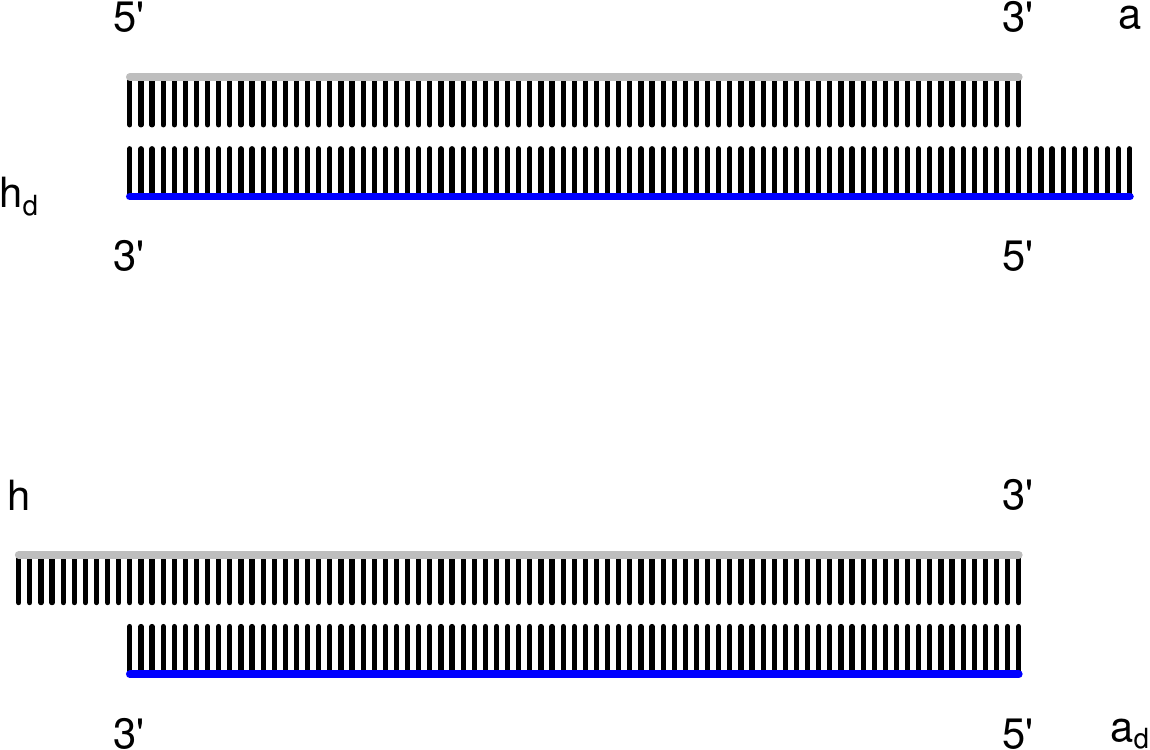}
\end{center}

During the next heating cycle the $h/a_d$ strand pair separate, as
does the $h_d/a$ strand pair.  The $h$ and $h_d$ strands behave as
above (as the $g$ and $g_d$ strands continue to do so as
well). Primers bind at the 5' ends of the $a$ and $a_d$ strands,

\begin{center}
  \includegraphics[scale=0.6]{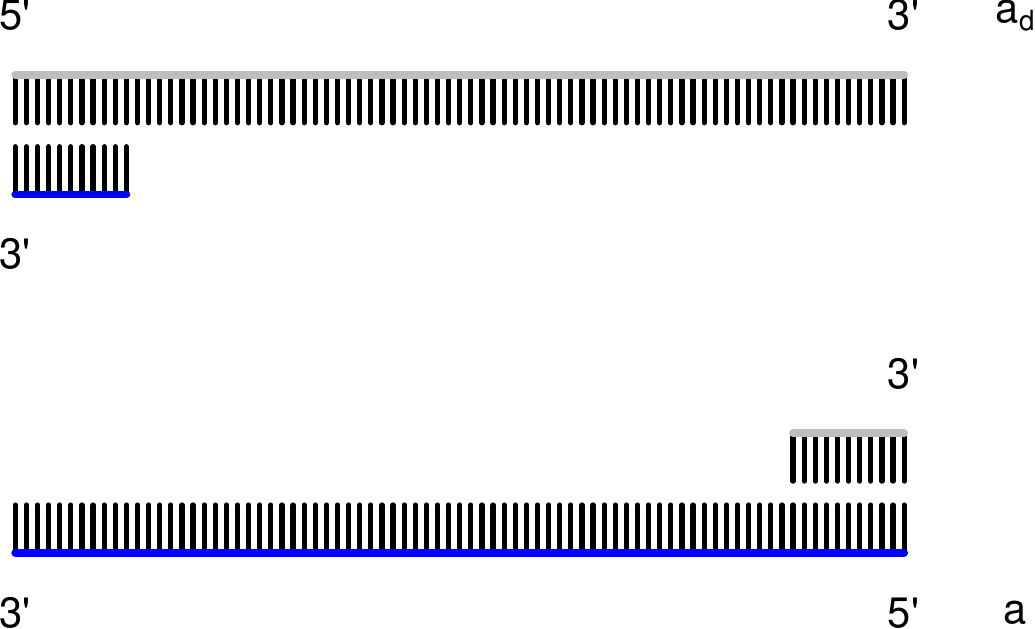}
\end{center}

\noindent
and the Taq polymerase completes the copy, so that the $a$ strand
makes a complimentary $a_d$ strand, and the $a_d$ strand makes a
complimentary $a$ strand.
\begin{center}
  \includegraphics[scale=0.6]{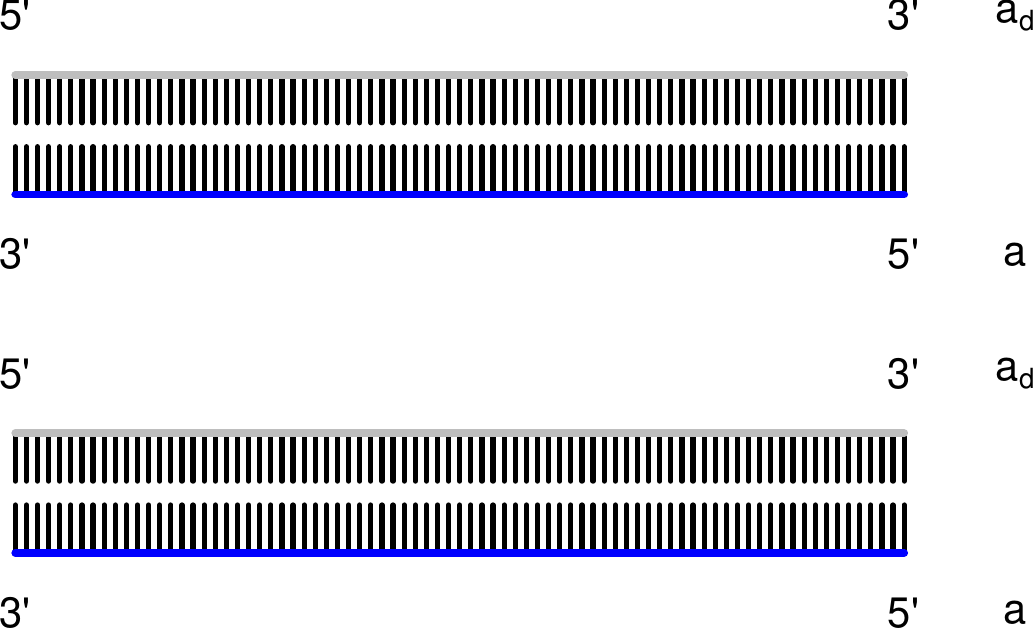}
\end{center}

Thus an amplified allele will consist of the allele word repeat
sequence region and two flanking regions, and will have a length
associated with it which is measured in the total number of base pairs
included in the word sequence and the flanking regions.  This is
called an \textit{amplicon}.
\begin{center}
  \includegraphics[scale=0.6]{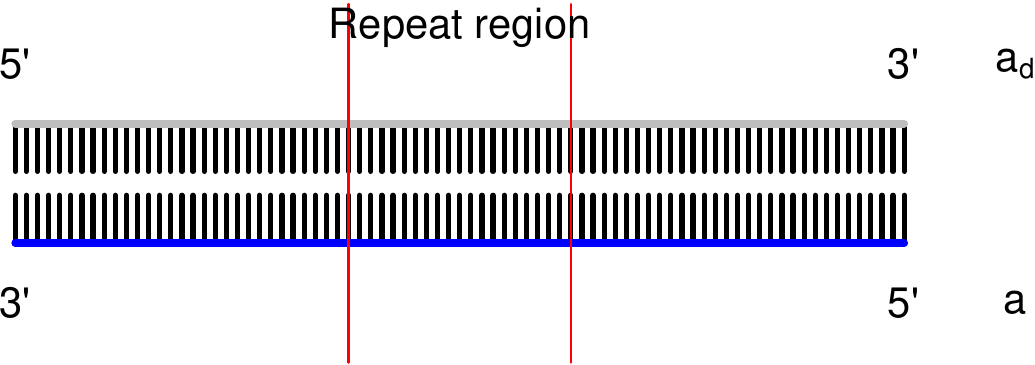}
\end{center}

Further heating cycles lead to an exponential growth in the number of
$a$ and $a_d$ strands with the number of cycles. After around 28 or 29
cycles there is a sufficient number of amplicons to measure their
amount.

For each marker, the DNA sequence (hence the size) forming each of the
two flanking regions is constant, but different across markers. Thus
quantifying a certain allele is equivalent to measuring how much DNA
is present of a certain size. This is carried out by the process of
\textit{electrophoresis}, as follows.

During the binding of primers a fluorescent dye is also attached. We
said earlier that the primers for the two ends of the strand are
different. Typically dyes are attached to only of  the two types or
primers: we can now reveal that the $d$ subscript used in the
description indicates the presence of a dye tag, thus $a_d$ represents
the strand in the complementary pair of strands making up the
dye-tagged amplicon, and $a$ other strand that is not tagged with a
dye. (Note that $h_d$ represents a dye-tagged half-strand, but that
$g_d$ strands are not dye tagged. However it simplifies notation to
treat the $g_d$ as if they are, rather than introduce another symbol
to distinguish the $g$ strand from its compliment.)  For brevity in
the remainder of the paper, we shall refer to $a_d$ as a dye-tagged
amplicon, and $a$ and an untagged amplicon.

Hence we wish to measure the number of dye-tagged amplicons,
$a_d$. The description above has assumed 100\% efficiency in the
amplification process. However, not every strand will get a primer
attached in each cycle, and the \PCR process will operate at less than
100\% efficiency. Typical amplification efficiencies are in the range
0.8-0.9.

\begin{figure}[htb]
  \begin{center}
    \includegraphics[width=0.9\textwidth]{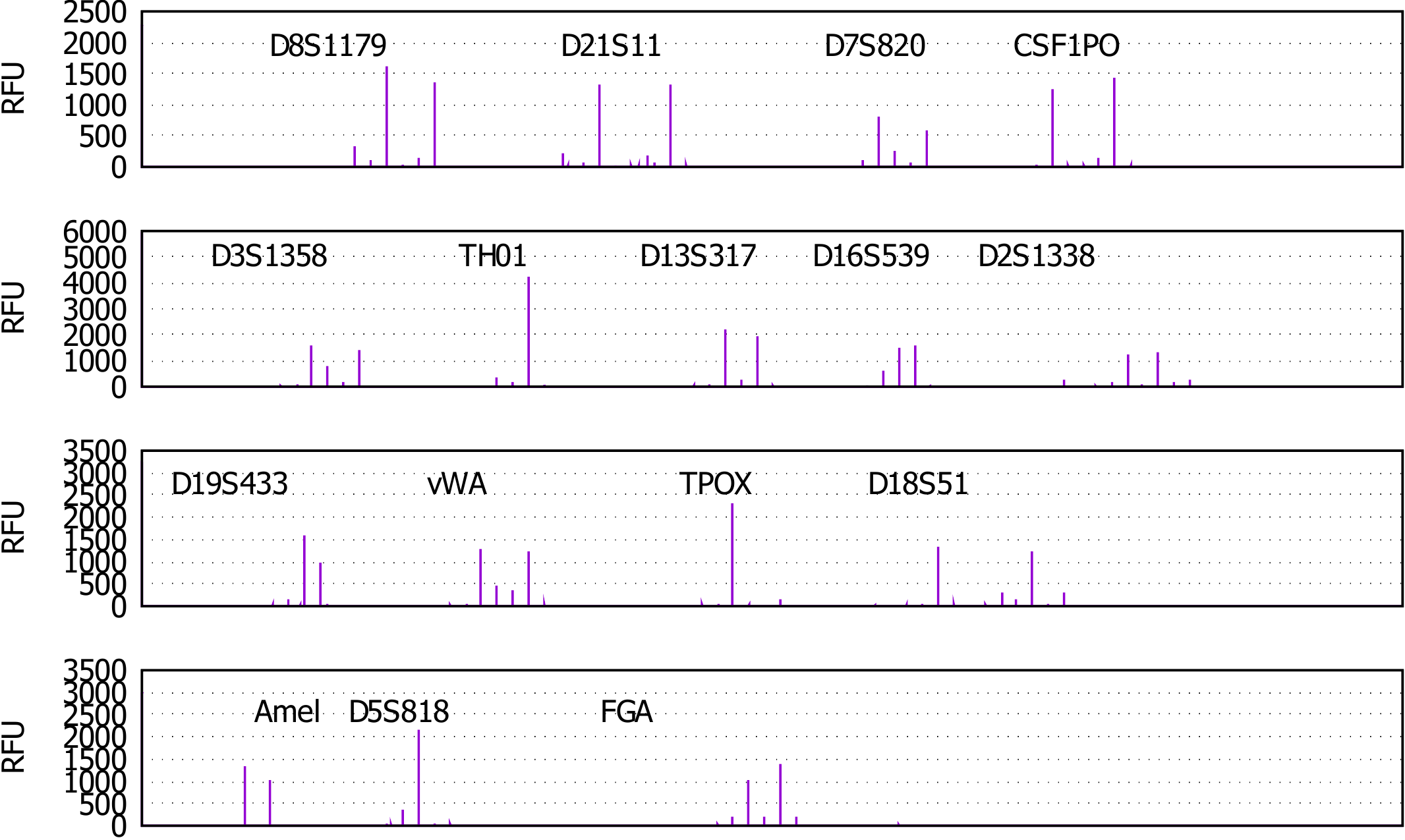}
  \caption{Schematic of an electropherogram plot for a simulated 3-person mixture, for the Identifiler Plus\textsuperscript{TM} STR kit of the Amelogenin locus and 15 autosomal loci. Each of the four panels represents a dye. On the left-hand side of each panel the scale of RFU is indicated. The horizontal scale (not shown) is in units of base-pairs, with a range of between 50 to 450 for the plots. Each vertical spike represents an allele.
    \label{fig:epg}}
\end{center}
\end{figure}

Several colours of fluorescent dyes are used to distinguish similarly
sized alleles from different markers.  The amplified DNA is drawn up
electro-statically through a fine capillary to pass through a light
detector, which illuminates the DNA with a laser and measures the
amount of fluorescence generated. The latter is then an indication of
the number of alleles tagged with the fluorescent marker. The longer
alleles are drawn up more slowly than the shorter alleles, however
alleles of the same length are drawn up together. This means that the
intensity of the detected fluorescence will sharply peak as a group of
alleles of the same length passes the light detector, and the value of
the intensity will be a measure of the number of alleles that
pass. The detecting apparatus thus measures a time series of
fluorescent intensity, but it converts the time variable into an
equivalent base pair length variable. The data may be presented to a
forensic scientist as an \emph{electropherogram} (\EPG) as illustrated
by the simulated \EPG shown in \figref{fig:epg}. Each panel in the EPG
corresponding to a different dye. The horizontal axes indicate the
base pair length, and the vertical axis the intensity.

In the absence of artefacts, a peak in the \EPG indicates presence of
an allele in the sample before amplification. The peak height is a
measure of the amount of the allele in the amplified sample expressed
in \textit{relative fluorescence units} (RFU). The area of the peak is
another measure of the amount, but this is highly correlated with the
height \citep{Tvedebrink2010}. Both peak height and peak area are
determined by software in the detecting apparatus.

We shall call the peak size information extracted from the \EPG the
\textit{profile of the DNA sample}, or more briefly, the {DNA
  profile}. Commonly, DNA profile also refers to the combined genotype
of a person across all markers.

In measuring the peak heights, low level noise give rise to small
spurious peaks. A \textit{peak amplitude threshold}, called the
\textit{analytic threshold}, may be set by the forensic analyst
whereby peaks below the analytic threshold level are ignored.  Thus an
allele present in the DNA sample will not be recorded as observed if
the peak it generates is below the analytic threshold; when this
happens a \textit{dropout} of the allele is said to have occurred. A
dropout of an allele can also occur if no genomic strands containing
the allele are selected for amplification.  Dropout is an artefact
that can make the analysis of DNA samples difficult. Another common
artefact is \textit{stutter}, where\-by an allele that is present in
the sample is mis-copied at some stage in the PCR amplification
process, and (for a tetraneucleotide marker) a four base pair word
segment is omitted.  This damaged copy itself takes part in the
amplification process, and so yields a peak located four base
pairs\footnote{For tetrameric loci.}  below the allele from which it
arose.  More rarely, two repeats are omitted during a \PCR cycle,
which is called \textit{double stutter}, or an extra repeat is
inserted, which is called \textit{forward stutter}; once formed these
artefacts can themselves replicate in subsequent cycles.

Another artefact is known as \textit{dropin}, referring to the
occurrence of small unexpected peaks in the EPG. This can for example
be due to sporadic contamination of a sample either at source or in
the forensic laboratory.

Finally, a mutation in the flanking region can result in the allele
not being picked up at all by the PCR process, in which case we say
that the allele is \emph{silent}. An allele can also be undetectable
and thus \emph{de facto} silent because its length is off-scale and
the peak does therefore not appear in the \EPG. Note that an allele
might be silent for a kit made by one manufacturer but not another;
this is because different manufacturers of kits for performing \PCR
use different primer binding regions.

\section{From sample to EPG}
\label{sec:steps}
In this section we give a description of the process of going from a
DNA sample, recovered for example from a crime scene, to the
electropherogram (\EPG) that is used to make inferences about the
constitution of the DNA in the sample. For a much more detailed
account of the laboratory processes please see
\citep{butler2011advancedmethod}.

We start with a descriptive summary, and follow this with a more
detailed mathematical specification.

\subsection{Steps in the process}
\begin{figure}
  \begin{center}

    \scalebox{0.7}{%
      \begin{tikzpicture}
        \node [aspect=2,style={cloud, cloud puffs=11},fill=red!80]
        (dd) at (10,34) {DNA sample/trace}; \node[style={draw, rounded
          rectangle,minimum size=35pt},rounded rectangle,fill=red!80 ]
        (d) at (10,31) {Small subsample};

        \node[style={draw, rounded rectangle,minimum
          size=35pt},rounded rectangle,fill=red!80 ](a) at (10,27)
        {\Large Aliquot};
        \node[style={draw, rounded rectangle,minimum
          size=35pt},rounded rectangle, fill=red!10 ] (c) at (10,24)
        {\Large Replicate(s)};

        \node[style={ rounded rectangle,minimum size=35pt},rounded
        rectangle] (th) at (10,21) {\Large Thermal Cycles (PCR)};

        \node[style={draw, rounded rectangle,minimum
          size=35pt},rounded rectangle, fill=red!40 ] (la) at (10,18)
        {\Large Amplified replicate(s)}; \node[style={draw, rounded
          rectangle,minimum size=35pt},rounded rectangle, ,
        fill=yellow!40] (f) at (10,15) {\Large Capillary
          electrophoresis}; \node[style={rounded rectangle,minimum
          size=35pt},rounded rectangle] at (10,12) (epg) {\LARGE
          Electropherogram(s)}; \draw node(b) at (5,25) {\Large
          Quantify DNA}; \draw node(bb) at (12,29) {\Large Extract
          DNA};

        \draw [->, line width=1mm] (a) to (b); \draw [->, line
        width=1mm] (d) to (a); \draw [->, line width=1mm] (dd) to (d);
        \draw [->, line width=1mm] (a) to (c); \draw [->, line
        width=1mm] (c) to (th);

        \draw [->, line width=1mm] (b) to (c);

        \draw [->, line width=1mm] (th) to (la); \draw [->, line
        width=1mm] (la) to (f); \draw [->, line width=1mm] (f) to
        (epg);

        \draw[ line width = 1mm] (3,9) -- (18,9); \draw [fill=blue]
        (4,9)--(4.2,9)--(4.1,10.5)--cycle; \draw [fill=blue]
        (5,9)--(5.2,9)--(5.1,9.5)--cycle; \draw [fill=blue]
        (7,9)--(7.2,9)--(7.1,10.2)--cycle; \draw [fill=blue]
        (6.7,9)--(6.9,9)--(6.8,10.0)--cycle; \draw [fill=blue]
        (12.6,9)--(12.8,9)--(12.7,9.6)--cycle;

        \draw [fill=blue] (14,9)--(14.2,9)--(14.1,9.8)--cycle; \draw
        [fill=blue] (13,9)--(13.2,9)--(13.1,10.6)--cycle;

\end{tikzpicture}
}
\caption{Overview of the sequence of steps to produce an
  electropherogram from a DNA sample.\label{fig:epgsteps}}
\end{center}
\end{figure}
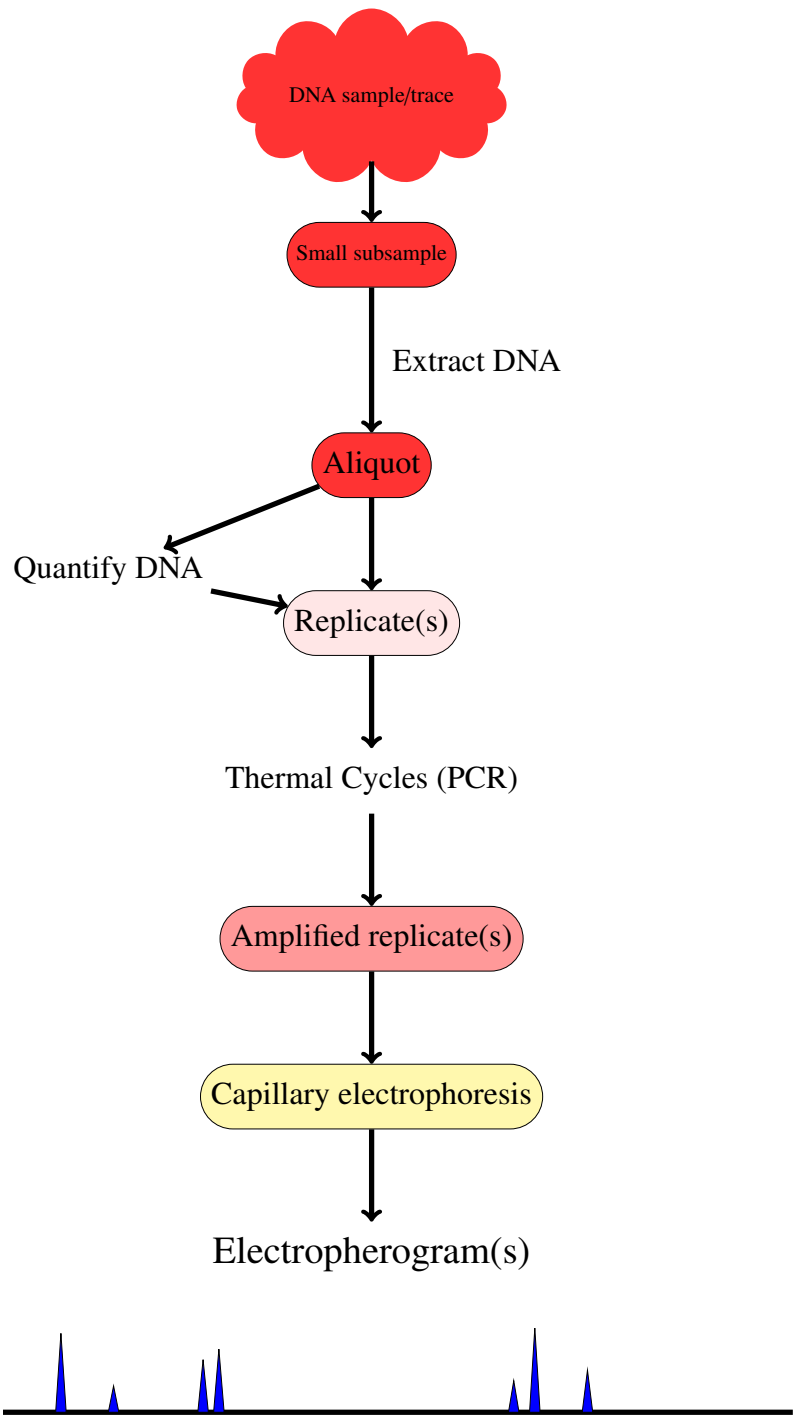

The following is a summary of the sequence of steps taken to obtain an
\EPG from a DNA sample, recovered for example at a crime scene;
\figref{fig:epgsteps} illustrates the sequence of steps involved.

\begin{enumerate}
\item Collect sample of DNA. The sample might be a single source
  trace, that is, it contains DNA from a single individual, or it
  could be a mixture, that is a trace with DNA originating from two or
  more people.
\item Take a sub-sample of the sample and using bi-chemical reagents
  extract nuclear DNA to create an aliquot (solution of extracted DNA-
  check this is the correct meaning). Typically this will be in a
  mini-tube of a volume of between \mcl{50} to \mcl{100}, depending on
  laboratory standard operating procedures.
\item Take a small extract of the aliquot, typically around \mcl{2},
  to determine the concentration of DNA in the aliquot. Usually this
  will be carried out using \textbf{qPCR}
  \citep{butler2011advancedmethod}, so that (if possible) an optimal
  amount DNA can be used for the \PCR step.

\item Based upon the estimated density of DNA in the aliquot, take
  enough of the aliquot for \PCR amplification so that there is an
  optimal amount of genomic DNA in the mini-tube for the \PCR
  amplification is to be carried. For \LTD samples, obtaining a
  sufficient amount might not be possible.  Typically a maximum volume
  of \mcl{20} is taken from the aliquot, so that there is sufficient
  volume left in the PCR mini-tubes for the PCR primers. In some cases
  the aliquot is split into several parts called \textit{replicates}
  and each is separately amplified via \PCR.
  
\item Carry out the \PCR amplification of each replicate using some
  kit and protocol.  Usually, but not always, the same manufacturer
  kit is used for all of the replicates when more than one replicate
  is made.

\item An \EPG is then produced from each replicate by capillary
  electrophoresis.
\end{enumerate}

\section{The objectives of an \EPG analysis}
\label{sec:objectives}
In the previous sections we have given a brief qualitative description
of how we obtain an \EPG from a DNA sample.  In this section we
describe the information a forensic scientist wishes to obtain from an
\EPG. In later sections we develop a mathematical framework to help
realise those aims.

A forensic scientist may wish to use the information in an \EPG to
calculate the \textit{weight of evidence} for two competing hypotheses
regarding the contributors to the DNA sample; another objective is to
try and identify the likely genotypes of contributors
\citep{article:Perlindata,wang2006least,tvedebrink:deconv:12}, a
process known as \emph{deconvolution}.

We consider each of these objectives in turn.
\paragraph{Weight of evidence}

In an adversarial court setting, we have two competing hypotheses, one
argued for by the prosecution, called the \emph{prosecution
  hypothesis} $\hyp_p$, the other called the \emph{defence hypothesis}
$\hyp_d$. Note that the defence is not obliged to propose an
alternative to the prosecution hypothesis $\hyp_p$; usually the
defence hypothesis is proposed by the prosecution for comparative
purposes.  The available \emph{evidence} $E$ in the case consists of
the peak heights as observed in the \EPG as well as the set of
genotypes of some known individuals; denote the known genotype
information by $G$. The prosecution and defence hypotheses differ in
their assumptions as to whose DNA is in the sample that produced the
\EPG. In most cases where a suspect $S$ is on trial, the prosecution
case would be that $S$ contributed to the sample and the defence case
that $S$ did not. However this is not always the case. For example, a
DNA sample might be recovered from a swab of an area of skin of the
suspect where physical contact is alleged to have been made during an
assault on a victim $V$. Under this scenario, the presence of $S$'s
DNA in the sample is not disputed, it is the presence of the victim's
DNA offered by the prosecution as proof of the assault that is
disputed.

The strength of the evidence \citep{good1950,lindley1977,balding:05}
is normally represented by the \emph{likelihood ratio}:
$$ LR=\frac{L(\hyp_p)}{L(\hyp_d)}=\frac{\Pr( E \cd \hyp_p)}{\Pr(E \cd \hyp_d)}.$$

This may be expressed on a base-10 $\log$ scale of units called the
\emph{ban}, introduced by Alan Turing \citep{good:79} so that one ban
represents a factor 10 on the likelihood ratio.  Then the \emph{weight
  of evidence} is $\mbox{WoE} =\log_{10}LR$ in bans
\citep{balding:13}.

The numerator and denominator in the likelihood ratio are calculated
based on models having the form
\begin{equation}
  \Pr( E \cd \hyp, G)= \sum_\mathbf{g}\Pr( E \cd \mathbf{g}) \Pr( \mathbf{g} \cd \hyp, G),\label{eq:likefn}
\end{equation}
so that the model for the conditional distribution
$\Pr( E \cd \mathbf{g})$ of the evidence given the genotypes
$\mathbf{g}$ of all contributors is the same for both hypotheses,
whereas the hypotheses differ concerning the distribution
$\Pr( \mathbf{g} \cd \hyp, G)$ of genotypes of the contributors. Note
that the genotypes $\mathbf{g}$ that are summed over may include those
of one or more \textit{unknown contributors}, that is, individuals
whose genetic profiles are not known (not included in $G$).

Note that $\Pr( E \cd \mathbf{g})$ has a dependence, not shown, on the
amount of DNA from each contributor in the sample, and other factors
such as sample degradation: this implies a similar implicit dependence
for the left hand side.

\paragraph{Deconvolution of DNA mixtures}

In the deconvolution of a DNA sample, typically a mixture, we assume
that there are one or more genetically untyped contributors to the DNA
sample. We then wish to find the genotypes of these untyped
individuals. Typically how this is done is that for each individual
separately a ranked list of genetic profiles ordered by their
likelihood is produced.  The potential profiles having high likelihood
could, for example, be checked against an offender database for a
match.  However, sometimes a ranking of the joint genotypes of two or
more individuals may also be of interest.

\clearpage
\part{Mathematical formulation}

In this part of the paper we give the mathematical notation
underpinning the framework of this paper.  From the right hand side of
\eqref{eq:likefn} we see that there are two components of the
framework that require specification.  One part is specifying the
(conditional) probability for the \EPG data, $\Pr( E \cd \mathbf{g})$,
the other specifying the genetic profile probabilities,
$ \Pr( \mathbf{g} \cd \hyp, G)$.  As the latter is relatively
uncontroversial, and is common to all models for the
$\Pr( E \cd \mathbf{g})$ distribution, we discuss that first. We then
discuss the simulation model of \cite{gill:etal:2005}, and show how
the full distribution may be obtained without simulation.  We then
extend the analysis to finding the distribution for the more realistic
model in which we start with genomic strands and we find the
distribution of dye-tagged amplicons. We then elaborate this in stages
leading to the general framework that includes background noise,
drop-in, forward and double-reverse stutters, degradation and
inhibition, for possibly multiple replicates analysed with possibly
multiple kits from possibly multiple independent samples.

\section{Specifying genetic profile probabilities}

Given the genotypes of contributors, the framework formulated in this
paper for specifying peak-height likelihoods is applicable to any type
of STR locus, that is for evaluating $\Pr( E \cd
\mathbf{g})$. However, if there are untyped contributors, evaluation
of $ \Pr( \mathbf{g} \cd \hyp, G)$ can be problematic for sex-linked
loci. In this paper we shall assume that the set of loci are all
autosomal loci plus, optionally, Amelogenin.  We shall also assume
that all contributors are unrelated. With these assumptions
\eqref{eq:likefn} simplifies to a product over the loci $L$, with
$\mathbf{g_l}$ denoting a genotype in the set $G_l$ of genotypes on
the locus $l\in L$:

\begin{equation}
  \Pr( E \cd \hyp, G)= \prod_{l \in L}\left(\sum_\mathbf{g_l}\Pr( E \cd \mathbf{g}) \Pr( \mathbf{g_l} \cd \hyp, G_l)\right),\label{eq:likefn2}
\end{equation}

We assume that the population from which each contributor comes from
is known, and that allele frequency estimates are available. We do not
assume that the contributors all come from the same population. It is
assumed that populations substructure correction parameters are known
for each population. Optionally, the finite size database correction
of \citep{cowell2016combining} may be applied. We do not have to be
concerned with linkage between autosomal loci because of the
assumption that contributors are unrelated. Relatedness amongst
contributors could be incorporated, but at the cost of complicating
the presentation presented here, especially if there are linked loci.
For a detailed discussion of these issues relating to populations and
genotype probability estimation, as they relate to forensic
applications, see
\citep{evett1998interpreting,balding2015weight,egeland2015relationship}.
A companion paper by the author is planned that proposes solutions to
the twin problems of Y-STR haplotype probability estimation and the
resolution of Y-STR mixtures.

\section{The simulation model of \cite{gill:etal:2005}: the amplicon
  model}

In this section we summarise the simulation model of
\cite{gill:etal:2005}, for further details please see the original
paper. We shall refer to this model as the \textit{amplicon model}.

The model presented in \cite{gill:etal:2005} has variants for diploid
cells and haploid cells. As mentioned earlier, their model fails to
take account of the fact that the process starts with genomic material
rather than amplicons, and it also ignores the dye-tagging of the
amplicons during the \PCR process. Despite these short-comings, the
model provides a convenient starting point for the more realistic
models presented later.  We start off with the case of no stutters;
\subsection{Simulating the process without stutters}

The model of \cite{gill:etal:2005} is a simulation model. It assumes
that initially we have a number $n_c$ of cells in our small subsample
of \figref{fig:epgsteps}.  We concentrate on just one allele from one
locus. If the cell is diploid, and the individual is homozygous, then
there will be $2n_c$ such amplicons within the cells to start with;
otherwise if the individual is heterozygous on the locus, or the cell
is haploid, then there will be $n_c$ such amplicons in the cells to
start with.  For whichever case holds, denote by $N$ the total number
of amplicons initially.

In the first stage the amplicons are extracted from the cells. This
process is not 100\% efficient, more typically only between 10-30\%
are extracted intact \citep{gill:etal:2005}. Following
\cite{gill:etal:2005}, let $\pi_{eff}$ denote the extraction
probability that an individual amplicon is released intact into the
aliquot. Assuming independence of the release of the distinct
amplicons, the total number of intact amplicons in the aliquot is
therefore binomially distributed with distribution
$\mbox{Binom}(N,\pi_{eff})$.

A fraction of the aliquot $\pi_{aliquot}$ is then taken for \PCR.  If
we let $\phi = \pi_{eff}\pi_{aliquot}$, then the total number of
amplicons intact and selected for amplification will be Binomially
distributed as
$\mbox{Binom}(N,\pi_{eff}\pi_{aliquot})\equiv \mbox{Binom}(N,\phi)$.
 
These are now subject to \PCR amplification, where it is assume that
in each cycle, each amplicon makes a copy of itself with probability
$\pi_{pcr}$ If we assume that there are $K$ cycles, then the total
number of amplicons can be simulated using the following algorithm,
starting with $N$ amplicons:

\begin{alg}[{\sc Simple amplicon \PCR simulation}]
  \label{alg-gillamplicon}
  \begin{itemize}
  \item Randomly sample $n$ from $\mbox{Binom}(N,\phi)$.
  \item For $K$ times do:
    \begin{itemize}
    \item Sample $m$ from $\mbox{Binom}(n,\pi_{pcr})$
    \item Update $n := n+m$
    \end{itemize}
  \end{itemize}
  \sbackup
\end{alg}

This algorithm is very simple to implement. If it is run many times, a
histogram, or kernel density estimate, plot may be made of the
distribution of the number of amplicons.

\begin{figure}[h]
  \begin{center}
    \includegraphics[scale=0.6]{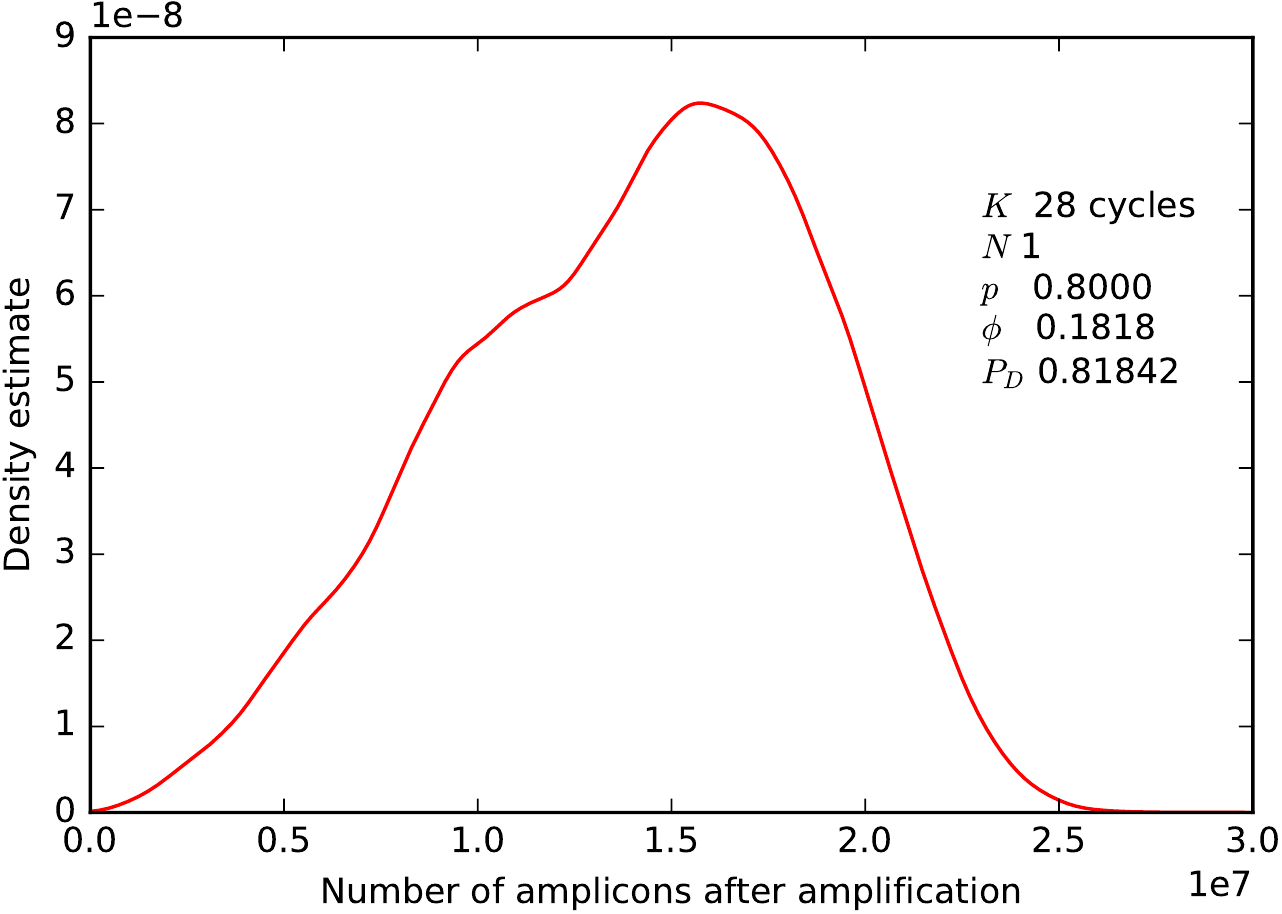}
    \caption{Kernel density estimate, based on 1 million simulations,
      of the final number of amplicons, starting with a single
      amplicon selected with probability $\phi = 2/11$ for \PCR, and
      amplified for 28 cycles with amplification probability of 0.8
      per cycle if selected. The estimated total drop-out probability
      is 0.81842, compared to the theoretical value of
      $9/11 = 0.81818\cdots$. Note that the zero amplicon values have
      been removed prior to the kernel density estimate, hence the
      kernel density estimate is for the conditional probability of
      the number of amplicons after the 28 amplification cycles,
      \textit{given} that one is selected for
      amplification.\label{fig:ampkd}}
  \end{center}
\end{figure}

Python code presented in \appref{sec:py:ampkd} produced the kernel
density plot shown in \figref{fig:ampkd}, based on a million
simulations.  The parameter values used were $\pi_{eff} = 0.6$,
$\pi_{aliquot} = 20/66$, $\pi_{PCR} = 0.8$ and $K=28$ cycles (values
taken from the \cite{gill:etal:2005} paper), for a single starting
amplicon, $N=1$.  Note that for each simulation there is a probability
$\phi = 0.6\times 20/66 = 2/11$ that the amplicon is selected for
amplification , so that we have a total dropout probability of
9/11. As this will dwarf the rest of the plot, these zeros have been
removed before the kernel density has been estimated.

\subsection{Generating the full distribution (without stutters)}

We now show how the simulation model can be represented by using
probability generating functions, following the approach of
\citep{good:1949} based on Galton-Watson cascade processes.  Consider
an initial single amplicon.  In the first PCR cycle it can either
amplify with probability $\pi_{PCR}$, or fail to amplify with
probability $1-\pi_{PCR}$. The probability distribution of the number
of alleles after one cycle can therefore be represented by the
\textit{probability generating function} (\PGF):

$$F_1(t) = f(t) = (1-\pi_{PCR})t + \pi_{PCR} t^2.$$
After two \PCR cycles the number of alleles has the \PGF
$$F_2(t) = f(f(t)),$$
and after $3$ cycles it has the \PGF
$$F_3(t) = f(f(f(t))).$$

More generally, we have

\begin{theorem}
  \label{thm:pgfs}
  The \PGF for the number of molecules after $r$ \PCR cycles, given
  that there is exactly one prior to any amplification cycle, is
  \begin{equation}
    F_r(t) = f(f(f(\ldots f(t) \ldots )))\label{eq:pgfr}
  \end{equation}
  where
  \begin{equation}
    F_1(t) = f(t),\,\,\, F_{s+1}(t) = f(F_s(t)), \hspace{0.2in} (s=1,2,3,\ldots).\label{eq:pgfs}
  \end{equation}

\end{theorem}
This is Theorem~1 of Good \cite{good:1949} (but here specialised to
$F(t)$ given above).
 
The above gives the \PGF for a single starting amplicon: if there are
$K$ cycles the \PGF will be a polynomial in $t$ of degree
$2^K$. However, for the simulation model we start not with a amplicon,
but with $N$ initial amplicons, that are each sampled independently
for amplification with probability $\phi$.  The \PGF taking into
account the initial number and the pre-\PCR sampling is
\citep{cowell:09}
$$
(1 - \phi + \phi F_K(t))^M
$$
a polynomial of degree $M\times 2^K$. This is simply the functional
composition of the Binomial \PGF $(1-\phi + \phi t)^N$ for the
pre-\PCR sampling of the amplicons with the \PGF of the number of
amplicons arising from a single amplicon in the \PCR branching
process.
 
Let us rewrite the recursion in \eqref{eq:pgfs} on substituting
$f(t)$; for later convenience we also replace $\pi_{PCR}$ by $p_t$,
thus the recurrence relation becomes:

$$
F_{s+1}(t) = (1- p_t)F_{s}(t) + p_t F_s(t)^2.
$$
 
with initial value $F_0(t) = t$.  It is in principle possible to find
the polynomial $F_K(t)$ quite simply by iteration using a computer
algebra package, however the growth in the numbers of terms in the
polynomials means that there is a relatively small limit to the number
of iterations that can be carried out before computer memory is
exhausted of the order of $K=10$ or so, even if numerical values for
$p_t$ are used. In addition in the later stages, the brute force
evaluation of the quadratic term $F_s(t)^2$ grows quadratically in
complexity, thus making the later stages of iterations slower and
slower.

\begin{figure}[h]
  \begin{center}
    \includegraphics[scale=0.6]{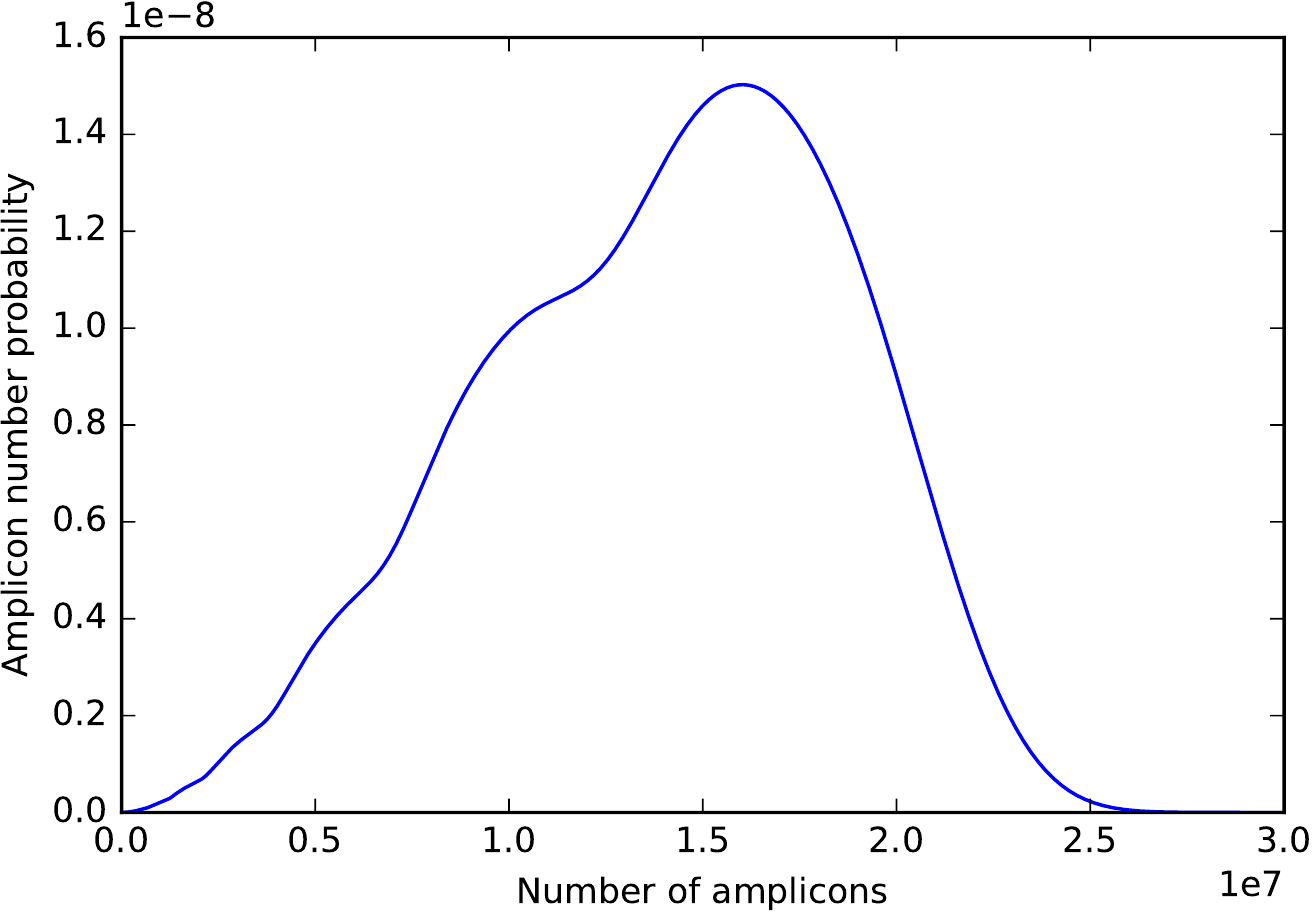}
\caption{%
Exact probability distribution for the total number of amplicons,  starting with a single  amplicon selected with probability $\phi = 2/11$ for \PCR, and amplified for 28 cycles with amplification probability of 0.8 per cycle if selected. The complete drop-out probability value of $9/11$ at 0 is off the scale of the plot. Unlike \figref{fig:ampkd}, the plot shown here is not conditional on selection, which is why the vertical scales of the two plots differ.
\label{fig:ampfft28}}
\end{center}
\end{figure}

There is however another way to evaluate the \PGF numerically, by
noting that the quadratic expression $F_s(t)^2$ is simply the \PGF of
the convolution of two (identical) probability distributions. The
convolution can be carried out using a \textit{Discrete Fourier
  Transform} (\DFT), which may be done efficiently using a
\textit{Fast Fourier Transform} (\FFT). Moreover, as we shall show,
the binomial sampling composition can also be carried out using the
\FFT.  Before given the details, the reader may care to look at
\figref{fig:ampfft28}, which shows the exact distribution for a single
starting amplicon evaluated using the \FFT.

\clearpage

\subsubsection*{The discrete Fourier transform}
Before presenting the algorithm for generating the full distribution
using the \FFT, we first give a brief review of the \DFT.  Suppose
that a sequence of (real) numbers $(x_0, x_1, x_2 , \ldots, x_{n-1})$ is
given, and an integer $N \ge n$.  Then the sequence of numbers
defined by
$$X_k = \sum_{j=0}^{N-1} x_je^{-2\pi i jk/N},$$
where $i = \sqrt{-1}$, is the discrete Fourier transform of the
sequence $(x_0, x_1, x_2 , \ldots, x_{n-1})$. There is an inversion
formula:
$$x_k = \frac{1}{N} \sum_{j=0}^{N-1} X_je^{2\pi ijk/N}$$

The key result that we use is that multiplication of two large
polynomials may be done efficiently using the \DFT, which may be
calculated efficiently using the \FFT. Suppose that $x(t)$ is a
polynomial of degree $n$ in $t$ with coefficients
$(x_0, x_1, x_2 , \ldots, x_{n})$, and that $y(t)$ is a polynomial of
degree $m$ in $t$ with coefficients $(y_0, y_1, y_2 , \ldots,
y_{m})$. Let $z(t) = x(t)y(t)$; this is a polynomial of degree $m+n$
with coefficients $(z_0, z_1, \ldots z_{n+m})$ where
$$ z_k = \sum_{j=0}^k x_{j}y_{k-j},$$
(defining $x_j=0$ for $j > n$ and $y_{k-j}=0$ for $k-j >m$).  Let $N$
be an integer such that $N \ge n+m+1$. if we extend the sequence
$(x_0, x_1, x_2 , \ldots, x_{n})$ with zeros to create a new sequence
$(x_0, x_1, x_2 , \ldots, x_{N-1})$, and similarly extend the sequence
$(y_0, y_1, y_2 , \ldots, y_{m})$ with zeros to create a new sequence
$(y_0, y_1, y_2 , \ldots, y_{N})$, then we may form the \DFT of these
extended sequences:
\begin{align*}
  X_k &= \sum_{j=0}^{N-1} x_je^{-2\pi i jk/N}\\
  Y_k &= \sum_{j=0}^{N-1} y_je^{-2\pi i jk/N}
\end{align*}
If the sequence $(z_0, z_1, \ldots z_{n+m})$, extended with zeros if
required to make the sequence $(z_0, z_1, \ldots z_{N-1})$, then the
\DFT of this sequence is
$$
Z_k = \sum_{j=0}^{N-1} z_je^{-2\pi i jk/N}
$$
and we have that
$$
Z_k = X_k Y_k \,\, \mbox{ for all } k \in \{0,N-1\}.
$$
so that
$$
z_k = \frac{1}{N}\sum_{j=0}^{N-1} X_jY_j e^{2\pi ijk/N}
$$
Hence to multiple the polynomials $x(t)$ and $y(t)$, we choose $N$
sufficiently large, take the {\DFT}s of the coefficients of $x(t)$ and
of $y(t)$, multiply the two {\DFT}s term-wise, and then take the
inverse \DFT. The computation of the forward and backward \DFT may be
carried out efficiently using the \FFT algorithm, an algorithm by
Gauss that dates back to 1805 (see \citep{heideman1984gauss} for an
interesting history of the \FFT , and \citep{rao2011fast} for a recent
monograph on \FFT algorithms.).

We can now present an algorithm for generating the full distribution
using the \DFT. It is given in \algref{alg-gillampliconfft} for the
case of a single starting amplicon and no binomial presampling.  Let
$F[]$ be a vector with indices starting from zero, such that $F[n]$
denotes the coefficient of $t^n$ in the \PGF. With $K$ cycles, the
number of amplicons will range up to $2^K$, hence $F[]$ must be a
vector of size at least $2^{K}+1$.

\begin{alg}[{\sc Single amplicon distribution using the \DFT}]
  \label{alg-gillampliconfft}
  \begin{itemize}
  \item Initialise $F[]$ to be a vector of size at least $2^{K}+1$,
    with all entries 0 except $F[1]=1$.
  \item Let ${\cal F}[]$ denote the \DFT of $F[]$ .

  \item For $K$ times do:
    \begin{itemize}
    \item for each element ${\cal F}[f]$ of $\cal{F}[]$ update
      ${\cal F}[f] := (1-p_t){\cal F}[f]+ p_t {\cal F}[f]^2$
    \end{itemize}
  \item Set $F[]$ equal to the inverse \DFT of $\cal{F}[]$.
  \end{itemize}
  \sbackup
\end{alg}

Note that taking the \DFT is a linear operator; denote it by $L$ and
its inverse by $L^{-1}$.  Hence we may write (with $\times$ denoting
element wise multiplication)
\begin{eqnarray*}
  F_{s+1}(t) &=& (1- p_t)F_{s}(t) + p_t F_s(t)^2\\
             &=& (1- p_t)F_{s}(t) + p_t L^{-1}( L(F_s(t)) \times  L(F_s(t)) )\\
             &=& (1- p_t)L^{-1}(L(F_{s}(t)) + p_t L^{-1}( L(F_s(t)) \times  L(F_s(t)) )\\
             &=& L^{-1}\left((1- p_t)L(F_{s}(t)) + p_t  L(F_s(t)) \times  L(F_s(t)) \right),
\end{eqnarray*}
thus justifying \algref{alg-gillampliconfft}.

Taking account of starting with $M$ amplicons, and binomially sampling
them with probability $\phi$ is almost trivial. We have, using the
linearity of $L$ and its inverse:
\begin{eqnarray*}
  (1 - \phi + \phi F_K(t))^M &=& L^{-1} L\left( (1 - \phi + \phi F_K(t))^M) \right)\\
                             &=&L^{-1} \left( \left(L(1 - \phi )+ \phi L(F_K(t)\right)^{\times M} \right)\\
\end{eqnarray*}
where the superscript $^{\times M}$ denotes taking the $M$-th power of
each element in the transform. Hence to take this into account we
modify \algref{alg-gillampliconfft} thus, (extending the initial size
of $F[]$ for the higher number of amplicons that could result):

\begin{alg}[{\sc Amplicon distribution with binomial pre-sampling
    using the \DFT}]
  \label{alg-gillampliconfftbin}
  \begin{itemize}
  \item Initialize $F[]$ to be a vector of size at least $M(2^{K}+1$),
    with all entries 0 except $F[1]=1$.
  \item Let ${\cal F}[]$ denote the \DFT of $F[]$ .

  \item For $K$ times do:
    \begin{itemize}
    \item for each element ${\cal F}[f]$ of $\cal{F}[]$ update
      ${\cal F}[f]:= (1-p_t){\cal F}[f]+ p_t {\cal F}[f]^2$
    \end{itemize}
  \item for each element ${\cal F}[f]$ of $\cal{F}[]$ update
    ${\cal F}[f]:= (1-\phi + \phi {\cal F}[f])^M$.
  \item Set $F[]$ equal to the inverse \DFT of $\cal{F}[]$.
  \end{itemize}
  \sbackup
\end{alg}

Now the extreme right-hand tail of the amplicon distribution has very
low probabilities, and for all practical intents and purposes can be
taken to be zero. Hence we may take the $F[]$ vector to have size
$M 2^K$ which makes finding the \DFT using the \FFT much more
computationally efficient, especially if $M$ itself is a power of
2. Using this approximation, \algref{alg-gillampliconfftbin} is
readily implemented.  \appref{r-alg-gillampliconfRcode} presents the
seven(!) lines of code $R$ code for this, repeated here.

\begin{verbatim}
N = M*2**K
F = rep(0,N) 
F[2] = 1	                      # F[] now corresponds to F(t) = t												
F = fft(F,inverse=FALSE)
for (k in 1:K) F = (1 - p)*F + p*F*F     # K amplifications cycles
F = (1-phi + phi*F)**M                 		 # binomial sampling
F = Re(fft(F,inverse=TRUE)) /N           # real part of inverse
\end{verbatim}

\begin{figure}[htb]
  \begin{center}
    \includegraphics[width=0.65\textwidth]{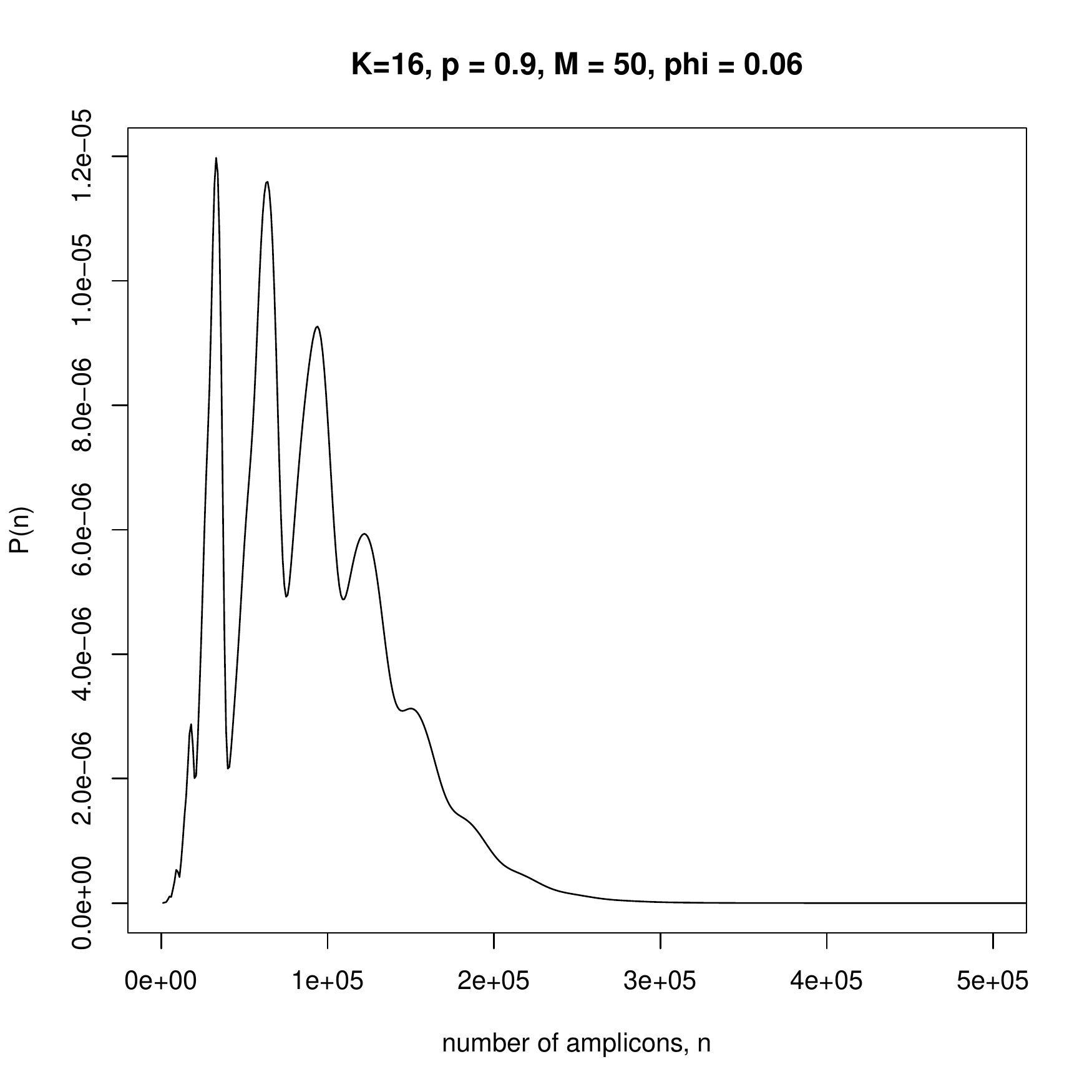}
  \caption{A marginal target distribution computed  with the R implementation
of  \algref{alg-gillampliconfftbin} 
given in \appref{r-alg-gillampliconfRcode} .
\label{fig:targetmargK16M50Phi06}}
\end{center}
\end{figure}

\clearpage
\figref{fig:targetmargK16M50Phi06} shows the amplicon distribution
obtained using the code given above for a low template amount, in
which multimodality is clearly present, and would be missed by current
probabilistic genotyping software based assuming a unimodal
probability distribution (such as lognormal or gamma) for peaks
heights.\footnote{Equivalent Python code, used for generating
  \figref{fig:ampfft28} is given in \appref{sec:py:ampfft}. Running
  the Python code for the full 28 cycles took around 3 minutes on a
  laptop with an Intel i7\textsuperscript{TM} processor, and used
  around 30Gb of ram.}

It is also worth emphasising that the model naturally includes
dropout, and that the full distribution calculated using
\algref{alg-gillampliconfftbin} enables the means to assess dropout
probabilities --there is no need to posit a separate additional model
for allelic dropout, such as the logistic regression model (or
variants thereof) of \cite{tvedebrink2009estimating}.  Indeed, in a
follow-up paper, \citep{tvedebrink2012allelic} refine their logistic
regression approach with a probit model based on and compared with the
amplicon model of \cite{gill:etal:2005}.

\begin{figure}[htb]
  \begin{center}
    \includegraphics[width=0.9\textwidth]{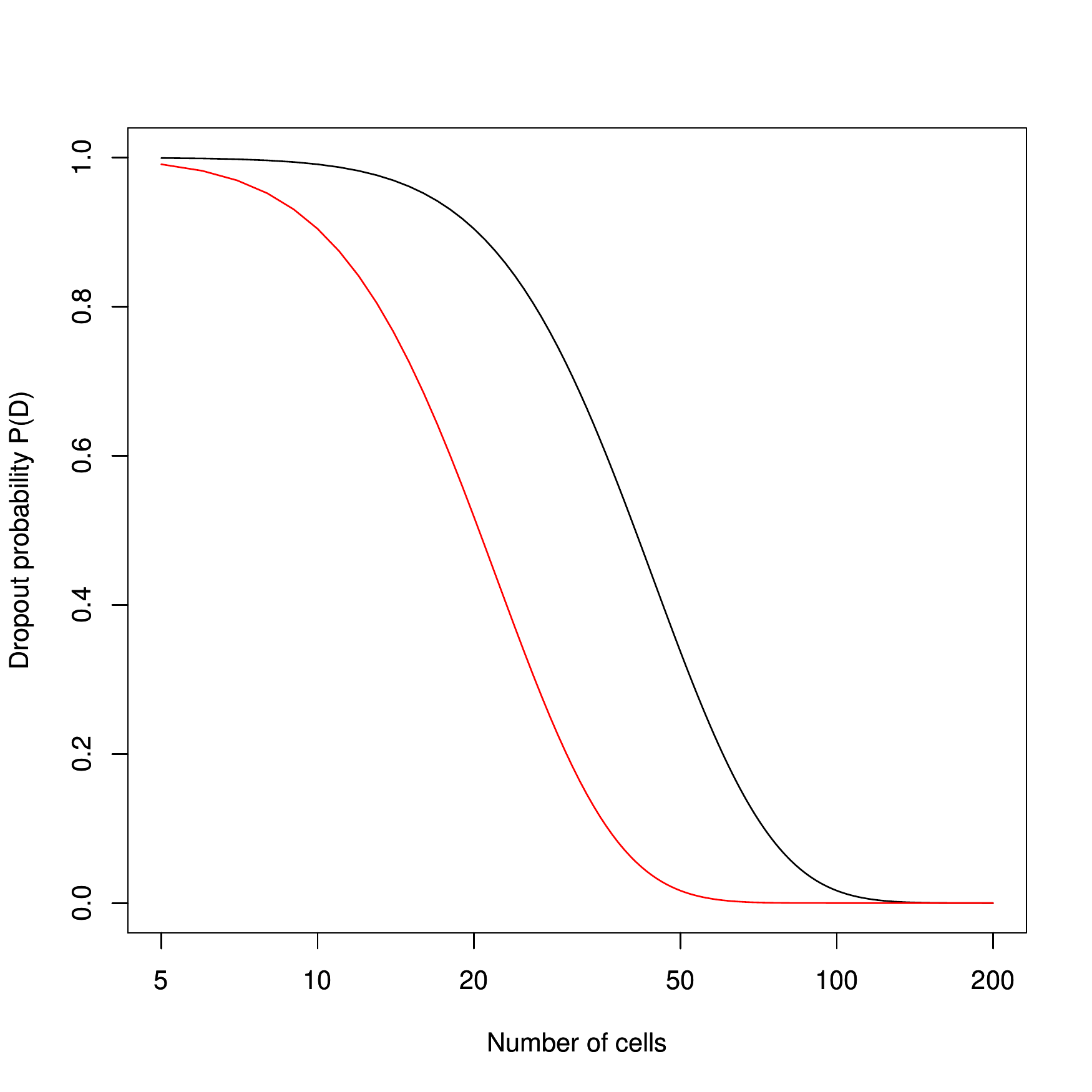}
  \caption{Dropout probabilities for homozygous (red) and heterozygous (black) individuals, using $K=15$ amplification cycles, pre-amplification sampling probability of $\phi = 0.1$, 
 an amplification probability of $p=0.85$ on each cycle, and a threshold of 40000 amplicons. 
 \label{fig:dropoutamp}}
\end{center}
\end{figure}

To illustrate this, we consider an amplification probability of
$p=0.85$. To keep computations manageable we take the number of cycles
to be $K=15$, take $\phi = 0.1$, and set the analytic threshold to be
40000 amplicons. \figref{fig:dropoutamp} shows the dropout
probabilities $P(D)$ as a function of the number of starting cells -
note that the horizontal axis is on a log-scale for comparison with
Figure~1 of \cite{tvedebrink2012allelic}. The plot shows two curves,
with the red corresponding to a homozygous individual, and the black
to a heterozygous individual. (R code to generate this plot is given
in \appref{r-fig:dropoutamp}.)

\clearpage

\subsection{Alternative derivation of the distribution probabilities}

There is another way to extract the probabilities from the probability
generating function $(1-\phi + \phi F_K(t))^M$. Note that this is a
finite degree polynomial in $t$. The coefficient of $t^n$ gives the
probability $F[n]$ of exactly $n$ amplicons. Instead of generating the
full distribution using the \DFT-based algorithms above, we can
extract this single value by using Cauchy's residue theorem with a
contour in the complex plane that contains the origin,
$$ F[n] = \frac{1}{2\pi i}\oint  \frac{(1-\phi + \phi F_K(t))^M}{t^{n+1}} dt.$$

If we take the contour to be the unit circle in the complex plane
centred at the origin, we may make a change of variable,
$t = \exp(- 2\pi i \theta)$, so that $\theta$ ranges over $[0,1]$, and
the contour integral becomes
$$ F[n] = \int_0^1  \left(1-\phi + \phi F_K(\exp( -2\pi i \theta))\right)^M  \exp( 2\pi n \theta)d\theta $$
We may evaluate this numerically by splitting the range up into a
large number $N$ of equal sized intervals (with size $d\theta = 1/N$)
and using the trapezoidal approximation. Note that the beginning and
the end point values are identical because we are evaluating the
closed circular contour, so that as the approximation we have
$$ F[n] \approx \frac{1}{N}\sum_{j=0}^{N-1} 
(1-\phi + \phi F_K\left(\exp( -2\pi i j/N)\right)^M\exp( 2\pi nj/N)
 $$
 
 \textbf{This is fully equivalent to using the \DFT above if the same
   value of $N$ is used, and hence is \underline{exact} if $N$ is
   sufficiently large}.
 
 The advantage of this formulation is that if only a single value from
 the distribution is required, it may be found without the large
 overhead in computer memory that using the \FFT incurs in storing the
 arrays $F[]$ and ${\cal F}[]$.

 There is another use for this approach. When evaluating a likelihood
 for a peak height of an allele, it may be that no peak is observed at
 the allele position in the \EPG, or one is observed but is not above
 the analytic threshold. In such cases we need to find the cumulative
 probability to the threshold.  Suppose that the threshold corresponds
 to $n$ amplicons. Then the cumulative probability is

\begin{align*}
  \sum_{m=0}^n F[n] &= \sum_{m=0}^n\frac{1}{2\pi i}\oint  \frac{(1-\phi + \phi F_K(t))^M}{t^{m+1}} dt\\
                    &= \frac{1}{2\pi i}\oint  (1-\phi + \phi F_K(t))^M
                      \left(
                      \sum_{m=0}^n \frac{1}{t^{m+1}}
                      \right) dt\\
                    &= \frac{1}{2\pi i}\oint  (1-\phi + \phi F_K(t))^M \frac{1 - t^{-n-1}}{t-1} dt
\end{align*}

Again, the trapezoidal rule may be used to evaluate this with line
integral on the unit circle in the complex plane.

\subsection{Simulating the process with stutters}
 
The paper of \cite{gill:etal:2005} also included a model for
stutters. There is a slight error in their paper that was pointed out
in \citep{cowell:09}. Their corrected model is as follows. During an
amplification an amplicon may make a copy of itself, with probability
$p_t$, or make a stutter copy with probability $p_s$. Thus neither a
stutter nor an exact copy is made with probability $1-p_t-p_s$.  It is
assumed that when a stutter amplicon is made it will make a copy of
itself in each subsequent cycle with probability $p_t$.  Let $n_k$
denote the number of amplicons and $m_k$ the number of stutter
amplicons after $k$ cycles. Assuming $N_0$ starting amplicons sampled
binomially with probability $\phi$ for amplification, a simulation for
the number of amplicons and stutters may be expressed as in
\algref{alg-gillstutt}. In this, the draw of $J$ simulates the total
number of new amplicons and stutters produced in the $k$-th cycle by
the existing amplicons. The $J_t$ draw then samples from these new
products to simulate how many are copies of the target amplicon, with
the remainder being stutters.  The $J_s$ draw simulates how many new
stutter amplicons are produced by currently existing stutter products.

\begin{alg}[{\sc Simple amplicon \PCR simulation with stutters}]
  \label{alg-gillstutt}
  \begin{itemize}
  \item Sample $n_0$ from $\mbox{Binom}(N_0,\phi)$, and set $m_0 = 0$.
  \item for k in 1 to K do
    \begin{itemize}
    \item Sample $J$ from $\mbox{Binom}(n_{k-1},p_s+p_t)$
    \item Sample $J_t$ from $\mbox{Binom}(J,p_t/(p_s+p_t))$
    \item Sample $J_s$ from $\mbox{Binom}(m_{k-1},p_s+p_t)$
    \item Set $n_k = n_{k-1}+ J_t$
    \item Set $m_k = m_{k-1} + (J-J_t) + J_s$
    \end{itemize}
  \end{itemize}
  \sbackup
\end{alg}

In \citep{gill:etal:2005} a value of $p_s = 0.002$ was estimated from
experimental data. However, after correcting their error (which was to
use $\mbox{Binom}(n_{k-1},p_t)$ for the number of copies of amplicons
generated, and $\mbox{Binom}(n_{k-1},p_s)$ for the number of stutter
products generated from amplicons, hence allowing the possibility for
an amplicon to generate both a copy of itself and also a stutter in a
single cycle) it appears that a value of around 0.004 would be more
appropriate.

\begin{figure}[htb]
  \begin{center}
    \includegraphics[width=0.8\textwidth]{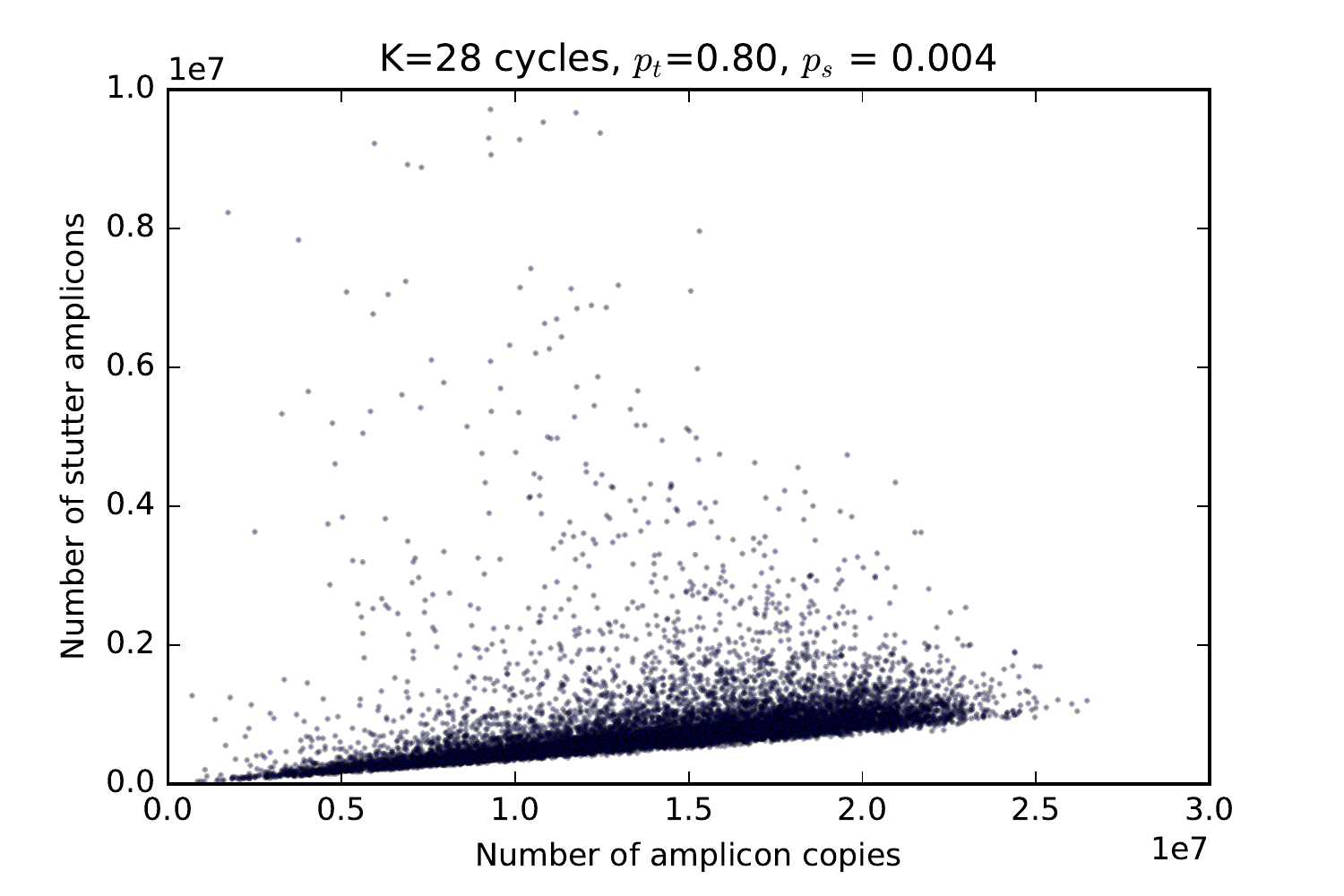}
  \caption{Scatterplot of main and stutter amplicon numbers for simple amplicon mode, for a simulation of size 10000.
    \label{fig:gillscatter}}
\end{center}
\end{figure}

\Figref{fig:gillscatter} shows a scatterplot of 10000 simulated
$(n_K, m_K)$ values for $K$ cycles starting from a single amplicon,
$n_0=1$ (so no pre-sampling) using $p_t = 0.8$ and $p_s = 0.004$. The
plot indicates some correlation between the two values --- for the
data in the plot the correlation coefficient is around 0.22, a
theoretical value will be given later.  \cite{weusten2012stochastic}
also found such correlations in their analysis.

\begin{figure}[htb]
  \begin{center}
    \includegraphics[width=0.8\textwidth]{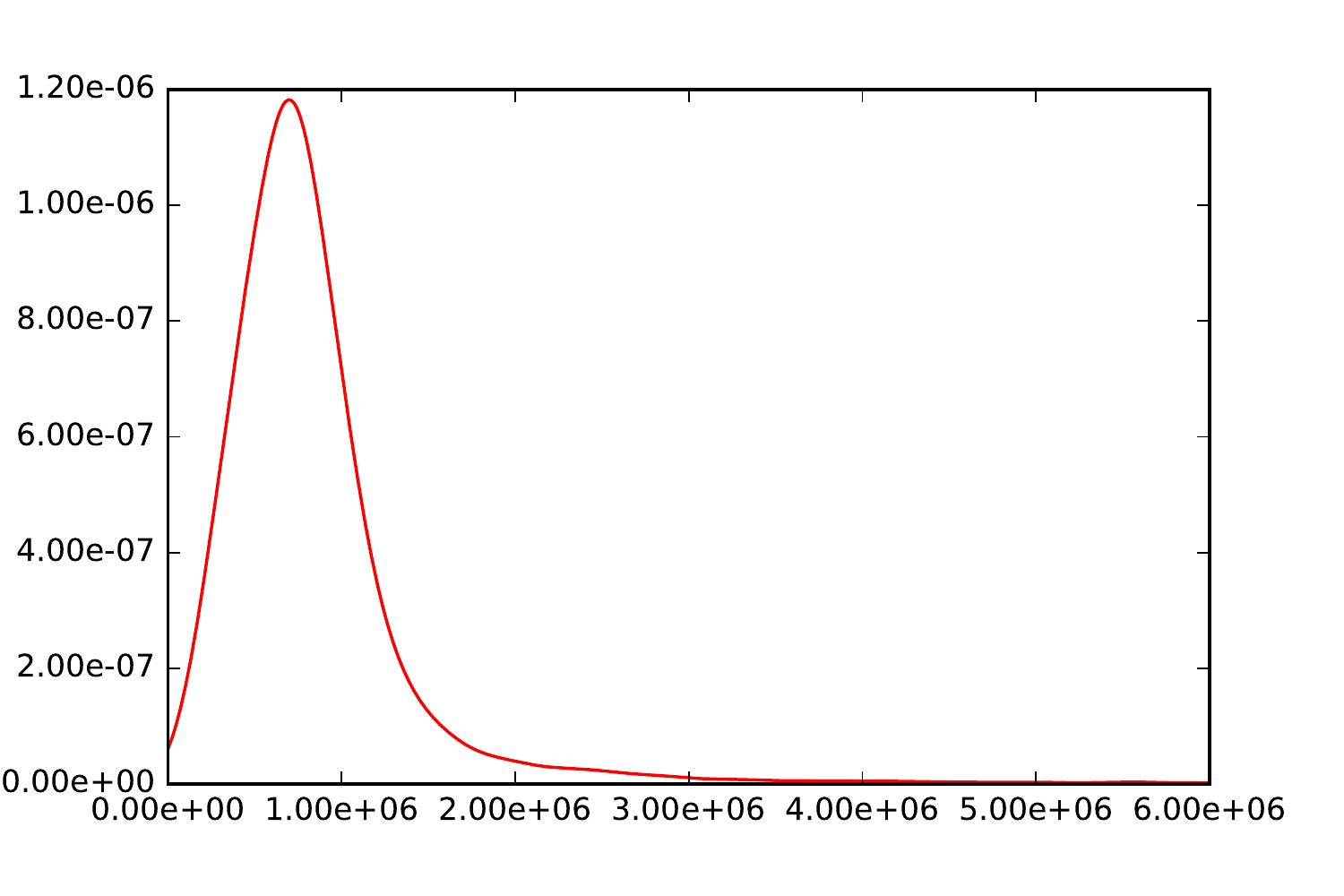}
    \caption{Kernel density estimate of the marginal distribution of the number of stutter
      amplicons from the simulation data  plotted in \figref{fig:gillscatter}.
    \label{fig:ampliconStutterMarg}}
\end{center}
\end{figure}

From the simulated data we may also estimate the marginal distribution
for stutter amplicons, a density estimate is shown in
\figref{fig:ampliconStutterMarg}.

\begin{figure}[htb]
  \begin{center}
    \includegraphics[width=0.8\textwidth]{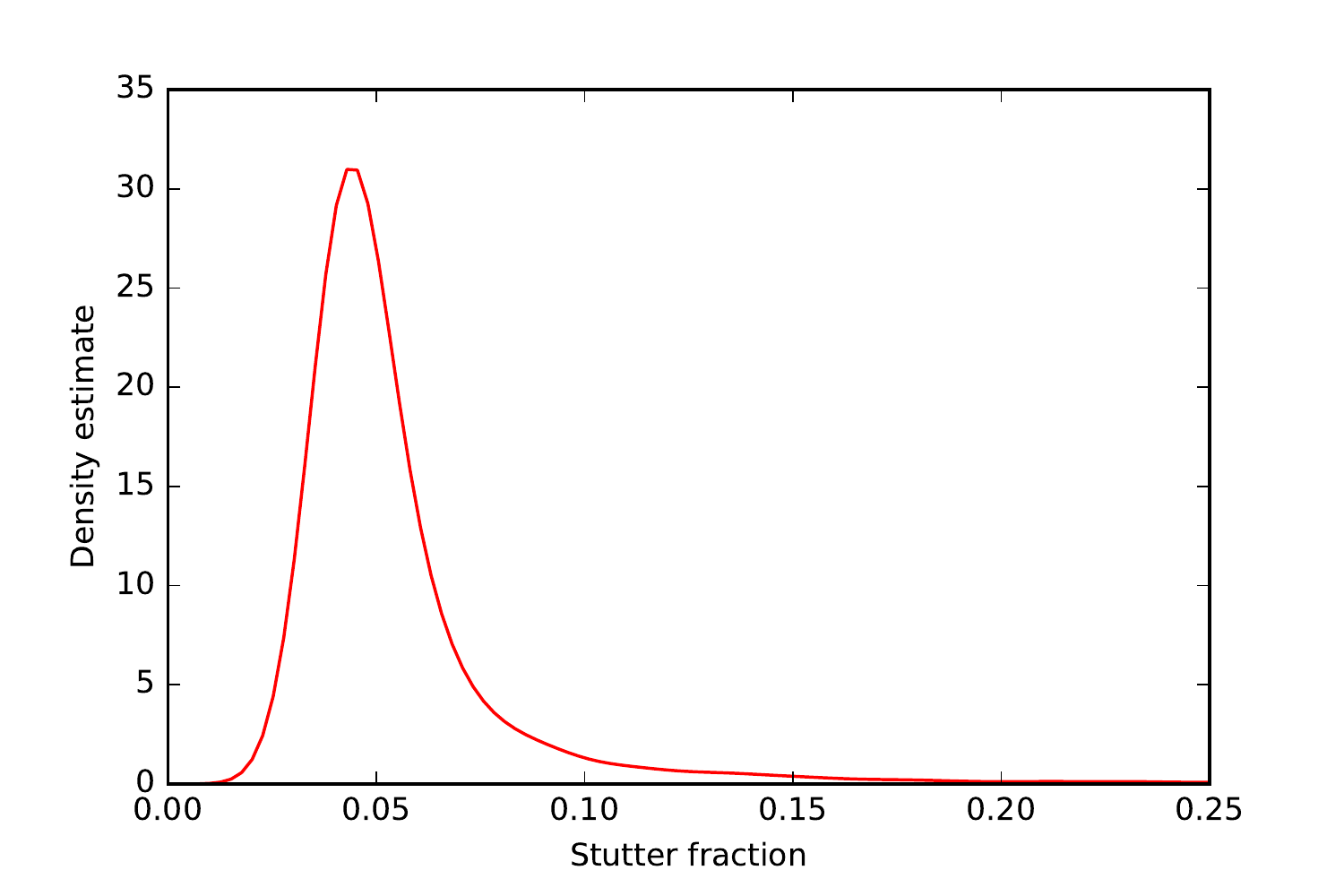}
  \caption{Kernel density estimate of the stutter fraction in  the simulation data  plotted in \figref{fig:gillscatter}.
    \label{fig:gillsratio}}
\end{center}
\end{figure}

\clearpage
We may also find the fraction of the number of amplicons generated
that are stutters, that is the ratio of the number of stutter
amplicons to the total number of amplicon and stutter amplicons, and
generate a kernel density estimate. (The values of this ratio are
therefore between 0 and 1, unlike the stutter proportion which is the
ratio of stutter amplicons to amplicons, which can grow very large.) A
kernel density estimate is shown in \figref{fig:gillsratio}.

\clearpage
\subsection{Generating the full joint distribution}
\label{sec:fullampstutdist}
We now show how to find the full joint- distribution of amplicon
copies and stutter amplicons, at least theoretically.  To do this we
combine a 2-dimensional \FFT analysis with bivariate probability
generating functions. A bivariate {\PGF} is appropriate here, as we
have both main amplicons and stutter amplicons, that is in the
amplification cycles we have a multi-type branching process of the
sort that was formulated by \cite{good1955joint} in terms of vectorial
generating functions.

We shall refer to the main, or initial, type of amplicon as a
\textit{target} amplicon, and use the symbol $t$ in the bivariate \PGF
to represent their number. We shall use the symbol $s$ to represent
the stutter amplicon number. As for the simple (no-stutter) model, we
may express the \PGF by functional iteration. However, because we have
two types of amplicons, we need to use coupled equations to express
the amplifications of single target and single stutter amplicons:

\begin{align*}
  t & \to  (1-p)t + p(1-\xi)t^2 + p\xi t s\\
  s &\to (1-p)s + ps^2\\
\end{align*} 

Here we let $p$ denote the probability that an amplicon (of either
type) makes a product in a cycle; $\xi$ is the conditional probability
that a target amplicon produces a stutter given either a target or
stutter is produced.  In terms of the previous notation we have
$p_s = p\xi$ and $p_t = p(1-\xi)$.  Let $F_0(s,t) = t$ and
$G_0(s) = s$. Let $F_{n}(t,s)$ denote the bivariate \PGF of the joint
distribution for the number of target and stutter amplicons after $n$
amplification cycles, arising from a single initial target amplicon,
and let $G_n(s)$ denote the number of stutter amplicons that (would)
arise from an initial single stutter amplicon.

Then for $n>0$ we have the iteration scheme:

\begin{align}
  F_{n}(t,s) &= (1-p)F_{n-1}(t,s) + p(1-\xi)F_{n-1}^2(t,s) + p\xi F_{n-1}(t,s)G_{n-1}(s) \label{eq:ampF}\\
  G_n(s) &= (1-p)G_{n-1}(s) + p G_{n-1}^2(s). \label{eq:ampG}
\end{align}

To see this is the case we argue as follows (the reader may find it
helpful to draw a probability tree). Initially we have one target
allele with \PGF $F_0(s,t) = t$. In the first cycle there are three
possible outcomes.

In the first outcome, with probability $(1-p)$ we still have just the
one target amplicon. Hence with a further $n-1$ cycles the bivariate
\PGF for the number target and stutter amplicons conditional on this
outcome is $F_{n-1}(t,s)$.

The second possible outcome from the first cycle is that the original
target makes a target copy.  This happens with probability $p(1-\xi)$,
and the \PGF conditional on this outcome is $t^2$ representing the two
targets. Now in the subsequent $(n-1)$ cycles each of these targets
will independently give rise to set of target and stutter amplicons.
The \PGF for the descendants from each target will each be
$F_{n-1}(t,s)$. Because the amplifications arising from each of the
two targets are independent, the \PGF representing the total number of
target and stutter amplicons from these two targets will be the
product of each of their {\PGF}s, that is $F_{n-1}^2(t,s)$.

The third outcome is that the initial target produces a stutter. This
happens with probability $p\xi$, and we have the joint \PGF
conditional on this outcome is $st$. In the subsequent $n-1$
amplifications the number of target and stutter amplicons generated by
the target will have \PGF $F_{n-1}(t,s)$. Independently the stutter
amplicon will generate further stutter amplicons with \PGF
$G_{n-1}(s)$. The bivariate \PGF for the total number of target and
stutter amplicons generated from the single target and amplicon is,
because of the independence of amplification, the product of their
{\PGF}s, that is $F_{n-1}(t,s)G_{n-1}(s)$.

Adding these three possibilities together with their probability
weights yields the coupled {\PGF} of \eqref{eq:ampF}: note that the
recurrence relation for $G_n(s)$ is that presented in
\Theoremref{thm:pgfs} for the non-stutter amplification process.

Now $F_{n}(t,s)$ given above is for a single starting amplicon. If we
start with $M$ amplicons that are pre-sampled binomially with
probability $\phi$, then the final distribution is given by

\begin{equation}
  \left( 1 - \phi + \phi F_{n}(t,s)\right)^M \label{eq:ampMF}
\end{equation}

\subsection{\FFT implementation of target and stutter distribution}

We now show how the recurrence relations \eqref{eq:ampF} and
\eqref{eq:ampG} may be evaluated numerically using \FFTs, at least in
principle. (In practice the memory requirements to carry this out will
be excessive for the number of cycles used in forensic \PCR analyses.)
The extension is to use a 2-dimensional
\DFT. \algref{alg-ampliconjointfft} gives the details (a Python
implementation is given in \appref{py:alg-ampliconjointfft}).

\begin{alg}[{\sc Joint distribution for target and stutter amplicons
    \DFT}]
  \label{alg-ampliconjointfft}
  \begin{itemize}
  \item Set $N = M 2^K$
  \item Initialize $F[,]$ to be a two dimensional $N\times N$ array,
    (with lowest index $[0,0]$) initialized such that all entries are
    zero except $F[1,0] = 1$.
  \item Initialize $G[]$ to be an $N$ dimensional array such that all
    entries are zero except $G[1] = 1$.
  \item Set ${\cal F}[,]$ equal to the 2-dimensional \DFT of $F[,]$ .
  \item Set ${\cal G}[]$ denote the one-dimensional \DFT of $G[]$ .

  \item For $K$ times do:
    \begin{itemize}
    \item for each element ${\cal F}[f,g]$ of ${\cal F}[f,g]$
      \begin{itemize}
      \item update
        ${\cal F}[f,g]:= (1-p){\cal F}[f,g] + p(1-\xi) {\cal F}[f,g]^2
        + p\xi{\cal F}[f,g]$.
      \end{itemize}
    \item for each element ${\cal G}[g]$ of ${\cal G}[]$ update
      ${\cal G}[g]:= (1-p){\cal G}[g] + p {\cal G}[g]^2 $.
    \end{itemize}
  \item for each element $(f,g)$ of $\cal{F}[]$ update
    ${\cal F}[f,g]:= (1 - \phi + \phi {\cal F}[f,g])^M$
  \item Set $F[,]$ equal to the inverse \DFT of ${\cal F}[]$.
  \end{itemize}
  \sbackup
\end{alg}

\begin{figure}[h]
  \begin{center}
    \includegraphics[width=0.99\textwidth]{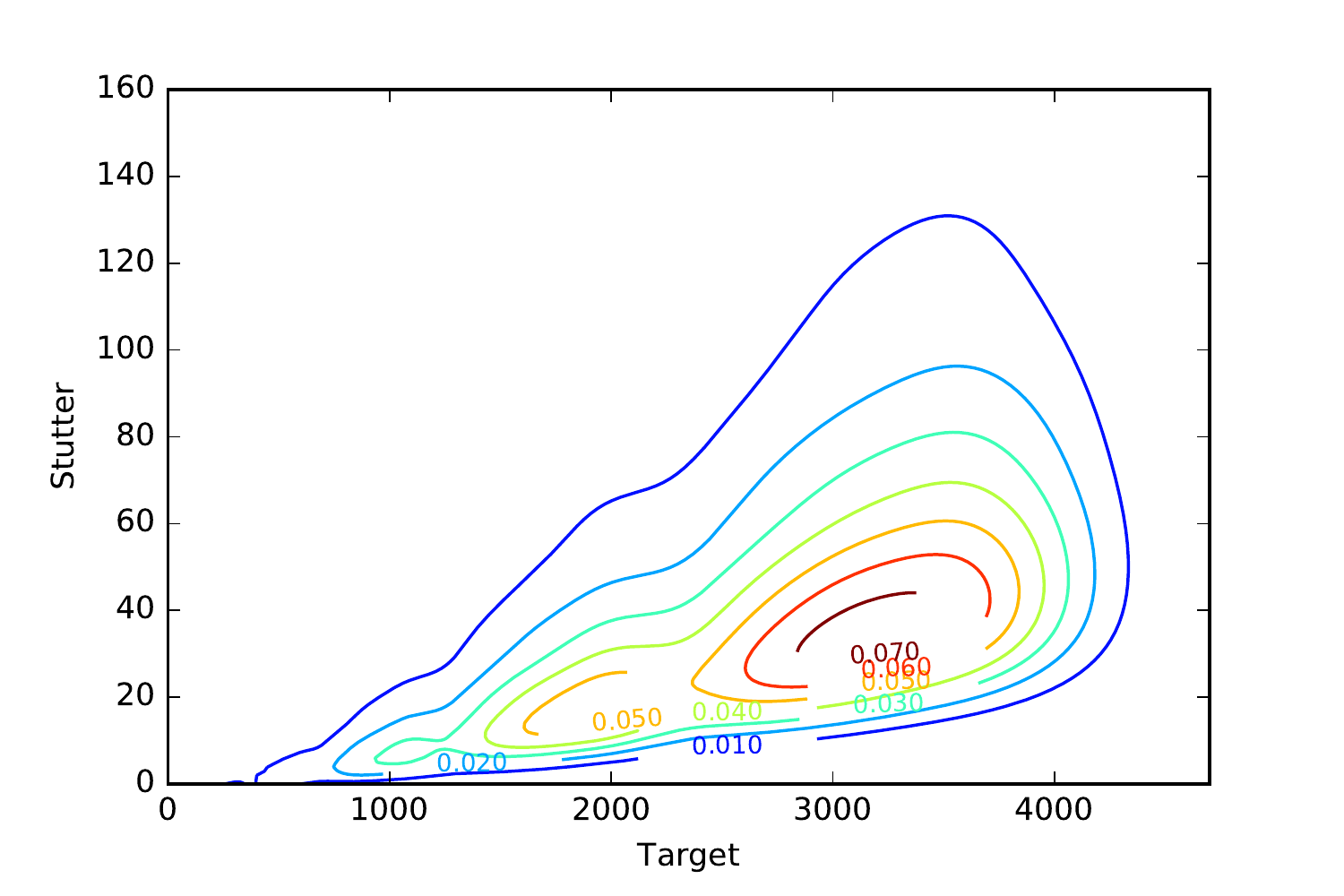}
  \caption{Contour plot of the joint distribution of the number of target and stutter amplicons, 
  arising from an single amplicon amplified for $K=13$ amplification cycles, with $p=0.85$ and 
  $\xi = 0.005$ on each cycle. Note that the numerical probabilities shown on the contours are the true values  
  multiplied by a factor of 10000.
  \label{fig:bivaramplicon}}
\end{center}
\end{figure}

\begin{figure}[h]
  \includegraphics[width=0.99\textwidth]{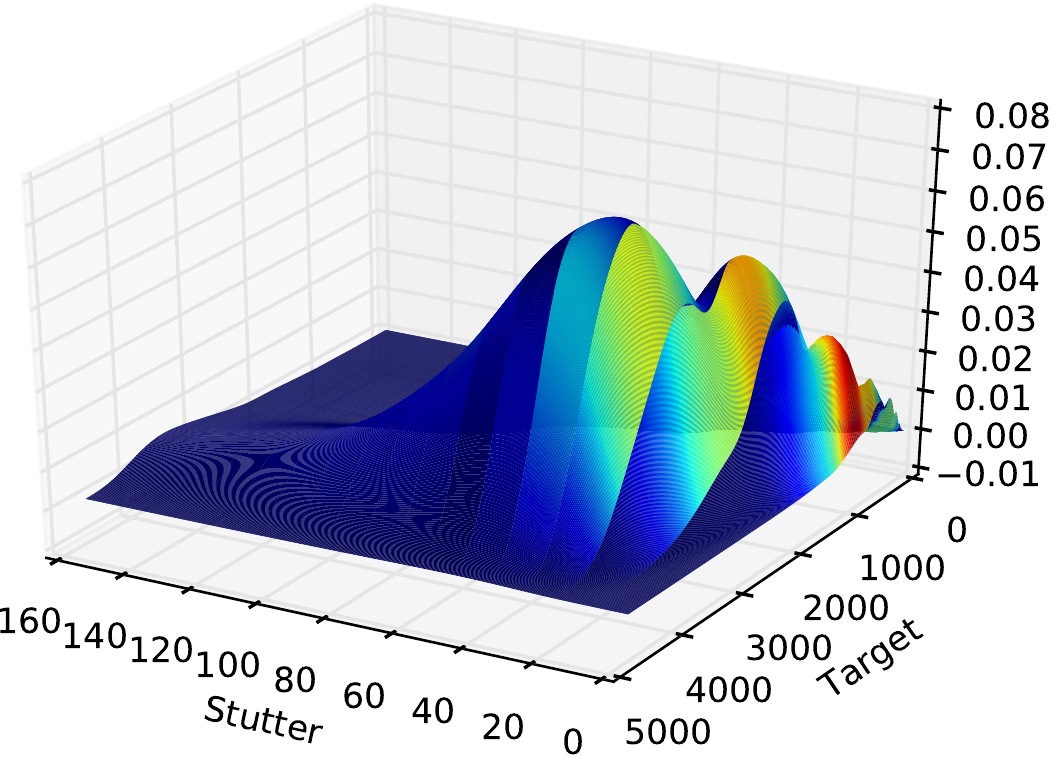}
  \caption{Surface plot of the joint distribution of the number of stutter and target amplicons, using the same values used for  \figref{fig:bivaramplicon}.
    \label{fig:bivarampsurface}}
\end{figure}

Although the use of the \algref{alg-ampliconjointfft} is not practical
for large $K$ and or $M$, it can be used for small $K$ and $M$ values.
\figref{fig:bivaramplicon} shows a contour plot of the joint
distribution for a single amplicon with subject to $K=13$
amplification cycles, with $p=0.85$ and $\xi = 0.005$ on each cycle,
and \figref{fig:bivarampsurface} a surface plot.  From the figures we
can see quite clearly the correlation between the number of target and
stutter amplicons.  Multi-modality of the distribution is also clearly
evident. The reader may care to compare these figures to
\figref{fig:gillscatter}.

\clearpage
\subsection{Stutter marginal distribution}

From the joint \PGF $F(t,s)$ of target and stutter amplicons, we may
obtain the marginal distribution of the target amplicon simply by
substituting $s=1$. Similarly we may obtain the \PGF of the marginal
distribution of the number of stutter amplicons by substituting
$t=1$. It is the latter we are interested in here. Recall that the
iterative equations finding the joint \PGF are

\begin{align*}
  F_{n}(t,s) &= (1-p)F_{n-1}(t,s) + p(1-\xi)F_{n-1}^2(t,s) + p\xi F_{n-1}(t,s)G_{n-1}(s) \\
  G_n(s) &= (1-p)G_{n-1}(s) + p G_{n-1}^2(s). 
\end{align*}
with initial values $F_0(t,s) = t$ and $G_0(s) = s$. Rather than
iterate to find $F_{n}(t,s)$ and then substituting $t=1$ to find the
stutter marginal, we may instead first make the substitution and then
iterate: the equations then simplify to

\begin{align*}
  F_{n}(s) &= (1-p)F_{n-1}(s) + p(1-\xi)F_{n-1}^2(s) + p\xi F_{n-1}(s)G_{n-1}(s) \\
  G_n(s) &= (1-p)G_{n-1}(s) + p G_{n-1}^2(s). 
\end{align*}
with initial values $F_0(s) = F_0(1,s) = 1$ and $G_0(s) = s$.

These can be evaluated numerically using a pair of 1-dimensional
\FFTs, the algorithm is almost identical to
\algref{alg-ampliconjointfft}, (a Python implementation is given in
\appref{py:alg-ampliconstuttermargfft}).

\begin{alg}[{\sc Joint distribution for target and stutter amplicons
    \DFT}]
  \label{alg-ampliconstuttermargfft}
  \begin{itemize}
  \item Set $N = M 2^K$
  \item Initialize $F[]$ to be an $N$ dimensional array such that all
    entries are zero except $F[0] = 1$.
  \item Initialize $G[]$ to be an $N$ dimensional array such that all
    entries are zero except $G[1] = 1$.
  \item Set ${\cal F}[,]$ equal to the \DFT of $F[,]$ .
  \item Set ${\cal G}[]$ equal to the \DFT of $G[]$ .

  \item for each index $g$ of ${\cal F}[]$
    \begin{itemize}
    \item For $K$ times do:
      \begin{itemize}
      \item update
        ${\cal F}[g]:= (1-p){\cal F}[g] + p(1-\xi) {\cal F}[g]^2 +
        p\xi{\cal F}[g]{\cal G}[g]$
      \item update
        ${\cal G}[g]:= (1-p){\cal G}[g] + p {\cal G}[g]^2 $.
      \end{itemize}
      update ${\cal F}[g]:= (1 - \phi + \phi {\cal F}[g])^M$
    \end{itemize}
  \item Set $F[]$ equal to the inverse \DFT of ${\cal F}[]$.
  \end{itemize}
  \sbackup
\end{alg}

\begin{figure}[ht]
  \begin{center}
    \includegraphics[width=0.8\textwidth]{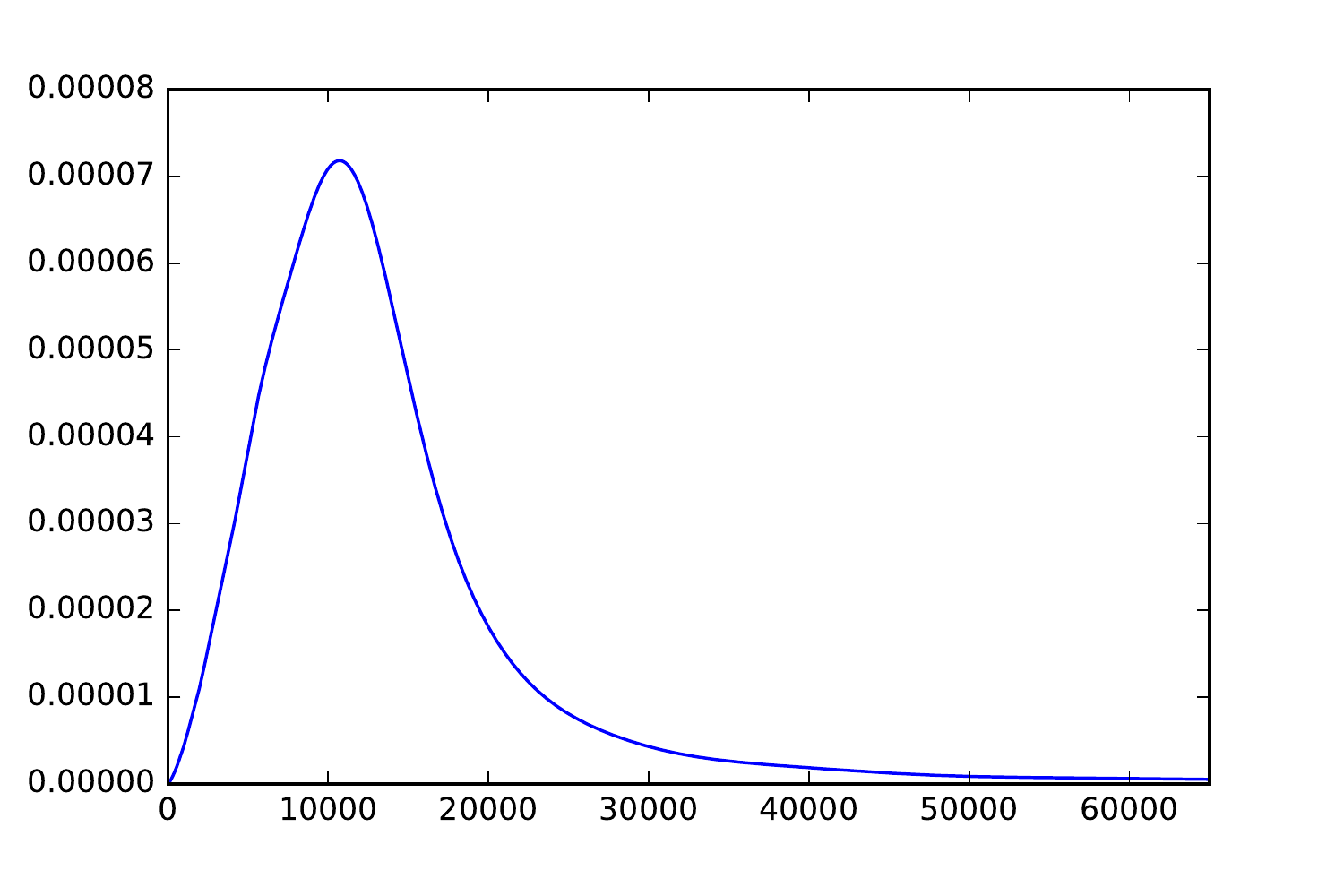}
    \caption{Marginal distribution of the number of stutter amplicons arising from a
      single target amplicon amplified for $K=22$ cycles,  with $p=0.80$ and 
  $\xi = 0.004$ on each cycle.
  \label{fig:ampliconstuttermarg}}
\end{center}
\end{figure}

\figref{fig:ampliconstuttermarg} shows the marginal distribution of
the number of stutter amplicons arising from a single initial target
amplicon, and amplified for $K=22$ cycles with $p=0.8$ and
$\xi = 0.004$ on each cycle. The long right-hand tail of the
distribution is evident.

\begin{figure}[ht]
  \begin{center}
    \includegraphics[width=0.95\textwidth]{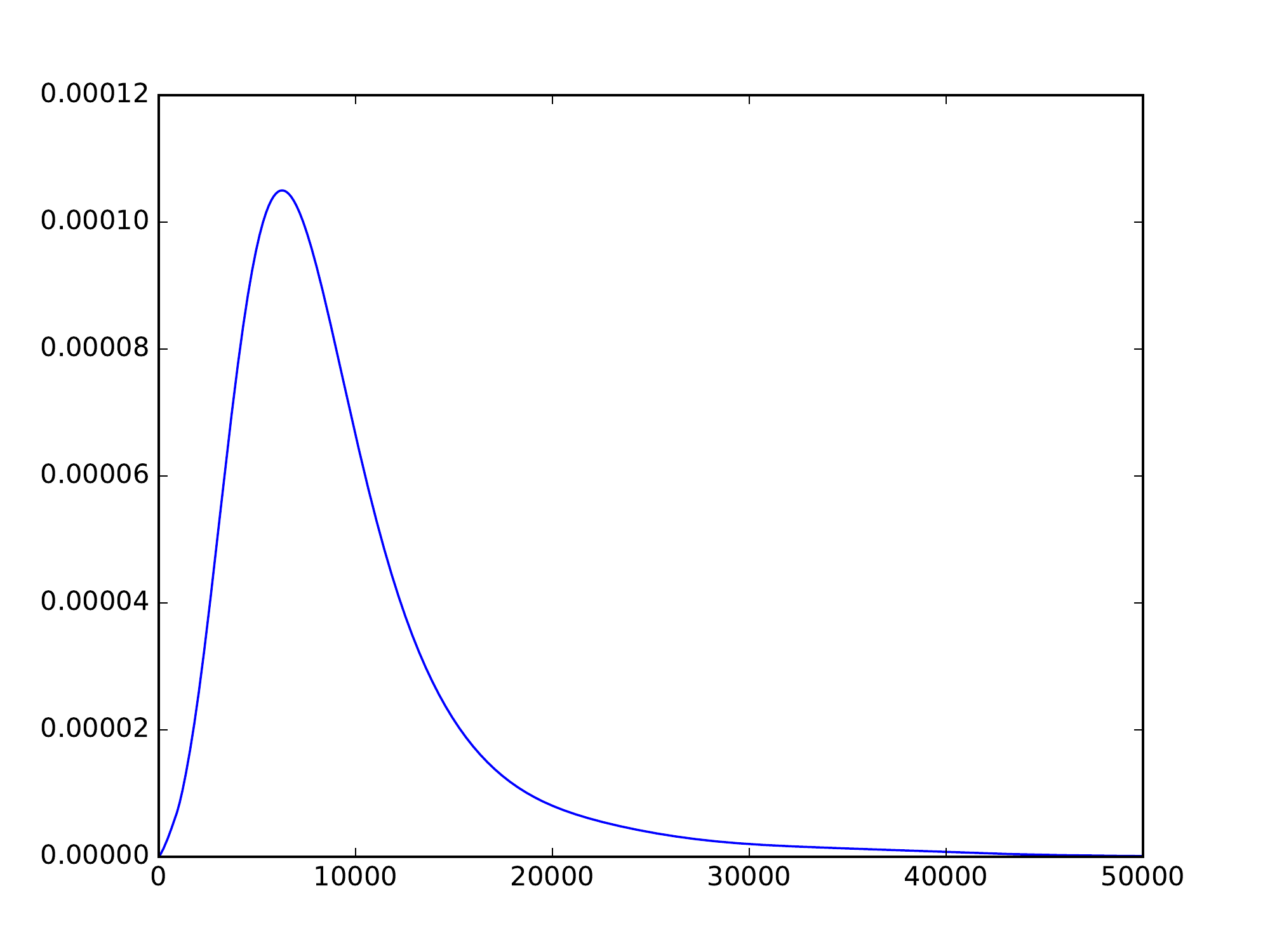}
  \caption{Marginal distribution of the number of stutter amplicons arising from $M=83$ amplicons
  pre-sampled with probability $\phi= 1/11$ prior to amplification, and  amplified for $K=18$ cycles,  with $p=0.8$ and 
  $\xi = 0.004$ on each cycle.
  \label{fig:stutterM83K18}}
\end{center}
\end{figure}

\figref{fig:stutterM83K18} shows the marginal distribution of the
number of stutter amplicons arising from an initial set of 83 target
amplicons, sampled binomially with selection probability $\phi = 2/11$
, and amplified for $K=18$ cycles with $p=0.8$ and $\xi = 0.004$ on
each cycle.

\clearpage

\subsection{Moment analysis}

The previous sections have shown how the full joint distribution of
target and stutter amplicons may be found numerically by evaluating
the \PGFs the \DFT. Up to now, probabilistic genotyping software
packages, lacking this evaluative ability, assume some simple
distributional assumption--- the most common distributions used are
the normal, lognormal of gamma. However, it is straightforward to
derive, from the \PGF, expressions for the means and variances of the
full distribution. These moments may be used for finding a `best
fitting' simple distribution based on moment matching.

Let $T$ denote the number of target amplicons, and $S$ the number of
stutter amplicons, arising from the amplification of a single target
amplicon. We have that
\begin{align*}
  \E T\cd n &= \frac{\partial F_n(t,s)}{\partial t}\vert_{t=1,s=1}\\
  \E T(T-1)\cd n &= \frac{\partial^2 F_n(t,s)}{\partial t^2}\vert_{t=1,s=1}\\
  \E S\cd n &= \frac{\partial F_n(t,s)}{\partial s}\vert_{t=1,s=1}\\
  \E S(S-1)\cd n &= \frac{\partial^2 F_n(t,s)}{\partial s^2}\vert_{t=1,s=1}\\
  \E TS\cd n &= \frac{\partial^2 F_n(t,s)}{\partial s\partial t}\vert_{t=1,s=1}
\end{align*}
from which the variances $\V T$ and $\V T$ and the correlation
$\cor(N,M)$ may be found.  (Derivations of algebraic solutions are
given in \appref{app:ampmoments} using different notation.) We have
(variance formulae are given in the appendix)

\begin{align*}
  \E T\cd n   &= (1 + p(1-\xi))^n\\
  \E S\cd n &=  (1+p)^n - ( 1 + p(1-\xi))^n\\\
  \E TS\cd n &=(1 + p*(1-x))^{n-1}\left( [1 - (1+p)^n]x + 2*[(1+p)^n - (1+p*(1-x))^n]\right)
  \\
  \cov (T,S\cd n) &=  (\E TS\cd n) - (\E T\cd n )(\E S\cd n )\\
              &= (1 + p(1-\xi))^{n-1}\left[ (1-p(1-\xi))[ (1+p)^n - (1 + p(1-\xi))^n] - \xi((1+p)^n-1)\right]
\end{align*}

\begin{figure}[ht]
  \begin{center}
    \caption{\label{fig:corr28}}
    \begin{minipage}{.5\textwidth}
      \centering
      \includegraphics[scale=0.4]{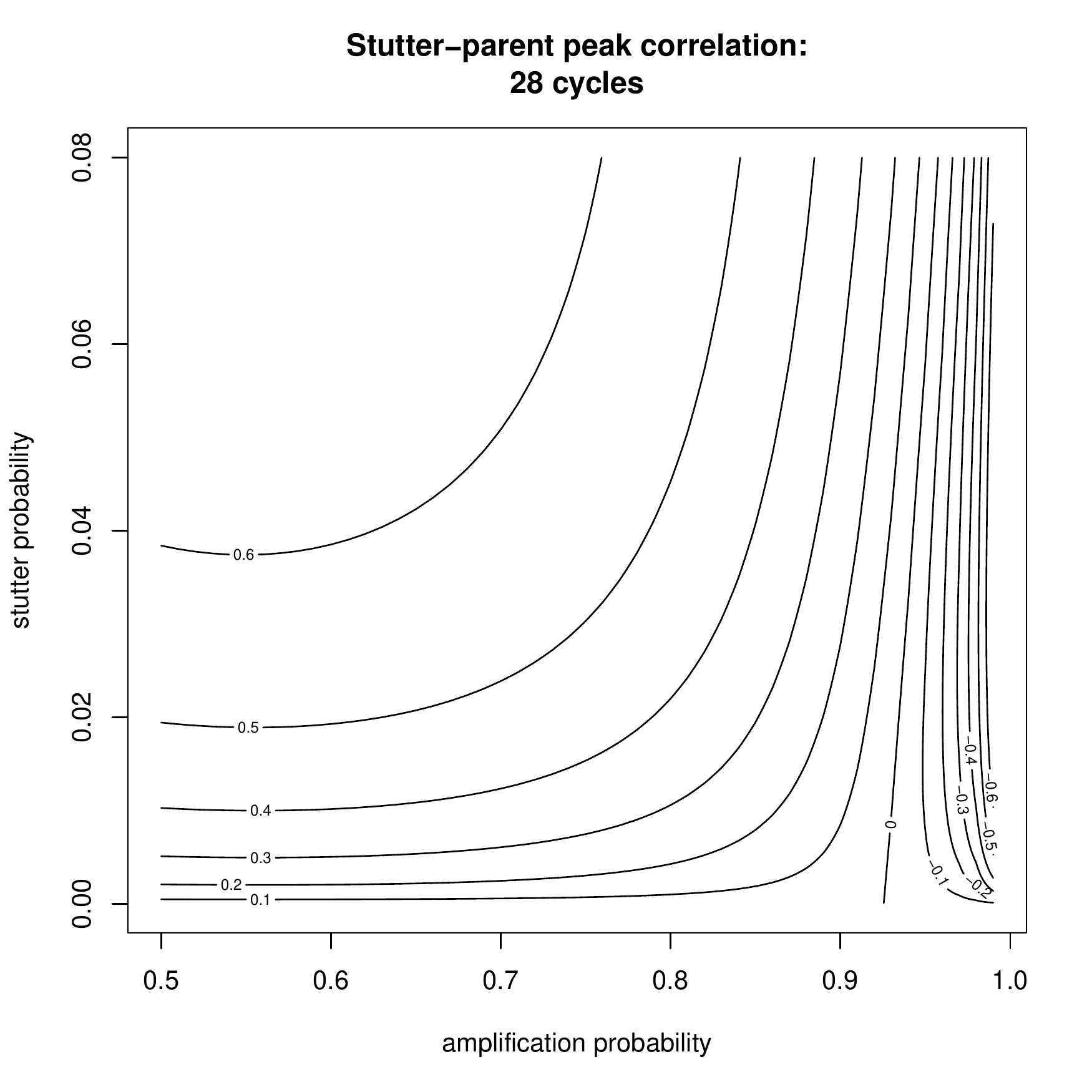}
    \end{minipage}%
    \begin{minipage}{.5\textwidth}
      \centering
      \includegraphics[scale=0.4]{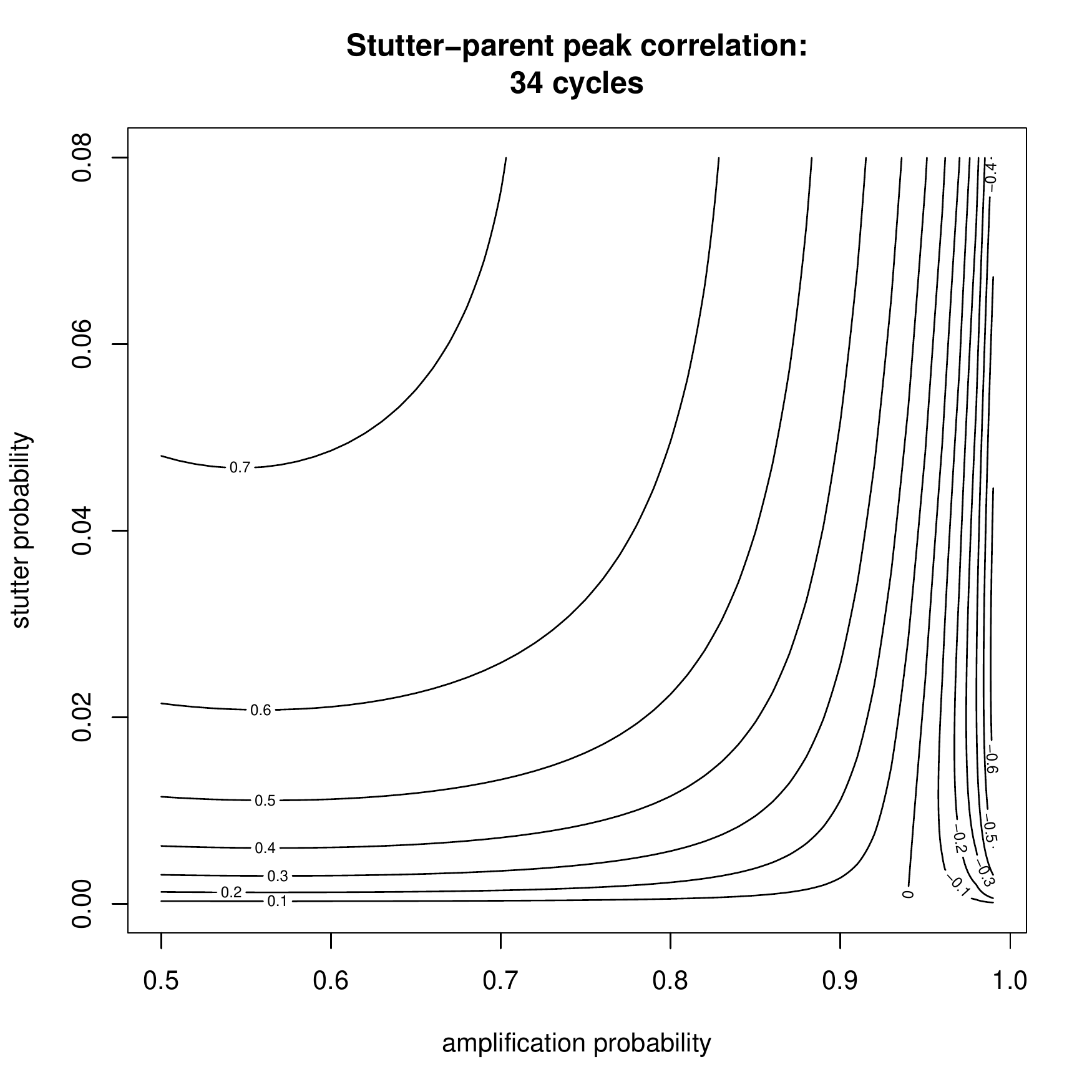}
    \end{minipage}%
  \end{center}
\end{figure}

As an alternative to using the analytic formula, instead one could fix
numerical values for $p$, $\xi$ and $K$ and solve the recurrence
relations numerically and very simply.  It is interesting to examine
how the correlation varies with $p$ and $\xi$ and the number of
cycles.  In \figref{fig:corr28} we show the dependence for 28 and 34
cycles, in the form of contour plots.
For forensic applications we expect that $p$ will be in the range
$0.75 - 0.95$, with $\xi$ in the range $0.004 - 0.01$. Thus we see
that there can be quite high correlation between the number of
amplicons (which is proportional to the peak heights in the \EPG) of
stutter and target alleles. This is apparently at variance with the
experimental results of \citep{bright2013developing}--- we shall
return to this point later.

The moment formulae above assume a single initial amplicon, but they are readily
extended to take account of binomial pre-sampling. There are two ways
to proceed. One way is to write down the \PGF to include the sampling;
this is given by

$$(1-\phi + \phi  F_n(t,s))^M.$$
This can then be differentiated to find the moments.

Another, simpler way, is to use conditional expectation:
\begin{align*}
  E Y &= E[E Y\cd X]\\
  V Y &= E[V Y\cd X] + V[E Y\cd X]
\end{align*}
for the means and variances of $T$ and $S$, where in our case
$X \sim \mbox{Binom}(M,\phi)$ having mean $M\phi$ and variance
$M\phi(1-\phi)$.  Let ${\cal T}$ denote the total number of amplicons
arising from pre-sampling $M$ target amplicons and then amplifying the
sample, and let ${\cal S}$ denote the number of stutter amplicon
products.

With this we obtain

\begin{align*}
  \E {\cal T} &= M\phi \E [T\cd n] \\
  \V {\cal T} &= M\phi (\V[T\cd n] + (1-\phi) E[T\cd n]^2)\\
  \E {\cal S} &= M\phi \E [S\cd n] \\
  \V {\cal S} &= M\phi (\V[S\cd n] + (1-\phi) E[S\cd n]^2)
\end{align*}

and for the covariance we have
$$
\cov({\cal T},{\cal S}) = M\phi\left(\cov (T,S\cd n) +(1-\phi)\E [T\cd
  n]\E [S\cd n]\right)
$$

from which we deduce that the correlation between $T$ and $S$ does not
depend on the initial number of target amplicons, but does depend on
the sampling probability $\phi$:
\begin{equation}
  \cor({\cal T},{\cal S}) \frac{\cov(T,S\cd n) + (1-\phi)(\E T\cd n)(\E S\cd n))}
  {\sqrt{(\V[T\cd n] + (1-\phi) (\E [T\cd n])^2}\sqrt{(\V[S\cd n]+ (1-\phi) (\E [S\cd n])^2}}\label{eq:bincor}
\end{equation}

In many forensic laboratory experiments DNA samples are prepared by
high dilution of large template DNA.  For such scenarios a Poisson
distribution would appear to be more appropriate.  Let us assume that
the total number of target amplicons selected for \PCR has a Poisson
distribution with rate $\lambda$ with $\lambda >0$.  If
$X\sim \mbox{Poisson}(\lambda)$, then the \PGF of $X$ has the form

$$\exp(\lambda(t-1))$$

To obtain the \PGF for the total number of amplicons we simply
substitute $F_n(t,s)$ for $t$ in this Poisson \PGF
$$\exp(\lambda(F_n(t,s)-1))$$
and take derivatives as appropriate, Alternatively, we may use
conditional expectation noting the mean and variance of a
Poisson($\lambda$) random variable is $\lambda$. Either way we obtain

\begin{equation}
  \cor({\cal T},{\cal S}) = \frac{\E [TS\cd n]}{\sqrt{(\E[T\cd n]^2)(\E[S\cd n]^2)}}\label{eq:poiscor}
\end{equation}

Surprisingly, the correlation $\cor({\cal T},{\cal S})$ does not
depend on $\lambda$ (see \appref{app:ampmoments}).  Note also that
\eqref{eq:bincor} reduces to \eqref{eq:poiscor} with $\phi= 0$, which
corresponds to the double limit $M\to \infty, \phi\to 0$ with
$ M\phi = \lambda$ fixed.

The correlations can get very high, for example with $p=0.8$,
$\xi = 0.005$ and $k=28$ cycles, the correlation is approximately
0.74.

\section{Amplifying genomic strands: a genomic model}
 
The previous section examined the simple amplicon model of
\citep{gill:etal:2005}, and showed how by using \PGFs and the \FFT is
it possible to extract, in a very simple manner, the full probability
distributions for the number of target and stutter amplicons. Two
simplifications in the model are that it starts from amplicons, and
not genomic strands, and it does not take into account the dye-tagging
of the amplicons. We shall now remove these limitations; we shall also
add in the artefacts of forward and double stutter, drop-in and
baseline noise, and show how they may all be expressed in a unified
manner using {\PGF}s, and evaluated using {\DFT}s. We being by
considering the genomic strand model without stutters.

\subsection{The basic model described}

The basic difference between the model developed here and the
simplified model of \cite{gill:etal:2005} is that the latter assumed
that we start the branching process with amplicons; in reality we
start the branching process with samples of genomic strands. As
pointed out by \cite{butler2011advancedmethod}, this means that
amplicons are not formed until at least the end of the second thermal
amplification cycle as we show in \secref{sec:mathformnostutt}.

Consider starting from a single genomic strand, and let us assume that
amplification is 100\% efficient in all stages. We are interested in a
particular locus.

\begin{itemize}
\item In the first thermal cycle, the genomic strand is melted, and
  primers attach to the flanking regions, one on each of the two
  strands of the melted genome. The TAQ polymerase lays down dNTRs
  along each strand from the primer to past the complimentary flanking
  region. Call these generated sequences \textit{half-genomes}.  Then
  on cooling we have two separate strands, each consisting of one of
  the complementary strands making up the genomic strand bound to a
  half-genome.
\item In the second thermal cycle these two hybrid genomes melt, and
  we have the two parts of the genome and also the two half-genome
  strands in solution. In this cycle the genomic strands behave as in
  the first thermal cycle, but the half-genome strands, which were not
  there then, attach primers to their complimentary flanking regions,
  and then the TAQ lays down dNTRs along it all the way to the (far)
  end of original flanking region, where the process stops because 
  that is the end of the molecule. Each half-genome strand now has
   a complementary amplicon strand attached to it.
\item In the third thermal cycle the two complementary amplicon
  strands separate from the half-genomes and make complementary copies
  of themselves, and so the exponential growth in the number of
  amplicons with further thermal cycles begins.
\end{itemize}

The above description is not quite complete, again as described by
\cite{butler2011advancedmethod}, in that in addition to making the
amplicons, the amplicons have attached to them a fluorescent dye at
one of the flanking regions, say the 5' region (so that it can be seen
in the capillary electrophoresis (CE) equipment). When measuring the
amplicons in the CE equipment, the amplicons are heated to separate
them out into individual complementary strands, only one each of each
complement has a dye attached.

\cite{weusten2012stochastic} included initial genomic strands and
amplicons in their analysis of \PCR, but did not include the
intermediate half-strands. They also did not consider direct double
stutter or forward stutter, or the tagging of amplicons with
fluorescent dyes.

\subsection{Initial mathematical formulation : no stutters}
\label{sec:mathformnostutt}

We may denote the original double-helix genomic strand by the pair $g$
and $g_d$ The half-genomic strands may be denoted by $h$ and $h_d$,
and the amplicons by $a$ and $a_d$.  The $d$ subscript denotes a
florescence dye attached. (The original genomic DNA does not have a
dye attached to it, but we use $g_d$ to avoid introducing a special
notation; the half-genomes and amplicons do have such dyes attached.)

The branching process is then summarized by the following processes on
the various components:
\begin{align*}
  g &\to g, h_d\\
  g_d &\to g_d, h\\
  h & \to h , a_d\\
  h_d &\to h_d, a\\
  a &\to a, a_d\\
  a_d& \to a_d, a
\end{align*}
and we are interested in the final number of tagged amplicons, $a_d$
after the $n$ cycles.

It is also simpler to break the process up into two independent
branching process, one that result in products from the $g$ strand,
and one from the $g$ strand:
\begin{align*}
  g&\to g, h_d\\
  h_d &\to h_d, a\\
  a &\to a, a_d\\
  a_d &\to a_d, a
\end{align*}
and from the $g_d$ strand:
\begin{align*}
  g_d&\to g_d, h \\
  h &\to h, a_d\\
  a_d &\to a_d, a\\
  a &\to a, a_d
\end{align*}

The following two tables show how the numbers of each type increase in
each cycle for these two processes treated independently, assuming
100\% efficiency in each cycle. The second table follows the pattern
of the first.

\begin{center}
  \begin{tabular}{c|cccc}
    cycle &$g$ &$h_d$ &$a$ &$a_d$ \\ \hline
    0 &1 &0 &0 &0 \\
    1 &1 &1 &0 &0 \\
    2 &1 &2 &1 &0 \\
    3 &1 &3 &3 &1 \\
    4 &1 &4 &7 &4 \\
    5 &1 &5 &15 &11 \\
    6 &1 &6 &31 &26 \\
    7 &1 &7 &63 &57 \\
    8 &1 &8 &127 &120 \\
    9 &1 &9 &255 &247 \\
    10 &1 &10 &511 &502 \\ \hline
  \end{tabular}
\end{center}

\begin{center}
  \begin{tabular}{c|cccc}
    cycle &$g_d$ &$h$ &$a_d$ &$a$ \\ \hline
    0 &1 &0 &0 &0 \\
    1 &1 &1 &0 &0 \\
    2 &1 &2 &1 &0 \\
    3 &1 &3 &3 &1 \\
    4 &1 &4 &7 &4 \\
    5 &1 &5 &15 &11 \\
    6 &1 &6 &31 &26 \\
    7 &1 &7 &63 &57 \\
    8 &1 &8 &127 &120 \\
    9 &1 &9 &255 &247 \\
    10 &1 &10 &511 &502 \\ \hline
  \end{tabular}
\end{center}

Of interest is the number of tagged amplicons, which is obtained from
adding up the row entries for $a_d$ in each of the two tables above:

\begin{center}
  \begin{tabular}{c|c}
    cycle $n$ &$a_d$ \\ \hline
    0 &0 \\
    1 &0 \\
    2 &1 \\
    3 &4 \\
    4 &11 \\
    5 &26 \\
    6 &57 \\
    7 &120 \\
    8 &247 \\
    9 &502 \\
    10 &1013 \\
  \end{tabular}
\end{center}

It is readily verified numerically that these totals are the numbers
$2^n-n-1$ for each number of cycles $n$.  (They form the sequence
A000225, called the \textit{Eulerian} numbers, in the
\href{https://oeis.org/}{The On-Line Encyclopedia of Integer
  Sequences}.)  We can verify this algebraically by induction as
follows. Let us consider the $g$ sequence. It clear that after $k$
cycles the number of $h_d$ is equal to $k$, as the $g$ has had $k$
opportunities to make and $h_d$.  We see that the values in the third
column are (for $k>0$) the integers $2^{k-1}-1$, and those in the
fourth column are the Eulerian numbers $k$ reduced by 1, that is
$2^{k-1}-k$. We can see this is true for all values given in the
table, we take this as a starting position on an inductive proof for
general $k>10$.
 
Thus at the $k$-th cycle, assume that there are $n_{h_d:k} = k$ copies
of $h_d$, $n_{a:k} = 2^k-1$ copies of $a$ and $n_{a_d:k} = 2^{k-1}-k$
copies of $a_d$. Then in the next cycle:

\begin{itemize}
\item The number of $h_d$ half-strands increases by 1, as the $g$
  makes a new copy and the previous copies are not destroyed,
  hence $$n_{h_d:k+1} = n_{h_d:k} +1 = k+1.$$
\item The number of $a$ amplicons increases by the number of $h_d$ and
  $a_d$ products that were present on the previous cycle:
$$n_{a:k+1} = n_{a:k} + n_{h_d:k} + n_{a_d:k} = 2^{k-1}-1 + k + 2^{k-1}-k = 2^k-1$$
\item The number of $a_d$ amplicons increases by the number of $a$
  amplicons that were present on the previous cycle:
$$n_{a_d:k+1} = n_{a_d:k} + n_{a_d:k} = 2^{k-1}-k +  2^{k-1}-1 = 2^k-(k+1)$$
\end{itemize}

Hence the inductive hypothesis is proved. Similar calculations go
through for the second table. From this we see that after $k+1$ cycles
the number of tagged amplicons is

$$2^k-1 +  2^k-(k+1) = 2^{k+1} - (k+1) -1,$$

which are the Eulerian numbers.  Note also that the simple amplicon
model would give $2^{k+1}$.  The above formulae correct the values in
Table~4.1 of \cite{butler2011advancedmethod}, who ignored the
persistent presence of the original genomic strands and the half
strands in the remaining cycles, assuming their contribution in
subsequent cycles would be negligible and concluding that after $n>2$
cycles there would be approximately $2^{n-2}$ amplicons.

\subsection{The basic model: PGF formulation}

The basic model can be formulated as a multivariate \PGF, derived
using vectorial generating functions.  For this we can consider the
{\PGF}s of the two types of initial genomic strands separately.

We start with the $g$ strand, which has the set of amplification
sequences:
\begin{align*}
  g&\to g, h_d\\
  h_d &\to h_d, a\\
  a &\to a, a_d\\
  a_d &\to a_d, a
\end{align*}
We introduce symbols $t_g$, $t_{h_d}$ $t_a$ and $t_{a_d}$ to be used
in the multivariate \PGF for the number each type of strand.  For a
single strand of each type, each amplifies in a single cycle according
to the {\PGF}s
\begin{align*}
  t_g & \to (1-p_g)t_g + p_gt_gt_{h_d} \\
  t_{h_d} & \to (1-p_{h_d}) t_{h_d}+ p_{h_d}t_{h_d}t_{a}\\
  t_{a} & \to(1-p_{a})  t_{a}+ p_{a}t_at_{a_d}\\
  t_{a_d} & \to(1-p_{a_d}) t_{a_d} + p_{a_d}t_{a_d}t_{a}
\end{align*}
where $p_g, p_{h_d}, p_a, p_{a_d}$ are branching process probabilities
for each of the types of strands.

Let $G_n(t_g, t_{h_d}, t_a, t_{a_d})$ denote the joint \PGF arising
from amplifying a single $g$ strand for $n$ cycles.

Let $H_{d;n}(t_{h_d}, t_a, t_{a_d})$ denote the joint \PGF arising
from amplifying a single $h_d$ strand for $n$ cycles.

Let $A_n( t_a, t_{a_d}))$ denote the joint \PGF arising from
amplifying a single $a$ strand for $n$ cycles.

Let $A_{d;n}(t_a, t_{a_d})$ denote the joint \PGF arising from
amplifying a single $a_d$ strand for $n$ cycles.

Then these {\PGF}s obey the following recurrence relations, which
follow the pattern of the single strand formulae above:
\begin{align*}
  G_{n+1} (t_g, t_{h_d}, t_a, t_{a_d})& = (1-p_g) G_n(t_g, t_{h_d}, t_a, t_{a_d})+
                                        p_gG_n(t_g, t_{h_d}, t_a, t_{a_d})H_{d;n}(t_{h_d}, t_a, t_{a_d}) ,\\
  H_{d;n+1}(t_{h_d}, t_a, t_{a_d}) & = (1-p_{h_d}) H_{d;n}(t_{h_d}, t_a, t_{a_d})+ p_{h_d}H_{d;n}(t_{h_d}, t_a, t_{a_d})A_n( t_a, t_{a_d}) ,\\
  A_{n+1}( t_a, t_{a_d}) & = (1-p_{a})A_n ( t_a, t_{a_d}))+ p_{a}A_n( t_a, t_{a_d})A_{d;n}( t_a, t_{a_d}),\\
  A_{d;n+1} ( t_a, t_{a_d})& = (1-p_{a_d})A_{d;n}( t_a, t_{a_d}) + p_{a_d}A_{d;n}( t_a, t_{a_d})A_n( t_a, t_{a_d}),
\end{align*}
with initial conditions
\begin{align*}
  G_0 (t_g, t_{h_d}, t_a, t_{a_d})&= t_g\\
  H_{d,0}(t_{h_d}, t_a, t_{a_d}) &= t_{h_d}\\
  A_{d}( t_a, t_{a_d}) &= t_{a}\\
  A_{d,0}( t_a, t_{a_d}) &= t_{a_d}
\end{align*}
It is the last two that give rise to the exponential growth of
amplicons in the \PCR process.  Similar equations arise when starting
from a genomic strand $g_d$, specifically:
\begin{align*}
  G_{d;n+1}(t_{g_d}, t_{h}, t_a, t_{a_d}) & = (1-p_{g_d}) G_{d;n}(t_{g_d}, t_{h}, t_a, t_{a_d}) + 
                                            p_{g_d}G_{d;n}(t_{g_d}, t_{h}, t_a, t_{a_d}) H_{n}(t_{h}, t_a, t_{a_d}) ,\\
  H_{n+1}(t_{h},(t_a, t_{a_d})  & = (1-p_{h})H_{n}(t_{h}, t_a, t_{a_d})  + p_{h}H_{n}(t_{h}, t_a, t_{a_d}) B_{d;n}( t_a, t_{a_d}) ,\\
  B_{d;n+1}( t_a, t_{a_d})  & = (1-p_{a_d}) B_{d;n}( t_a, t_{a_d}) + p_{a_d}B_n( t_a, t_{a_d}) B_{d;n}( t_a, t_{a_d}) ,\\
  B_{n+1}( t_a, t_{a_d})  & = (1-p_{a}) B_{n}( t_a, t_{a_d}) + p_{a}B_{d;n}( t_a, t_{a_d}) B_n( t_a, t_{a_d}) ,
\end{align*}
with initial conditions
\begin{align*}
  G_{d;0}(t_{g_d}, t_{h}, t_a, t_{a_d})  &= t_{g_d}\\
  H_{0} (t_{h}, t_a, t_{a_d}) &= t_{h}\\
  B_{0}(t_a, t_{a_d})  &= t_{a}\\
  B_{d,0} (t_a, t_{a_d}) &= t_{a_d}
\end{align*}
where we introduce functions $B$ and $B_d$ in place of $A$ and $A_d$
as these are to be considered as iterating (amplifying) independently.
The justification of the recurrence relations is similar to that given
in \secref{sec:fullampstutdist} for the joint distribution of target
and stutter amplicons in the amplicon model, and is omitted.

In general, this is a multivariate polynomial in six $t$-parameters,
as well as depending on six branching probabilities.  However, because
the genomic strand itself does not duplicate, we see that $G$ and
$G_d$ are proportional to $t_g$ and $t_{g_d}$ respectively, which is a
simplification that effectively reduces the number of $t$ parameters
to four.

We are particularly interested in the final number of dye-tagged
amplicons. This is the sum of their number arising from the $g$ and
$g_d$ strands, and therefore (because of independence of the branching
process amplifications) is the coefficient of $t_{a_d}$ in the product
of the {\PGF}s of each strand:
\begin{equation}
  F_n(t_g, t_{g_d}, t_{h},t_{h_d}, t_a, t_{a_d}) = G_n(t_g, t_{h_d}, t_a, t_{a_d})G_{d;n}(t_{g_d}, t_{h}, t_{a}, t_{a_d})
  \label{eq:fullpgf}
\end{equation}
in which we may set all of the $t$'s except $t_{a_d}$ to unity to
obtain the marginal \PGF for the tagged amplicons, that is:
$$F_n(1,1,1,1, t_{a_d}) = G_n(1,1,1, t_{a_d})G_{d;n}(1,1, 1,t_{a_d})$$

\subsection{Moment analysis}

Of interest also are the various moments, in particular the mean and
variance, which may be found from the marginal \PGF by differentiation
with respect to $t_{a_d}$ and then setting this to 1.  Algebraic
derivations maybe found in \appref{app:genomicmoments}, where it is
shown, for example, that the mean number of tagged amplicons is given
by
$$
\frac{p_{g}p_{h_d}}{p_ap_{a_d}}\left(
  \sqrt{\frac{p_a}{p_{a_d}}}\frac{(1+\sqrt{p_ap_{a_d}})^n-(1-\sqrt{p_ap_{a_d}})^n}{2}-np_a
\right) + \frac{p_{g_d}p_{h}}{p_ap_{a_d}}\left(
  \frac{(1+\sqrt{p_ap_{a_d}})^n+(1-\sqrt{p_ap_{a_d}})^n}{2}-1 \right)
$$
If $p_a = p_{a_d} = p$, this simplifies to:
$$
\frac{p_{g}p_{h_d}}{p^2}\left( \frac{(1+p)^n-(1-p)^n}{2}-np \right) +
\frac{p_{g_d}p_{h}}{p^2}\left( \frac{(1+p)^n+(1-p)^n}{2}-1 \right)
$$
and if in addition $p_g = p_{g_d} = p_{h} = p_{h_d} = p$ this reduces
to
$$ (1+p)^n - np -1.$$
When $p=1$ this yields $2^n-n-1$, the Eulerian numbers given earlier.

There are far too many parameters to explore the behaviour of the
variance in detail here, however one special case could be of
interest, that in which all the amplification parameters take the same
value.  We can compare the behaviour for the mean and variance to that
of the amplicon model.
We see the mean values will be close, viz $ (1+p)^n - np -1$ compared
to $(1+p)^n$. Somewhat surprisingly, the variance of the genomic
amplification model turns out to be almost exactly half that of the
Gill amplification model for sufficient number of cycles (the ratio is
0.49997 with N=20 and p=0.7, and 0.50000 if we increase to n=24
cycles) and appears to asymptote to almost 0.5 as the number of cycles
increases.  Thus, although the exact formula for the variance is quite
lengthy, for $n \ge 24$ cycles taking the variance formula from the
simplified amplicon model, and dividing the result by 2, appears to
give an excellent approximation to it.

\subsection{Full distribution from vectorial PGF}

The numerical derivation of the full distribution follows a similar
pattern to the derivation for amplicon model with stutter. Using a
6-dimensional \DFT one could, in principle, derive the full joint
distribution for the six types of strand. However the number of $g$
and $g_d$ strands does not change in each cycle, thus a 4-dimensional
\DFT suffices (for amplification of a single genome). The number of
each of the $h$ and $h_d$ strands increases by at most 1 in each
cycle, with at most $n-1$ of each after $n$ cycles. The numbers of $a$
and $a_d$ strands will have a maximum of $2^n-n-1$, hence the size of
the \DFT required is approximately $n^2 2^{2n}$, too large for
forensic applications with $n$ between 27 and 34 cycles.

For $M$ genomic strands pre-sampled binomially with probability
$\phi$, the joint \PGF is given by

\begin{equation}
  (1-\phi + \phi F_n(t_g, t_{g_d}, t_{h},t_{h_d}, t_a, t_{a_d}) )^M
  \label{eq:fullpfgM}
\end{equation}

However, obtaining the marginal distribution of tagged amplicons can
be found by setting $t_g =t_{g_d} =t_{h}=t_{h_d}=t_a= 1$, and using 1-dimensional
{\DFT}s of size at most $2^n$.  The algorithm is given in
\algref{alg:gemomemarg}, with an R implementation in
\appref{sec:r-fullgill09}.

\begin{alg}[{\sc Marginal target amplicon distribution for the genomic
    model}]
  \label{alg:gemomemarg}
  \begin{itemize}
  \item Set $N = M 2^K$
  \item Initialize $G[]$ ,$G_D[]$ ,$H[]$ , $H_D[]$ and $A[]$ to be an
    $N$ dimensional arrays a such that all entries are zero except the
    first, eg, $G[0] = 1$.
  \item Initialize $A_D[]$ to be an $N$ dimensional array such that
    all entries are zero except the second, $AD[1] = 1$.
  \item Find the \DFT of all the arrays, for example, ${\cal G}[]$.
  \item Initialize ${\cal F}[]$ to be an $N$ dimensional array.

  \item for each index $g$ of ${\cal G}[]$
    \begin{itemize}
    \item For $K$ times do:
      \begin{itemize}
      \item update
        ${\cal G}[g]:= (1-p_g){\cal G}[g] + p_g{\cal G}[g]{\cal
          H_D}[g]$
      \item update
        ${\cal G_D}[g]:= (1-p_{g_d}){\cal G_D}[g] + p_{g_d}{\cal
          G_D}[g]{\cal H}[g]$
      \item update
        ${\cal H}[g]:= (1-p_h){\cal H}[g] + p_h{\cal H}[g]{\cal
          A_D}[g]$
      \item update
        ${\cal H_D}[g]:= (1-p_{h_d}){\cal H_D}[g] + p_{h_d}{\cal
          H_D}[g]{\cal A}[g]$
      \item set $a = {\cal A}[g]$ and $a_d = {\cal A_D}[g]$
      \item update ${\cal A}[g]:= (1-p_a)a+ p_a aa_d$
      \item update ${\cal A_D}[g]:= (1-p_{a_d})a_d+ p_{a_d}aa_d$
      \end{itemize}
      update
      ${\cal F}[g]:= (1 - \phi + \phi {\cal G}[g]{\cal G_D}[g])^M$
    \end{itemize}
  \item Set $F[]$ equal to the inverse \DFT of ${\cal F}[]$.
  \end{itemize}
  \sbackup
\end{alg}

\begin{figure}[ht]
  \begin{center}
    \caption{\label{fig:genomeamp}}
    \begin{minipage}{.5\textwidth}
      \centering
      \includegraphics[scale=0.4]{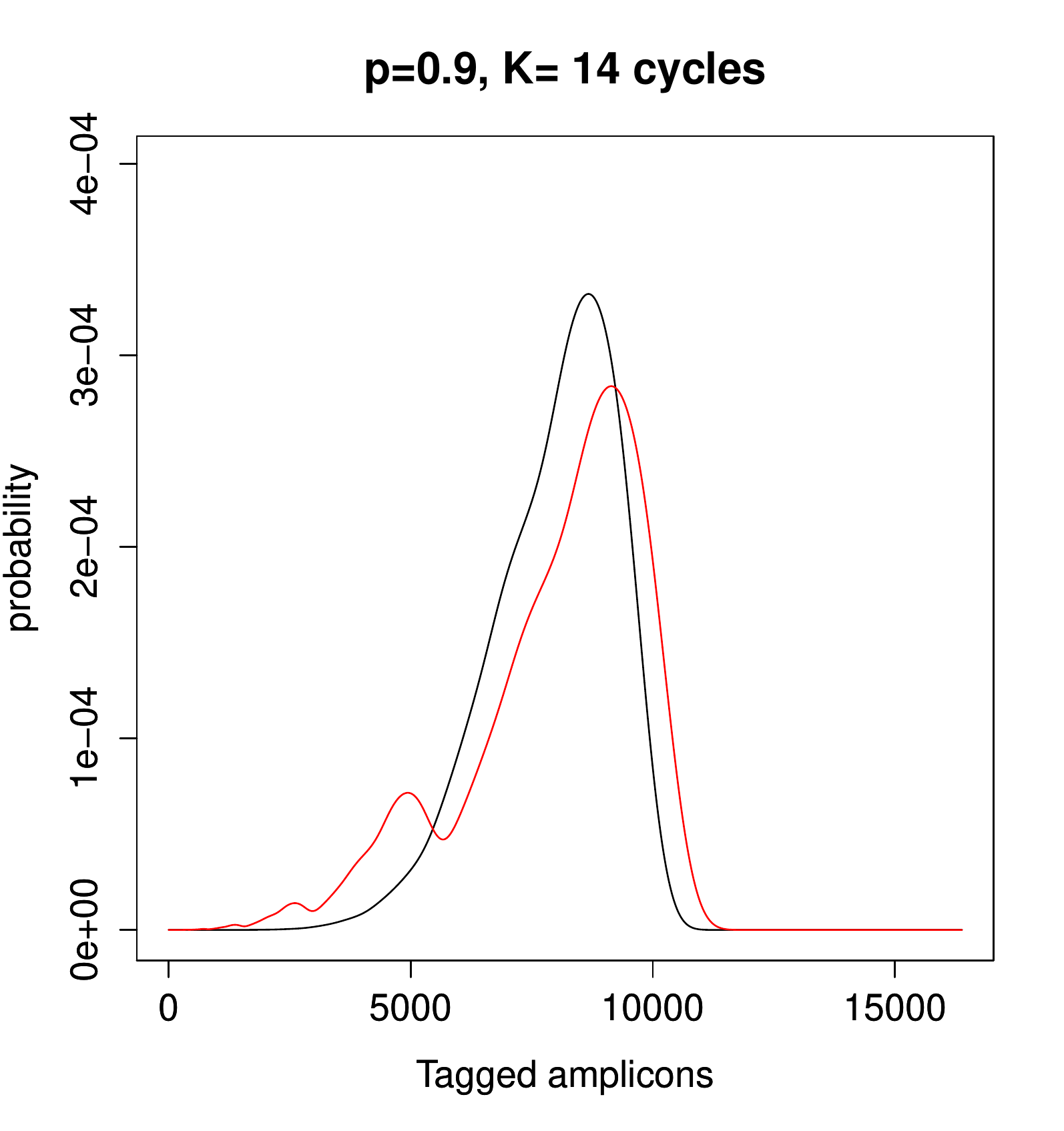}
    \end{minipage}%
    \begin{minipage}{.5\textwidth}
      \centering
      \includegraphics[scale=0.4]{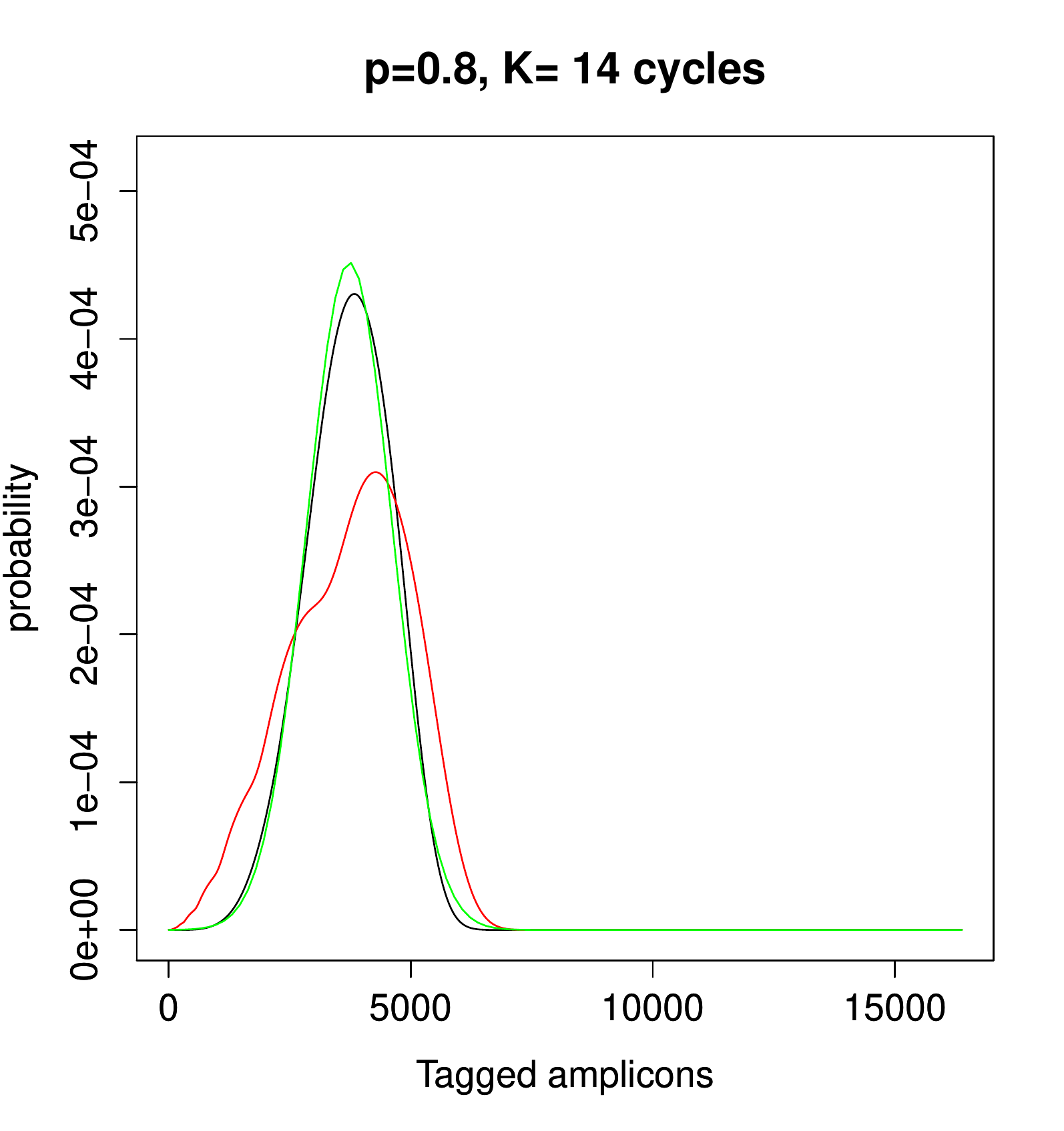}
    \end{minipage}%
  \end{center}
\end{figure}

\figref{fig:genomeamp} shows the tagged amplicon distribution for the
genomic strand model, with all amplification probabilities equal to
$p=0.9$, for $K=14$ cycles ($M= 1$ and $\phi = 1$); in
\textcolor{red}{red} is shown the corresponding amplicon model
distribution
We see that the genomic plot is somewhat smoother than the amplicon
model, and more peaked as well, confirming visually the lower variance
discussed above. The plot on the right shows the results on lowering
$p \to 0.8$. The additional \textcolor{green}{green} curve shows a
normal distribution with mean and variance matching the genomic model
- we see the fit is appears to be quite good. However it is the
vertical distances at each $x$ value that are important for evaluating
likelihoods) and in this regard the ratios of values are quite
different from 1 in places. For example in the range [1000,5000] the
ratio of the normal approximation to full distribution value ranges
from 0.777 to 1.07.  Multiplying many such ratios together for many
observed peaks could lead cumulatively to a gross over-estimation or
under-estimation of the likelihood, if there is a systematic bias.

\subsection{Including single step backward stutter}

We now include stutter into the genomic model.  For simplicity we do
not consider either forward stutter or double stutter, nor allow a
stutter to stutter, at this stage.  The iterative equations for
amplifying a $g$-strand without stutter is:

\begin{align*}
  t_g & \to t_g(1-p_g) + p_gt_gt_{h_d}\\
  t_{h_d} & \to t_{h_d}(1-p_{h_d}) + p_{h_d}t_{h_d}t_{a}\\
  t_{a} & \to t_{a}(1-p_{a}) + p_{a}t_at_{a_d}\\
  t_{a_d} & \to t_{a_d}(1-p_{a_d}) + p_{a_d}t_{a_d}t_{a}
\end{align*}

To include stutter we allow there to be stutter variants of $h_d$, $a$
and $a_d$ strands, which we denote with an extra $s$ suffix. We need
extra $t$'s to corresponding to each of these variants, and also
additional amplification probabilities for the new amplification
possibilities.  We could introduce many stutter probabilities as well,
but will keep for simplicity to a single conditional stutter
probability.  Our non-stutter equations become extended to:

\begin{align*}
  t_g & \to t_g(1-p_g) + p_g(1-\xi)t_gt_{h_d} + p_g\xi t_gt_{h_{sd}}\\
  t_{h_d} & \to t_{h_d}(1-p_{h_d}) + p_{h_d}(1-\xi)t_{h_d}t_{a}+ p_{h_d}\xi t_{h_d}t_{{a_s}}\\
  t_{h_{sd}} & \to t_{h_{sd}}(1-p_{h_{sd}}) + p_{h_{sd}} t_{h_{sd}}  t_{a_s} \\
  t_{a} & \to t_{a}(1-p_{a}) + p_{a}(1-\xi)t_at_{a_d} + p_{a}\xi t_{a}t_{a_{sd}} \\
  t_{a_d} & \to t_{a_d}(1-p_{a_d}) + p_{a_d}(1-\xi) t_{a_d}t_{a} + p_{a_d}\xi t_{a_d}t_{a_s}\\
  t_{a_s} & \to t_{a_s}(1-p_{a_s}) + p_{a_s}t_{a_s}t_{a_{sd}}\\
  t_{a_{sd}} & \to t_{a_{sd}}(1-p_{a_{sd}}) + p_{a_{sd}}t_{a_{sd}}t_{a_s}
\end{align*}

For the initial $g_d$ strand we have to introduce the additional
stutter variant $h_{s}$ and symbols $t_{g_d}$ and $t_{h_{s}}$.  The
$g_d$ equations are thus:
\begin{align*}
  t_{g_d} & \to t_{g_d}(1-p_{g_d}) + p_{g_d}(1-\xi)t_{g_d}t_{h} + p_{g_d}\xi t_{g_d}t_{h_{s}}\\
  t_{h} & \to t_{h}(1-p_{h}) + p_{h}(1-\xi)t_{h}t_{a_d}+ p_{h}\xi t_{h}t_{{a_{sd}}}\\
  t_{h_{s}} & \to t_{h_{s}}(1-p_{h_{s}}) + p_{h_{s}} t_{h_{s}} t_{a_s} \\
  t_{a} & \to t_{a}(1-p_{a}) + p_{a}(1-\xi)t_at_{a_d} + p_{a}\xi t_{a}t_{a_{sd}} \\
  t_{a_d} & \to t_{a_d}(1-p_{a_d}) + p_{a_d}(1-\xi) t_{a_d}t_{a} + p_{a_d}\xi t_{a_d}t_{a_s}\\
  t_{a_s} & \to t_{a_s}(1-p_{a_s}) + p_{a_s}t_{a_s}t_{a_{sd}}\\
  t_{a_{sd}} & \to t_{a_{sd}}(1-p_{a_{sd}}) + p_{a_{sd}}t_{a_{sd}}t_{a_s}
\end{align*}

Hence we need to keep track of 10 $t$'s corresponding to the 2 genomic
strands, the 4 half strands and 4 amplicon strands. For the marginal
stutter distribution we are interested in the tagged stutter
amplicons, $t_{a_{sd}}$.

\begin{figure}[htb]
  \begin{center}
    \includegraphics[scale=0.5]{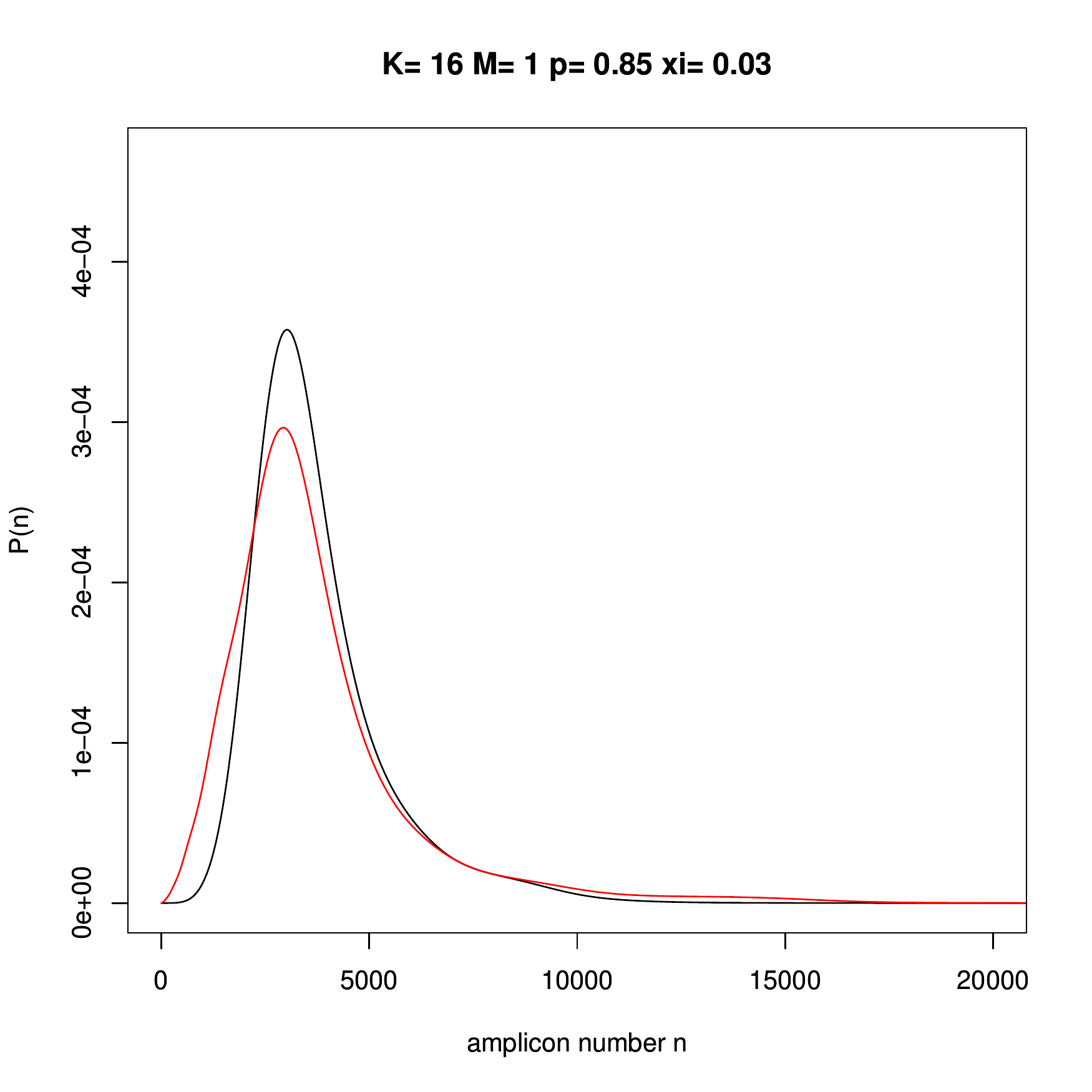}
    \caption{Marginal stutter distributions for the genomic (black)
      and amplicon model (red), with all amplifications probabilities
      equal to 0.85, and conditional stutter probability equal to
      0.03, for a single strand pair amplified for 16
      cycles.\label{fig:genampstutt}}
  \end{center}
\end{figure}

\begin{figure}[htb]
  \begin{center}
    \includegraphics[scale=0.6]{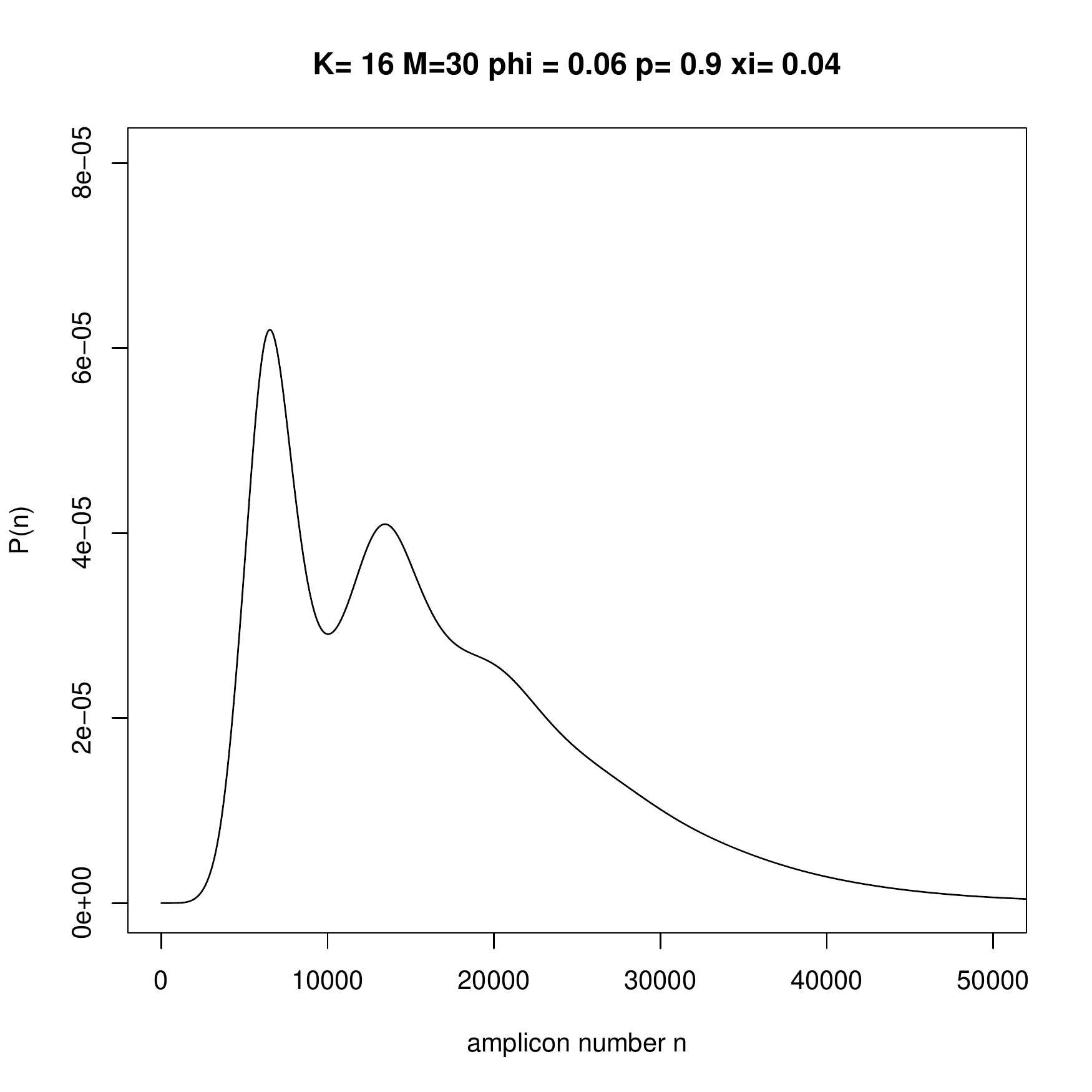}
    \caption{Marginal stutter distribution for the genomic model, with
      all amplifications probabilities $p=0.9$, conditional stutter
      probability $\xi = 0.04$, $M=30$ genome strands selected with
      probability $\phi=0.06$ and amplified for $K=16$ cycles. The
      complete dropout probability value $0.94^{30} = 0.1562556$ at
      $n=0$ is not shown. \label{fig:genamp2}}
  \end{center}
\end{figure}

These single-strand equations can be `lifted' to iterative equations
for the \PGFs, giving generalizations of \algref{alg:gemomemarg} that
are omitted, for obtaining the marginal distributions of target and
stutter amplicons.  \figref{fig:genampstutt} shows the marginal
distribution of stutter amplicons for the genomic model, with the red
curve showing the marginal from the amplicon model.
We see that the genomic stutter distribution is narrower and more
peaked, thus having a lower variance than the amplicon model. For the
parameters in the plot, the means and variances are shown in
\tabref{tab:gneampmv}. The means are approximately equal, whereas the
variances are about a factor of 2 different. The derivation of moments
is deferred to \appref{app:gmom2}. \figref{fig:genamp2} shows an
example of a marginal stutter distribution with binomial pre-sampling
of strands.
\begin{table}
  \caption{Means and variances of the amplicon and genomic model
    curves shown in \figref{fig:genampstutt}. \label{tab:gneampmv}}
  \begin{center}
    \begin{tabular}{l|cc}
      & mean &variance\\ \hline
      Amplicon model& 3749.002 &  5330275\\
      Genomic  model& 3748.594 & 2664897\\
    \end{tabular}
  \end{center}
\end{table}

\clearpage

\subsection{Extension to forward stutter and double stutter}

The extension to include forward stutter and double back-stutter is
straightforward, and follows the path of stutter.  We need two extra
$h$ variables for the forward stutter and double stutter products, and
4 extra $a$ variables, two for forward stutter and two for the double
stutter. Thus we have 16 variables altogether.

Let the subscript $f$ denotes forward stutter, and $r$ double reverse
stutter. Let $\xi_r$ denote the conditional probability of double
stutter in a cycle, and $\xi_f$ that for forward stutter.  Then the
individual branching equations are given by:
\begin{align*}
  t_g & \to t_g(1-p_g) + p_g(1- \xi_r -\xi_s - \xi_f )t_gt_{h_d}+ 
        p_g\xi_r t_gt_{h_{rd}} + p_g\xi_s t_gt_{h_{sd}}+
        p_g\xi_f t_gt_{h_{fd}}\\
  t_{g_d} & \to t_{g_d}(1-p_{g_d}) + p_{g_d}(1- \xi_r -\xi_s -\xi_f)t_{g_d}t_{h} +
            p_{g_d}\xi_r t_{g_d}t_{h_{r}}+  p_{g_d}\xi_s t_{g_d}t_{h_{s}}+ p_{g_d}\xi_f t_{g_d}t_{h_{f}}\\
  t_{h} & \to t_{h}(1-p_{h}) + p_{h}(1- \xi_r -\xi_s-\xi_f)t_{h}t_{a_d}+
          p_{h}\xi_r t_{h}t_{{a_{rd}}}+ p_{h}\xi_s t_{h}t_{{a_{sd}}}+ p_{h}\xi_f t_{h}t_{{a_{fd}}}\\
  t_{h_r} & \to t_{h_r}(1-p_{h_r}) + p_{h_r}(1-\xi_f)t_{h_r}t_{a_{rd}}
            + p_{h_r}\xi_f t_{h_r}t_{a_{sd}}\\
  t_{h_{s}} & \to t_{h_{s}}(1-p_{h_{s}}) + p_{h_{s}} (1-\xi_s-\xi_f)t_{h_{s}} t_{a_{sd}}
              +  p_{h_{s}}\xi_s t_{h_{s}} t_{a_{rd}}
              +  p_{h_{s}}\xi_f t_{h_{s}} t_{a_d}\\
  t_{h_{f}} & \to t_{h_{f}}(1-p_{h_{f}}) + p_{h_{f}} (1-\xi_s)t_{h_{f}} t_{a_{fd}}+ p_{h_{f}} \xi_s t_{h_{f}} t_{a_d}\\
  t_{h_d} & \to t_{h_d}(1-p_{h_d}) + p_{h_d}(1-\xi_r -\xi_s-\xi_f)t_{h_d}t_{a}+
            p_{h_d}\xi_r t_{h_d}t_{{a_r}}+ p_{h_d}\xi_s t_{h_d}t_{{a_s}}+ p_{h_d}\xi_f t_{h_d}t_{{a_f}}\\
  t_{h_{rd}} & \to t_{h_{rd}}(1-p_{h_{rd}}) + p_{h_{rd}}(1-\xi_f)t_{h_{rd}}t_{a_r}
               + p_{h_{rd}}\xi_f t_{h_{rd}}t_{a_s}\\
  t_{h_{sd}} & \to t_{h_{sd}}(1-p_{h_{sd}}) + p_{h_{sd}}(1-\xi_f) t_{h_{sd}}  t_{a_s} + p_{h_{sd}}\xi_f t_{h_{sd}}  t_{a}\\
  t_{h_{fd}} & \to t_{h_{fd}}(1-p_{h_{fd}}) + p_{h_{fd}} (1-\xi_s)t_{h_{fd}} t_{a_f}+ p_{h_{fd}} \xi_s t_{h_{fd}} t_{a}\\
  t_{a} & \to t_{a}(1-p_{a}) + p_{a}(1-\xi_s - \xi_f)t_at_{a_d} + p_{a}\xi_s t_{a}t_{a_{sd}} + p_{a}\xi_f t_{a}t_{a_{fd}}\\
  t_{a_d} & \to t_{a_d}(1-p_{a_d}) + p_{a_d}(1-\xi_s-\xi_f) t_{a_d}t_{a} + p_{a_d}\xi t_{a_d}t_{a_s}+ p_{a_d}\xi_f t_{a_d}t_{a_f}\\
  t_{a_s} & \to t_{a_s}(1-p_{a_s}) + p_{a_s}(1-\xi_s-\xi_f)t_{a_s}t_{a_{sd}}
            + p_{a_s}\xi_s t_{a_s}t_{a_{rd}}
            + p_{a_s}\xi_f t_{a_s}t_{a_{d}}
  \\
  t_{a_{sd}} & \to t_{a_{sd}}(1-p_{a_{sd}}) + p_{a_{sd}}(1-\xi_s-\xi_f)t_{a_{sd}}t_{a_s}
               + p_{a_{sd}}\xi_s t_{a_{sd}}t_{a_r}
               + p_{a_{sd}}\xi_f t_{a_{sd}}t_{a}\\
  t_{a_r} & \to t_{a_r}(1-p_{a_r}) + p_{a_r}(1-\xi_f)t_{a_r}t_{a_{rd}}
            + p_{a_r}\xi_r t_{a_r}t_{a_{sd}}\\
  t_{a_{rd}} & \to t_{a_{rd}}(1-p_{a_{rd}}) + p_{a_{rd}}(1-\xi_f)t_{a_{rd}}t_{a_r}
               + p_{a_{rd}}\xi_f t_{a_{rd}}t_{a_s}\\
  t_{a_f} & \to t_{a_f}(1-p_{a_f}) + p_{a_f}(1-\xi_s)t_{a_f}t_{a_{fd}} + p_{a_f}\xi_s t_{a_f}t_{a_{d}}\\
  t_{a_{fd}} & \to t_{a_{fd}}(1-p_{a_{fd}}) + p_{a_{fd}}(1-\xi_s)t_{a_{fd}}t_{a_f}
               + p_{a_{fd}}\xi_s t_{a_{fd}}t_{a}\\
\end{align*}

The corresponding recurrence relations for the multivariate \PGF in
these 18 variables is left to the reader.

With the introduction of forward and double stutter we have the
possibility that over several cycles a stutter product may amplify by
forward stuttering to make target amplicon, or could itself stutter to
make a double stutter. Similarly a forward stutter product from a
target could amplify and stutter thus producing a target, or could
double stutter and product a stutter. Likewise a double stutter
product could amplify by forward stuttering and create a stutter
product. This means that there is a cyclic dependency in the equations
above, so care is required in their computer implementation.

Note that the above equations do not describe the possibility that a
double-stutter strand could itself stutter or double stutter when
amplified, or a forward stutter strand could itself forward
stutter. These more general possibilities are accounted
in \partref{part:newframework}, in which we give a general modelling
framework, and also include the additional artefacts of drop-in and
background noise, and DNA degradation, that we have not dealt with so
far. But first we look at some other mathematical computational
aspects of the branching process.

\section{Further Mathematical and computational aspects}
\label{part:maths}
In this section, further mathematical properties of the branching
process as revealed by the probability generating functions are
explored.  Readers interested in forensic applications may wish to
skip this part and go onto \partref{part:newframework}.

\subsection{Finding  marginal distributions without a full \DFT}
\begin{figure}[htb]
  \begin{center}
    \includegraphics[scale=0.6]{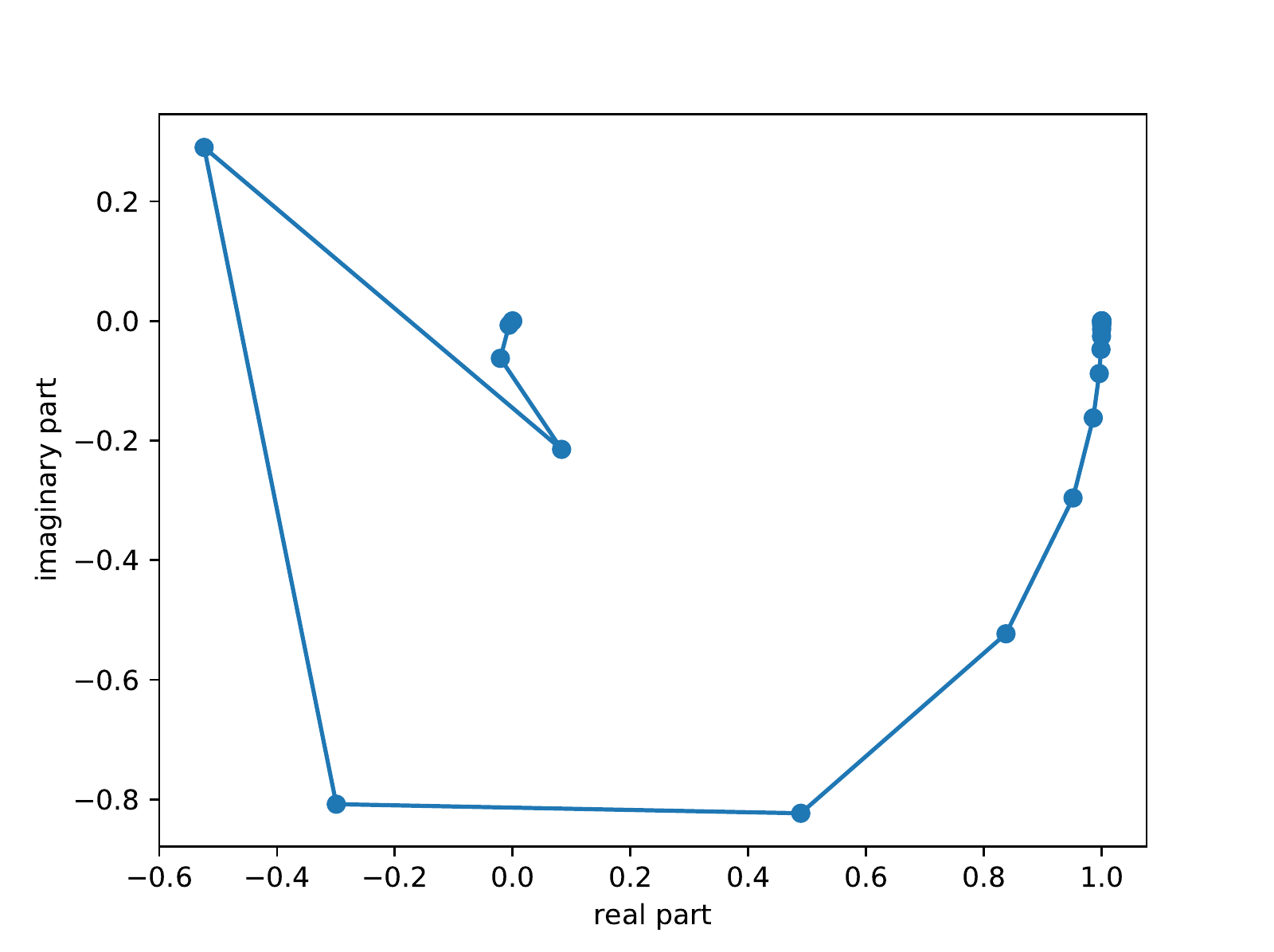}
\caption{Convergence of iterates to 0 in the complex plane for the simple amplicon model, of $F_k(e^{-i2\pi j/N}): k = 0,1,\ldots K$, for  $j=201$ 
in  \eqref{eq:pnsum}, for $p=0.85$ and up to $K=28$ iterations. The initial value is 
$  F_0(e^{-i2\pi j/N}) = 0.9999999999890425 - i4.681337853637813\times 10^{-6}$, and the 
the final iterate is at  $F_{28}(e^{-i2\pi j/N}) = -2.084\times 10^{-5} - i2.223\times 10^{-5}$ in the complex plane.
\label{fig:cxampiters}}
\end{center}
\end{figure}

We consider again the simple amplicon model, in which we start with a
single amplicon with \PGF recurrence relation
\begin{equation}
  F_{k+1}(t) = (1-p)F_{k}(t)  + pF_{k}(t) ^2
  \label{eq:amprecu}
\end{equation}
and initial condition $F_0(t) = t$.  If we let $N=2^K$, then we can
find the full distribution using the \DFT, or alternatively, for a
particular value $n$ we can find the probability of exactly $n$
amplicons via Cauchy's residue theorem

\begin{align}
  F[n] &= \frac{1}{2\pi i}\oint \frac{ F_{K}(t)}{t^{n+1}}dt  \nonumber \\
       &= \frac{1}{N}\sum_{j=0}^{N-1}F_K(e^{-i2\pi j/N})e^{i2\pi jn/N}\label{eq:pnsum}
\end{align}
To evaluate this sum, we have to evaluate each term
$A_K(e^{-i2\pi j/N})$ for points located on the unit circle in the
complex plane. This can be done the recurrence equation
\eqref{eq:amprecu}. If one views \eqref{eq:amprecu} as a nonlinear
discrete dynamical system, the we see that it has fixed points at
$t=0$ and $t=1$. The fixed point at $t=0$ is stable, whilst the fixed
point at $t=1$ is unstable: for real values of $t\in [0,1)$, iterates
of \eqref{eq:amprecu} converges to zero, whilst for real $t>1$
iterates diverge to infinity. It turns out that, apart from the $j=0$
term which corresponds to the stable point $t=1$, the iteration scheme
$ t \to (1-p)t + p*t^2$ on the unit circle forms a contraction mapping
so that the $F_k(e^{-i2\pi j/N})$ iterates converge to the origin as
$k\to \infty$. This is illustrated in \figref{fig:cxampiters}. (See
also \cite{stolovitzky1996efficiency}.)

Not only do the $F_k(e^{-i2\pi j/N})$ converge to the origin, they do
so rapidly with increasing $j$. Hence, we may approximate
\eqref{eq:pnsum} by a truncated series with a point symmetrically
located around $j=0$ (ensuring the sum total is real valued), thus:

\begin{equation}
  F[n] \approx \frac{1}{N}\sum_{j=-L}^{L}F_K(e^{-i2\pi j/N})e^{i2\pi jn/N} \label{eq:trunc}
\end{equation}

\begin{figure}[htb]
  \begin{center}
    \includegraphics[scale=0.6]{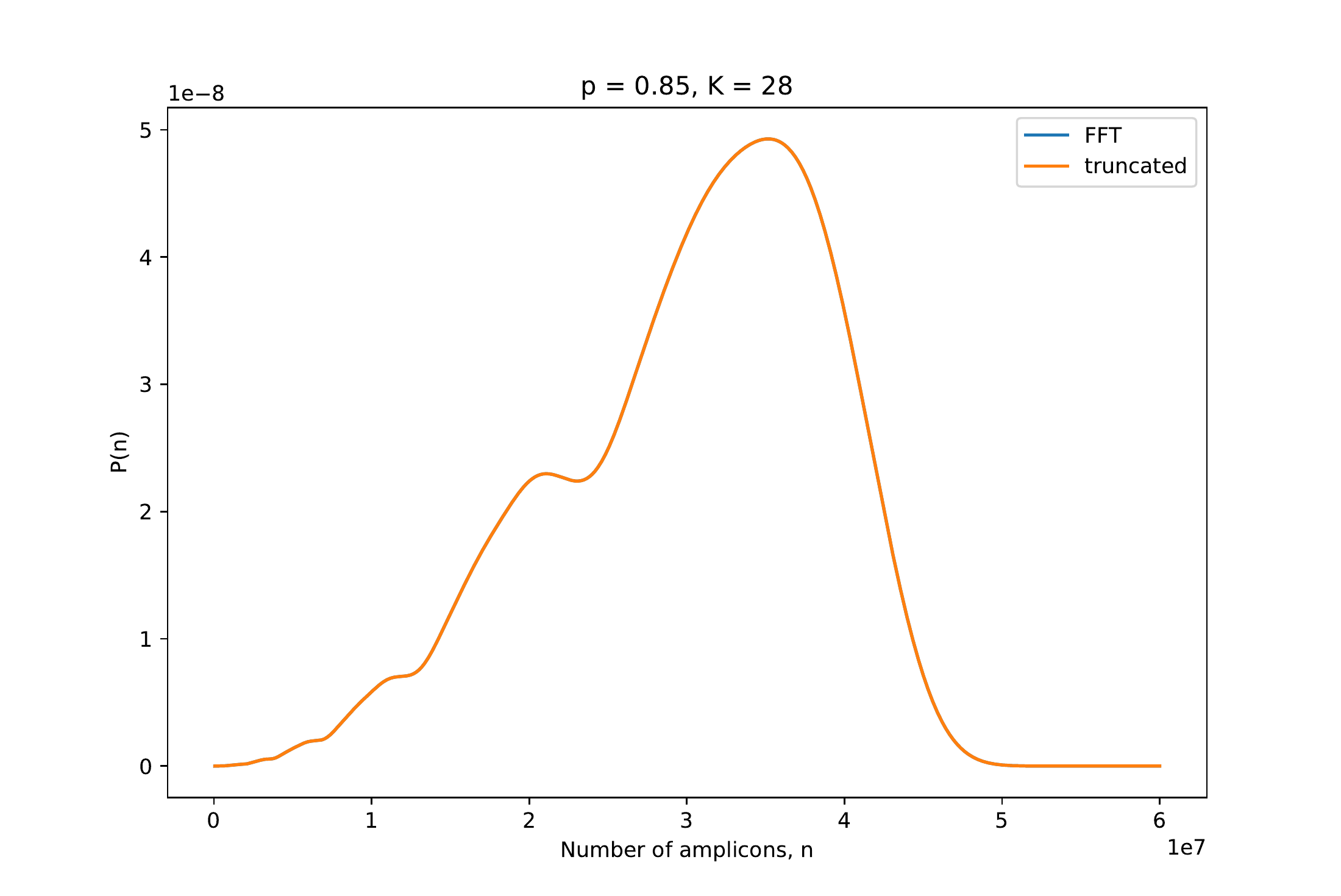}
    \caption{Distribution of the number of amplicons for the simple
      amplicon model, starting with a single amplicon, amplified for
      28 cycles with amplification probability 0.85 per cycle. The
      figure shows plots of 1000 evenly spaced $n$ values, generated
      using the Julia code in \appref{jl:full trunc} for the exact
      distribution calculated using the a \FFT analysis, and using the
      truncation approximation of \eqref{eq:trunc} with $L=1024$. The
      exact \FFT took several minutes to evaluate, and required
      approximately 10Gb of ram. In contrast, evaluating all 1000
      points using the truncated approximation took around 0.25
      seconds with minimal memory overhead.  The two curves are
      visually indistinguishable.\label{fig:fulltraunc}}
  \end{center}
\end{figure}

\figref{fig:fulltraunc} overlays the exact distribution obtained from
a full \FFT analysis, and the distribution for the truncated
approximation of \eqref{eq:trunc}. The curves are visually
indistinguishable.

\clearpage

Taking account of binomial pre-sampling with truncation requires a
little care. The full series expansion, for $M$ starting amplicons, is

$$
\frac{1}{N}\sum_{j=0}^{N-1}(1-\phi + \phi F_K(e^{-i2\pi j/N}))^M
e^{i2\pi jn/N}
$$

To carry out the truncation, we set $F_K(e^{-i2\pi j/N}))=0$ on the
$j$ terms on the circle further than $\pm L$ from the $j=0$ term, to
obtain
$$
\frac{1}{N}\sum_{j=-L}^{L}(1-\phi + \phi F_K(e^{-i2\pi j/N}))^M
e^{i2\pi jn/N} + \frac{1}{N}\sum_{j=-L*}^{L*}(1-\phi )^M e^{i2\pi
  jn/N}
$$
where the second summation is over terms complementary to the first
summation. We now use that $e^{i2\pi n/N}$ is an $N$-th root of unity,
for which $\sum_{j=0}^{N-1}e^{i2\pi jn/N} = 0$, to obtain
$$
\sum_{j=-L*}^{L*}(1-\phi )^M e^{i2\pi jn/N} =
-\sum_{j=-L}^{L}(1-\phi)^M\ e^{i2\pi jn/N}
$$
so that
\begin{equation}
  P[n] \approx \frac{1}{N}
  \sum_{j=-L}^{L}
  \left(  
    (1-\phi + \phi F_K(e^{-i2\pi j/N})^M - (1-\phi)^M
  \right)
  e^{i2\pi jn/N} \label{eq:truncbinom}
\end{equation}

Drop-in can be handled in a similar manner, as can finding stutter
marginal distribution values.  Obtaining good convergence typically
requires more terms than for \eqref{eq:trunc}. Details are left to the
interested reader to investigate (or assign to a research student if
available).  Note that if the series is truncated prematurely it could
lead to poor, and even negative, probabilities.

\begin{figure}[htb]
  \begin{center}
    \includegraphics[scale=0.6]{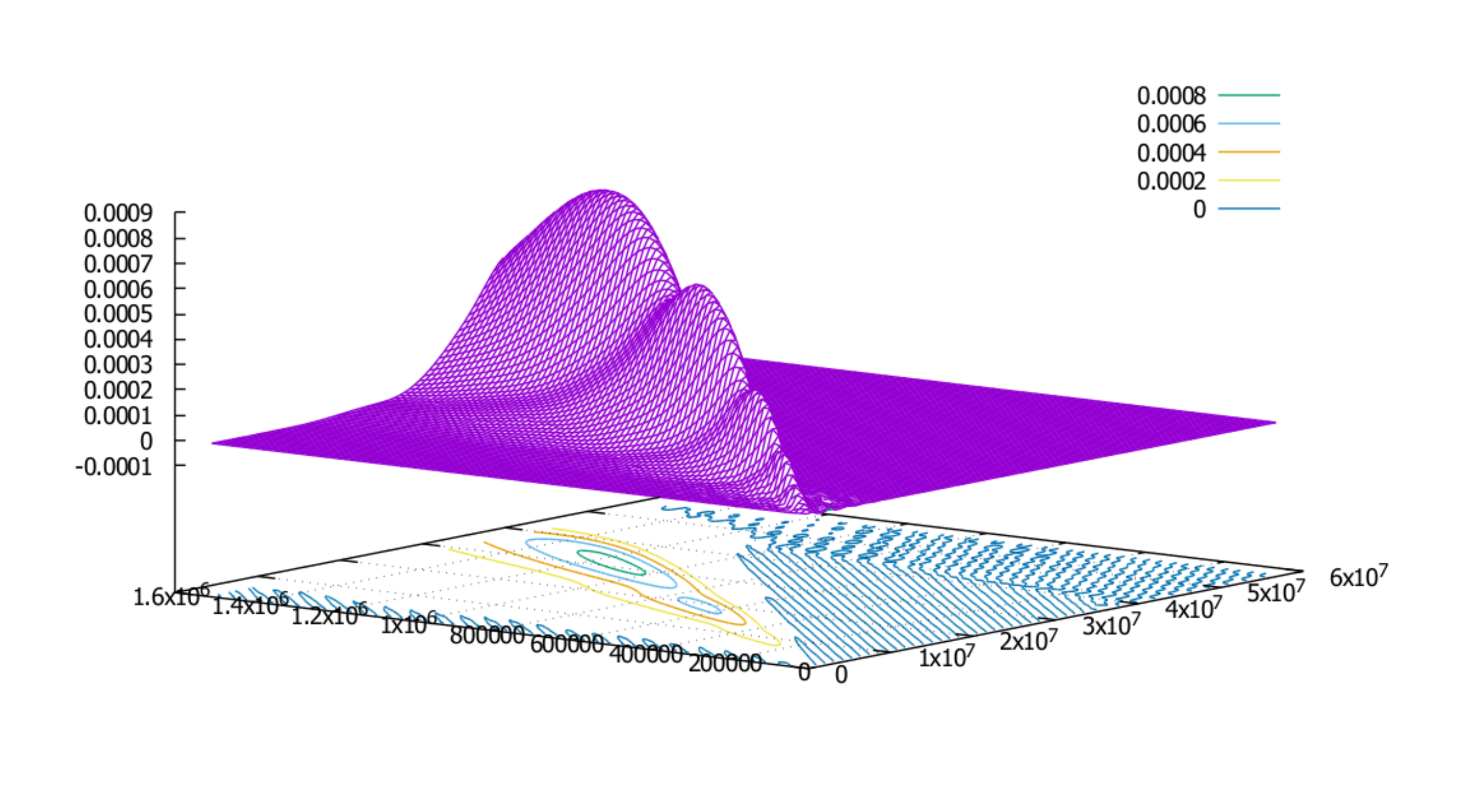}
    \caption{Bivariate distribution for target and stutter amplicons,
      starting from a single amplicon, for $K=28$, $p=0.85$ and
      $\xi = 0.004$. Note that the vertical scale (i.e., the
      probability values) have been scaled upwards by a factor of
      $598684 *15822 = 9472378248$ so that the legend for the contour
      levels show non-zero values. \label{fig:bivargnupt} The
      projected contour plot is similar to the plot in
      \figref{fig:bivarampsurface}.  (The values 598684 and 15822 are
      the intervals in the target and stutter amplicon numbers for the
      grid of points used to generate the data-points in the plot. In
      each direction 96 values were used, thus 9126 evaluations in
      total.)}
  \end{center}
\end{figure}

Bivariate distributions may also be found using similar truncated
series. As pointed out in \appref{sec-ampliconjointfftPy}, to do a
full \FFT analysis to obtain the bivariate distribution of a target
and stutter amplicons numbers would require an impossible amount of
computer memory and time. However, by writing the 2-D\FFT function out
as a double series and truncating to a smaller set of terms near the
origin, efficient and accurate evaluation of individual bivariate
probabilities may be carried out. Accuracy can be gauged by seeing how
the values change with using an increasing number of terms in the
double series. The details are left to the reader to fill in:
\figref{fig:bivargnupt} shows two views of the bivariate distribution
for a single starting amplicon, of the number of target and stutter
amplicons, using summation limits of $L = \pm 10^{12}$ for each
evaluated point (thus around $2^{26}\approx $67 million terms
altogether).

\clearpage

\part{New Framework in detail}
\label{part:newframework}

The previous sections have given some indication of the possible
models that can be formulated, and how they may be analyzed using the
mathematical tools of multivariate {\PGF}s combined with {\DFT}s.  In
this part in \secref{sec:modgen} we present a general framework of the
process leading to an \EPG, or set of {\EPG}s in the case of replicate
analyses.  Specialization of this framework to a specific model is
then presented in \secref{sec:particularmodel}, and its performance is
illustrated using simulated in \secref{sec:simdata} and publicly
available real data in \secref{sec:budata}.

\section{Modelling the \EPG generation process}
\label{sec:modgen}
\subsection{Contributor DNA}

We assume that we have a sample of $C$ DNA cells in a volume $V$. The
DNA is assumed to come from $I$ individuals, with the $i$-th
individual denoted by $K_i$, who is a contributor of $c_i$ cells to
the sample. Thus $C = \sum_{i=1}^I c_i$.  Let $L$ denote the set of
STR loci under investigation, and let $A_l$ denote the set of alleles
for a given locus $l \in L$.  The total number of genomic strands of
type $a_l \in A_l$ within the cells contributed by $K_i$ depends on
the genotype of the individual. Let $n_{ia_l}$ denote the number of
alleles of type $a_l\in A_l$ of individual $K_i$.  The values that
$n_{ia_l}$ can take depends on the locus $l$. With a few exceptions,
the possibilities are as follows.

\begin{itemize}
\item If $l$ is autosomal, then $n_{ia_l} \in \{ 0,1,2\}$.
\item if $l$ is Amelogenin, then $n_{iX} \in \{ 1,2\}$ and
  $n_{iY} \in \{ 0,1\}$. If $K_i$ is male then $n_{iX} = n_{iY} = 1$;
  otherwise $K_i$ is female with $n_{iX} = 2$ and $n_{iY} = 0$.
\item If $l$ is a Y-linked locus, then either
  $n_{ia_l} \in \{ 0,1,2\}$ or $n_{ia_l} \in \{ 0,1\}$, depending on
  the precise locus.
\item if If $l$ is an X-linked locus, then $n_{ia_l}\in \{ 0,1\}$ if
  $K_i$ is male, and $n_{ia_l} \in \{ 0,1,2\}$ if $K_i$ is female.
\end{itemize}

Let $g_{a_l}$ and $g_{a_l:d}$ denote the two complimentary strands for
the genome of type $a_l \in A_L$. We use $t_{g_{a_l}}$ and
$t_{g_{a_l:d}}$ for symbols in their generating functions. It is not
until the \PCR process begins that the $g_{a_l}$ and $g_{a_l:d}$
strands become separated. Prior to \PCR they are combined, hence the
\PGF of the number of both of these genomic strands in the DNA sample
of allele type $a_l$ from $K_i$ is
$(t_{g_{a_l}}t_{g_{a_l:d}})^{n_{ia_l}c_i}$, and the multivariate \PGF
of all alleles in $A_l$ from person $K_i$ is
\begin{equation}
  \prod_{a_l\in A_l}(t_{g_{a_l}}t_{g_{a_l:d}})^{n_{ia_l}c_i}.\label{eq:kiknown}
\end{equation}
If, for example, $l$ is autosomal and $K_i$ is homozygous with
genotype $(a_l,a_l)$ then the \PGF is
$(t_{g_{a_l}}t_{g_{a_l:d}})^{2c_i}$.

Using enzymes, DNA is extracted from the cells in the volume $V$. Let
$\pi_{e:i,a_l}$ denote the extraction efficiency, that is, the
probability that a genome segment containing the allele $a_l \in A_l$
from $K_i$ is extracted in a state suitable for amplification.  The
efficiency $\pi_{e:i,a_l}$ will depend on the process of the DNA
extraction, and also the allele via its length. The extraction
process, as well as removing genomes from cells, breaks them up into
small pieces. The location of the breakage points along a genome can
be considered a Poisson process, so that the distance between breaks
has an exponential distribution. If a break occurs in or between the
flanking regions of a genomic strand pair, then it cannot be
amplified, hence the dependence of $\pi_{e:i,a_l}$ on the
allele.\footnote{With the Poisson process just described, i
  $\pi_{e:i,a_l}$ will decay exponentially at a rate proportional to
  the total base-pair size of the between and including the flanking
  regions. This is one possible mechanism of \textit{preferential
    amplification} \citep{walsh1992preferential}.}  The \PGF for
extraction of the (single) pair is thus
$$ E_x( t_{g_{a_l}},t_{g_{a_l:d}}) =  1 - \pi_{e:i,a_l} + \pi_{e:i,a_l} t_{g_{a_l}}t_{g_{a_l:d}} .$$

However, prior to extraction the DNA may be degraded by age,
environmental or other factors. Such degradation could also be person
specific (for example, depending on the cell type of the DNA from the
person).

One form of degradation leads to a break in the genome-pair strand in
or between the flanking regions. If we let $\lambda$ denote the
probability of there not being such a break, then the \PGF for
extraction, of amplify-able (single) genome pair, becomes
$$E_x( t_{g_{a_l}},t_{g_{a_l:d}}) =  1 - \pi_{e:i,a_l}\lambda + \pi_{e:i,a_l} \lambda t_{g_{a_l}}t_{g_{a_l:d}}.$$

\begin{figure}[htb]
  \begin{center}
    \includegraphics[scale=0.6]{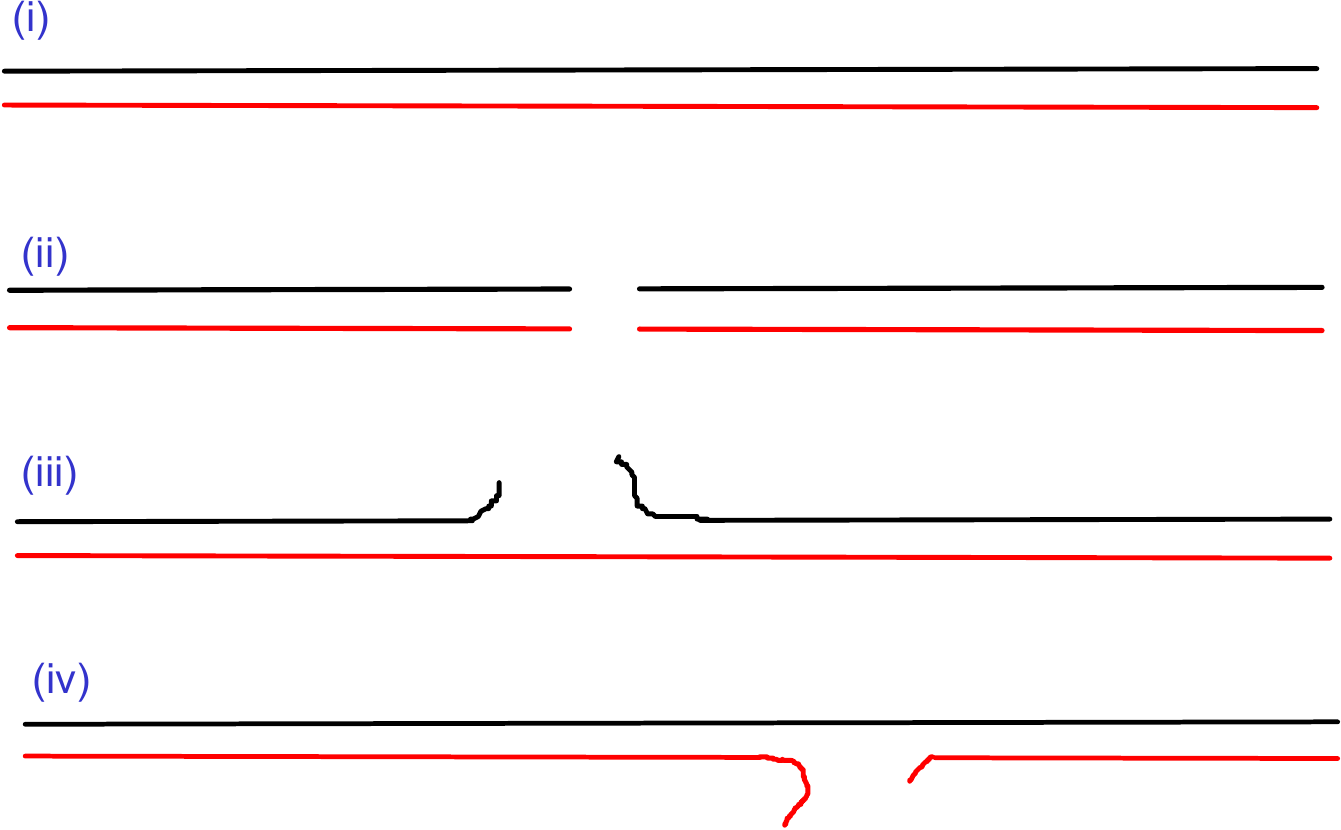}
    \caption{Genomic strand breakage possibilities for a degradation
      model. (i) An intact strand pair; (ii) a simultaneous breakage
      in both strands at a common point; (iii) breakage in one strand;
      (iv) breakage in the complementary strand. \label{fig:degrade2}}
  \end{center}
\end{figure}

Alternatively degradation could lead to one of the complimentary
strand pairs having a break and the other not, as illustrated in
\figref{fig:degrade2}.  If we assume such breaks occur independently
for each if the complimentary strands, then the joint \PGF for
extraction of the (single) genome pair strands becomes
$$ Ex( t_{g_{a_l}},t_{g_{a_l:d}}) = 1 - \pi_{e:i,a_l}(1 - (1-\lambda)(1-\lambda_d))+ \pi_{e:i,a_l} (1-\lambda +\lambda t_{g_{a_l}})(1-\lambda_d + \lambda_d t_{g_{a_l:d}}).$$
where we allow for the possibility of different breakage
probabilities, $\lambda$ and $\lambda_d$, for the two strands (arising
from their different, but complimentary, chemical base composition).

A fraction $\pi_f$ of the volume $V$ is taken for amplification. Hence
the joint \PGF for successful extraction and selection of the genome
strands from a single pair is
$$ 1 -\pi_f  + \pi_f E_x( t_{g_{a_l}},t_{g_{a_l:d}})$$

The total number of such strands from person $K_i$ extracted and
selected for amplification and are amplify-able therefore has the \PGF
\begin{subequations}
  \begin{equation}
    \left(1 -\pi_f  + \pi_f E_x( t_{g_{a_l}},t_{g_{a_l:d}})\right) ^{n_{ia_l}c_i}.\label{eq:degrad1}\\
  \end{equation}
\end{subequations}
where $E_x( t_{g_{a_l}},t_{g_{a_l:d}}) $ depends on the breakage model
chosen.

The full multivariate \PGF for the contributors' strands from all loci
under consideration ready for amplification is therefore

\begin{equation}
  \prod_{l\in  L} \prod_{a \in A_l} \prod_{i \in I}\left(1 -\pi_f  + \pi_f E_x( t_{g_{a_l}},t_{g_{a_l:d}}) \right)^{n_{ia_l}c_i}.
  \label{eq:prepcrpgf}
\end{equation}

\subsection{Including Drop-in}

In addition to the extracted alleles, there is a possibility that
spurious \textit{drop-in} alleles may get into the minitube in which
the aliquot is ready for the \PCR process.  We follow
\cite{puch2014dropin} and assume that such drop-in events occur by a
Poisson process. Denote the drop-in rate for allele $a_l \in A_l$ by
$\lambda_{a_l}$. Then the \PGF for the total number of genome strand
pairs that drop-in is given by

\begin{equation}
  D_r(\ t_{g_{a_l}},t_{g_{a_l:d}}) =  \exp\left( \lambda_{a_l} (  t_{g_{a_l}}t_{g_{a_l:d}}-1)\right)
  \label{eq:genomicdropin}
\end{equation}

Drop-ins occur independently for each locus, and independently of the
genotypes of the $I$ individuals. Hence the \PGF for the genomic
strands from drop-in and contributors prior to amplification is
\begin{equation}
  \prod_{l\in  L} \prod_{a \in A_l} \prod_{i \in I}(1 -\pi_f  + \pi_f E_x( t_{g_{a_l}},t_{g_{a_l:d}}) )^{n_{ia_l}c_i}
  D_r(\ t_{g_{a_l}},t_{g_{a_l:d}})
  \label{eq:dropinandK}
\end{equation}

Within the genomic strand framework presented here, there is an
alternative model for drop-in, in which amplicon-pair strands, rather
than genomic-pair strands, fall into the minitube that the \PCR is
carried out in. One reason amplicons could drop-in 
is that there is a build-up over time of
contaminating amplicons from the previous \PCR analyses carried out in
 the forensic
laboratory. This
possibility is supported by the observation that there are lower
levels of drop-in in samples amplified after a laboratory has been
deep-cleaned, compared to just before a deep-cleaning has been carried
out, and that as time goes on dropin-rates increase until the
laboratory is cleaned again\footnote{Sue Pope, personal
  communication.}

To model this alternative, we use a Poisson drop-in model for the
number of amplicons that drop-in, given by the \PGF
\begin{equation}
  D_r(\ t_{a_{a_l}},t_{a_{a_l:d}})= \exp\left( \lambda_{a_l} (  t_{a_{a_l}}t_{a_{a_l:d}}-1)\right),
  \label{eq:amplicondropin}
\end{equation}
instead of the genomic strand drop-in model \eqref{eq:genomicdropin}.
Hence the \PGF for the genomic strands and amplicons prior to
amplification is now
\begin{equation}
  \prod_{l\in  L} \prod_{a \in A_l} \prod_{i \in I}(1 -\pi_f  + \pi_f E_x( t_{g_{a_l}},t_{g_{a_l:d}}) )^{n_{ia_l}c_i}
  D_r(\ t_{a_{a_l}},t_{a_{a_l:d}}).
  \label{eq:dropinandK2}
\end{equation}

One could even consider both Poisson processes occurring
simultaneously so that we allow drop-in of both genomic strands and
amplicon strands, leading to

\begin{equation}
  \prod_{l\in  L} \prod_{a \in A_l} \prod_{i \in I}(1 -\pi_f  + \pi_f E_x( t_{g_{a_l}},t_{g_{a_l:d}}) )^{n_{ia_l}c_i}
  D_r(\ t_{g_{a_l}},t_{g_{a_l:d}}) D_r(\ t_{a_{a_l}},t_{a_{a_l:d}}).
  \label{eq:dropinandK3}
\end{equation}
in which we can allow a different drop-in rate $\lambda_{a_l} $ for
the two processes.

\subsection{\PCR amplification}
During the \PCR process we may generate strands of type
$g_{a_l}, g_{a_ld}, h_{a_l}, h_{a_ld}, a_{a_l}$ and $ a_{a_ld}$.  For
any of these strands to make a successful copy (or copy of some
stutter type) a primer has to bind at the primer binding site. The
probability of this happening will depend on various factors in how
the \PCR is carried out (e.g., the kit, temperature etc.). It could
also depend on the type of strand considered. For example the $g$
strands could be quite long and coil up to create a barrier preventing
a primer from binding. It also depends on the binding energy of the
primer to the strand, and each end of the flanking region generally
has different base-pair composition, so we should expect the binding
energies to be different. We thus introduce primer binding
probabilities specific to each complementary-strand
$$p_{l,g}, p_{l,gd}, p_{l,h}, p_{l,hd}, p_{l,a} \mbox{ and } p_{l,ad}.$$
Binding a primer is a pre-requisite for a successful copy. However
during the remainder of the duplication process a copying error could
occur, leading to a stutter variant. The probability that this
happens, and the resulting stutter artefact, will depend on the
initial allele $a_l$ and the resulting allele $b_l$.  We thus
introduce $\xi_{b_l\cd a_l}$ to denote the conditional probability
that a strand of allele type $a_l$ creates strand of allele type $b_l$
given that some product is produced We have that for all $a_l\in A_l$,
$$\sum_{b_l \in A_l} \xi_{b_l\cd a_l} = 1$$

Note that we have the conditional probability for copying without
stuttering, given some sort of copy if made, is
$$
\xi_{a_l\cd a_l} = 1 - \sum_{b_l \in A_l: b_l \ne a_l} \xi_{b_l\cd
  a_l}.
$$
We shall assume that these probabilities remain constant throughout
all of the cycles of the \PCR process, though this assumption could be
relaxed.

The set of {\PGF}s for single-strand single-cycle duplication is then
given by
\begin{subequations}
  \begin{align}
    g_{a_l} &: (1 - p_{l,g}) t_{g_{a_l}} +   p_{l,g}\sum_{b_l \in A_l}\xi_{b_l\cd a_l} t_{g_{a_l}}t_{h_{a_ld}}\\
    g_{a_ld} &: (1 - p_{l,gd}) t_{g_{a_l}} +  p_{l,gd}\sum_{b_l \in A_l} \xi_{b_l\cd a_l} t_{g_{a_ld}}t_{h_{a_l}}\\
    h_{a_l} &: (1 - p_{l,h}) t_{h_{a_l}} +  p_{l,h} \sum_{b_l \in A_l}\xi_{b_l\cd a_l} t_{h_{a_l}}t_{a_{a_ld}}\\
    h_{a_ld} &: (1 - p_{l,hd}) t_{h_{a_ld}} +  p_{l,hd}\sum_{b_l \in A_l} \xi_{b_l\cd a_l} t_{h_{a_ld}}t_{a_{a_l}}\\
    a_{a_l} &: (1 - p_{l,a}) t_{a_{a_l}} +  p_{l,a} \sum_{b_l \in A_l}\xi_{b_l\cd a_l} t_{a_{a_l}}t_{a_{a_ld}}\\
    a_{a_ld} &: (1 - p_{l,ad}) t_{a_{a_ld}} +   p_{l,ad}\sum_{b_l \in A_l}\xi_{b_l\cd a_l} t_{a_{a_ld}}t_{a_{a_l}}
  \end{align}
\end{subequations}
These single strand {\PGF}s are `lifted' to the full vectorial \PGF of
the \PCR process, as follows. Let $F_n(\mathbf{t}_l\cd s )$ denote the
multivariate \PGF for all the possible types of strands that are
generated from a single strand of types
$$s \in \{ g_{a_l}, g_{a_ld}, h_{a_l}, h_{a_ld}, a_{a_l},a_{a_ld}: a_l \in A_l\}.$$ The $\mathbf{t}_l$ represents all the possible symbols for all the possible alleles and their strand types (there will be $6 \cd A_l\cd $ such symbols). Then the recurrence relations for the \PCR branching process have the form:

\begin{subequations}
  \begin{align}
    F_{n+1}(\mathbf{t}_l\cd g_{a_l})&= (1 - p_{l,g}) F_{n}(\mathbf{t}_l\cd g_{a_l} ) + 
                                      p_{l,g}\sum_{b_l \in A_l}\xi_{b_l\cd a_l}F_{n}(\mathbf{t}_l\cd g_{a_l} ) F_{n}(\mathbf{t}_l\cd h_{a_ld})\\
    F_{n+1}(\mathbf{t}_l\cd g_{a_ld}) &=(1 - p_{l,gd}) F_{n}(\mathbf{t}_l\cd g_{a_ld}) +  
                                        p_{l,gd}\sum_{b_l \in A_l} \xi_{b_l\cd a_l} F_{n}(\mathbf{t}_l\cd g_{a_ld}) F_{n}(\mathbf{t}\_l\cd  h_{a_l})\\
    F_{n+1}(\mathbf{t}_l\cd h_{a_l}) &= (1 - p_{l,h}) F_{n}(\mathbf{t}_l\cd h_{a_l}) +  
                                       p_{l,h}\sum_{b_l \in A_l}\xi_{b_l\cd a_l} F_{n}(\mathbf{t}_l\cd h_{a_l}) F_{n}(\mathbf{t}_l\cd a_{a_ld}) \\
    F_{n+1}(\mathbf{t}_l\cd h_{a_ld}) &= (1 - p_{l,hd}) F_{n}(\mathbf{t}_l\cd h_{a_ld}) +  
                                        p_{l,hd}\sum_{b_l \in A_l} \xi_{b_l\cd a_l}F_{n}(\mathbf{t}_l\cd h_{a_ld}) F_{n}(\mathbf{t}_l\cd h_{a_l}) \\
    F_{n+1}(\mathbf{t}_l\cd a_{a_l}) &= (1 - p_{l,a}) F_{n}(\mathbf{t}_l\cd a_{a_l}) +  
                                       p_{l,a}\sum_{b_l \in A_l} \xi_{b_l\cd a_l} F_{n}(\mathbf{t}_l\cd a_{a_l}) F_{n}(\mathbf{t}_l\cd a_{a_ld}) \\
    F_{n+1}(\mathbf{t}_l\cd a_{a_ld}) &= (1 - p_{l,ad})F_{n}(\mathbf{t}_l\cd a_{a_ld}) +  
                                        p_{l,ad}\sum_{b_l \in A_l} \xi_{b_l\cd a_l} F_{n}(\mathbf{t}_l\cd a_{a_ld}) F_{n}(\mathbf{t}_l\cd a_{a_l})
  \end{align}
\end{subequations}
with initial conditions
\begin{subequations}
  \begin{align}
    F_{0}(\mathbf{t}_l\cd g_{a_l})=& \,t_{g_{a_l}} \\
    F_{0}(\mathbf{t}_l\cd g_{a_ld}) =&\,t_{g_{a_ld}}\\
    F_{0}(\mathbf{t}_l\cd h_{a_l}) =& \,t_{h_{a_l}}\\
    F_{0}(\mathbf{t}_l\cd h_{a_ld}) =& \,t_{h_{a_ld} }\\
    F_{0}(\mathbf{t}_l\cd a_{a_l}) =&\, t_{a_{a_l}}\\ 
    F_{0}(\mathbf{t}_l\cd a_{a_ld}) =& \,t_{a_{a_ld} }
  \end{align}
\end{subequations}

The joint \PGF of all types of strand products after $n$ cycles
generated from a single initial pair strand $(g_{a_l},g_{a_ld})$ is
\begin{equation}
  F_{n}(\mathbf{t}_l\cd g_{a_l} ) F_{n}(\mathbf{t}_l\cd g_{a_ld} ) \label{eq:singleGGD}
\end{equation}

\subsection{The joint \PGF after \PCR}

If using the genomic drop-in model \eqref{eq:genomicdropin}, one now
substitutes $t_{g_{a_l}}\to F_{n}(\mathbf{t}_l\cd g_{a_l} ) $ and
$t_{g_{a_ld}}\to F_{n}(\mathbf{t}_l\cd g_{a_ld} ) $ into
\eqref{eq:dropinandK} to obtain the joint \PGF of all types of \PCR
product:

\begin{equation}
  \prod_{l\in  L} \prod_{a \in A_l} \prod_{i \in I}
  (1 -\pi_f  + \pi_f E_x(  F_{n}(\mathbf{t}_l\cd g_{a_l} )  ,F_{n}(\mathbf{t}_l\cd g_{a_ld} ) ) )^{n_{ia_l}c_i}
  D_r( F_{n}(\mathbf{t}_l\cd g_{a_l} )F_{n}(\mathbf{t}_l\cd g_{a_ld} ) )
  \label{eq:precepgf}
\end{equation}

Alternatively, if using the amplicon drop-in model
\eqref{eq:amplicondropin}, one substitutes
$t_{g_{a_l}}\to F_{n}(\mathbf{t}_l\cd g_{a_l} ) $ and
$t_{g_{a_ld}}\to F_{n}(\mathbf{t}_l\cd g_{a_ld} ) $ into
\eqref{eq:dropinandK2} together with the additional substitutions of
$t_{a_{a_l}}\to F_{n}(\mathbf{t}_l\cd a_{a_l} ) $ and
$t_{a_{a_ld}}\to F_{n}(\mathbf{t}_l\cd a_{a_ld} ) $ :

\begin{equation}
  \prod_{l\in  L} \prod_{a \in A_l} \prod_{i \in I}
  (1 -\pi_f  + \pi_f E_x(  F_{n}(\mathbf{t}_l\cd g_{a_l} )  ,F_{n}(\mathbf{t}_l\cd g_{a_ld} ) ) )^{n_{ia_l}c_i}
  D_r( F_{n}(\mathbf{t}_l\cd a_{a_l} )F_{n}(\mathbf{t}_l\cd a_{a_ld} ) )
  \label{eq:precepgfb}
\end{equation}

Depending on the drop-in model chosen, either \eqref{eq:precepgf} or
\eqref{eq:precepgfb} is the multivariate \PGF for all the possible
\PCR products. For the capillary electrophoresis we require only the
multivariate \PGF for the tagged amplicons. This is obtained by
setting all of the $\mathbf{t}_l$ components to 1 except for the
symbols representing the tagged amplicons, that is, the set
$\{t_{a_{a_ld} }: l\in L, a_l\in A_l\}$.  Let $\mathbf{t}_{ld}$ denote
$\mathbf{t}_{l}$ with this substitution made. Then \eqref{eq:precepgf}
with this substitution becomes
\begin{equation}
  \prod_{l\in  L} \prod_{a \in A_l} \prod_{i \in I}
  (1 -\pi_f  + \pi_f E_x(  F_{n}(\mathbf{t}_l\cd g_{a_l} )  ,F_{n}(\mathbf{t}_{ld}\cd g_{a_ld} ) ) )^{n_{ia_l}c_i}
  D_r( F_{n}(\mathbf{t}_{ld}\cd g_{a_l} )F_{n}(\mathbf{t}_l\cd g_{a_ld} ) )
  \label{eq:precepgf2}
\end{equation}
with a similar equation for substitution into
\eqref{eq:precepgfb}. Hence, taking into account the two variations of
$E_x(\cdot, \cdot)$ given earlier, we have four modelling
possibilities.

\subsection{RFU scaling and baseline noise}
During the capillary electrophoresis phase, a thin tube is dipped into
the amplified product and a high voltage is applied. This forces a
fraction of the product into the tube where it is carried along by a
voltage differential. At a specific point along the capillary tube a
lasers excite the dyes on the tagged amplicons, and the amount that
fluoresces is recorded as the RFU reading. The RFU value is
proportional to the number of tagged amplicons.

There is, therefore, some scale factor which we denote by $\rho$, that
relates the final number of tagged amplicons of each type to the RFU
reading. The factor will depend upon the machinery, but may also be
expected to be dye dependent with all the other factors constant. Thus
if we consider a single locus $l$, a given peak height RFU value
$r_{a_l}>0$ will correspond to the range
$[\rho(r_{a_l}-1/2), \,\rho(r_{a_l}+1/2)]$ of tagged amplicons of that
allele, and conversely.

However this does not take account the baseline noise that is usually
present. If we denote the discrete probability distribution of the
baseline noise distribution by $\eta_l$ for the range $[0,w]$, say,
then we may form the \PGF of the distribution as
$$
\sum_{j=0}^w \eta_l[j]z^j
$$
Given the locus, the noise distribution will be independent of the
allele type (it may depend on the dye lane that the locus is in). Thus
for each $a_l\in A_l$ we may form a``tagged-amplicon equivalent''
noise distribution given by
$$
\eta_{l,a_l} := \sum_{j=0}^w \eta_l[j] t_{a_{a_ld}}^{j\rho}
$$
These {\PGF}s may then be multiplied into \eqref{eq:precepgf2} and the
resulting multivariate \PGF used for assessing peak height likelihoods
for observed data.

Alternatively one could derive peak height distributions from
\eqref{eq:precepgf} and convolute them with the baseline noise
distributions, and use the convolved distribution to evaluate peak
height likelihoods.

\subsection{Multiple replicates from a sample}

Sometimes more than one \PCR amplification is carried out on a sample
from which DNA has been extracted, each amplified sub-sample is called
a \textit{replicate}.  This maybe modelled using a multinomial \PGF as
follows. Let there be $R$ replicates, with $\pi_{f_r}$ the fraction of
the sample used for replicate $r$.

If we ignore breakage through degradation, then for a single genomic
pair strand $g_{a_l}g_{a_ld}$ in the sample, the multivariate \PGF for
the number of strands (0 or 1) of the pair being in each replicate is
$$
1 - \sum_{r=1}^R\pi_{f_r} + \sum_{r=1}^R\pi_{f_r}
t_{g_{a_l}:r}t_{g_{a_ld}:r}
$$
where we introduce new symbols to represent the specific replicate the
pair get selected for, indicated by the additional `$:r$' subscripts.
New symbols for half-strand and amplicons are also required to specify
the replicate they belong to.

To take into account the extraction efficiency, we multiply each
$\pi_{f_r} $ by $\pi_{e:i,a_l}$:
\begin{equation}
  \left(1 - \pi_{e:i,a_l}\sum_{r=1}^R\pi_{f_r}\right)  + \pi_{e:i,a_l}\sum_{r=1}^R\pi_{f_r} t_{g_{a_l}:r}t_{d_{a_ld}:r}\, . 
  \label{eq:replicate}
\end{equation}

To take account of degradation breakage, replace
$t_{g_{a_l}:r}t_{h_{a_ld}:r}$ by $E_x( t_{g_{a_l}:r},t_{g_{a_ld}:r})$
in \eqref{eq:replicate}:

\begin{equation}
  \left(1 - \pi_{e:i,a_l}\sum_{r=1}^R\pi_{f_r}\right)  + \pi_{e:i,a_l}\sum_{r=1}^R\pi_{f_r}E_x( t_{g_{a_l}:r},t_{g_{a_ld}:r})\, . 
  \label{eq:replicate2}
\end{equation}
so that for person $K_i$,
\begin{equation}
  \left(\left(1 - \pi_{e:i,a_l}\sum_{r=1}^R\pi_{f_r}\right)  + \pi_{e:i,a_l}\sum_{r=1}^R\pi_{f_r}E_x( t_{g_{a_l}:r},t_{g_{a_ld}:r})\right)^{n_{ia_l}c_i}\, . 
  \label{eq:replicate3}
\end{equation}

\subsection{Untyped contributors}

Equation \eqref{eq:kiknown} assumes that the genotypes of all
contributors are known. For a contributor $U_j$ whose genotype is not
known, we replace \eqref{eq:kiknown}, for autosomal loci, by the \PGF
\begin{equation}
  \left( \sum_{a_l\in A_l} p_{a_l}\,(t_{g_{a_l}}t_{g_{a_l:d}})^{c_j}\right )^2\label{eq:unknown}
\end{equation}
where $p_{a_l}$ is the allele frequency of allele $a_l$ in the locus
$l$ for the population that $U_j$ comes from.  For a set of $J$
\textit{unrelated} untyped individuals we have
\begin{equation}
  \prod_{j=1}^J \left( \sum_{a_l\in A_l} p_{a_l}\,(t_{g_{a_l}}t_{g_{a_l:d}})^{c_j}\right )^2.\label{eq:unknownm}
\end{equation}

This represents intact genomic pair strands in the sample prior to
extraction and without degradation or splitting a fraction into
$r \ge 1$ replicates. To take all these into consideration we replace
the $t_{g_{a_l}}t_{g_{a_l:d}}$ product pairs thus:

\begin{equation}
  \prod_{j=1}^J \left( \sum_{a_l\in A_l} p_{a_l}\,
    \left(\left(1 - \pi_{e:i,a_l}\sum_{r=1}^R\pi_{f_r}\right)  + \pi_{e:i,a_l}\sum_{r=1}^R\pi_{f_r}E_x( t_{g_{a_l}:r},t_{g_{a_ld}:r})\right)
    ^{c_j}\right )^2.\label{eq:unknownmr}
\end{equation}

\subsection{Another variation of the framework}

\cite{grisedale2014method} proposed a method for improving the
detection of alleles as described in the Method Summary of their
paper:
\begin{quote}
  DNA template is first divided into two aliquots. One aliquot is used
  as template for a \PCR using a primer mix containing all forward
  primers for loci targeted in the PowerPlex ESI 16 kit (Promega),
  while the other aliquot is amplified with all reverse
  primers. Amplification products are then pooled for use as template
  in a standard \PCR with the STR kit primer mix. The forward and
  reverse primer reactions result in a linear amplification of the
  target sequences to boost the amount of template available for \PCR,
  thus reducing the stochastic effects commonly seen with low template
  DNA analysis.
\end{quote}

The framework presented so far does not apply to their experimental
set-up; however it is readily adapted to it as follows.  Assume that
the DNA template is split into two equal aliquot parts for the first
$\kappa$ pre-cycles, and then put back together for $n$ amplifications
cycles.  Consider a single genome-pair $(g ,g_d)$ (dropping the $a_l$
indexing from earlier notation for simplicity) in the aliquot prior to
splitting the sample.  Let $S$ and $S_d$ denote the two split samples.
If the sample is split into two equal volumes, then the $(g,g_d)$ pair
will be in $S$ with probability 0.5, and $S_d$ with probability 0.5.
Suppose that the sample $S$ has primers that bind to the $g$ strand;
then the number of $h_d$ strand generated by the $\kappa$ pre-cycles
will be binomially distributed as $\mbox{Bin}(\kappa, p_g)$ if the
stand pair is in $S$ This is because the $g$ strand produces at most
one $h_d$ strand per cycle. Similarly the number of $h$ strands
generated in $S_d$ in the $\kappa$ pre-cycles will be binomially
distributed as $\mbox{Bin}(\kappa, p_{g_d})$ if the strand pair is in
$S_d$.

Hence when the two samples are re-combined for the main $n$ \PCR
cycles that generate the amplicons, the \PGF for the starting number
of $g, g_d, h$ and $h_d$ strands will be

$$ t_g t_{g_d} \frac{(1+t_h)^\kappa + (1+t_{h_d})^\kappa}{2}$$

One then replaces $t_g\to F_n(\mathbf{t}\cd t_g)$, and so on for the
joint \PGF after the $n$ cycles of \PCR. Other factors such as an
unequal splitting of the DNA template, extraction efficiency,
degradation, multiple replicates, number of cells and genotypes of
contributors, and modification to the drop-in model, can be taken into
account in a straightforward manner. For example, for the genomic
drop-in model we make an adjustment from
  $$D_r(\ t_{g},t_{g_{d}}) =  \exp\left( \lambda (  t_g t_{g_d}-1)\right)$$
  to
  $$D_r(\ t_{g},t_{g_{d}}) =  \exp\left( \lambda\left(  t_{g}t_{g_{d}} \frac{(1+t_h)^\kappa + (1+t_{h_d})^\kappa}{2}-1\right)\right).$$

  \section{A particular model realisation}
  \label{sec:particularmodel}

  In this section we present a model that specialises the framework
  given earlier, and which introduces approximations that make the
  model computationally tractable. A satisfying feature of this
  approach is that the mathematical nature of the model approximations
  are clearly specified, and they can be judged on their merits. The
  author has implemented the model in a computationally efficient and
  accurate system, and this implementation is used in the performance
  analysis of the model applied to real and simulated data presented
  below.

\subsection{Model assumptions and approximations}
In the model presented we shall assume that we have a single DNA
sample from which a single replicate has been produced.  First the
sample is described, then the assumptions regarding the \PCR process
are given, and how the tagged amplicon number is interpreted as \
measurement in terms of \RFU units. We shall assume that the kit being
used has autosomal loci, and perhaps also Amelogenin but no other
sex-linked loci are among the loci of the kit. We also assume that all
contributors are unrelated.

\subsubsection{The sample}

We shall assume that we have a mini-tube of volume $V_s$, that
contains the extracted DNA of $I$ contributors, in which the $i$th
contributor has contributed $c_i$ cells.  The DNA has been extracted
with an efficiency $\psi \in [0,1]$, and may have degradation
characterised by a parameter $\delta$ which has units of inverse
length (measured in base pairs).  The values of the cell amounts $c_i$
and the degradation parameter are taken as unknown, and to be
estimated from the \EPG.  The extraction efficiency is taken as known.

A very small amount of the sample in $V_s$ is taken to quantify the
concentration of DNA in $V_s$, and based on this a fraction fraction
$\pi_f$ of the volume $V_s$ is taken is put into a mini-tube of volume
$V$, to which is added primers \etc. \PCR is carried out with this
aliquot.

Note that quantification of DNA is usually carried out using qPCR, so
that the amount of DNA estimated to be in $V$ is the amplify-able
amount of DNA. If the estimated amount of DNA in $V$ is $\gamma$, then
an estimate of the amount of DNA in the sample volume $V_s$ is given
by $\gamma_s = \gamma/(\psi\pi_f)$, if we ignore degradation.

We shall assume that degradation has the effect of breaking a pair
strand in two or more pieces, according to a Poisson process.  For a
given locus $l \in L$, the total number of genomic pair strands of an
allele $a_l \in A_L$, having total size in base-pairs between and
including flanking regions denoted by $b_l$, that are in the aliquot
volume $V$ and are amplify-able is given by the binomial distribution,
 taken to be
$$ \mbox{Binom}\left( \sum _{i=1}^I n_{ia_l}c_i, \psi\pi_f \exp(-\delta b_l)\right).$$

We shall assume the genomic drop-in model, with a Poisson model for
the number of drop-in genome strands with drop-rates for every allele
assumed known. We ignore stutter products that may be generated during
the \PCR process by drop-in strands.

\subsubsection{The \PCR}
We shall use the genomic model described earlier. We assume that all
amplification probabilities, conditional and unconditional, are known
for all allele and loci.

We shall assume that in an amplification cycle, forward stutter,
no-stutter, single reverse stutter and double reverse stutter products
may be \textit{directly} produced from a strand of type of the allele
$a_l$. Note that we allow triple stutters arising indirectly via two
amplification cycles. Some alleles might not form one or more of these
stutter products, for example because they are at the low or high end
of the allelic range of $A_l$; additionally Amelogenin is assumed not
to form stutters.

We only allow stuttering for multiples of the base-pair repeat size of
each locus. Thus while we include a stutter of $9.3\to 8.3$, for TH01,
we exclude the stutter possibility $9.3\to 9$; this is purely for
computational efficiency, to make the likelihood evaluation tractable.

\subsubsection{Converting amplicon numbers to peak height {\RFU}s}

At the conclusion of the \PCR, the volume $V$ will have a large number
of tagged amplicons for the various alleles in the amplification
kit. A small number of these are drawn up electrostatically into the
capillary-electrophoresis machine. We shall assume that the number of
tagged amplicons of a specific allelic type drawn up is proportional
to their number (the proportionality depends on other factors such as
the voltage applied; see \cite{butler2011advancedmethod} pp.144-145),
and that the \RFU peak height is proportional to the number drawn up,
and that the proportionality constant may depend upon the particular
dye.  Hence we introduce dye-lane proportionality factors so that the
\RFU peak height generated by the tagged amplicons is proportional to
the total number in the volume $V$.

To this must be added a random value arising from the baseline
noise. It is assumed that the prior baseline noise distribution is
known, and may be dye-lane dependent.

\subsubsection{Likelihood evaluation}

In order to evaluate the likelihood function requires evaluation of
the {\PGF}s $F_{n}(\mathbf{t}_{ld}\cd g_{a_l} )$ and
$F_{n}(\mathbf{t}_{ld}\cd g_{a_{ld}} )$, based on the assumed known
amplification probabilities and binomial sampling rates and genome
counts, for all the alleles.  Let us write this as
\begin{equation}
  \prod_{a_l \in A_L} F_{n}(\mathbf{t}_{ld}\cd g_{a_l} )  = 
  \prod_{a_l \in A_L}  \prod_{b_l \in A_L} ( 1 -\phi_{b_{l}} + \phi_{b_{l}} G(t_{b,a_{ld}}))^{k_{b_l}},
  \label{eq:likeexact}
\end{equation}
where $G(t_{b,a_{ld}})$ is the \PGF of the joint distribution of all
allelic products arising from a single genomic strand of type $b_l$,
$\phi_{b_{l}}$ is the binomial sampling probability for such alleles,
of which there are ${k_{b_l}}$ in the sample.

This is computationally intractable for two reasons. The first is in
evaluating the individual {\PGF}s $G(t_{b,a_{ld}})$. We therefore make
the approximation in which each $ G(t_{a_{ld}})$ factorizes as
\begin{equation}
  G(t_{a_{ld}}) =  G_{-2}(t_{a_{ld}}) G_-{1}(t_{a_{ld}}) G_{0}(t_{a_{ld}}) G_{+1}(t_{a_{ld}}) 
  = \prod_{j=-2}^1G_{j}(t_{a_{ld}}) 
  \label{eq:likeapprox1}
\end{equation}
in which $G_{-2}(t_{a_{ld}}) $ is the \PGF of the number of tagged
amplicons in double-stutter position, $G_{-1}(t_{a_{ld}}) $ in stutter
position, $G_{0}(t_{a_{ld}}) $ copies of the target allele, and
$G_{-1}(t_{a_{ld}}) $ forward stutter.

Even with this approximation, the computations are still intractable,
we therefor make the following further approximation:
\begin{equation}
  \left( 1 -\phi_{b_{l}} + \phi_{b_{l}}\prod_{j=-2}^1G_{j}(t_{a_{ld}})  \right)^{k_{b_l}}
  \to
  \prod_{j=-2}^1\left (1 -\phi_{b_{l}} + \phi_{b_{l}}G_{j}(t_{a_{ld}}) \right)^{k_{b_l}}.
  \label{eq:likeapprox2}
\end{equation}
Under these two approximations, the total likelihood, given the
genotypes of contributors, will factorise into a product of functions,
each of which depend on a single allele type.  For a particular allele
$a$ , there will be up to four factors arising from alleles 2 repeats
higher (which double stutter to make $a$, one repeat higher (which
stutter to form $a$) , alleles of the same type $a$, and alleles one
repeat lower (which forward stutter to produce $a$ product).

This needs to be multiplied by {\PGF}s from each allele drop-in \PGF,
and combined with baseline noise distribution on each allele, to get
the marginal peak-height distributions for each allele.  Hence for
each allele we have a product of univariate {\PGF}s that can be
convoluted using the \FFT to find the marginal distribution on each
allele, which can then be used directly to find the likelihoods on
each allele given the peak height.

The peak height likelihood from all peaks is then found from these
individual allele likelihoods by multiplication---given the genotypes
of the contributors---and such products averaged using population
genotype probabilities over the possible genotypes of the untyped
contributors.

\section{Model performance with simulated data}
\label{sec:simdata}
One of the useful features of the framework developed in this paper is
that, because the model is an idealisation of the DNA extraction and
amplification process, not only can realistic simulated single source
and mixtures {\EPG}s be produced, the simulated data can be analysed
by the model using the same parameters as used in simulating the data.
The performance of the model can then be compared to the `gold
standard' simulated values. Additionally, one can tweak the parameters
of the model so that they differ from those used to simulate the data,
so that robustness of the model can be gauged.  It is also possible to
use unrealistic model parameters to isolate and analyse specific
effects, for use with software validation and regression tests.  All
these can be used to judge the limitations of the model. Here we
present a set of simulations of increasing complexity, beginning with
unrealistic single contributor sample simulations.  All the
simulations use population Caucasian data from \cite{butler:etal:03}.
All allele frequencies in a locus were increased so that the minimum
allele count of any was 5. (The total allele count was also increased
to yield population probabilities that add to 1, so this is not quite
the same as the $5/2N$ adjustment of \citep{bodner2016recommendations}
which does not adjust the normalisation). An Fst value of 0.02 was
used in all analyses.  Following \citep{puch2014dropin} we use a
locus-wide drop-in rate of 0.021, with allele specific rates equal to
this multiplied by the relative allele frequency in the
\textit{unadjusted} population counts. Contributor genotypes were
taken from those in the dataset of contributors from the
PROVEDIt Initiative \citep{alfonse2016development,alfonse2018large},
which we shall return to in \secref{sec:budata} when we examine the
model performance on experimental data.

All simulations are based on the Identifiler\texttrademark\ kit, in
which we assume that all strand amplification probabilities are
$p_g = p_{g_d} = p_h = p_{h_d} = p_a = p_{a_d} = 0.85$ for all loci.

\subsection{Single contributor simulations}

\subsubsection{Simulations with no stutter and no noise }
In this set of simulations we take all the conditional stutter
probabilities $\xi$ to be zero on all alleles and loci.  With this
assumption, we will have
$G_{-2}(t_{a_{ld}}) = G_-{1}(t_{a_{ld}}) = G_{+1}(t_{a_{ld}}) = 1$,
and \eqref{eq:likeapprox2} is no longer an approximation, with only
the $j=0$ terms surviving on both sides of the equation to give an
equality. We are thus, in these simulations, testing the performance
of the model under the approximation of \eqref{eq:likeapprox1}.

We shall simulate data the genotype data of subject
\texttt{RD14-0003-01} taken from the PROVEDit dataset, and assume that
the profile is known when estimating cell counts.  We take as further
idealisations that there is no drop-in (the drop-in rate is zero), and
that there is no baseline noise.  We use a factor of 800,000 to scale
post-\PCR tagged-amplicon numbers to \RFU\ values, and 28 cycles for
the \PCR amplification.

We use maximum likelihood to estimate the cell counts of the
contributor, a value known at the time of simulation. We can thus
examine the predictive accuracy and variability of the cell count
estimates.

In the first set of simulations, we shall take the degradation
parameter to be zero, and shall assume this value when estimating the
cell counts.  We compare the predictions made using the model based on
the \FFT analysis and normal, lognormal and gamma distribution models
based on matching the means and variances of the marginal
distribution.  We take the binomial sampling probability $\phi$ to be
the same for all loci and alleles, and see how the models behave as we
vary the number of cells, and the values of $\phi$. We include in the
simulations the unrealistic value $\phi=1$, so that
\eqref{eq:likeapprox1} is exact, and we can compare the (now exact)
\FFT model to the moment approximation models.

The following table shows maximized log-likelihoods and estimated
cells counts from 10 simulations, with 500 cells, $\phi =1$, and an
analytic threshold of 1 \RFU in all simulations; we see that the
estimates are very close to the true value of 500 cells used to
generate the simulations.  We also see that the log-likelihood maxima
are all of a similar value.

\begin{center}
  \begin{tabular}{|cc|cc|cc|cc|}
    \hline
    \multicolumn{2}{|c|}{Normal} & \multicolumn{2}{c|}{Logormal} & \multicolumn{2}{c|}{Gamma} & \multicolumn{2}{c|}{\FFT} \\
    $\widehat{LL}_{max}$ & cells &$\widehat{LL}_{max}$ & cells &$\widehat{LL}_{max}$ & cells &$\widehat{LL}_{max}$ & cells \\ \hline
    -171.356 &499 &-171.341 &499 &-171.345 &499 &-171.367 &499 \\
    -168.651 &499 &-168.658 &499 &-168.655 &499 &-168.648 &499 \\
    -170.734 &500 &-170.757 &500 &-170.749 &500 &-170.719 &500 \\
    -169.804 &499 &-169.802 &499 &-169.802 &499 &-169.806 &499 \\
    -171.148 &500 &-171.176 &500 &-171.166 &500 &-171.129 &500 \\
    -173.965 &500 &-174.017 &500 &-173.999 &500 &-173.931 &500 \\
    -173.689 &498 &-173.770 &498 &-173.743 &498 &-173.635 &498 \\
    -170.764 &499 &-170.665 &499 &-170.697 &499 &-170.830 &499 \\
    -172.017 &500 &-172.036 &500 &-172.029 &500 &-172.008 &500 \\
    -172.890 &499 &-172.851 &499 &-172.863 &499 &-172.919 &499 \\ \hline
  \end{tabular}
\end{center}

Reducing the initial number of cells to 10 , we obtain the following
table from 10 simulations, in which every model estimated the actual
number of cells used. Again the log-likelihood values are very similar
in each simulation across the models.
\begin{center}
  \begin{tabular}{|cc|cc|cc|cc|}
    \hline
    \multicolumn{2}{|c|}{Normal} & \multicolumn{2}{c|}{Logormal} & \multicolumn{2}{c|}{Gamma} & \multicolumn{2}{c|}{\FFT} \\
    $\widehat{LL}_{max}$ & cells &$\widehat{LL}_{max}$ & cells &$\widehat{LL}_{max}$ & cells &$\widehat{LL}_{max}$ & cells \\ \hline
    -110.371 &10 &-110.321 &10 &-110.322 &10 &-110.468 &10 \\
    -106.731 &10 &-106.708 &10 &-106.716 &10 &-106.746 &10 \\
    -116.895 &10 &-116.288 &10 &-116.455 &10 &-117.502 &10 \\
    -105.7 &10 &-105.645 &10 &-105.663 &10 &-105.767 &10 \\
    -107.805 &10 &-107.955 &10 &-107.896 &10 &-107.753 &10 \\
    -111.267 &10 &-111.747 &10 &-111.564 &10 &-111.012 &10 \\
    -112.118 &10 &-113.147 &10 &-112.767 &10 &-111.622 &10 \\
    -110.533 &10 &-111.237 &10 &-110.984 &10 &-110.134 &10 \\
    -106.879 &10 &-106.954 &10 &-106.926 &10 &-106.842 &10 \\
    -119.358 &10 &-120.664 &10 &-120.166 &10 &-118.811 &10 \\
    \hline
  \end{tabular}
\end{center}

We now reduce the $\phi$ value to 0.7.  For 500 cells, we obtain the
following estimates:

\begin{center}
  \begin{tabular}{|cc|cc|cc|cc|}
    \hline
    \multicolumn{2}{|c|}{Normal} & \multicolumn{2}{c|}{Logormal} & \multicolumn{2}{c|}{Gamma} & \multicolumn{2}{c|}{\FFT} \\
    $\widehat{LL}_{max}$ & cells &$\widehat{LL}_{max}$ & cells &$\widehat{LL}_{max}$ & cells &$\widehat{LL}_{max}$ & cells \\ \hline
    -200.822 &495 &-200.725 &495 &-200.749 &495 &-200.853 &495 \\
    -193.161 &502 &-193.307 &502 &-193.256 &502 &-193.13 &502 \\
    -203.948 &500 &-204.114 &500 &-204.047 &500 &-203.927 &500 \\
    -192.587 &504 &-192.632 &504 &-192.616 &504 &-192.576 &504 \\
    -191.827 &498 &-192.032 &498 &-191.962 &498 &-191.773 &497 \\
    -194.707 &500 &-195.243 &500 &-195.057 &500 &-194.598 &499 \\
    -195.202 &498 &-195.243 &498 &-195.227 &498 &-195.194 &498 \\
    -195.921 &496 &-195.927 &496 &-195.921 &496 &-195.924 &496 \\
    -195.392 &498 &-195.356 &498 &-195.365 &498 &-195.4 &498 \\
    -192.545 &498 &-192.6 &498 &-192.58 &498 &-192.532 &498 \\
    \hline
  \end{tabular}
\end{center}

and for 10 cells
\begin{center}
  \begin{tabular}{|cc|cc|cc|cc|}
    \hline
    \multicolumn{2}{|c|}{Normal} & \multicolumn{2}{c|}{Logormal} & \multicolumn{2}{c|}{Gamma} & \multicolumn{2}{c|}{\FFT} \\
    $\widehat{LL}_{max}$ & cells &$\widehat{LL}_{max}$ & cells &$\widehat{LL}_{max}$ & cells &$\widehat{LL}_{max}$ & cells \\ \hline
    -139.783 &11 &-147.468 &10 &-143.755 &11 &-139.270 &11 \\
    -137.910 &10 &-142.610 &10 &-140.283 &10 &-137.284 &10 \\
    -142.794 &10 &-146.055 &9 &-144.733 &9 &-142.067 &10 \\
    -133.883 &9 &-134.231 &10 &-134.058 &10 &-133.530 &9 \\
    -140.499 &10 &-142.933 &10 &-141.646 &10 &-140.285 &10 \\
    -130.732 &10 &-129.973 &10 &-130.279 &10 &-131.016 &10 \\
    -133.887 &10 &-133.595 &10 &-133.623 &10 &-134.086 &10 \\
    -135.397 &9 &-133.938 &9 &-134.267 &9 &-136.617 &9 \\
    -139.761 &10 &-141.720 &10 &-140.478 &10 &-140.018 &10 \\
    -142.768 &10 &-147.142 &9 &-145.116 &9 &-142.328 &10 \\\hline
  \end{tabular}
\end{center}
Again the estimates are good for all models, and similar
log-likelihoods are obtained.  Reducing $\phi$ to 0.07, we obtain for
500 cells

\begin{center}
  \begin{tabular}{|cc|cc|cc|cc|}
    \hline
    \multicolumn{2}{|c|}{Normal} & \multicolumn{2}{c|}{Logormal} & \multicolumn{2}{c|}{Gamma} & \multicolumn{2}{c|}{\FFT} \\
    $\widehat{LL}_{max}$ & cells &$\widehat{LL}_{max}$ & cells &$\widehat{LL}_{max}$ & cells &$\widehat{LL}_{max}$ & cells \\ \hline
    -177.336 &495 &-177.035 &495 &-177.027 &495 &-177.141 &495 \\
    -175.239 &477 &-174.969 &479 &-175.003 &479 &-175.107 &478 \\
    -176.421 &497 &-176.359 &498 &-176.295 &498 &-176.324 &497 \\
    -171.300 &460 &-171.109 &466 &-171.179 &464 &-171.251 &462 \\
    -176.719 &537 &-178.073 &538 &-177.447 &538 &-176.984 &538 \\
    -184.978 &501 &-187.879 &489 &-186.498 &494 &-185.504 &498 \\
    -177.330 &523 &-178.051 &524 &-177.673 &524 &-177.434 &523 \\
    -183.672 &510 &-185.782 &501 &-184.699 &504 &-183.973 &507 \\
    -171.51 &523 &-171.445 &532 &-171.511 &529 &-171.532 &526 \\
    -176.99 &516 &-175.459 &518 &-175.868 &517 &-176.388 &516 \\ \hline
  \end{tabular}
\end{center}

and for 10 cells
\begin{center}
  \begin{tabular}{|cc|cc|cc|cc|}
    \hline
    \multicolumn{2}{|c|}{Normal} & \multicolumn{2}{c|}{Logormal} & \multicolumn{2}{c|}{Gamma} & \multicolumn{2}{c|}{\FFT} \\
    $\widehat{LL}_{max}$ & cells &$\widehat{LL}_{max}$ & cells &$\widehat{LL}_{max}$ & cells &$\widehat{LL}_{max}$ & cells \\ \hline
    -90.0015 &12 &-112.940 &4 &-96.5969 &6 &-74.8928 &10 \\
    -90.1253 &13 &-112.852 &4 &-96.6292 &7 &-77.1458 &11 \\
    -78.6269 &8 &-96.6019 &4 &-83.227 &5 &-63.8753 &8 \\
    -88.6604 &12 &-108.518 &4 &-93.0871 &7 &-73.4382 &10 \\
    -80.5898 &11 &-93.9211 &3 &-80.5046 &4 &-66.668 &9 \\
    -86.3204 &9 &-107.53 &5 &-93.3741 &7 &-81.5557 &9 \\
    -69.2068 &8 &-80.6773 &2 &-68.6397 &3 &-52.4727 &7 \\
    -77.0180 &9 &-92.9245 &3 &-79.659 &5 &-61.3999 &8 \\
    -81.5496 &8 &-108.38 &4 &-92.4763 &6 &-74.1947 &9 \\
    -91.0582 &11 &-106.786 &4 &-92.8809 &6 &-78.1598 &10 \\ \hline
  \end{tabular}
  \label{page:cell10tab}
\end{center}

We see that variability in the cell estimates is increasing, and that
for the 500-cell simulations the log-likelihoods are close amongst the
four models in each simulation. However for the 10-cell simulations we
see that the \FFT based model has higher maximized likelihoods in all
the simulations, indicating a better overall fit to the simulated peak
heights. We can also see that the lognormal and gamma models appear to
be underestimating the true cell-count value, whilst the normal and
\FFT models are less biased in their estimates, but the normal model
having a greater variability. The following boxplot of cell counts
estimated for 10 cells, and $\phi = 0.07$, is based on 200
simulations, and confirms the small sample behaviour in the table
above:

\begin{center}
  \includegraphics[scale=0.8 ]{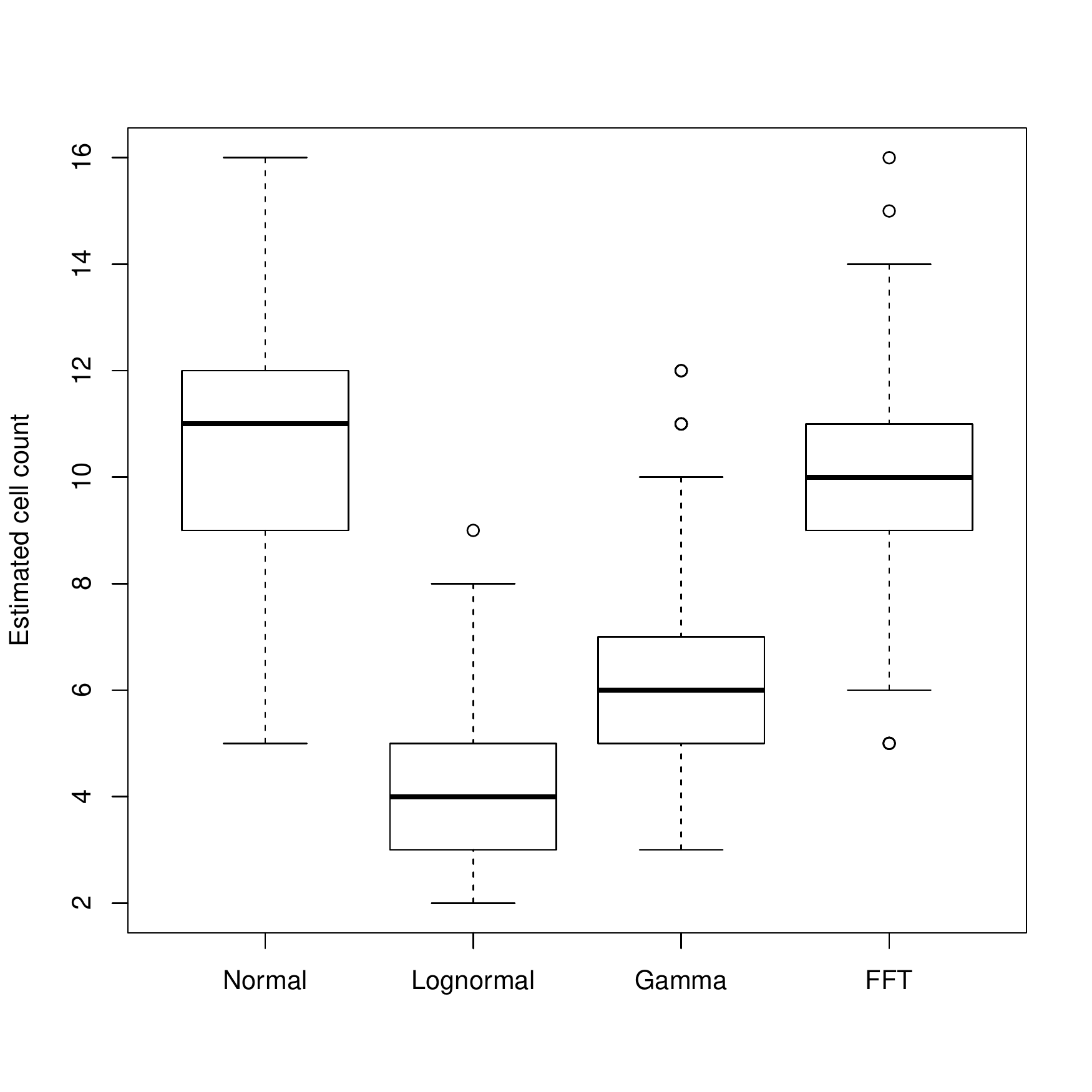}
\end{center}

\subsubsection{Simulations with stutter but no noise }

We now include stutter products in the simulations. For all
simulations we set the conditional probability of stutter on each
allele to be 0.004, and for forward and double stutter conditional
probabilities are all set to 0.001. We do not include background
noise, but we set the analytic threshold to be 30 \RFU.

For $\phi = 1$ and 500 cells, we obtain the following estimates from
10 simulations. We see that all models give similar cell count
estimates and log likelihoods. However we see that there is much more
variation amongst the maximized likelihoods between the four models in
each simulation.
\begin{center}
  \begin{tabular}{|cc|cc|cc|cc|}
    \hline
    \multicolumn{2}{|c|}{Normal} & \multicolumn{2}{c|}{Logormal} & \multicolumn{2}{c|}{Gamma} & \multicolumn{2}{c|}{\FFT} \\
    $\widehat{LL}_{max}$ & cells &$\widehat{LL}_{max}$ & cells &$\widehat{LL}_{max}$ & cells &$\widehat{LL}_{max}$ & cells \\ \hline
    -398.925 &500 &-397.548 &500 &-397.95 &500 &-400.347 &501 \\
    -401.835 &500 &-400.62 &500 &-400.966 &500 &-400.65 &500 \\
    -390.493 &501 &-389.57 &501 &-389.845 &501 &-396.683 &502 \\
    -401.541 &500 &-401.002 &500 &-401.14 &500 &-403.778 &500 \\
    -399.756 &500 &-399.765 &500 &-399.731 &500 &-401.508 &500 \\
    -390.409 &499 &-390.135 &499 &-390.196 &499 &-397.262 &500 \\
    -403.813 &500 &-402.911 &500 &-403.161 &500 &-400.463 &501 \\
    -394.292 &499 &-393.968 &499 &-394.046 &499 &-399.208 &500 \\
    -393.24 &498 &-392.758 &498 &-392.896 &498 &-396.412 &499 \\
    -392.309 &500 &-392.337 &500 &-392.311 &500 &-401.011 &500 \\
    \hline
  \end{tabular}
\end{center}

Reducing the number of cells to 200 we obtain the following table. We
see that all models are making accurate estimates of the numbers of
cells, but we also see that the \FFT model likelihoods are, with one
simulation exception, lower than the moment based models.
\begin{center}
  \begin{tabular}{|cc|cc|cc|cc|}
    \hline
    \multicolumn{2}{|c|}{Normal} & \multicolumn{2}{c|}{Logormal} & \multicolumn{2}{c|}{Gamma} & \multicolumn{2}{c|}{\FFT} \\
    $\widehat{LL}_{max}$ & cells &$\widehat{LL}_{max}$ & cells &$\widehat{LL}_{max}$ & cells &$\widehat{LL}_{max}$ & cells \\ \hline
    -344.81 &200 &-343.656 &200 &-344.009 &200 &-354.253 &200 \\
    -349.42 &199 &-346.64 &199 &-347.456 &199 &-353.245 &200 \\
    -350.053 &199 &-348.586 &199 &-348.993 &199 &-354.592 &199 \\
    -365.401 &200 &-364.874 &200 &-364.998 &200 &-372.019 &200 \\
    -343.929 &200 &-342.793 &200 &-343.114 &200 &-351.199 &200 \\
    -369.769 &200 &-366.598 &200 &-367.482 &200 &-366.688 &200 \\
    -342.514 &200 &-341.61 &200 &-341.877 &200 &-350.469 &200 \\
    -335.855 &201 &-335.204 &201 &-335.411 &201 &-347.73 &201 \\
    -344.183 &200 &-343.186 &200 &-343.482 &200 &-349.88 &200 \\
    -358.427 &199 &-357.657 &199 &-357.856 &199 &-363.16 &200 \\
    \hline
  \end{tabular}
\end{center}

Reducing the number of cells down to 100 we obtain the following
table.  Cell estimates are again good for all models, but we see that
divergence between the \FFT log-likelihood values and the moment
models is becoming more pronounced.

\begin{center}
  \begin{tabular}{|cc|cc|cc|cc|}
    \hline
    \multicolumn{2}{|c|}{Normal} & \multicolumn{2}{c|}{Logormal} & \multicolumn{2}{c|}{Gamma} & \multicolumn{2}{c|}{\FFT} \\
    $\widehat{LL}_{max}$ & cells &$\widehat{LL}_{max}$ & cells &$\widehat{LL}_{max}$ & cells &$\widehat{LL}_{max}$ & cells \\ \hline
    -233.411 &100 &-233.38 &100 &-233.384 &100 &-240.193 &100 \\
    -236.636 &101 &-235.228 &101 &-235.645 &101 &-235.647 &101 \\
    -237.502 &100 &-236.287 &100 &-236.657 &100 &-240.525 &100 \\
    -234.364 &99 &-234.535 &99 &-234.442 &99 &-236.879 &99 \\
    -223.076 &100 &-223.137 &100 &-223.106 &100 &-229.342 &100 \\
    -228.312 &101 &-227.795 &101 &-227.945 &101 &-231.916 &101 \\
    -230.257 &99 &-229.443 &99 &-229.69 &99 &-232.93 &99 \\
    -228.276 &100 &-227.977 &100 &-228.071 &100 &-234.852 &100 \\
    -224.315 &100 &-223.785 &100 &-223.957 &100 &-231.222 &100 \\
    -221.219 &100 &-221.08 &100 &-221.128 &100 &-229.573 &100 \\
    \hline
  \end{tabular}
\end{center}

Reducing the number of cells down to 10 we obtain the following table,
in which all models correctly estimate the number of cells in each
simulations, but the divergence between the \FFT log-likelihood values
and the moment models is still apparent.
\begin{center}
  \begin{tabular}{|cc|cc|cc|cc|}
    \hline
    \multicolumn{2}{|c|}{Normal} & \multicolumn{2}{c|}{Logormal} & \multicolumn{2}{c|}{Gamma} & \multicolumn{2}{c|}{\FFT} \\
    $\widehat{LL}_{max}$ & cells &$\widehat{LL}_{max}$ & cells &$\widehat{LL}_{max}$ & cells &$\widehat{LL}_{max}$ & cells \\ \hline
    -108.134 &10 &-108.002 &10 &-108.038 &10 &-115.534 &10 \\
    -111.576 &10 &-111.361 &10 &-111.413 &10 &-119.017 &10 \\
    -110.092 &10 &-110.252 &10 &-110.185 &10 &-117.129 &10 \\
    -111.932 &10 &-111.352 &10 &-111.528 &10 &-119.666 &10 \\
    -110.431 &10 &-111.194 &10 &-110.915 &10 &-117.359 &10 \\
    -107.527 &10 &-108.362 &10 &-108.054 &10 &-114.481 &10 \\
    -106.965 &10 &-106.724 &10 &-106.799 &10 &-114.368 &10 \\
    -113.55 &10 &-113.298 &10 &-113.352 &10 &-121.412 &10 \\
    -109.083 &10 &-108.913 &10 &-108.956 &10 &-116.539 &10 \\
    -116.432 &10 &-117.402 &10 &-117.013 &10 &-123.514 &10 \\
    \hline
  \end{tabular}
\end{center}

We now set $\phi = 0.2$.  For 500 cells, we obtain the following
table, in which we are now seeing the effects of the pre-sampling
variability in the wider range of cell estimates in all models, which
are giving similar estimates.

\begin{center}
  \begin{tabular}{|cc|cc|cc|cc|}
    \hline
    \multicolumn{2}{|c|}{Normal} & \multicolumn{2}{c|}{Logormal} & \multicolumn{2}{c|}{Gamma} & \multicolumn{2}{c|}{\FFT} \\
    $\widehat{LL}_{max}$ & cells &$\widehat{LL}_{max}$ & cells &$\widehat{LL}_{max}$ & cells &$\widehat{LL}_{max}$ & cells \\ \hline
    -279.819 &496 &-280.204 &497 &-280.053 &497 &-284.162 &498 \\
    -313.331 &497 &-307.348 &493 &-308.797 &494 &-302.683 &495 \\
    -279.83 &496 &-279.6 &498 &-279.678 &497 &-283.113 &498 \\
    -279.828 &504 &-279.664 &506 &-279.717 &505 &-283.564 &506 \\
    -288.832 &496 &-288.937 &496 &-288.811 &496 &-290.632 &497 \\
    -302.379 &497 &-298.563 &495 &-299.36 &495 &-297.065 &496 \\
    -286.874 &504 &-287.079 &504 &-286.944 &504 &-289.078 &505 \\
    -286.703 &507 &-286.583 &507 &-286.567 &507 &-289.549 &509 \\
    -301.076 &502 &-301.912 &500 &-301.295 &500 &-299.415 &500 \\
    -292.959 &500 &-293.018 &499 &-292.885 &500 &-294.815 &501 \\
    \hline
  \end{tabular}
\end{center}

Reducing the number of cells to 200, and keeping $\phi = 0.2$ we
obtain

\begin{center}
  \begin{tabular}{|cc|cc|cc|cc|}
    \hline
    \multicolumn{2}{|c|}{Normal} & \multicolumn{2}{c|}{Logormal} & \multicolumn{2}{c|}{Gamma} & \multicolumn{2}{c|}{\FFT} \\
    $\widehat{LL}_{max}$ & cells &$\widehat{LL}_{max}$ & cells &$\widehat{LL}_{max}$ & cells &$\widehat{LL}_{max}$ & cells \\ \hline
    -227.224 &199 &-228.354 &199 &-227.737 &199 &-228.009 &198 \\
    -237.356 &196 &-233.14 &195 &-234.183 &195 &-233.538 &195 \\
    -229.561 &205 &-231.812 &206 &-230.937 &206 &-232.314 &207 \\
    -233.089 &196 &-237.701 &193 &-235.661 &194 &-234.839 &195 \\
    -223.573 &197 &-222.67 &199 &-222.915 &198 &-224.836 &198 \\
    -233.505 &205 &-233.881 &205 &-233.493 &205 &-234.295 &205 \\
    -232.002 &198 &-230.505 &197 &-230.724 &197 &-231.164 &196 \\
    -225.549 &195 &-225.824 &196 &-225.692 &195 &-227.969 &196 \\
    -216.563 &192 &-218.336 &194 &-217.713 &193 &-219.421 &193 \\
    -220.548 &199 &-221.047 &200 &-220.84 &199 &-222.54 &200 \\
    \hline
  \end{tabular}
\end{center}

Reducing the number of cells to 100, and keeping $\phi = 0.2$ we
obtain similar behaviour,

\begin{center}
  \begin{tabular}{|cc|cc|cc|cc|}
    \hline
    \multicolumn{2}{|c|}{Normal} & \multicolumn{2}{c|}{Logormal} & \multicolumn{2}{c|}{Gamma} & \multicolumn{2}{c|}{\FFT} \\
    $\widehat{LL}_{max}$ & cells &$\widehat{LL}_{max}$ & cells &$\widehat{LL}_{max}$ & cells &$\widehat{LL}_{max}$ & cells \\ \hline
    -161.866 &91 &-163.224 &92 &-162.538 &92 &-162.197 &91 \\
    -174.273 &102 &-174.564 &99 &-174.032 &100 &-173.428 &100 \\
    -164.162 &99 &-163.554 &100 &-163.703 &100 &-164.374 &99 \\
    -166.503 &100 &-168.923 &100 &-167.691 &101 &-167.244 &100 \\
    -165.579 &96 &-165.367 &96 &-165.226 &96 &-165.526 &96 \\
    -161.76 &99 &-161.274 &101 &-161.398 &100 &-161.841 &100 \\
    -169.946 &106 &-171.175 &105 &-170.44 &105 &-170.32 &106 \\
    -164.081 &98 &-164.207 &99 &-164.096 &99 &-164.48 &98 \\
    -165.631 &98 &-167.698 &97 &-166.658 &98 &-166.109 &98 \\
    -174.81 &103 &-173.074 &101 &-173.319 &102 &-174.481 &102 \\
    \hline
  \end{tabular}
\end{center}

However as we go below 40 cells the behaviours of the \FFT and moment
based models start to change.  For 40 cells we obtain the following
table, in which all the models are in broad agreement:

\begin{center}
  \begin{tabular}{|cc|cc|cc|cc|}
    \hline
    \multicolumn{2}{|c|}{Normal} & \multicolumn{2}{c|}{Logormal} & \multicolumn{2}{c|}{Gamma} & \multicolumn{2}{c|}{\FFT} \\
    $\widehat{LL}_{max}$ & cells &$\widehat{LL}_{max}$ & cells &$\widehat{LL}_{max}$ & cells &$\widehat{LL}_{max}$ & cells \\ \hline
    -147.227 &38 &-151.059 &38 &-149.056 &39 &-147.405 &39 \\
    -149.565 &44 &-149.253 &44 &-149.071 &44 &-148.65 &46 \\
    -144.248 &39 &-145.536 &40 &-145.022 &40 &-144.202 &41 \\
    -149.696 &40 &-159.711 &38 &-154.874 &39 &-150.288 &41 \\
    -148.827 &41 &-148.863 &41 &-148.599 &41 &-148.133 &42 \\
    -149.489 &41 &-149.389 &41 &-149.102 &41 &-148.598 &42 \\
    -151.896 &41 &-153.712 &40 &-152.28 &40 &-150.926 &41 \\
    -142.965 &33 &-143.202 &35 &-142.985 &35 &-143.77 &37 \\
    -145.254 &36 &-149.028 &36 &-146.878 &36 &-145.521 &37 \\
    -153.346 &37 &-149.71 &35 &-150 &36 &-151.28 &37 \\ 
    \hline
  \end{tabular}
\end{center}

However the further reduction to 35 cells produces the following table
in which we start to see the \FFT model overestimating the number of
cells in a biased manner, although the likelihoods of the four models
are in broad agreement.
\begin{center}
  \begin{tabular}{|cc|cc|cc|cc|}
    \hline
    \multicolumn{2}{|c|}{Normal} & \multicolumn{2}{c|}{Logormal} & \multicolumn{2}{c|}{Gamma} & \multicolumn{2}{c|}{\FFT} \\
    $\widehat{LL}_{max}$ & cells &$\widehat{LL}_{max}$ & cells &$\widehat{LL}_{max}$ & cells &$\widehat{LL}_{max}$ & cells \\ \hline
    -144.129 &36 &-148.72 &36 &-146.337 &37 &-144.584 &37 \\
    -146.785 &36 &-149.664 &36 &-147.998 &37 &-146.792 &37 \\
    -150.221 &33 &-151.583 &32 &-149.991 &33 &-151.201 &37 \\
    -142.658 &34 &-143.072 &36 &-142.915 &36 &-143.154 &37 \\
    -152.339 &33 &-153.707 &30 &-152.106 &31 &-153.466 &35 \\
    -147.424 &37 &-144.977 &38 &-145.683 &38 &-146.49 &39 \\
    -152.856 &36 &-156.567 &34 &-154.021 &35 &-152.478 &37 \\
    -152.798 &33 &-154.999 &28 &-152.755 &30 &-155.032 &35 \\
    -146.274 &34 &-150.299 &34 &-148.062 &34 &-147.571 &37 \\
    -144.768 &36 &-145.235 &37 &-144.882 &37 &-144.767 &37 \\
    \hline
  \end{tabular}
\end{center}
Reducing the number of cells to 30 yields divergence in the likelihood
estimates too, as in the following table:
\begin{center}
  \begin{tabular}{|cc|cc|cc|cc|}
    \hline
    \multicolumn{2}{|c|}{Normal} & \multicolumn{2}{c|}{Logormal} & \multicolumn{2}{c|}{Gamma} & \multicolumn{2}{c|}{\FFT} \\
    $\widehat{LL}_{max}$ & cells &$\widehat{LL}_{max}$ & cells &$\widehat{LL}_{max}$ & cells &$\widehat{LL}_{max}$ & cells \\ \hline
    -140.094 &28 &-142.829 &29 &-141.302 &29 &-145.536 &34 \\
    -140.487 &29 &-143.953 &29 &-142.178 &29 &-146.174 &34 \\
    -137.771 &27 &-138.535 &29 &-138.321 &28 &-145.971 &33 \\
    -142.205 &30 &-145.612 &28 &-143.452 &29 &-146.512 &34 \\
    -141.715 &30 &-146.407 &29 &-143.991 &30 &-146.539 &34 \\
    -147.524 &30 &-146.787 &29 &-146.315 &29 &-151.487 &34 \\
    -145.765 &30 &-146.033 &30 &-145.324 &30 &-149.674 &34 \\
    -139.777 &30 &-143.263 &30 &-141.447 &30 &-144.618 &34 \\
    -146.34 &32 &-153.725 &28 &-149.89 &29 &-150.328 &35 \\
    -148.841 &31 &-147.083 &30 &-146.882 &31 &-150.594 &35 \\
    \hline
  \end{tabular}
\end{center}

Reducing the number of cells to 10 we obtain a stark difference
between the \FFT and moment based models, as shown in the following
table:

\begin{center}
  \begin{tabular}{|cc|cc|cc|cc|}
    \hline
    \multicolumn{2}{|c|}{Normal} & \multicolumn{2}{c|}{Logormal} & \multicolumn{2}{c|}{Gamma} & \multicolumn{2}{c|}{\FFT} \\
    $\widehat{LL}_{max}$ & cells &$\widehat{LL}_{max}$ & cells &$\widehat{LL}_{max}$ & cells &$\widehat{LL}_{max}$ & cells \\ \hline
    -73.3783 &10 &-76.5299 &10 &-75.1702 &10 &-138.845 &23 \\
    -69.2668 &10 &-69.0933 &10 &-68.4735 &10 &-129.504 &23 \\
    -66.276 &10 &-63.7138 &10 &-64.0525 &10 &-130.571 &23 \\
    -83.0595 &11 &-84.5144 &13 &-84.3105 &12 &-138.676 &24 \\
    -67.5142 &9 &-68.668 &10 &-68.2279 &9 &-137.273 &3 \\
    -81.0544 &12 &-81.9969 &12 &-81.2933 &12 &-132.806 &24 \\
    -80.4326 &11 &-79.8365 &11 &-79.5108 &11 &-137.78 &23 \\
    -66.6831 &9 &-67.4642 &11 &-67.4318 &10 &-130.458 &23 \\
    -76.19 &11 &-78.2126 &11 &-77.0102 &11 &-135.338 &23 \\
    -65.9317 &9 &-65.3378 &10 &-65.1538 &10 &-130.649 &23 \\
    \hline
  \end{tabular}
\end{center}

It appears that the moment based models are giving close to unbiased
estimates for the number of cells, whereas the \FFT model is giving
quite biased estimates. The problem with the \FFT model can be traced
to factorisation approximation of \eqref{eq:likeapprox2}.  Recall that
this is a double approximation here: one in which the joint \PGF is
replaced with a factorised product, given in \eqref{eq:likeapprox1},
and the second factorisation approximation made in going to
\eqref{eq:likeapprox2}. Now consider an allele having $k$ genomes. The
probability that it completely drops out will be
$(1-\phi)^k$. However, if the allele completely drops out then it
cannot produce any stutter products, hence the joint probability of
dropout for that allele and its stutter products will be
$(1-\phi)^k$. However the factorisation approximation
\eqref{eq:likeapprox2} means that the factor $(1-\phi)^k$ is counted
for the allele and each of it stutter product, hence will give a total
joint probability of dropout of $(1-\phi)^{4k}$. For large $\phi$ and
$k$, both $(1-\phi)^k$ and $(1-\phi)^{4k}$ will be very close to zero,
and so the factorisation approximation will be good, but if $k$ and/or
$\phi$ are small, then the terms will have a large effect on the
likelihood calculations, and their differences will also grow.  It
appears from the results in the tables above that the transition point
where the approximation starts becoming bad is when the number of
cells is around 35-40: with $\phi=0.2$ this corresponds to around 7-8
amplify-able cells, approximately 45-50\textit{pg} by weight, of DNA
subject to \PCR.

It is curious that the moment models appear to give approximately
unbiased estimates for the number of cells.  If we compare to the
10-cell, $\phi=1$ table on \pref{page:cell10tab}, in which the
lognormal and gamma models gave underestimates for the number of
cells, it appears that the factorisation approximation has had a
correcting effect on these models. The normal model appears to have
less variability in its log-likelihood values when compared to to the
lognormal and gamma models.

Now $\phi=0.2$ is a high value for forensic samples. Recall that
$\phi$ is the product of the extraction efficiency and the fraction of
extracted DNA that is taken for \PCR amplification.  A more typical
extraction efficiency value would be around 0.2, and similarly the
fraction of aliquot taken for amplification would also be around 0.2,
so that $\phi$ values of around $\phi = 0.04$ would be more realistic
for forensic casework \cite{gill:etal:2005}. Hence, if the transition
for the \FFT model happens when expected number of amplify-able cells
is around 7-8, this corresponds to around 200 cells in the sample.

The following table shows estimated cell counts and log-likelihoods
for 200 cells and $\phi = 0.04$, corresponding to around 8 cells of
DNA. We see that all the models are quite comparable in their
estimates of cell counts and their likelihoods values.

\begin{center}
  \begin{tabular}{|cc|cc|cc|cc|}
    \hline
    \multicolumn{2}{|c|}{Normal} & \multicolumn{2}{c|}{Logormal} & \multicolumn{2}{c|}{Gamma} & \multicolumn{2}{c|}{\FFT} \\
    $\widehat{LL}_{max}$ & cells &$\widehat{LL}_{max}$ & cells &$\widehat{LL}_{max}$ & cells &$\widehat{LL}_{max}$ & cells \\ \hline
    -153.806 &218 &-158.376 &212 &-155.638 &217 &-154.016 &222 \\
    -158.168 &213 &-160.208 &199 &-158.317 &206 &-157.783 &208 \\
    -156.432 &212 &-158.452 &204 &-156.755 &208 &-156.127 &208 \\
    -154.188 &195 &-155.784 &189 &-154.183 &193 &-153.931 &195 \\
    -153.257 &201 &-152.824 &199 &-152.443 &201 &-152.747 &202 \\
    -145.644 &177 &-143.678 &190 &-144.625 &187 &-146.097 &190 \\
    -157.511 &206 &-157.945 &190 &-156.17 &196 &-155.666 &202 \\
    -149.454 &197 &-149.449 &203 &-149.159 &202 &-148.911 &202 \\
    -146.514 &183 &-152.516 &182 &-149.646 &185 &-148.237 &192 \\
    -156.928 &227 &-157.76 &220 &-156.721 &224 &-155.927 &224 \\
    \hline
  \end{tabular}
\end{center}

The following boxplots for estimated cell counts and log-likelihoods,
based on 200 simulations confirms this:

\begin{center}
  \begin{minipage}{.5\textwidth}
    \centering
    \includegraphics[scale=0.4]{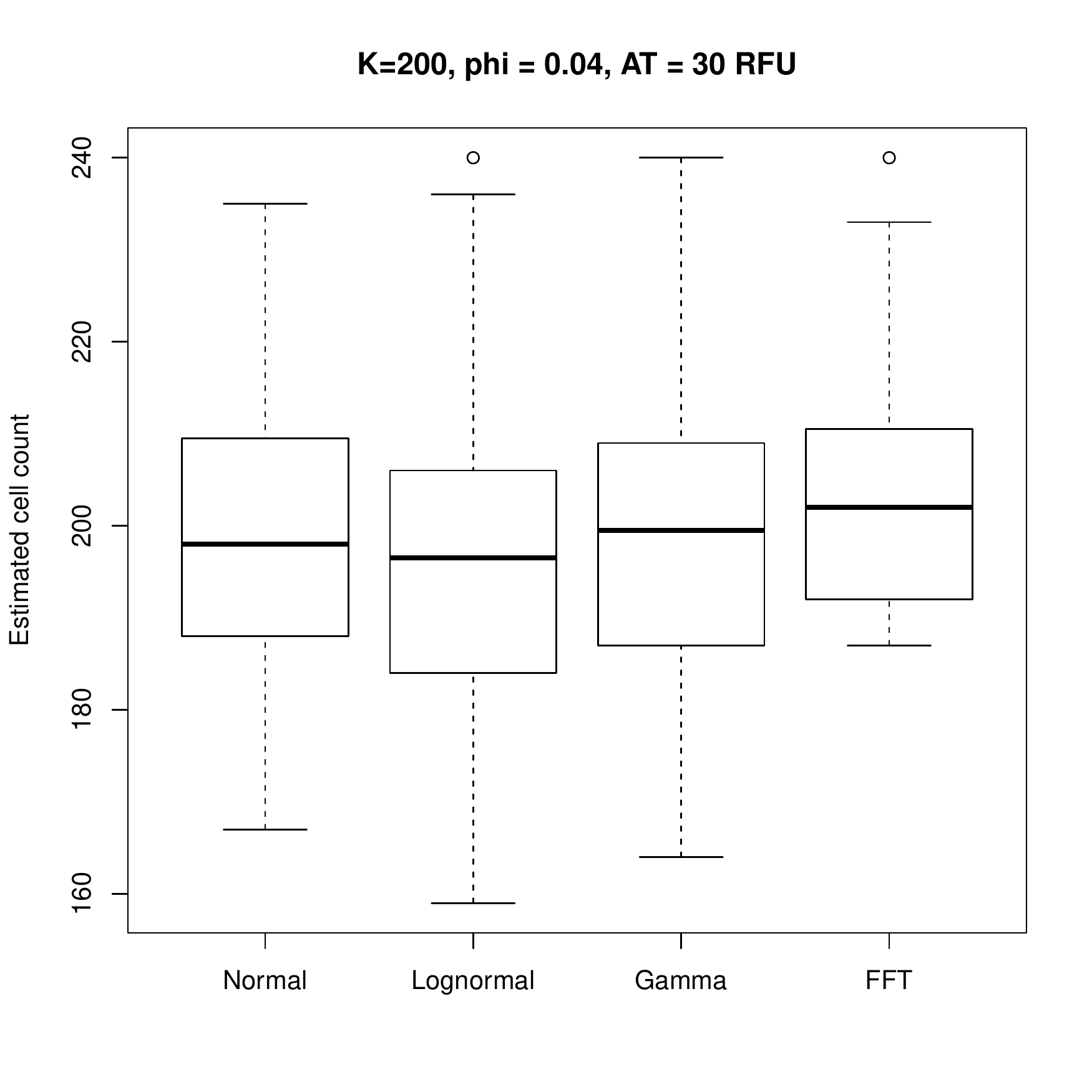}
  \end{minipage}%
  \begin{minipage}{.5\textwidth}
    \centering
    \includegraphics[scale=0.4]{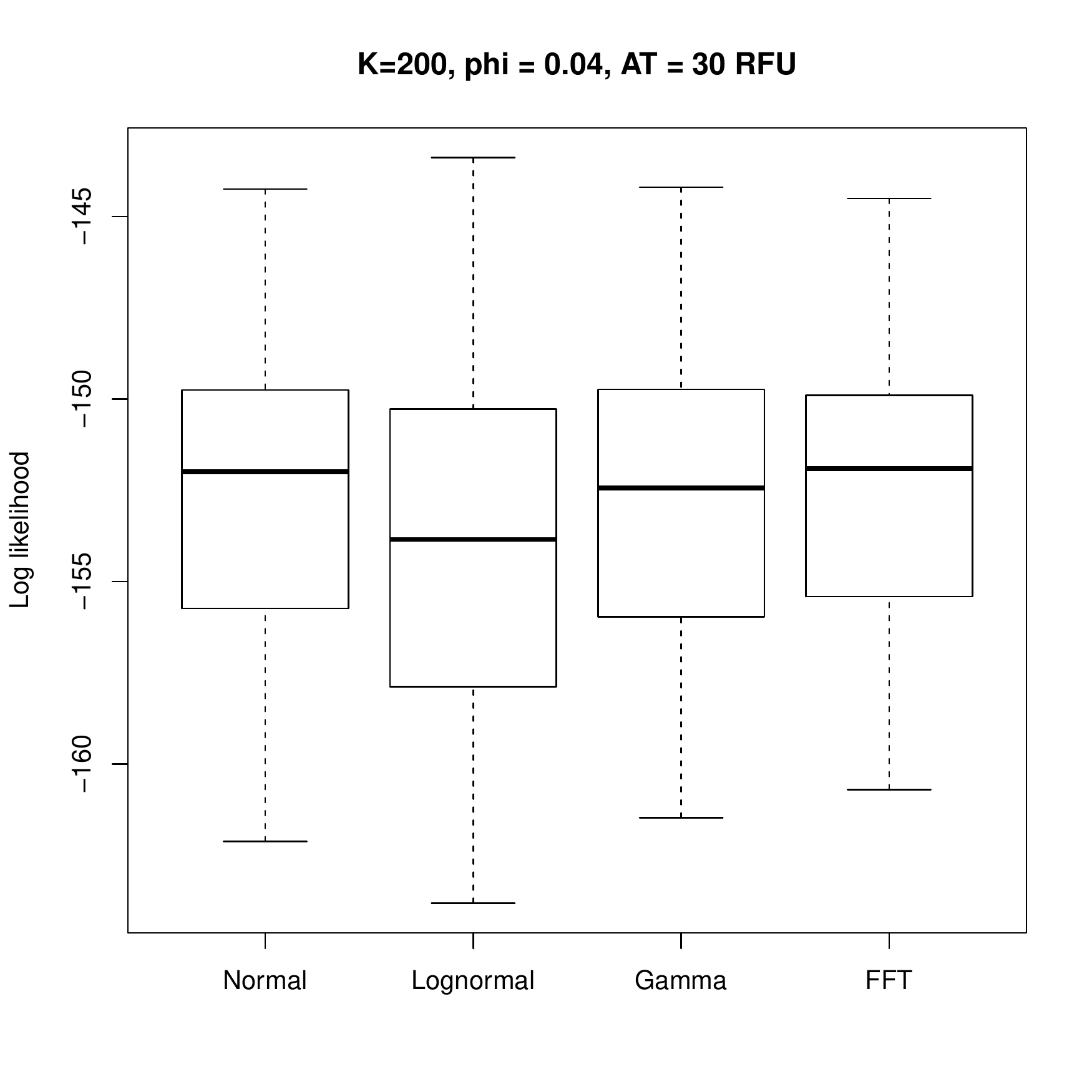}
  \end{minipage}%
\end{center}

However, reducing the number of cells to 150 we obtain the following
table
\begin{center}
  \begin{tabular}{|cc|cc|cc|cc|}
    \hline
    \multicolumn{2}{|c|}{Normal} & \multicolumn{2}{c|}{Logormal} & \multicolumn{2}{c|}{Gamma} & \multicolumn{2}{c|}{\FFT} \\
    $\widehat{LL}_{max}$ & cells &$\widehat{LL}_{max}$ & cells &$\widehat{LL}_{max}$ & cells &$\widehat{LL}_{max}$ & cells \\ \hline
    -144.025 &136 &-142.352 &143 &-142.807 &142 &-151.579 &170 \\
    -137.712 &141 &-144.199 &131 &-140.921 &136 &-146.336 &170 \\
    -140.71 &141 &-142.032 &147 &-141.36 &147 &-147.833 &173 \\
    -150.072 &171 &-151.597 &164 &-150.225 &167 &-151.937 &188 \\
    -145.085 &138 &-142.67 &143 &-143.295 &143 &-151.939 &171 \\
    -140.398 &135 &-142.12 &134 &-140.728 &136 &-148.936 &168 \\
    -144.896 &146 &-146.134 &145 &-144.956 &147 &-151.04 &175 \\
    -142.788 &140 &-145.464 &127 &-143.368 &133 &-151.46 &169 \\
    -137.54 &131 &-141.125 &134 &-139.346 &135 &-147.239 &167 \\
    -148.65 &156 &-146.808 &158 &-147.003 &158 &-151.945 &181 \\
    \hline
  \end{tabular}
\end{center}
and these corresponding plots based on 200 simulations:
\begin{center}
  \begin{minipage}{.5\textwidth}
    \centering
    \includegraphics[scale=0.4]{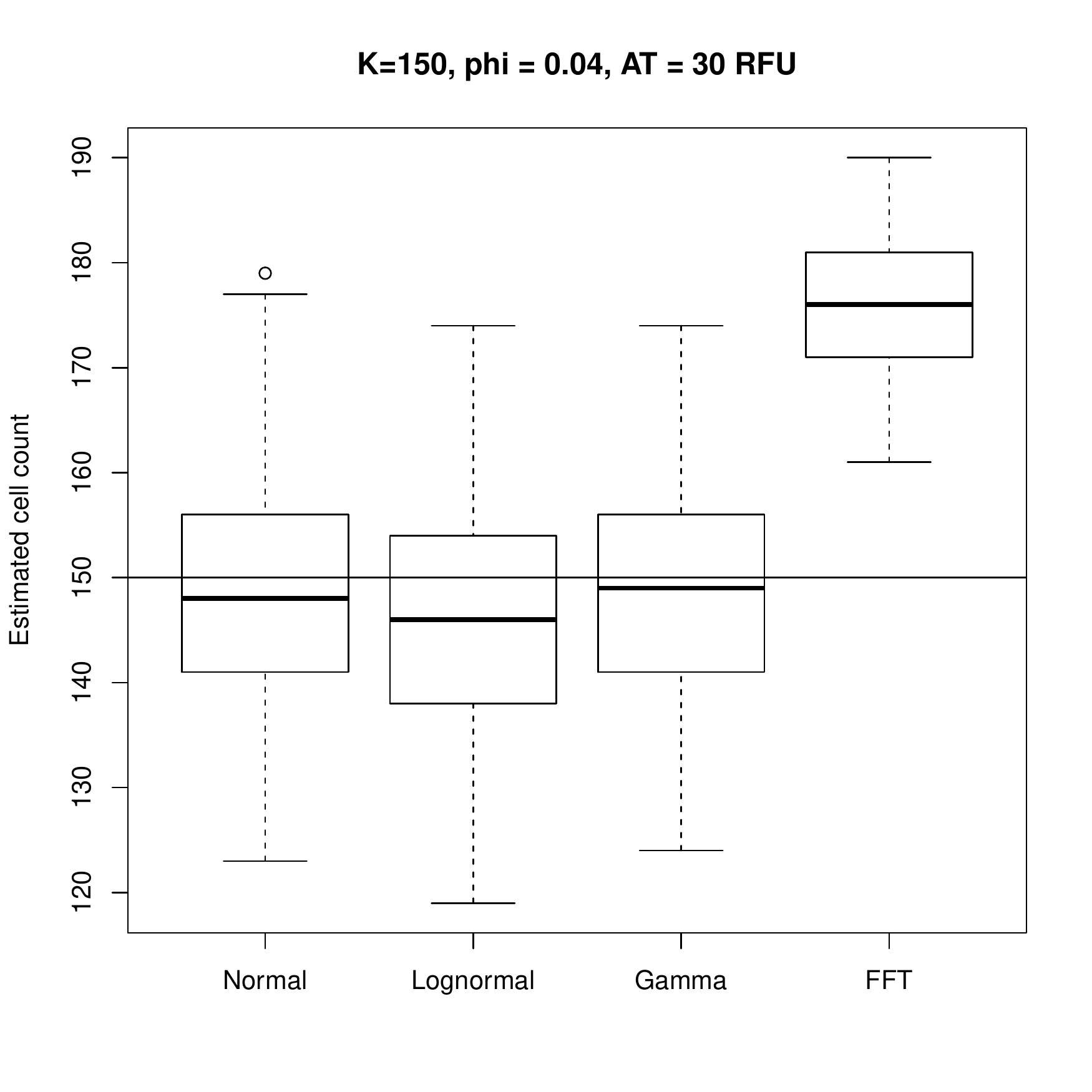}
  \end{minipage}%
  \begin{minipage}{.5\textwidth}
    \centering
    \includegraphics[scale=0.4]{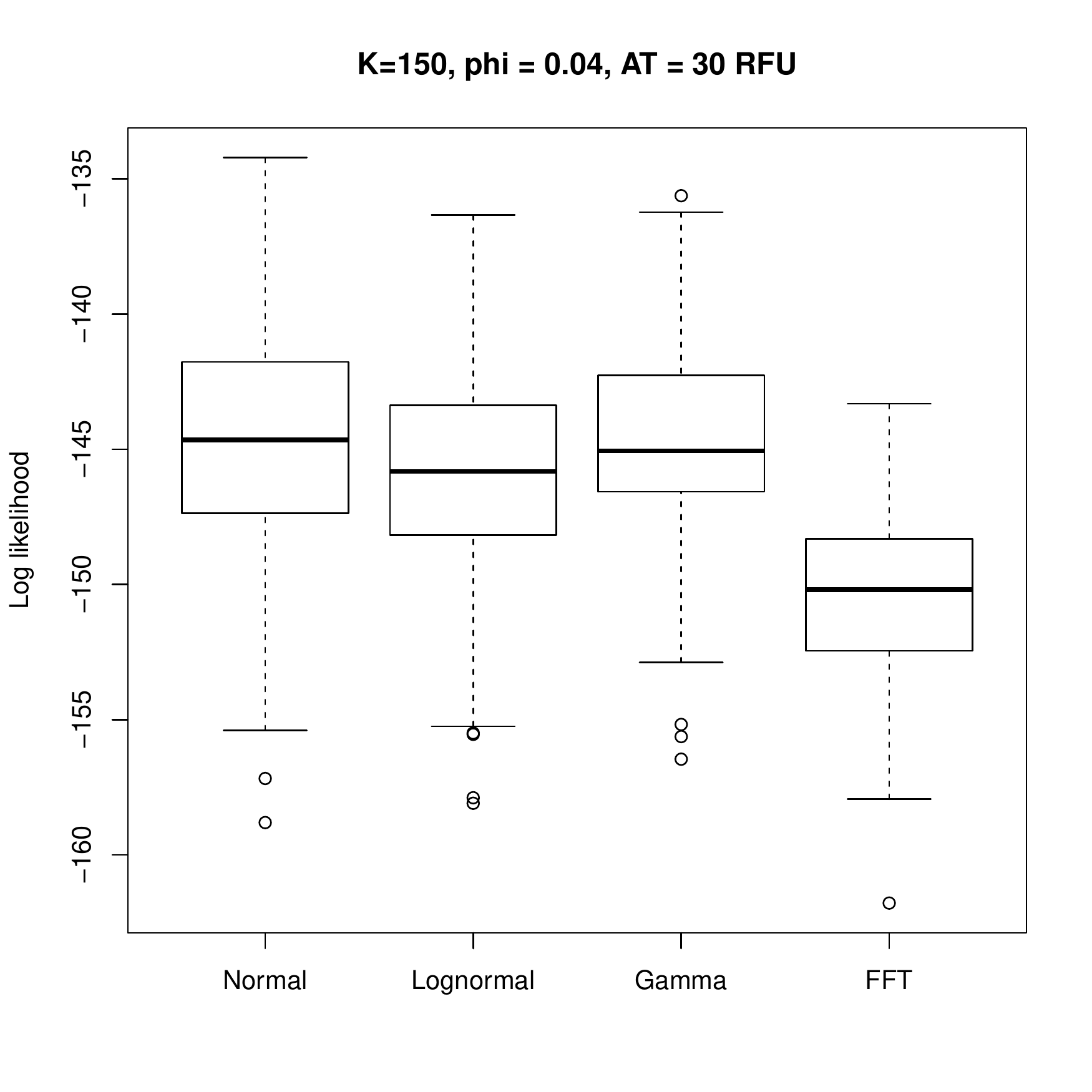}
  \end{minipage}%
\end{center}

Reducing the number of cells to just 30, (so giving an average of just
1.2 cells worth of DNA for \PCR) we obtain the following:

\begin{center}
  \begin{tabular}{|cc|cc|cc|cc|}
    \hline
    \multicolumn{2}{|c|}{Normal} & \multicolumn{2}{c|}{Logormal} & \multicolumn{2}{c|}{Gamma} & \multicolumn{2}{c|}{\FFT} \\
    $\widehat{LL}_{max}$ & cells &$\widehat{LL}_{max}$ & cells &$\widehat{LL}_{max}$ & cells &$\widehat{LL}_{max}$ & cells \\ \hline
    -26.6781 &26 &-26.6658 &27 &-26.226 &24 &-51.4625 &4 \\
    -31.7871 &30 &-31.8301 &32 &-31.5047 &29 &-64.4055 &6 \\
    -29.1594 &27 &-29.4184 &33 &-29.4628 &29 &-61.8782 &5 \\
    -35.5867 &28 &-37.4363 &39 &-37.6121 &34 &-72.5157 &6 \\
    -25.1085 &25 &-24.6105 &28 &-24.5907 &24 &-51.6145 &4 \\
    -20.4599 &24 &-20.3479 &25 &-20.191 &22 &-44.142 &4 \\
    -27.5168 &25 &-28.6467 &30 &-28.5487 &26 &-53.8391 &4 \\
    -35.9236 &31 &-36.4142 &35 &-36.2378 &32 &-74.6161 &7 \\
    -17.6097 &21 &-17.9368 &25 &-18.0696 &21 &-38.5006 &3 \\
    -38.5072 &33 &-37.1138 &34 &-36.9363 &31 &-75.1298 &7 \\
    \hline
  \end{tabular}
\end{center}

and these corresponding plots:
\begin{center}
  \begin{minipage}{.5\textwidth}
    \centering
    \includegraphics[scale=0.4]{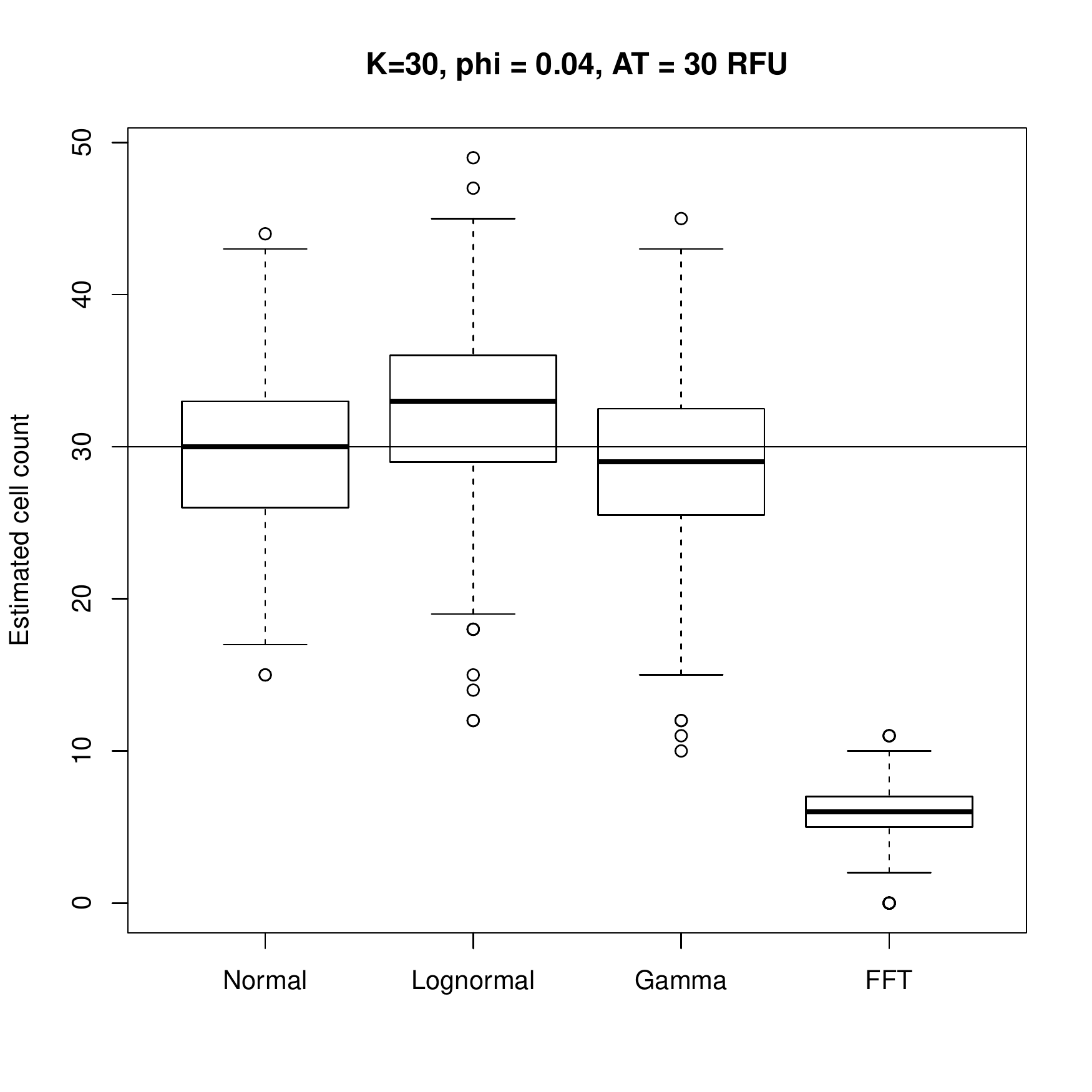}
  \end{minipage}%
  \begin{minipage}{.5\textwidth}
    \centering
    \includegraphics[scale=0.4]{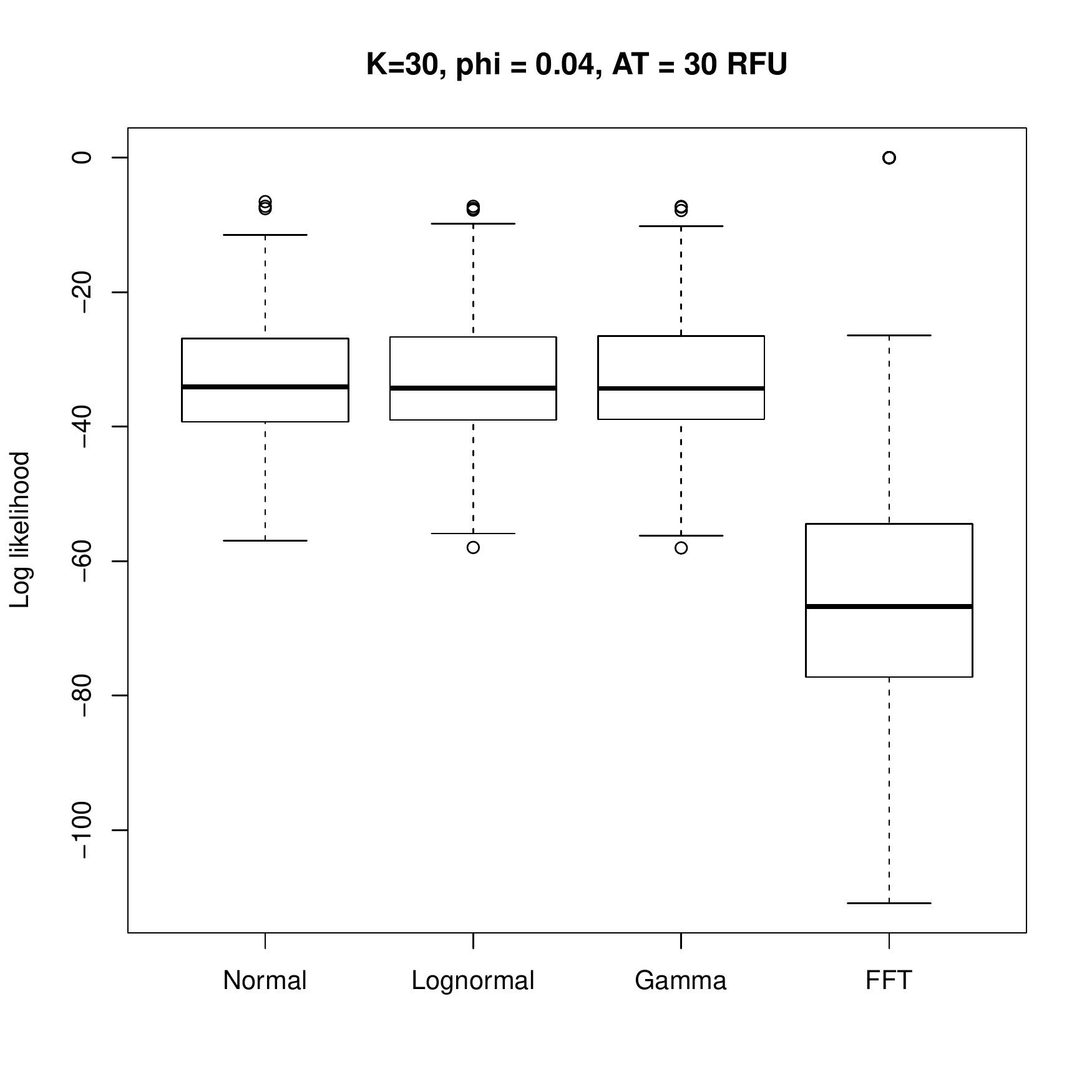}
  \end{minipage}%
\end{center}

However, if we look at the simulated data for these latter
simulations, we find that they have complete drop-out on most of the
markers, and just one peak on each of the other markers. For such
samples it is not necessary to include double and forward stutters in
the model.  If we remove these model components, so that we include
only the target and single back-stutter allele distributions, we
obtain the following fits for the models (using the same simulated
data).

\begin{center}
  \begin{tabular}{|cc|cc|cc|cc|}
    \hline
    \multicolumn{2}{|c|}{Normal} & \multicolumn{2}{c|}{Logormal} & \multicolumn{2}{c|}{Gamma} & \multicolumn{2}{c|}{\FFT} \\
    $\widehat{LL}_{max}$ & cells &$\widehat{LL}_{max}$ & cells &$\widehat{LL}_{max}$ & cells &$\widehat{LL}_{max}$ & cells \\ \hline
    -15.3989 &21 &-14.676 &18 &-14.4406 &16 &-20.3993 &11 \\
    -26.7267 &26 &-26.7212 &27 &-26.2684 &24 &-32.4348 &25 \\
    -31.7885 &30 &-31.8266 &32 &-31.4997 &29 &-37.5075 &31 \\
    -29.1259 &28 &-29.3837 &33 &-29.4362 &29 &-35.7485 &31 \\
    -35.5782 &28 &-37.4145 &39 &-37.5979 &34 &-42.0806 &33 \\
    -25.0822 &25 &-24.588 &28 &-24.5746 &24 &-31.628 &26 \\
    -20.4623 &24 &-20.3585 &25 &-20.1993 &22 &-27.3342 &23 \\
    -27.4902 &25 &-28.6197 &30 &-28.5286 &26 &-33.9025 &25 \\
    -35.9469 &31 &-36.4683 &35 &-36.2816 &32 &-41.3409 &35 \\
    -17.6271 &21 &-17.9428 &25 &-18.0737 &21 &-24.178 &19 \\
    \hline
  \end{tabular}
\end{center}

We see that the \FFT estimate are now much better. The lower
log-likelihoods are explained by the inclusion of modelling
stutters. However the few observed peaks heights in the simulations
are within the range 30-100, so that any stutter peaks would not be
observed with the threshold 30. So, if we also remove the modelling of
stutter, then we obtain the following fits, in which we see that the
log-likelihoods of all four models are now comparable.

\begin{center}
  \begin{tabular}{|cc|cc|cc|cc|}
    \hline
    \multicolumn{2}{|c|}{Normal} & \multicolumn{2}{c|}{Logormal} & \multicolumn{2}{c|}{Gamma} & \multicolumn{2}{c|}{\FFT} \\
    $\widehat{LL}_{max}$ & cells &$\widehat{LL}_{max}$ & cells &$\widehat{LL}_{max}$ & cells &$\widehat{LL}_{max}$ & cells \\ \hline
    -15.366 &21 &-14.6613 &18 &-14.4301 &16 &-14.3498 &15 \\
    -26.6653 &27 &-26.6806 &27 &-26.2362 &24 &-26.1813 &22 \\
    -31.7059 &30 &-31.7632 &32 &-31.4469 &29 &-31.785 &27 \\
    -29.2184 &28 &-29.4225 &33 &-29.4644 &29 &-30.134 &27 \\
    -35.5742 &29 &-37.4283 &39 &-37.6174 &34 &-36.6476 &29 \\
    -25.388 &26 &-24.8448 &28 &-24.7658 &24 &-25.9578 &23 \\
    -20.4145 &24 &-20.3252 &25 &-20.1738 &22 &-21.0171 &21 \\
    -27.437 &25 &-28.5662 &30 &-28.4889 &26 &-27.6955 &23 \\
    -35.856 &32 &-36.3818 &36 &-36.2137 &32 &-36.0407 &31 \\
    -17.5942 &21 &-17.9095 &25 &-18.0512 &21 &-17.7821 &19 \\
    \hline
  \end{tabular}
\end{center}

\subsubsection{Simulations with degradation, but no baseline noise}

In the simulations above, we have simulated samples with a degradation
parameter equal to zero, and have used that value in estimating the
cell counts. We now consider simulations in which the degradation
parameter is non zero, and is estimated from the data along with the
cell counts.

Our first simulation set will have $\phi=0.04$, and degradation
parameter $\delta = 0.01$. This means that for an allele of size $l$
in base-pairs, that its binomial selection probability if
$\phi \exp(-\delta l)$.  So an an allele of size 100 will have a
selection probability of around 0.0147, whilst a longer allele of
length 400 will have a selection probability of around 0.0007.

We take the initial number of cells to be 1000. Including all possible
forms of stutter in fitting the model we obtain the following
estimates from 10 simulations (the degradation for only the FFT model
is shown);

\begin{center}
  \begin{tabular}{|cc|cc|cc|ccl|}
    \hline
    \multicolumn{2}{|c|}{Normal} & \multicolumn{2}{c|}{Logormal} & \multicolumn{2}{c|}{Gamma} & \multicolumn{2}{c}{\FFT}  & $\hat{\delta}_{FFT}$\\
    $\widehat{LL}_{max}$ & cells &$\widehat{LL}_{max}$ & cells &$\widehat{LL}_{max}$ & cells &$\widehat{LL}_{max}$ & cells & \\ \hline
    -119.379 &1055 &-120.355 &1003 &-119.455 &1029 &-147.95 &1172 &0.00995539 \\
    -110.236 &1002 &-111.833 &1038 &-111.143 &1030 &-140.249 &1146 &0.0100925 \\
    -118.145 &1218 &-116.529 &1127 &-115.724 &1154 &-147.429 &1320 &0.0111535 \\
    -123.923 &1033 &-123.257 &1113 &-123.683 &1092 &-149.485 &1172 &0.00954747 \\
    -113.382 &1020 &-111.865 &1012 &-111.872 &1017 &-142.634 &1147 &0.0101153 \\
    -128.499 &1108 &-127.712 &1028 &-126.872 &1059 &-153.195 &1204 &0.00958187 \\
    -133.743 &1002 &-133.587 &954 &-132.82 &965 &-161.607 &1098 &0.00973301 \\
    -116.822 &1038 &-117.803 &1036 &-116.58 &1037 &-146.971 &1165 &0.0103188 \\
    -124.35 &1025 &-122.464 &1058 &-122.859 &1049 &-149.219 &1145 &0.0093805 \\
    -117.821 &1082 &-114.408 &1066 &-114.787 &1066 &-148.147 &1207 &0.0108218 \\
    \hline
  \end{tabular}
\end{center}

Give the quite high level of degradation we would not expect to see
double or forward stutters, even amongst the smaller alleles.  So
removing these from the model, but retaining the stutter model, we
obtain (for the same simulated data) much better concordance between
the \FFT model and the moment-based models.

\begin{center}
  \begin{tabular}{|cc|cc|cc|ccl|}
    \hline
    \multicolumn{2}{|c|}{Normal} & \multicolumn{2}{c|}{Logormal} & \multicolumn{2}{c|}{Gamma} & \multicolumn{2}{c}{\FFT}  & \multicolumn{1}{c|}{$\hat{\delta}_{FFT}$}\\
    $\widehat{LL}_{max}$ & cells &$\widehat{LL}_{max}$ & cells &$\widehat{LL}_{max}$ & cells &$\widehat{LL}_{max}$ & cells & \\ \hline
    -119.262 &1057 &-120.176 &1007 &-119.301 &1033 &-120.526 &1058 &0.00995539 \\
    -110.268 &1004 &-111.917 &1040 &-111.211 &1032 &-111.825 &1027 &0.0100925 \\
    -118.006 &1220 &-116.363 &1132 &-115.572 &1159 &-117.445 &1190 &0.0111535 \\
    -123.927 &1035 &-123.266 &1116 &-123.691 &1095 &-124.538 &1070 &0.00954747 \\
    -113.252 &1023 &-111.715 &1016 &-111.741 &1021 &-113.954 &1027 &0.0101153 \\
    -128.584 &1111 &-127.824 &1030 &-126.974 &1061 &-128.595 &1094 &0.00958187 \\
    -133.737 &1005 &-133.594 &957 &-132.831 &968 &-133.919 &984 &0.00973301 \\
    -116.784 &1040 &-117.731 &1039 &-116.525 &1040 &-117.751 &1046 &0.0103188 \\
    -124.403 &1028 &-122.545 &1060 &-122.924 &1052 &-124.78 &1043 &0.0093805 \\
    -117.763 &1084 &-114.387 &1069 &-114.757 &1070 &-117.423 &1076 &0.0108218 \\
    \hline
  \end{tabular}
\end{center}

The following plot shows the peak height data used in the last
simulation of the previous table. We see that there is complete
allelic drop-out on the loci CSF1PO and D18S51.

\begin{flushleft}
  \includegraphics[scale=0.7]{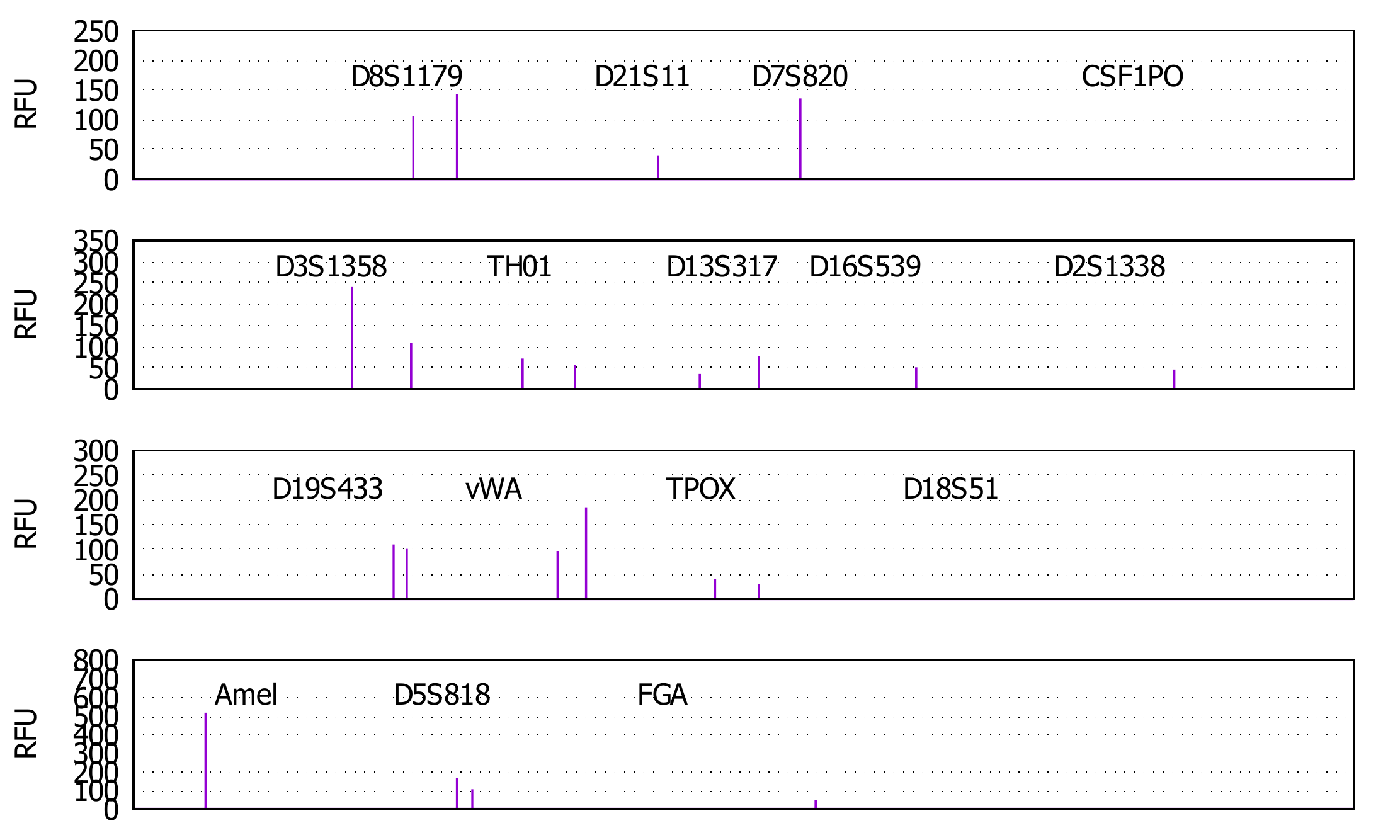}
\end{flushleft}
 
The following plot shows the peak height probability distribution
obtained from the \FFT model, for allele 25 of the locus FGA, in which
the distribution is conditional on the values all of the other peak
heights and the profile of the contributor. The smaller vertical line
at $(30,0)$ locates the analytic threshold, the larger vertical line
locates the observed peak. The complete dropout probability value
point at $(0, 0.0569)$, is off the scale of the plot. The plot
suggests that, most likely, either 3 or 4 intact genomic strands were
randomly selected for the simulated \PCR amplification.

\begin{center}
  \includegraphics[scale=0.5]{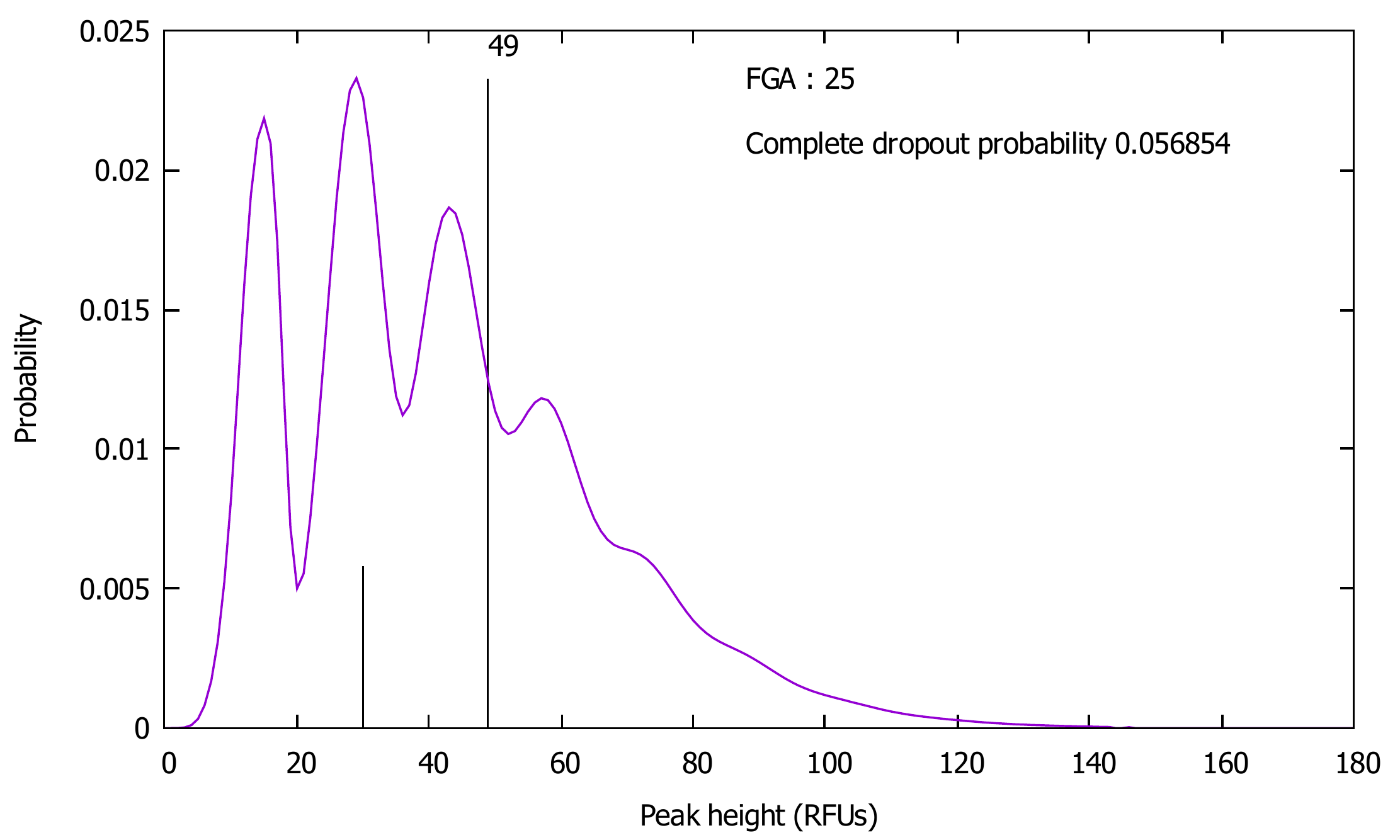}
\end{center}

\subsubsection*{Summary for single contributor simulations}
For moderate to large numbers of cells, all three moment models and
the \FFT based model yield comparable estimates for estimated cell
counts and maximized log-likelihood values. However when the expected
number of cells drops below 8 or so, the \FFT model exhibits bias in
overestimating the number of cells, and produces more extreme (lower)
likelihoods.  Overall the lognormal and gamma models appear to give
the better estimates. However, for such low amounts it seems sensible
to simplify the model by omitting the forward and double backward
stutter model components; doing so it appears that the \FFT model is
concordant with the moment based models.

\subsection{Two person simulations}

We now look at a few two-person simulations, concentrating on
low-template scenarios with various relative initial amounts of DNA.

\subsubsection{ Two person mixtures}
\label{sec:sim2pers}
In the first simulation we have no degradation, that is $\delta = 0$,
set $\phi = 0.04$, and we simulate from contributors
\texttt{RD14-0003-01} ($C_1=200$ cells) and \texttt{RD14-0003-2}
($C_2=200$ cells).  With $\phi=0.04$ this corresponds to an average of
16 amplify-able cells worth of DNA, around 150$pg$.  Using all four
stutter distributions, we obtain this set of simulated fits, assuming
that both contributor profiles are known. All models are giving
similar maximized likelihood and cell estimates.

\begin{center}
  \begin{tabular}{|ccc|ccc|ccc|ccc|}
    \hline
    \multicolumn{3}{|c|}{Normal} & \multicolumn{3}{c|}{Logormal} & \multicolumn{3}{c|}{Gamma} & \multicolumn{3}{c|}{\FFT}  \\
    $\widehat{LL}_{max}$ & $\hat{C}_1$ & $\hat{C}_2$  &$\widehat{LL}_{max}$ &  $\hat{C}_1$ & $\hat{C}_2$ &$\widehat{LL}_{max}$ &  $\hat{C}_1$ & $\hat{C}_2$ &$\widehat{LL}_{max}$ &  $\hat{C}_1$ & $\hat{C}_2$  \\ \hline
    -242.429 &193 &194 &-241.955 &200 &201 &-242.155 &198 &200 &-242.31 &200 &200 \\
    -244.799 &202 &196 &-241.404 &210 &198 &-242.383 &208 &198 &-243.576 &208 &200 \\
    -248.631 &184 &181 &-248.005 &182 &180 &-247.48 &184 &181 &-249.832 &201 &196 \\
    -246.318 &175 &207 &-244.447 &187 &206 &-244.682 &182 &206 &-246.326 &194 &206 \\
    -249.793 &207 &224 &-247.785 &207 &226 &-247.889 &208 &226 &-248.492 &208 &224 \\
    -240.07 &180 &214 &-239.761 &185 &225 &-239.775 &185 &222 &-240.727 &190 &222 \\
    -253.473 &204 &212 &-251.879 &200 &209 &-251.465 &203 &210 &-251.749 &204 &213 \\
    -252.464 &197 &207 &-251.199 &192 &206 &-250.913 &194 &208 &-251.412 &195 &208 \\
    -246.45 &198 &214 &-252.804 &210 &197 &-249.38 &206 &205 &-247.265 &205 &206 \\
    -245.234 &191 &209 &-243.718 &200 &209 &-243.995 &198 &210 &-244.59 &200 &212 \\
    \hline
  \end{tabular}
\end{center}

Repeating the simulation but with $C_1= 50$ and $C_2=200$ cells, (a
1:4 mixture), we see that the three moment models are in broad
agreement, but that the FFT model is diverging in both the cell
estimates for \texttt{RD14-0003-01} and the maximized likelihoods,
which are much lower than the moment models in comparison.

\begin{center}
  \begin{tabular}{|ccc|ccc|ccc|ccc|}
    \hline
    \multicolumn{3}{|c|}{Normal} & \multicolumn{3}{c|}{Logormal} & \multicolumn{3}{c|}{Gamma} & \multicolumn{3}{c|}{\FFT}  \\
    $\widehat{LL}_{max}$ & $\hat{C}_1$ & $\hat{C}_2$  &$\widehat{LL}_{max}$ &  $\hat{C}_1$ & $\hat{C}_2$ &$\widehat{LL}_{max}$ &  $\hat{C}_1$ & $\hat{C}_2$ &$\widehat{LL}_{max}$ &  $\hat{C}_1$ & $\hat{C}_2$  \\ \hline
    -181.783 &41 &239 &-180.183 &40 &237 &-179.908 &39 &238 &-215.229 &9 &251 \\
    -193.674 &50 &193 &-192.84 &47 &193 &-191.837 &46 &195 &-239.032 &16 &204 \\
    -186.171 &42 &209 &-190.008 &51 &212 &-188.724 &47 &213 &-232.002 &13 &224 \\
    -193.501 &50 &199 &-190.686 &54 &200 &-191.185 &52 &200 &-234.826 &111 &189 \\
    -188.659 &48 &200 &-191.771 &46 &204 &-189.964 &47 &204 &-235.36 &15 &204 \\
    -201.354 &56 &202 &-200.354 &60 &196 &-199.818 &57 &200 &-240.015 &115 &192 \\
    -172.509 &34 &201 &-173.196 &41 &215 &-173.62 &38 &212 &-205.595 &9 &208 \\
    -177.388 &41 &189 &-176.959 &42 &196 &-176.441 &40 &195 &-213.534 &11 &204 \\
    -190.686 &44 &201 &-194.281 &48 &188 &-191.67 &45 &195 &-233.393 &13 &204 \\
    -194.286 &52 &196 &-200.514 &54 &192 &-197.259 &53 &195 &-237.177 &112 &187 \\
    \hline
  \end{tabular}
\end{center}

However if we take out the double and forward stutter models, we see
that balance is restored:

\begin{center}
  \begin{tabular}{|ccc|ccc|ccc|ccc|}
    \hline
    \multicolumn{3}{|c|}{Normal} & \multicolumn{3}{c|}{Logormal} & \multicolumn{3}{c|}{Gamma} & \multicolumn{3}{c|}{\FFT}  \\
    $\widehat{LL}_{max}$ & $\hat{C}_1$ & $\hat{C}_2$  &$\widehat{LL}_{max}$ &  $\hat{C}_1$ & $\hat{C}_2$ &$\widehat{LL}_{max}$ &  $\hat{C}_1$ & $\hat{C}_2$ &$\widehat{LL}_{max}$ &  $\hat{C}_1$ & $\hat{C}_2$  \\ \hline
    -181.742 &42 &240 &-180.308 &42 &238 &-179.996 &41 &239 &-183.262 &45 &237 \\
    -193.528 &51 &194 &-192.87 &49 &193 &-191.85 &47 &195 &-195.064 &52 &193 \\
    -186.354 &43 &209 &-190.206 &52 &213 &-188.941 &49 &213 &-191.354 &50 &212 \\
    -193.559 &51 &199 &-190.807 &55 &201 &-191.313 &53 &201 &-194.136 &55 &200 \\
    -188.622 &49 &201 &-191.907 &48 &204 &-190.039 &48 &204 &-191.171 &53 &201 \\
    -201.419 &57 &203 &-200.684 &61 &196 &-200.109 &59 &200 &-202.325 &60 &202 \\
    -172.836 &35 &202 &-173.856 &43 &215 &-174.24 &40 &212 &-176.951 &39 &207 \\
    -177.56 &42 &189 &-177.187 &44 &196 &-176.674 &41 &196 &-179.332 &44 &192 \\
    -190.541 &44 &202 &-194.199 &49 &189 &-191.58 &46 &196 &-192.827 &49 &199 \\
    -194.748 &53 &196 &-200.858 &56 &192 &-197.676 &55 &195 &-197.303 &57 &196 \\
    \hline
  \end{tabular}
\end{center}

We now look at the data from the final simulation. We refit the set of
models under three more scenarios, making the following four
scenarios, in which the K1K2 scenario is that use in the previous
table.

\begin{itemize}
\item[K1K2 scenario] \texttt{RD14-0003-01} genotype is treated as
  known, and \texttt{RD14-0003-02} genotype is treated as known
\item[U1K2 scenario] \texttt{RD14-0003-01} genotype is treated as
  unknown, and \texttt{RD14-0003-02} genotype is treated as known
\item[K1U2 scenario] \texttt{RD14-0003-01} genotype is treated as
  known, and \texttt{RD14-0003-02} genotype is treated as unknown
\item[U1U2 scenario] \texttt{RD14-0003-01} genotype is treated as
  unknown, and \texttt{RD14-0003-02} genotype is treated as unknown
\end{itemize}

\begin{figure}
  \begin{center}
    \includegraphics[scale=0.5]{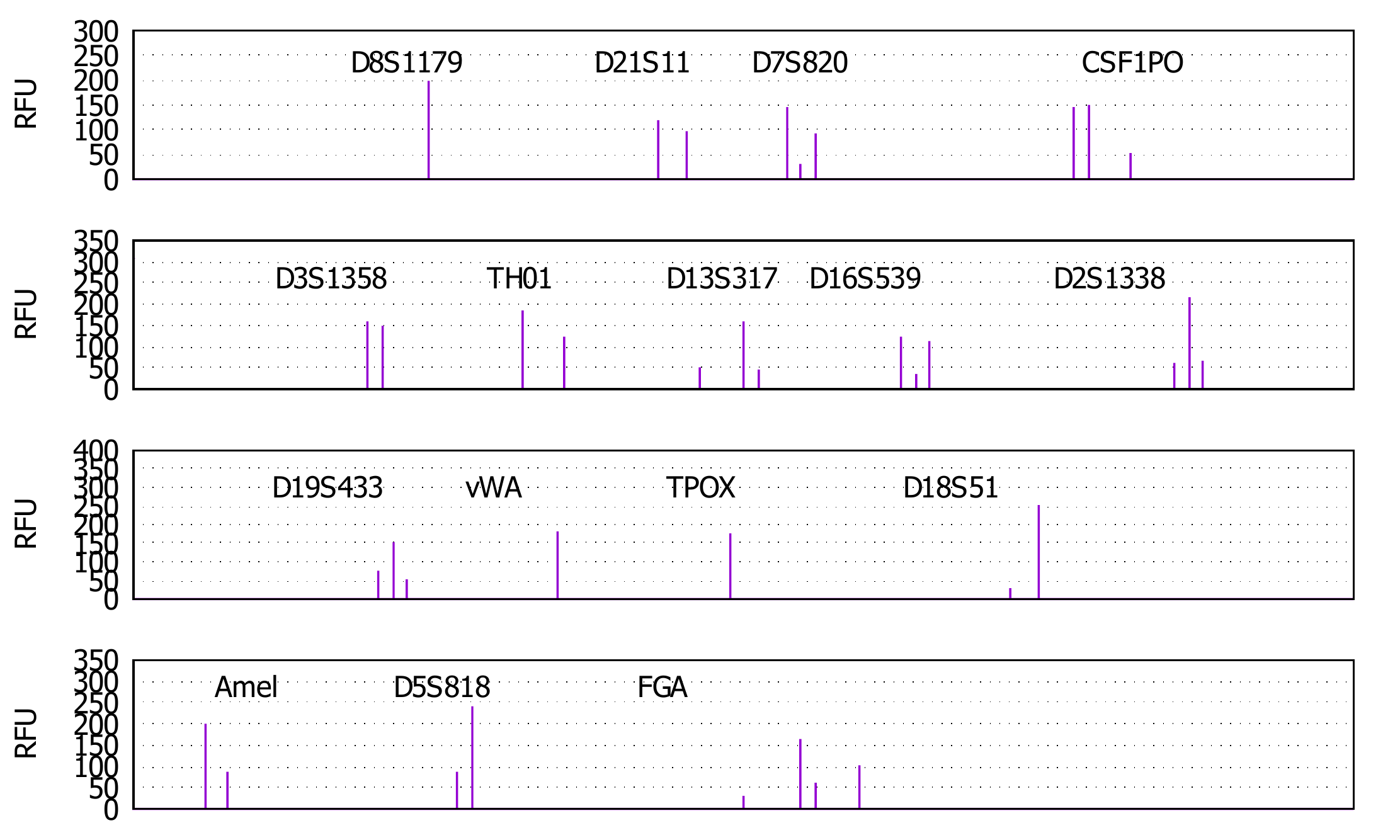}
    \caption{Simulated \EPG for the 1:4 mixture of two persons of
      \tabref{tab:sim2data}. \label{fig:sim2pepg}}
  \end{center}
\end{figure}

The genotypes of the individuals, and the observed peaks using the
analytic threshold of 30 RFUs, are shown in \tabref{tab:sim2data}, and
the simulated \EPG is plotted in \figref{fig:sim2pepg}.  From
\tabref{tab:sim2data} we can see that a number of alleles have dropped
out, and that for some loci there is complete drop-out of the
low-donor contributor.

\begin{table}[htbp]
  \begin{center}
\caption{Genotypes and peaks heights in a two person simulation, with \texttt{RD14-0003-01} having 50 cells, and \texttt{RD14-0003-02}  200 cells\label{tab:sim2data}.}
{\scriptsize
\begin{tabular}{|lcc|llll|} \hline
Locus &\texttt{RD14-0003-01} &\texttt{RD14-0003-02} & Allele/height  & Allele/height  & Allele/height  & Allele/height 
\\ \hline
Amelogenin 	&	 X / X	&	 X / Y	&	X     199	&	Y     88	&		&		\\
CSF1PO 	&	 11 / 12	&	 7 / 8	&	7     147	&	8     148	&	11    55	&		\\
D13S317 	&	 8 / 12	&	 11 / 11	&	8     53	&	11    160	&	12    46	&		\\
D16S539 	&	 12 / 13	&	 11 / 13	&	11    123	&	12    34	&	13    114	&		\\
D18S51 	&	 13 / 15	&	 17 / 17	&	15    31	&	17    251	&		&		\\
D19S433 	&	 14 / 15	&	 13 / 14	&	13    77	&	14    154	&	15    54	&		\\
D21S11 	&	 29 / 31	&	 29 / 31	&	29    118	&	31    96	&		&		\\
D2S1338 	&	 20 / 22	&	 21 / 21	&	20    61	&	21    219	&	22    66	&		\\
D3S1358 	&	 14 / 18	&	 15 / 16	&	15    161	&	16    149	&		&		\\
D5S818 	&	 11 / 12	&	 11 / 12	&	11    89	&	12    240	&		&		\\
D7S820 	&	 9 / 9	&	 8 / 10	&	8     145	&	9     33	&	10    93	&		\\
D8S1179 	&	 12 / 15	&	 13 / 13	&	13    198	&		&		&		\\
FGA 	&	 20 / 25	&	 24 / 28	&	20    33	&	24    163	&	25    60	&	28    103	\\
TH01 	&	 6 / 9.3	&	 6 / 9	&	6     188	&	9     125	&		&		\\
TPOX 	&	 8 / 11	&	 9 / 9	&	9     178	&		&		&		\\
vWA 	&	 17 / 19	&	 15 / 17	&	17    185	&		&		&		\\
\hline
\end{tabular}
}
\end{center}
\end{table}

\begin{table}
  \caption{Fitting each of the four scenarios to each of the four
    models.\label{tab:sim2results}}
\begin{center}
{\scriptsize
\begin{tabular}{|c|ccc|ccc|ccc|ccc|}
\hline
\multicolumn{1}{|c|}{Scenario}&
\multicolumn{3}{|c|}{Normal} & \multicolumn{3}{c|}{Logormal} & \multicolumn{3}{c|}{Gamma} & \multicolumn{3}{c|}{\FFT}  \\
&$\widehat{LL}_{max}$ & $\hat{C}_1$ & $\hat{C}_2$  &$\widehat{LL}_{max}$ &  $\hat{C}_1$ & $\hat{C}_2$ &$\widehat{LL}_{max}$ &  $\hat{C}_1$ & $\hat{C}_2$ &$\widehat{LL}_{max}$ &  $\hat{C}_1$ & $\hat{C}_2$  \\ \hline
K1K2 &-194.748 &53 &196 &-200.858 &56 &192 &-197.676 &55 &195 &-197.303 &57 &196 \\
U1K2 &-209.557 & 57 & 189 & -215.185 & 67 & 185 &  -212.8 & 62 & 189 & -212.005 & 63 & 191 \\
K1U2 &-237.658 & 55 & 193 & --239.057& 55 & 200 &  -238.359 & 54 & 199 &  -239.642 & 58 & 198\\
U1U2 &-250.112 &  54 & 192 & -250.986 & 58 & 195 &  -250.658 & 55 & 196 & -251.555 &  59 & 195\\
\hline
\end{tabular}
}
\end{center}
\end{table}

The results of fitting each of the models to each of the four
scenarios is shown in \tabref{tab:sim2results}. We see that for each,
the models give similar results, even with the considerable allelic
drop-out that is taking place.  Note also that the table may be used
to find for prosecution to defence likelihood ratios.  Suppose that
the prosecution case is that \texttt{RD14-0003-01} is a contributor,
and the defence hypothesis that \texttt{RD14-0003-01} is not. Then the
log-likelihood ratio in favour of the prosecution hypothesis will be
$( -194.748 - (-209.557))/\log(10) = 6.43$ Bans.
\tabref{tab:sim2ratioresults} shows various combinations of
log-likelihood ratios.  Notice that the values are higher for the last
two rows, which are hypotheses concerning the presence of the major
contributor, than equivalents hypothesis comparisons in the first two
rows concerning the presence of the minor contributor.  This is to
anticipated. What is perhaps surprising are the quite high values in
the first two rows, concerning the presence of the minor contributor,
given the very low template DNA from  the minor contributor.

\begin{table}
  \caption{Various log-likelihood ratios for possible prosecution vs
    defence hypothesis concerning the presence of one of the
    individuals, for each of the four models: values given are
    log-likelihood ratios expressed in
    Bans.\label{tab:sim2ratioresults}}
  \begin{center}
    \begin{tabular}{|c|cccc|}\hline
      Hypotheses & Normal & Lognormal & Gamma & \FFT \\ \hline
      K1K2 vs  U1K2	&6.43			&6.22		&	6.57		&	6.38\\
      K1U2 vs U1U2	&5.41			&5.18	&		5.34	&		5.17\\
      K1K2 vs K1U2		&18.64			&16.59	&		17.67	&		18.39\\
      U1K2 vs U1U2	&17.61			&15.55	&		16.44	&		17.18\\
      \hline
    \end{tabular}
  \end{center}
\end{table}

\section{Application to sample data}
\label{sec:budata}

This section describes the performance of the model above to the
publicly available DNA dataset from the PROVEDIt Initiative
\citep{alfonse2016development,alfonse2018large}. This is a very large
dataset of laboratory controlled single source and mixed DNA samples,
with amplifications carried out for the Identifiler
Plus\texttrademark, PowerPlex16H\texttrademark\ and
Globalfiler\texttrademark\ kits. From it we shall use the sets of
samples for the Identifiler Plus\texttrademark\ kit that were
amplified for 28 cycles. We shall use the processed data in the EXCEL
files available from the PROVEDIt Initiative website, rather than the
raw fsa files that are also available. The EXCEL files contain allele
calls (including O/L designations) in Genemapper output format, in
which an analytic threshold has been set to 1 \RFU. Although these
files contain artefacts such as split-peaks and dye-blobs, and the
PROVEDIt Initiative provides an EXCEL spreadsheet to help remove
these, they have been left in so that the data can be processed in
batch automatically. Hence our analyses presented here are using quite
noisy data.

\subsection{Calibration of model parameters to the data}

The samples from the PROVEDIt Initiative were prepared from the
dilution of high-density extracted DNA. Hence a Poisson model for the
amount of DNA would be appropriate. However, we use a binomial model
as presented so far, as this is more appropriate for real life
samples.\footnote{Essentially we are approximating Poisson
  distributions by binomial distributions, usually the approximation
  is the other way around.} We assume an extraction efficiency of
$\psi = 0.3$, a sample volume of 25$\mu$L of which 10$\mu$L is used in
the amplification, hence $\pi_f = 0.06$.  We also used a drop-in rate
per locus of 0.021.

\begin{figure}
  \begin{center}
    \includegraphics[scale=0.75]{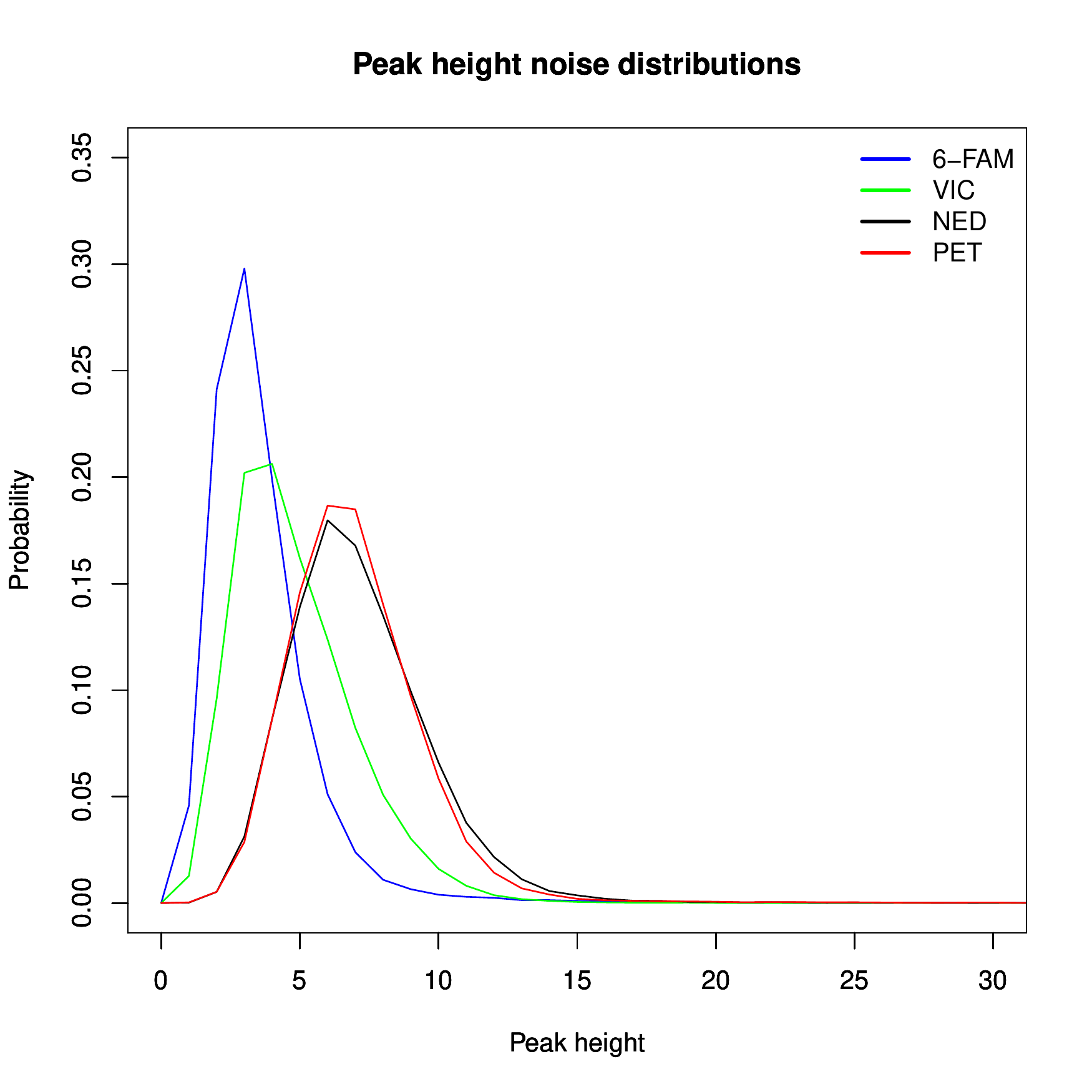}
    \caption{Empirical noise distributions for each dye lane for
      analyzing the PROVEDIt Initiative data for the
      Identifiler-Plus\texttrademark\ kit. The long tails going out to
      100 {\RFU}s is omitted.\label{fig:idnoise}}
  \end{center}
\end{figure}

To estimate the noise distribution, the peak heights of all
\textit{off-ladder} (O/L) designated alleles in the high DNA single
source samples were extracted and treated as empirical distributions
for each lane - each lane distribution is based on over 40,000
values. A cut-off of 100 {\RFU}s was used for the noise distribution.
\figref{fig:idnoise} shows the noise distributions for each of the
four dyes, omitting the long tail up to 100 {\RFU}s.

To form the \EPG data files, we retained for each sample all of the
alleles that were called, that is, those that were given an allele
designation that was not given as O/L.

Amplification probabilities were assigned as follows. It is assumed
that all strand probabilities in any given locus are the same, that is
$p_g = p_{g_d} = p_h = p_{h_d} = p_a = p_{a_d} = p$.  From the
high-template DNA single source, non-degraded, samples, in each sample
the peak heights of called alleles were added together. These total
peak height values were scaled by the estimated amount of initial DNA
in each sample (this information is provided in the name of each
sample). The mean peak heights of every locus in the kit was then
found. It was found that TPOX had the largest mean value.  This locus
was - somewhat arbitrarily - assigned an amplification probability of
0.85 for all types of strand. Using the fact that the number of cycles
was 28, and that the mean number of amplicons from a single strand is
approximately $(1+p)^{28}$, amplification probabilities for the other
loci were found so that their mean values scaled correctly according
to their empirical mean when compared to the mean peak height of TPOX.

Conditional stutter probabilities were either assigned values of 0.001
for forward and double stutter and 0.004 for single stutter, except
for the loci in the allelic ladder of the kit where mean stutter
ratios were available from the manufacturer's literature,
\citep{scientific2012ampflstr} in which case somewhat crude estimates
were obtained from the plots of stutter ratio versus allele lengths to
obtain stutter probabilities that matched in mean; forward and double
stutter probabilities for the alleles were obtained from these by
division by 4.

A scale factor of two million was used for all lanes to convert the
number of tagged amplicons to an \RFU equivalent value.

We shall focus on the samples that were amplified for 5 seconds, and
for these use an analytic threshold of 15 {\RFU}s.

With all of these parameters set, and using the laboratory estimated
amount in each sample, maximum likelihood estimation was used to
estimate the degradation and initial cell counts of each hypothesised
contributor.

\subsection{Analysis of a  two person sample}
From the two-person data the author extracted the profiles of 189 samples.
These two person mixtures were treated as 3 person mixtures in
analysis.  The profiles are discussed here, and in later sections use
the naming convention of the PROVEDIt Initiative datasets.

\noindent
\textbf{Sample
  \texttt{E02\_RD14-0003-42\_43-1\_9-M2S10-0.15IP-Q1.0\_001.5sec}}

\begin{figure}
  \begin{center}
    \includegraphics[scale=0.75]{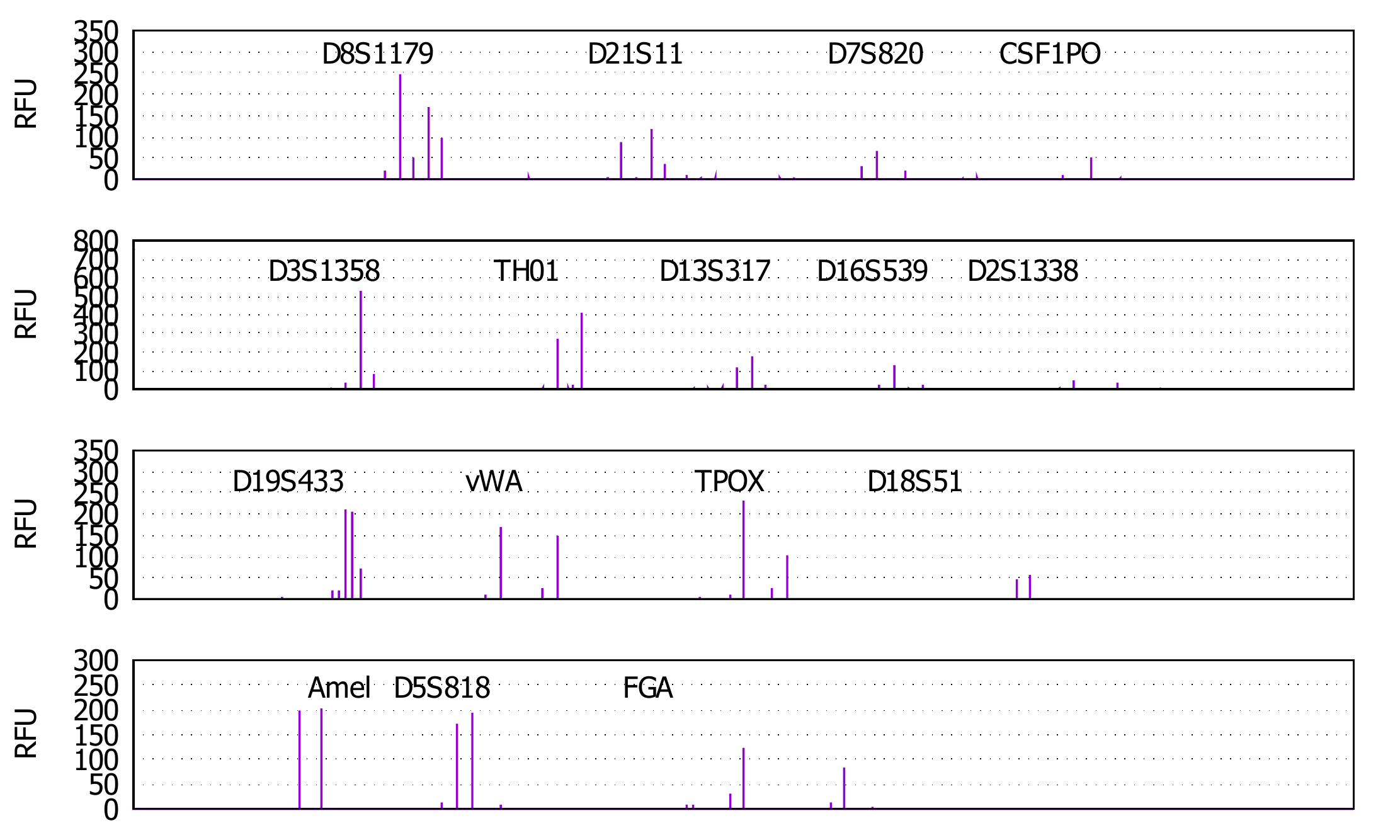}
    \caption{Schematic \EPG plot for the
      \texttt{E02\_RD14-0003-42\_43-1\_9-M2S10-0.15IP-Q1.0\_001.5sec}
      sample.\label{fig:buexample1}}
  \end{center}
\end{figure}

This two sample mixture was prepared in 1 ratio of 1:9, and subject to
10 seconds of sonification to simulate degradation. The target amount
of DNA was 0.15ng. With our use of $\phi =0.06$, this corresponds to
approximately $150/0.06 = 2500pg$ of DNA initially, that is around 378
cells in total, with around 22 cells worth in the amplified
sample. This is a highly degraded very low template sample, we
therefore omit the double and forward stutter components when fitting
the models: the \EPG plot is shown in \figref{fig:buexample1};

We first analyse this sample under the hypothesis that both of the
actual contributors are contributors, and an additional untyped
contributor is assumed to be present as well. Maximized likelihoods
and estimated cell amounts are shown in \tabref{tab:buexample1}. The
degradation is estimated on a discretized scale, and in this case all
four models have estimated the same degradation value.

Of particular note is that all models have estimated the cell amount
for the extra untyped contributor to be 0 - which is in agreement with
the manner of the preparation of the two-person mixture. The forensic
science literature contain much discussion concerning how to estimate
the number of contributors to a mixture(see for example
\citep{haned2011estimating,lauritzen2002bounding,egeland2003estimating,
  swaminathan2015nocit}).  In the models presented here the maximum
likelihood estimates can return 0 for cell counts, because the cell
amounts beign estimated are discrete integers. This is in marked contrast to other
continuous peak height models that model contributor amounts or
relative amounts by continuous variables, for which the nesting of
models by adding extra hypothesised contributors will lead to
increasing maximized likelihoods. \citep{cowell2015analysis}

\begin{table}
  \begin{center}
    \caption{Maximum likelihood estimates for the
      \texttt{E02\_RD14-0003-42\_43-1\_9-M2S10-0.15IP-Q1.0\_001.5sec}
      sample.\label{tab:buexample1}}
    \begin{tabular}{|l|ccccc|}
      \hline
      Model & $\widehat{LL}_{max}$ & $\hat{C}_1$ & $\hat{C}_2$  & $\hat{C}_3$ &  $\hat{\delta}$\\ \hline
      Normal & -236.85 & 72 & 343 & 0 &  0.00504874\\
      Lognormal &-240.971 & 64 & 337 & 0 &  0.00504874\\
      Gamma & -237.857 &67 & 343 & 0 &  0.00504874\\
      \FFT &-235.605 & 71 & 347 & 0 &  0.00504874 \\
      \hline
    \end{tabular}
  \end{center}
\end{table}

\begin{figure}
  \begin{center}
    \includegraphics[scale=0.75]{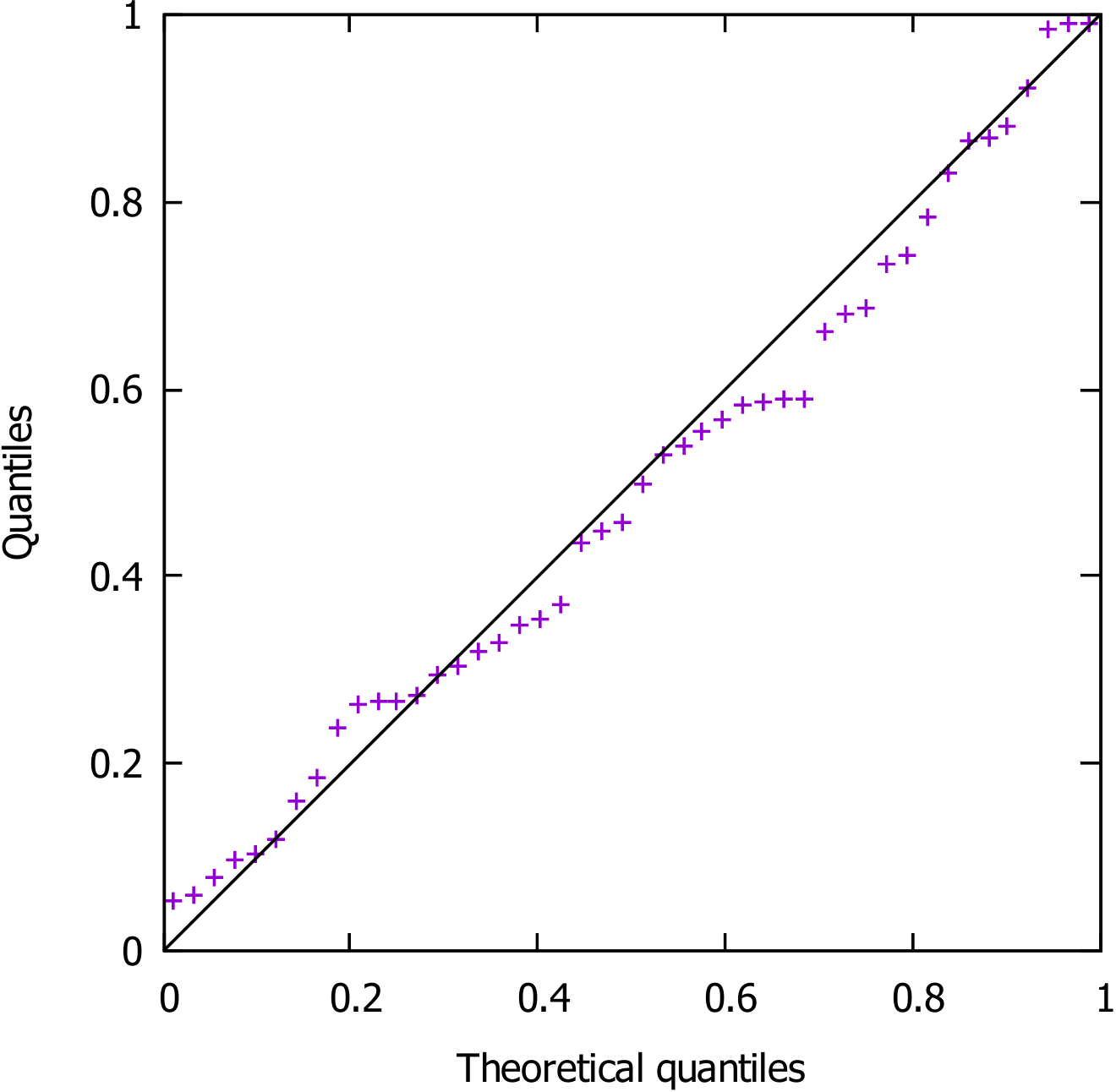}
    \caption{QQ-plot for the \FFT model for the
      \texttt{E02\_RD14-0003-42\_43-1\_9-M2S10-0.15IP-Q1.0\_001.5sec}
      sample.\label{fig:buexample1qqplot}}
  \end{center}
\end{figure}

\citep{graversen2015computational,cowell2015analysis} introduced
various statistically based diagnostics to check how well a model fits
the profile data. One such diagnostic is a QQ-plot in which for each
allelic peak $h_a$ in the set $\cal H$ of peaks observed above the
threshold $T$ the quantities
$P( H< h_a\cd H > T, h_b\in {\cal H}, b\ne a)$ are calculated. If the
models is `true', then these values should follow a uniform
distribution, so that a plot of these sorted values should follow a
straight line when plotted against the quantiles of the uniform
distribution. \figref{fig:buexample1qqplot} shows the QQ-plot for the
\FFT model, which is the model with the highest likelihood in
\tabref{tab:buexample1}: we see that the fit is very good.

\begin{figure}
  \begin{center}
    \includegraphics[scale=0.55]{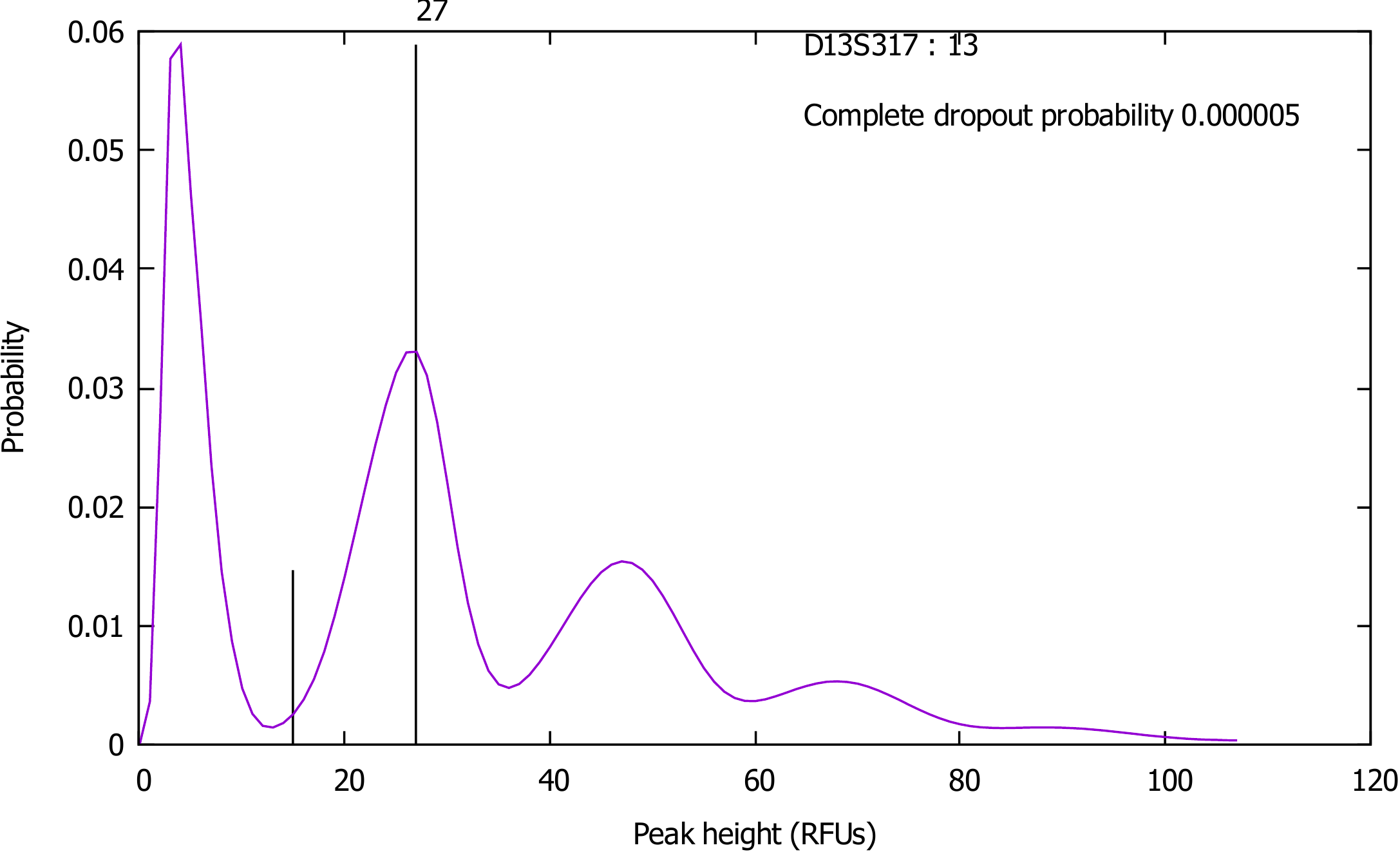}
    \includegraphics[scale=0.55]{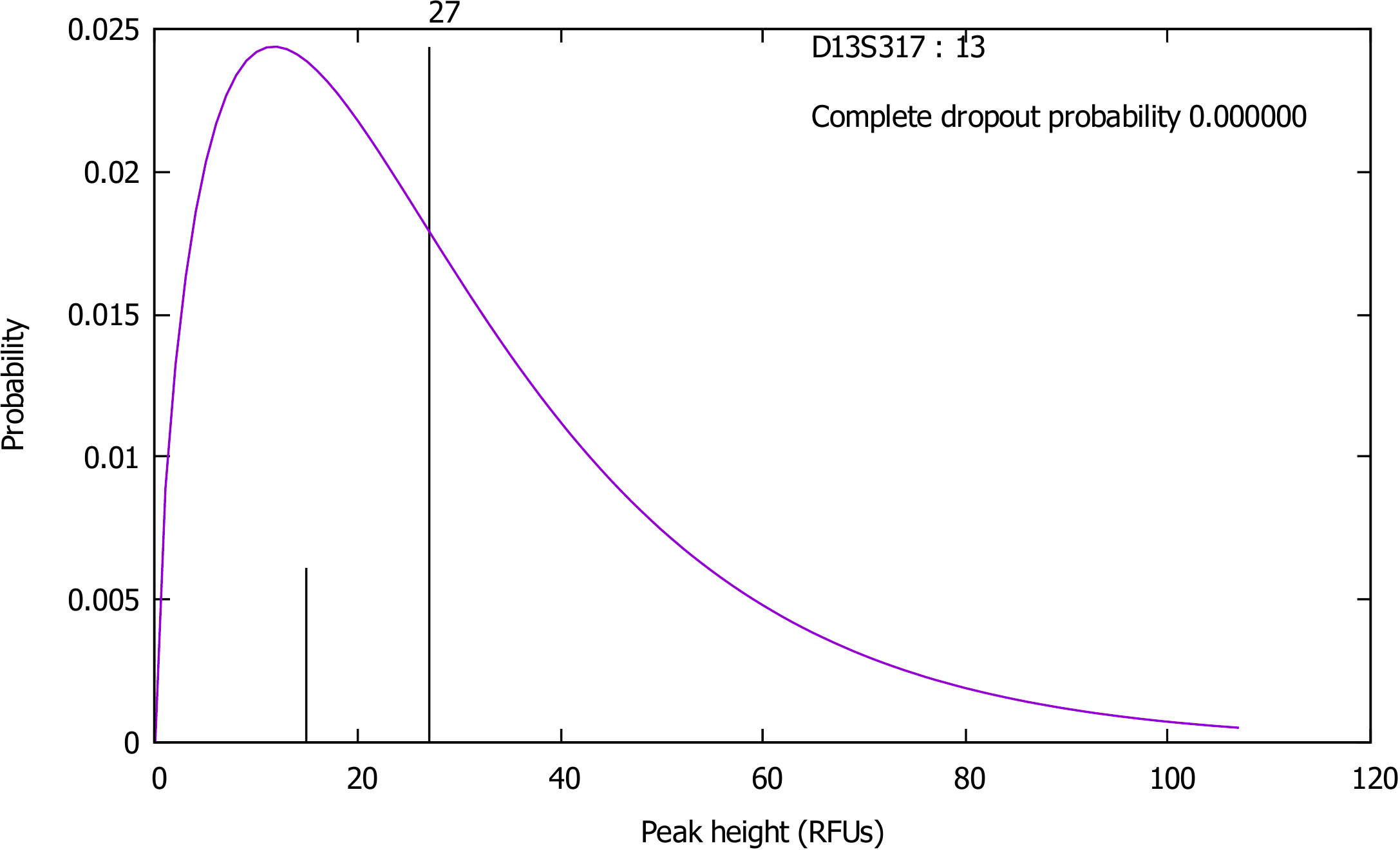}
    \caption{Predictive distributions for allele 13 of locus D13S317,
      for the
      \texttt{E02\_RD14-0003-42\_43-1\_9-M2S10-0.15IP-Q1.0\_001.5sec}
      sample assuming the genotypes of the true contributors and a
      third untyped contributor. Top plot, the \FFT model, bottom plot
      the (moment matching) Gamma model. The first vertical line at 15 RFUs
      is the location of the analytic threshold, the second
      at 27 RFUs is the observed peak height.\label{fig:buexample1dist}}
  \end{center}
\end{figure}

As an example of the predictive distribution of a peak, we take the
allele 13 of locus D13S317. For that locus, three peaks were observed
above the threshold of 15 {\RFU}s: allele 11 with a peak height of
117, allele 12 with a height of 177 and allele 13 with a height of 27.
Contributor \texttt{RD14-0003-42} has genotype $(12/13)$ on locus
D13S317, and \texttt{RD14-0003-43} has genotype $(11/12)$. Thus the 13
peak can arise from the minor contributor \texttt{RD14-0003-42}
(recall that we are not including forward stutter in this analysis) or
drop-in.  From the \FFT estimates in \tabref{tab:buexample1}, there are an estimated 71 initial
genomic strands of this allele type, sampled intact with probability
0.06 for amplification, hence an estimated mean of around 4.26
amplify-able strands .  \figref{fig:buexample1dist} shows the
predictive distribution for this allele conditional on the observed
peak heights and estimated cell amounts. The first peak on the left is
the baseline-noise distribution for VIC lane (compare it to
\figref{fig:idnoise}).  The plot is giving a clear signal that just
one amply-able strand of this allele was in the aliquot minitube when
the \PCR was carried out. For comparison, the distribution obtained
from using the Gamma model is also shown in
\figref{fig:buexample1dist}.

\begin{figure}
  \begin{center}
    \includegraphics[scale=0.55]{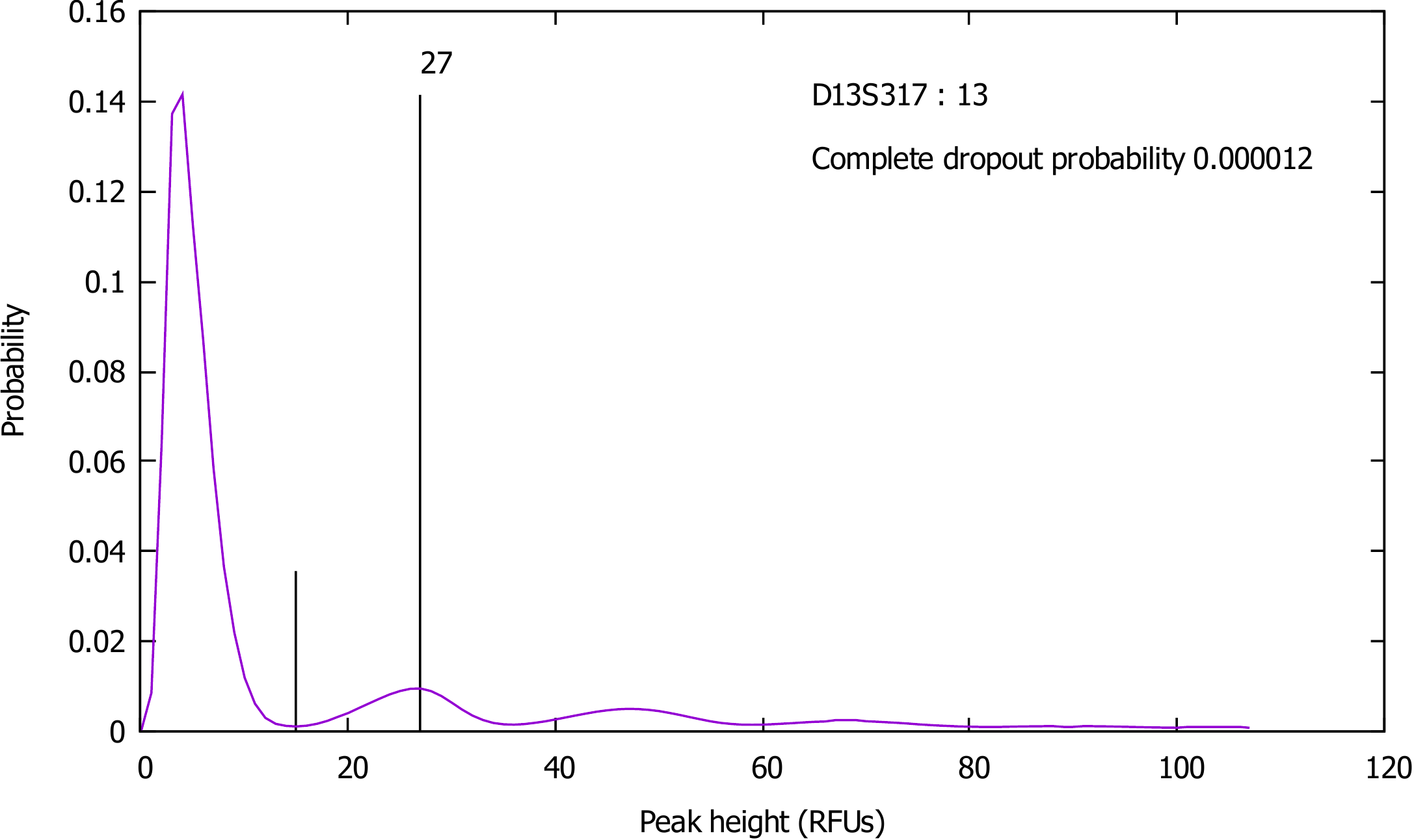}
    \includegraphics[scale=0.55]{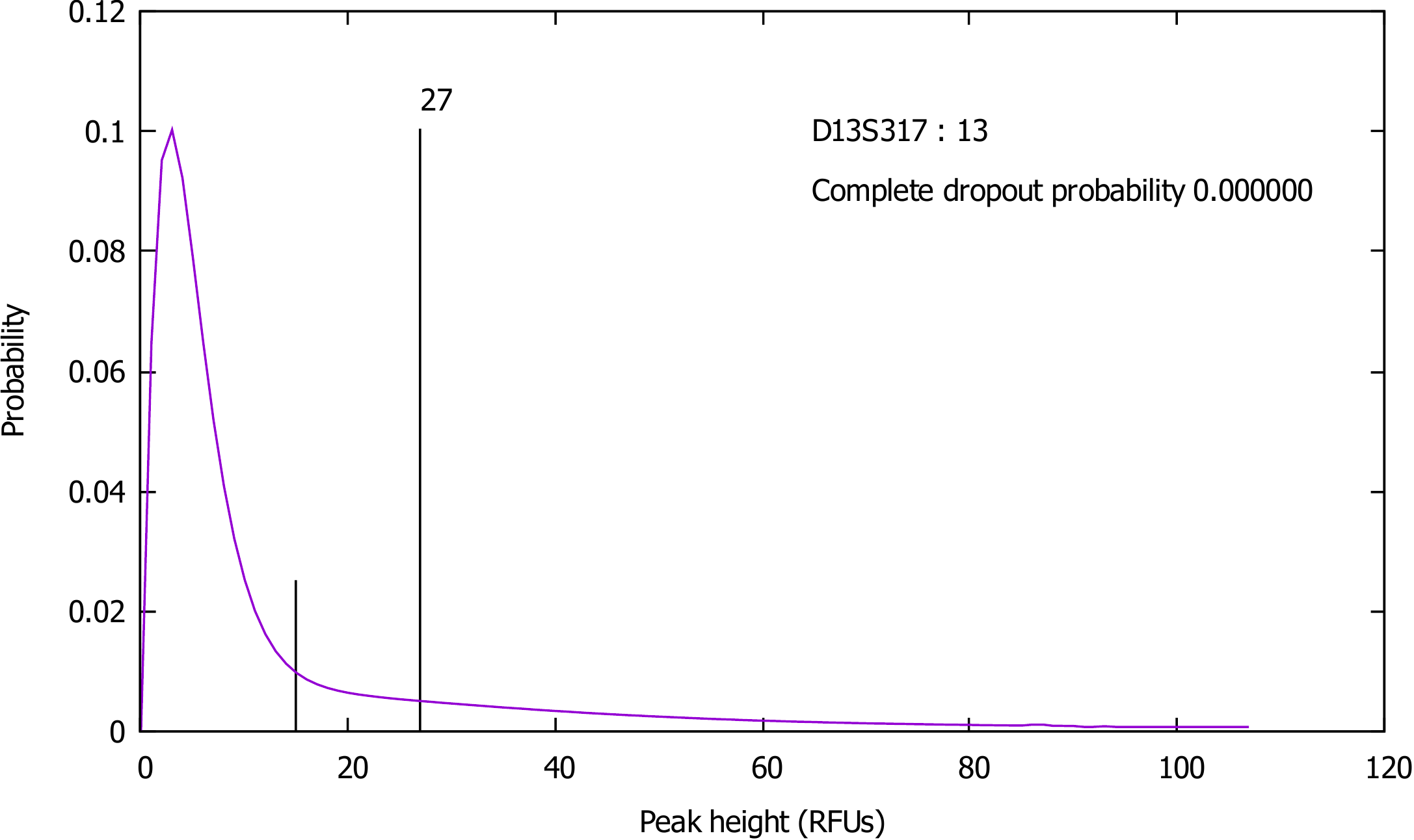}
    \caption{Predictive distributions for allele 13 of locus D13S317,
      for the
      \texttt{E02\_RD14-0003-42\_43-1\_9-M2S10-0.15IP-Q1.0\_001.5sec}
      sample assuming three contributors of unknown genotype. Top
      plot, the \FFT model, bottom plot the  (moment matching) Gamma
      model.\label{fig:buexample1dist3u}}
  \end{center}
\end{figure}

\tabref{tab:buexample1w3u} shows estimates for each model analyzed
under the assumption that the mixture is of three untyped persons. The
gamma model has the highest likelihood, and now the normal model is
giving a non-zero cell count estimate for all three persons.
Distribution plots for allele 13 of locus D13S317 are given in
\figref{fig:buexample1dist3u}.

\begin{table}
  \begin{center}
    \caption{Maximum likelihood estimates for the
      \texttt{E02\_RD14-0003-42\_43-1\_9-M2S10-0.15IP-Q1.0\_001.5sec}
      sample assuming 3 untyped
      contributors.\label{tab:buexample1w3u}}
    \begin{tabular}{|l|ccccc|}
      \hline
      Model & $\widehat{LL}_{max}$ & $\hat{C}_1$ & $\hat{C}_2$  & $\hat{C}_3$ &  $\hat{\delta}$\\ \hline
      Normal &   -297.752 & 90 & 370 & 9 & 0.00567983 \\
      Lognormal & -297.842 &  79 & 388 & 0 & 0.00567983\\
      Gamma &  -297.56 & 78 & 391 & 0 & 0.00567983\\
      \FFT & -298.815 & 71 &  347 & 0 & 0.00567983\\
      \hline
    \end{tabular}
  \end{center}
\end{table}

\clearpage
\subsection{Analysis of a three person sample}
\label{sec:bu3person}
\begin{figure}[htbp]
  \begin{center}
    \includegraphics[scale=0.55]{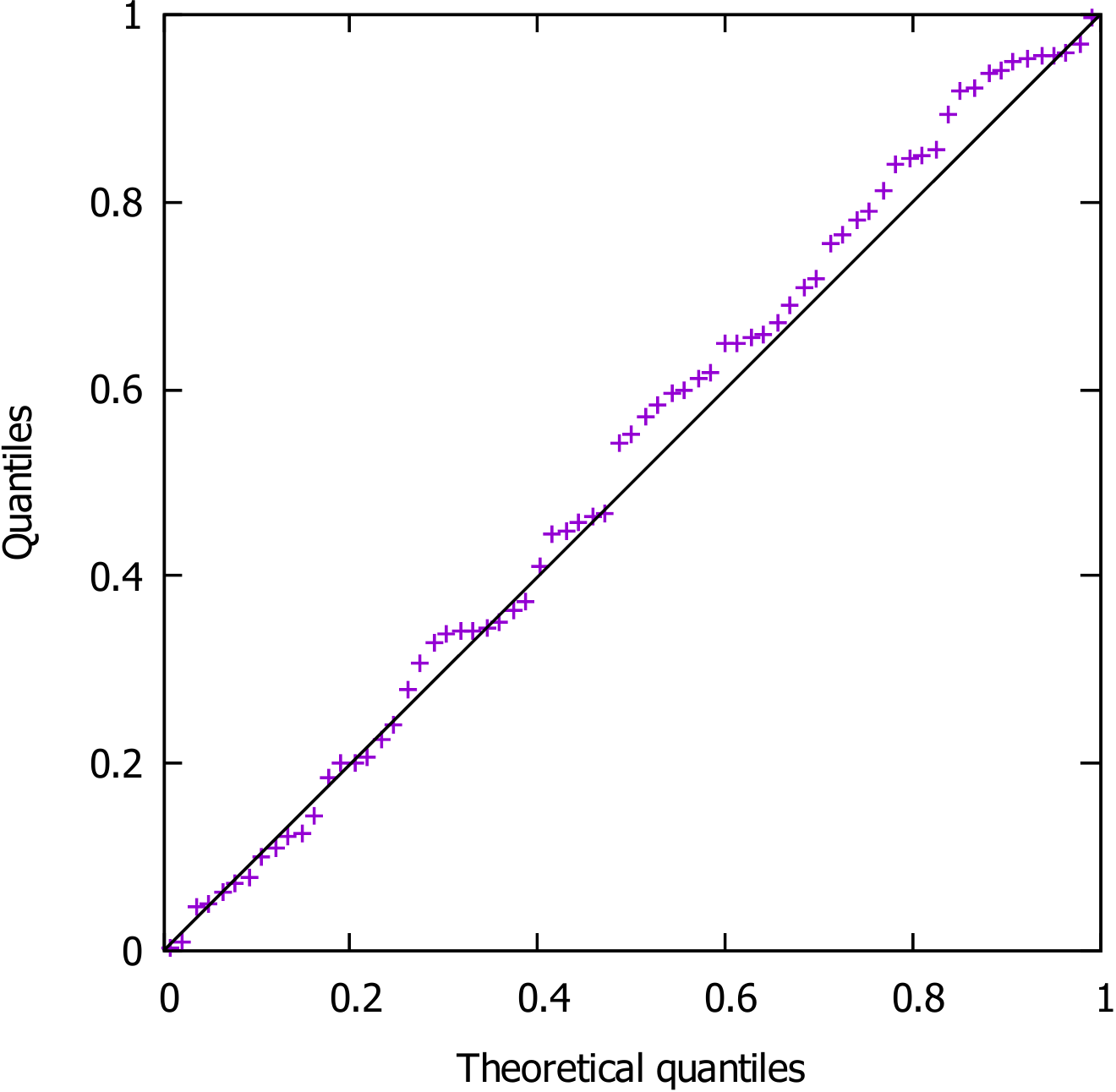}
    \caption{QQ-plot for the \FFT model for the
      \texttt{A06\_RD14-0003-47\_48\_49-1\_9\_9-M2c-0.589IP-Q1.1\_001.5sec}
      sample, with stutter but no double or forward
      stutter.\label{fig:buexample2fftqqplot}}
  \end{center}
\end{figure}

The sample
\texttt{A06\_RD14-0003-47\_48\_49-1\_9\_9-M2c-0.589IP-Q1.1\_001.5sec}
has contributor amounts in the proportions of 1:9:9, with a medium
level of degradation cause by rDNase I enzyme added to the
extract. Analyzing as a four person mixture, assuming as knowns the
genotypes of the three contributors and one extra untyped individual,
and ignoring forward and double stutter, we obtain the estimates in
\tabref{tab:buexample3w3k}.  The
\FFT model correctly estimate zero cells for the untyped contributor,
and the \FFT model gives the highest likelihood.  The estimated
mixture proportions are in line with the experimentally prepared
ratios for all four models.  The QQ-plot for the \FFT model is shown
in \figref{fig:buexample2fftqqplot}.

\begin{table}[htbp]
  \begin{center}
    \caption{Maximum likelihood estimates for the
      \texttt{A06\_RD14-0003-47\_48\_49-1\_9\_9-M2c-0.589IP-Q1.1\_001.5sec}
      sample assuming the genotypes of the three contributors, plus a
      fourth of unknown genotype.\label{tab:buexample3w3k}}
    \begin{tabular}{|l|cccccc|}
      \hline
      Model & $\widehat{LL}_{max}$ & $\hat{C}_1$ & $\hat{C}_2$  & $\hat{C}_3$ &  $\hat{C}_4$ & $\hat{\delta}$\\ \hline
      Normal &    -384.74148962 & 73 & 687 & 718 & 2 &  0.00490963394042 \\
      Lognormal &   -386.801221118 & 67 & 693 & 682 &  2&0.00490963394042\\
      Gamma &   -384.573053649 & 68 & 693 & 695 & 2 & 0.00490963394042\\
      \FFT &  -382.407751392 & 62 & 622 & 639 & 0 & 0.00490963394042\\
      \hline
    \end{tabular}
  \end{center}
\end{table}

\begin{figure}[ht]
  \begin{center}
    \begin{minipage}{.5\textwidth}
      \centering
      \includegraphics[scale=0.4]{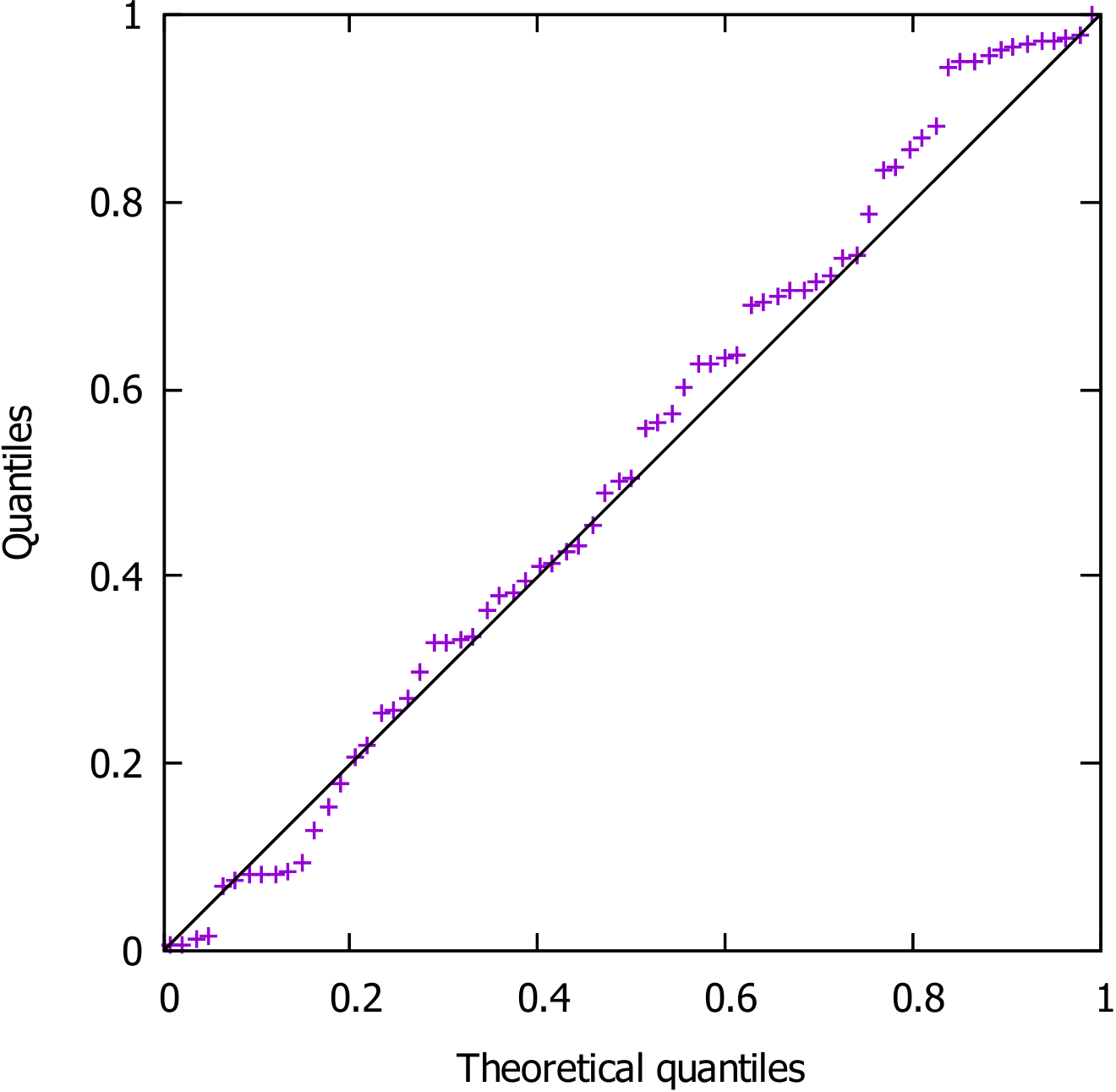}
    \end{minipage}%
    \begin{minipage}{.5\textwidth}
      \centering
      \includegraphics[scale=0.4]{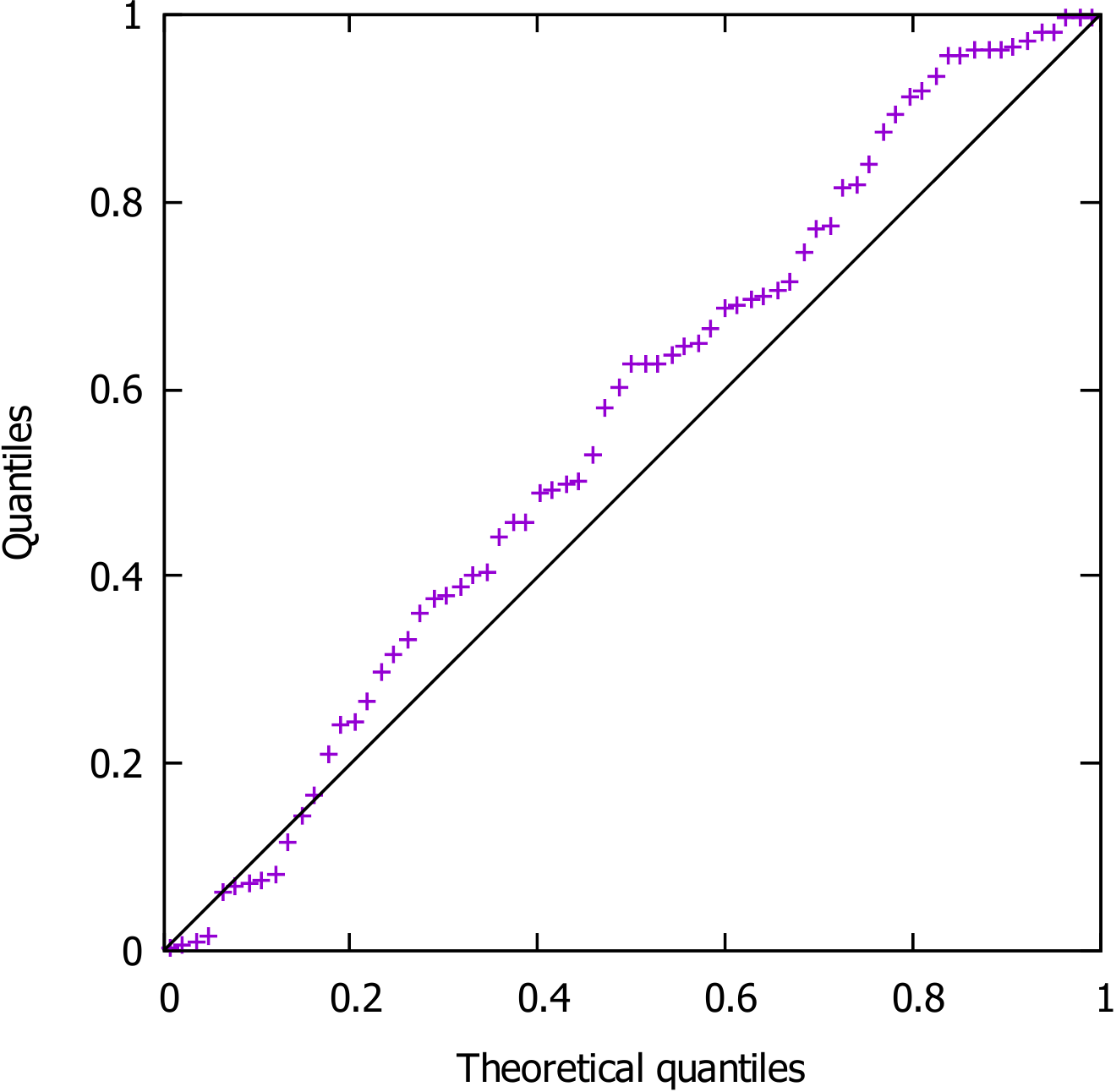}
    \end{minipage}%
    \caption{QQ-plots
      the\texttt{A06\_RD14-0003-47\_48\_49-1\_9\_9-M2c-0.589IP-Q1.1\_001.5sec}
      sample, when forward and double stutter are included. The Normal
      model (left), and the \FFT model
      ((right).\label{fig:buexample2normalqqplot}}
  \end{center}
\end{figure}

\clearpage
If we include forward and double stutter in the models, we obtain the
estimates obtained in \tabref{tab:buexample3w3kfd}. We see that the
\FFT model has become out-of-line with the other models in having a
much lower likelihood, and much lower estimates for the number of
cells from the minor contributor.  The normal model is now the one
having the highest likelihood: QQ-plots for the Normal and \FFT model
are shown in \figref{fig:buexample2normalqqplot}.
\begin{table}
  \begin{center}
    \caption{Maximum likelihood estimates for the
      \texttt{A06\_RD14-0003-47\_48\_49-1\_9\_9-M2c-0.589IP-Q1.1\_001.5sec}
      sample assuming the genotypes of the three contributors, plus a
      fourth of unknown genotype, with estimation include stutter,
      double stutter and forward stutter
      distributions.\label{tab:buexample3w3kfd}}
    \begin{tabular}{|l|cccccc|}
      \hline
      Model & $\widehat{LL}_{max}$ & $\hat{C}_1$ & $\hat{C}_2$  & $\hat{C}_3$ &  $\hat{C}_4$ & $\hat{\delta}$\\ \hline
      Normal &     -395.225614561 & 59 & 591 & 615 & 0 &  0.00436411905815\\
      Lognormal &  -401.337705571 & 45 & 587 & 584 & 4 & 0.00436411905815\\
      Gamma &  -394.142509492 & 53 & 595 & 604 & 3 &  0.00436411905815\\
      \FFT &  -411.894302982 & 17 & 599 & 625 & 0 &  0.00436411905815\\
      \hline
    \end{tabular}
  \end{center}
\end{table}

\section{A modification of the \FFT model to correct for correlations}

We have seen that in some cases we can improve the \FFT model by
removing the forward and double back-stutters; however this will not
be appropriate for all cases.  Here we introduce an adjustment to the
\FFT model that puts back some of the correlation that was discarded
in the factorization approximations of \eqref{eq:likeapprox1} and
\eqref{eq:likeapprox2}; we denote the new model by \mFFT.

Recall that the point of making those approximations was purely to
make the computations tractable, by enabling the peak height
likelihood to factorize into a product of convolutions of univariate
probability distributions of individual alleles. However, as pointed
out earlier, a side effect of this approximation is that the complete
dropout probability for an allele is included as a factor multiple
times in the likelihood. The modification here aims to reduce such
multiple factors.

For any given allele $a$, there will up to 6 distributions to convolve
in finding its peak height likelihood contribution:

\begin{enumerate}
\item The base-line noise distribution
\item The drop-in distribution
\item The peak height distribution arising from genomes of type $a$
\item The peak height distribution arising from stutter from genomes
  of type one repeat larger than $a$
\item The peak height distribution arising from double stutter from
  genomes of type two repeats larger than $a$
\item The peak height distribution arising from forward stutter from
  genomes of type one repeat smaller than $a$
\end{enumerate}

In the new approximation we always retain the first three
distributions. Now consider the contribution from say stutter from one
repeat higher. Suppose that no peak above threshold is observed at the
one-repeat higher position. This means that any genomic material of
allelic type one-repeat higher than $a$ has dropped out. In this case,
we can therefore omit the marginal stutter peak height distribution in
the convolution as it is extremely likely that any stutter product
would be too low to be observed.  However, given that stutter peaks
are typically around 5-15\% of the target peak producing the stutter,
this means that even if a peak is observed at one-repeat higher than
$a$, but is below a small multiple of the analytic threshold then we
can assume that any stutter contribution it makes to the allele $a$
will not be seen.  A similar argument applies also to the double
stutter and forward stutter distributions---more so as these tend to
have smaller stutter ratios than single stutter peaks.

Hence the modification is as follows. In evaluating the peak height
likelihood for an allele $a$, when forming the marginal peak height
distribution:
\begin{itemize}
\item Always include the base-line noise distribution.
\item Always include the drop-in distribution.
\item Always include the peak height distribution arising from genomes
  of type $a$.
\item Only include the peak height distribution arising from stutter
  from genomes of type one repeat larger than $a$ if there is a peak
  above three times the threshold at that position.
\item Only include the peak height distribution arising from double
  stutter from genomes of type two repeats larger than $a$ if there is
  a peak above three times the threshold at that position.
\item Only include the peak height distribution arising from forward
  stutter from genomes of type one repeat smaller than $a$ if there is
  a peak above three times the threshold at that position.
\item Convolve all of the included distributions to obtain the
  marginal peak height distribution to use for finding the likelihood
  for the peak of the allele $a$.
\end{itemize}

A factor different to three for the multiple of the analytic threshold
could be used---further research is required to find the best factor
to use.

\subsection{Revisiting the two person simulation of
  \secref{sec:sim2pers}}

The following table shows the model fits obtained using the \mFFT
model for the simulation with $C_1 = 50$ cells and $C_2 = 200$, which
should be compared to the second table in \secref{sec:sim2pers}: we
see that the \mFFT model is now much more in line with the other
models both in its maximum likelihood values and the cell estimates of
the two contributors.

\begin{center}
  \begin{tabular}{|ccc|ccc|ccc|ccc|}
    \hline
    \multicolumn{3}{|c|}{Normal} & \multicolumn{3}{c|}{Logormal} & \multicolumn{3}{c|}{Gamma} & \multicolumn{3}{c|}{\mFFT}  \\
    $\widehat{LL}_{max}$ & $\hat{C}_1$ & $\hat{C}_2$  &$\widehat{LL}_{max}$ &  $\hat{C}_1$ & $\hat{C}_2$ &$\widehat{LL}_{max}$ &  $\hat{C}_1$ & $\hat{C}_2$ &$\widehat{LL}_{max}$ &  $\hat{C}_1$ & $\hat{C}_2$  \\ \hline
    -181.783 &41 &239 &-180.183 &40 &237 &-179.908 &39 &238 &-180.334 &38 &237 \\
    -193.674 &50 &193 &-192.84 &47 &193 &-191.837 &46 &195 &-191.74 &47 &199 \\
    -186.171 &42 &209 &-190.008 &51 &212 &-188.724 &47 &213 &-188.69 &44 &215 \\
    -193.501 &50 &199 &-190.686 &54 &200 &-191.185 &52 &200 &-193.364 &52 &202 \\
    -188.659 &48 &200 &-191.771 &46 &204 &-189.964 &47 &204 &-188.564 &47 &204 \\
    -201.354 &56 &202 &-200.354 &60 &196 &-199.818 &57 &200 &-205.226 &61 &202 \\
    -172.509 &34 &201 &-173.196 &41 &215 &-173.62 &38 &212 &-172.733 &34 &210 \\
    -177.388 &41 &189 &-176.959 &42 &196 &-176.441 &40 &195 &-176.003 &39 &195 \\
    -190.686 &44 &201 &-194.281 &48 &188 &-191.67 &45 &195 &-190.566 &44 &200 \\
    -194.286 &52 &196 &-200.514 &54 &192 &-197.259 &53 &195 &-195.177 &51 &199 \\
    \hline
  \end{tabular}
\end{center}

\subsection{Revisiting the three person mixture of
  \secref{sec:bu3person} from the PROVEDit Initiative}

Re-analyzing
\texttt{A06\_RD14-0003-47\_48\_49-1\_9\_9-M2c-0.589IP-Q1.1\_001.5sec},
with both forward and double stutter included using the \mFFT model, we
obtain a maximized log-likelihood of -405.944809925, and estimated
cell counts of $\hat{C}_1 = 31$, $\hat{C}_2 = 596$, $\hat{C}_3 = 617$
and $\hat{C}_4 = 0$, less extreme than the estimates in
\tabref{tab:buexample3w3kfd}. The corresponding QQ-plot is little
changed from \figref{fig:buexample2normalqqplot} and is omitted.

\subsection{Another 2-person mixture from the PROVEDit datatset}

\begin{figure}[htbp]
  \begin{center}
    \includegraphics[scale=0.75]{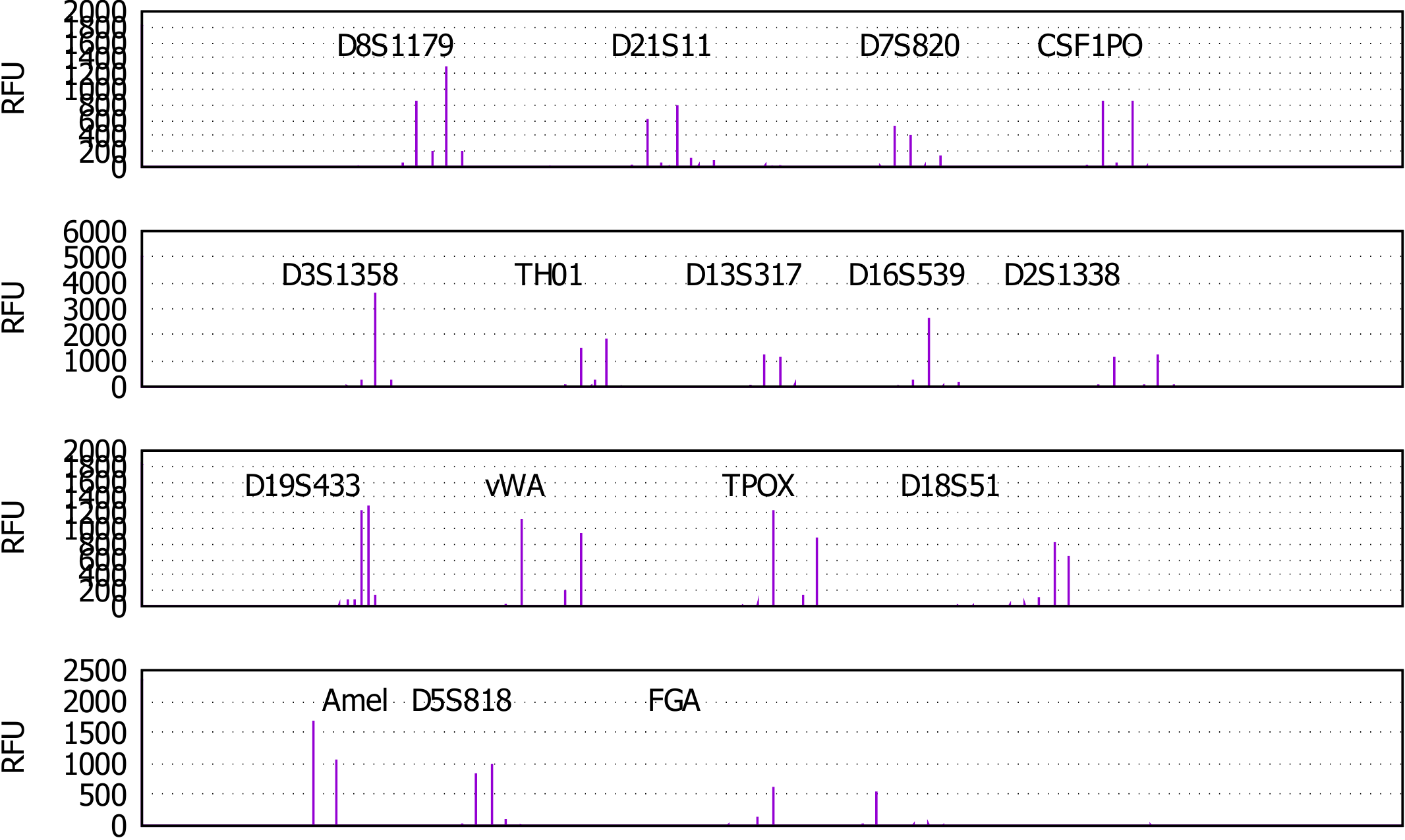}
    \caption{\EPG plot for the
      \texttt{C04\_RD14-0003-42\_43-1\_9-M1U105-0.54IP-Q0.6\_003.5sec}
      sample.\label{fig:bugoodepg}}
  \end{center}
\end{figure}

In all the examples seen so far, we have seen broad agreement between
the various models. We now look at an example from PROVEDit datatset
in which the models give quite different model fits. The dataset we
shall look at is \newline
\texttt{C04\_RD14-0003-42\_43-1\_9-M1U105-0.54IP-Q0.6\_003.5sec}. This
is a two-person mixture of 540pg total DNA template in the ratio of
1:9 from the two contributors. The U105 indicates that the sample was
subject to 105 minutes exposure to ultra-violet radiation (to degrade
the sample) before \PCR.  \tabref{tab:bugood2epg} shows the peak
heights of the sample, above the analytic threshold of 15 {\RFU}s, and
the profiles of the two contributors. A plot of the \EPG is shown in
\figref{fig:bugoodepg}.

\begin{sidewaystable}[htbp]
  \begin{center}
\caption{Peaks heights and the genotypes of constributors \texttt{RD14-0003-42}  \texttt{RD14-0003-43} \label{tab:bugood2epg}.}
{\scriptsize
\begin{tabular}{lcc|llllllll} \hline
Locus &\texttt{RD14-0003-42} &\texttt{RD14-0003-43} & Allele/height  & Allele/height  & Allele/height  & Allele/height & Allele/height  & Allele/height  & Allele/height  & Allele/height 
\\ \hline
Amelogenin	&	X	/	X			&	X	/	Y	&	X	/	1686	&	Y	/	1049	&				&				&				&				&					\\ 
CSF1PO	&	10	/	12			&	10	/	12	&	9	/	33	&	10	/	848	&	10.2	/	17	&	11	/	53	&	12	/	867					&					\\
D13S317	&	12	/	13			&	11	/	12	&	10	/	67	&	11	/	1269	&	12	/	1125	&	
13	/	79					\\ 
D16S539	&	9	/	12			&	10	/	10	&	9	/	297	&	10	/	2682	&	11	/	22	&	12	/	183				\\ 
D18S51	&	17	/	18			&	18	/	19	&	15	/	15	&	16	/	18	&	17	/	130	&	18	/	834	&	19	/	633			\\
D19S433	&	14	/	14.2			&	13.2	/	14	&	12	/	18	&	12.2	/	92	&	13	/	88	&	13.2	/	1250	&	14	/	1295	&	14.2	/	141					\\
D21S11	&	30	/	31.2			&	27	/	29	&	26	/	29	&	27	/	616	&	28	/	61	&	29	/	786	&	30	/	122	&	31.2	/	100	&	35	/ 19	&36	/ 23	\\ 
D2S1338	&	22	/	23			&	19	/	22	&	18	/	92	&	19	/	1148	&	21	/	127	&	22	/	1230	&	23	/	109										\\ 
D3S1358	&	15	/	16			&	15	/	15	&	13	/	20	&	14	/	252	&	15	/	3613	&	16	/	266				\\
D5S818	&	12	/	13			&	11	/	12	&	10	/	49	&	11	/	848	&	12	/	989	&	13	/	124		\\
D7S820	&	11	/	11			&	8	/	9	&	8	/	523	&	9	/	426	&	11	/	138			\\
D8S1179	&	12	/	14			&	11	/	13	&	10	/	44	&	11	/	858	&	12	/	207	&	13	/	1309	&	14	/	204	\\ 
FGA	&	20	/	21			&	21	/	28	&	20	/	134	&	21	/	617	&	27	/	40	&	28	/	557	&	31.2	/	23				\\
TH01	&	9	/	9.3			&	8	/	9.3	&	7	/	90	&	8	/	1508	&	8.3	/	22	&	9	/	230	&	9.3	/	1896					\\
TPOX	&	8	/	10			&	8	/	11	&	7	/	23	&	8	/	1234	&	10	/	141	&	11	/	880						\\ 
vWA	&	16	/	17			&	13	/	17	&	12	/	37	&	13	/	1123	&	16	/	215	&	17	/	932					\\ \hline
\end{tabular}
}
\end{center}
\end{sidewaystable}

We shall analyse this mixture under the assumption of two
contributors. If we assume the true genotypes of the contributors, and
include forward and double stutter, then obtain a maximized
log-likelihood of -756.241357805 using the normal model. Omitting the
forward and double stutter, the maximized log-likelihood increases
significantly to -651.9280003. Hence in all analyses we shall include
only stutter, and omit both forward and double stutter model
components.

\tabref{tab:bu2goodresults} shows the results of fitting  
models to four possible two-contributor scenarios.  The first line of
the table is the scenario in which we assume as known the profiles of
the (true) contributors \texttt{RD14-0003-42} and
\texttt{RD14-0003-43}.  In the second line we replace
\texttt{RD14-0003-42} with an untyped contributor labelled
\texttt{U1}.  In the third line we replace \texttt{RD14-0003-43} with
an untyped contributor labelled \texttt{U2}.  In the last line we
treat both contributors as having unknown genotypes.

We see that cell estimates are broadly in line for all models except
the normal model, which has higher major contributor estimates, and in
agreement with the 1:9 preparation ratio of the sample. There is quite
a lot of variation between the models in their log-likelihood
estimates for each scenario, with the \mFFT model outperforming the
moment based models, by a considerable margin. The log-likelihood
estimates are also out of line with the other models, in that for the
second and fourth scenarios in which the \texttt{RD14-0003-42} has
been replaced by an unknown person, the likelihoods are lower than the
scenario in which both contributor genotype are assumed known,
indicating evidence against \texttt{RD14-0003-42} being a contributor.

\begin{table}
  \caption{Fitting each of the four scenarios to each of the four
    models.\label{tab:bu2goodresults}}
\begin{center}
{\scriptsize
\begin{tabular}{|c|ccc|ccc|ccc|ccc|}
\hline
\multicolumn{1}{|c|}{Scenario}&
\multicolumn{3}{|c|}{Normal} & \multicolumn{3}{c|}{Logormal} & \multicolumn{3}{c|}{Gamma} & \multicolumn{3}{c|}{\mFFT}  \\
&$\widehat{LL}_{max}$ & $\hat{C}_1$ & $\hat{C}_2$  &$\widehat{LL}_{max}$ &  $\hat{C}_1$ & $\hat{C}_2$ &$\widehat{LL}_{max}$ &  $\hat{C}_1$ & $\hat{C}_2$ &$\widehat{LL}_{max}$ &  $\hat{C}_1$ & $\hat{C}_2$  \\ \hline
42-43 &-651.92&154 & 1267 &-528.83 &162& 1108 &-543.31&161 & 1158 &-502.20 & 158 & 1158 \\ 
U1-43 & -599.47 & 138 & 1233 &-549.52  & 142 & 1094&-553.55 & 136 & 1145& -534.37 & 140 & 1161 \\ 
42-U2 & -700.63  & 154 & 1267&-577.53  & 162 & 1108 &-592.01 & 161 & 1158&-550.90 & 158 & 1158 \\ 
U1-U2 &-648.809 & 138 & 1223& -598.54 & 143 & 1094&-602.67& 136 & 1145& -583.39& 141 & 1161 \\  
\hline
\end{tabular}
}
\end{center}
\end{table}

\tabref{tab:bugoodratioresults} shows various log-likelihood ratios
(expressed in bans) obtained by comparing pairs of scenarios from
\tabref{tab:bu2goodresults}. We see that all models are giving similar
log-likelihood ratios in favour of the major contributor, compared to
the maximum of 24.71 Bans (the inverse log profile probability of
\texttt{RD14-0003-43)}, and the \mFFT model gives much larger values
in favour of the minor contributor compared to the other models, with
the normal model suggesting very strong evidence \textit{against}
\texttt{RD14-0003-42} being a contributor (the inverse log profile
probability of \texttt{RD14-0003-42} is 20.22 Bans).
\begin{table}
\caption{Various scenario log-likelihood  ratios expressed in Bans.
  \label{tab:bugoodratioresults}}
\begin{center}
  \begin{tabular}{|l|cccc|}\hline
    Hypotheses & Normal & Lognormal & Gamma & \mFFT \\ \hline
    42-43 vs U1-43	&	-22.78	&		8.99	&		4.45	&		13.97	\\
    42-U2 vs U1-U2	&	-22.51	&		9.12	&		4.63	&		14.11	\\
    42-43 vs 42-U2	&	21.15	&		21.15	&		21.15	&		21.15	\\
    U1-43 vs U1-U2	&	21.43	&		21.29	&		21.33	&		21.29	\\
    \hline
  \end{tabular}
\end{center}
\end{table}

\begin{figure}[t]
  \begin{center}
    \includegraphics[width=0.4\textwidth]{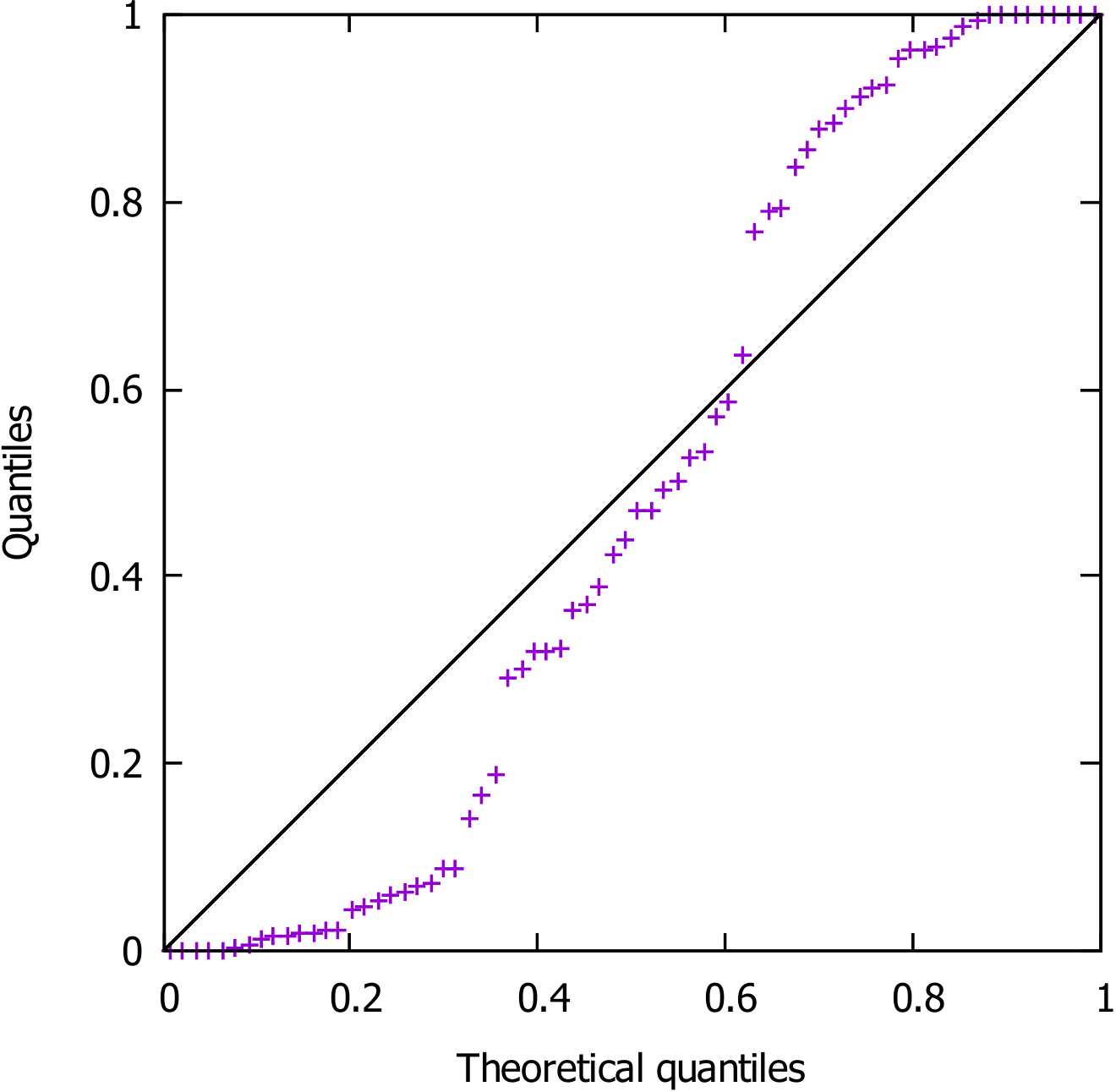}
    \hfill
    \includegraphics[width=0.4\textwidth]{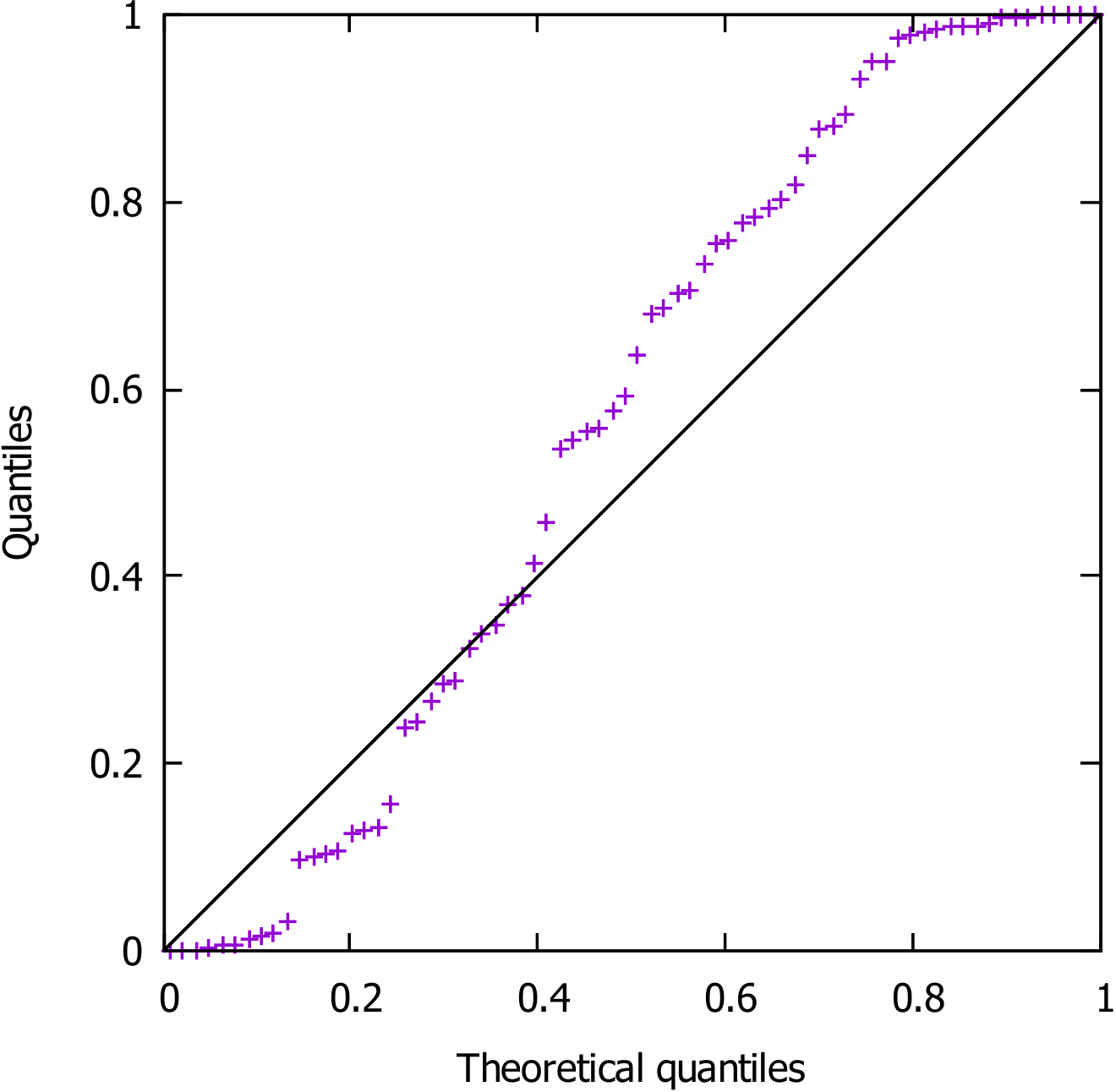}
    \vskip\baselineskip
    \includegraphics[width=0.4\textwidth]{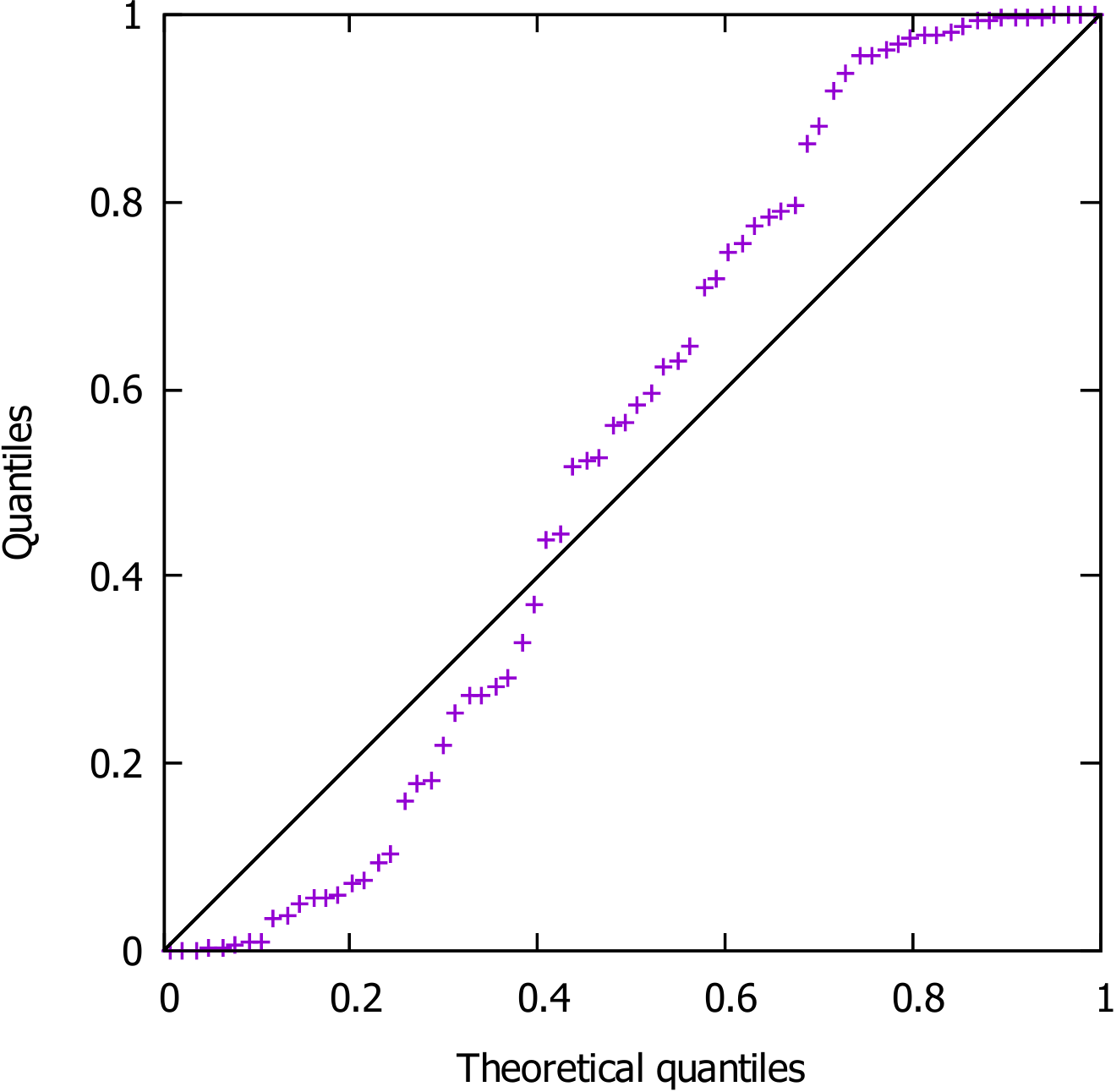}
    \hfill
    \includegraphics[width=0.4\textwidth]{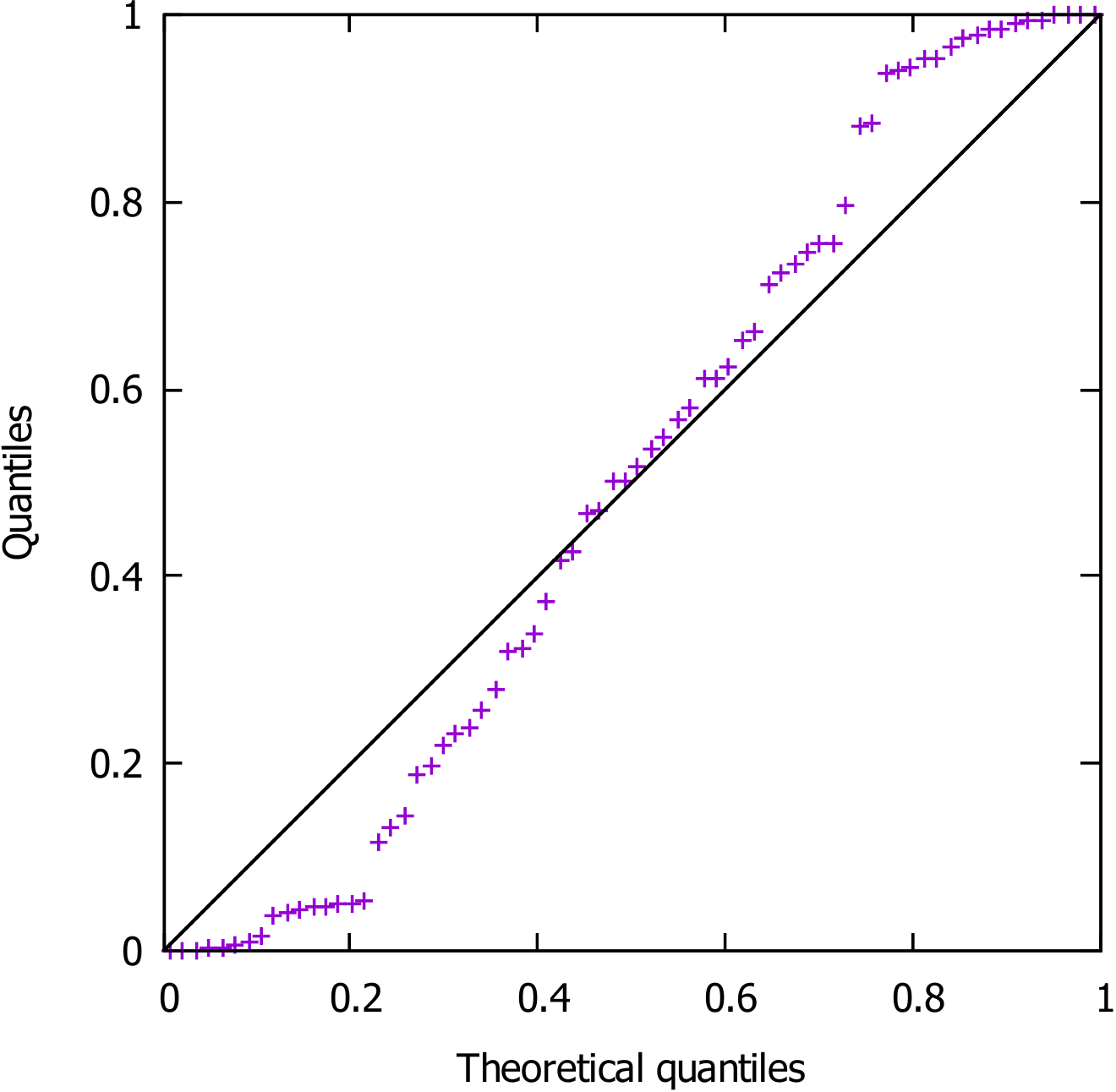}
  \end{center}
  \caption{Top-left: normal model; top-right: lognormal model; bottom
    left: Gamma model; bottom right: \mFFT
    model.\label{fig:bugoodqqplots}}
\end{figure}

\figref{fig:bugoodqqplots} shows the QQ-plots for each of the models
under the scenario in which both contributors are known.
\clearpage

If we add a third un-profiled contributor, then all the models give
similar results. Here are the values under the scenario of the two
known contributors and a third untyped contributor.\newline

\begin{center}
  \begin{tabular}{l|cccc}
    Model &$\widehat{LL}_{max}$ & $\hat{C}_1$ & $\hat{C}_2$ & $\hat{C}_3$ \\ \hline
    Normal &-489.24 &  153 & 1204 & 16 \\
    Lognormal & -500.16 & 153 & 1081 & 26 \\
    Gamma &-496.16 & 153 & 1121& 27 \\
    \mFFT & -491.83 & 152 & 1140 & 17 \\ \hline
  \end{tabular}
\end{center}

If we add another untyped contributor, so that we analyse the sample
assuming the profiles of the two (true) contributors and two untyped
contributors, we obtain the following estimates in which the \mFFT
model gives an estimate of no cells to the second untyped contributor,
unlike the moment based models:\newline
\begin{center}
  \begin{tabular}{l|ccccc}
    Model &$\widehat{LL}_{max}$ & $\hat{C}_1$ & $\hat{C}_2$ 
    & $\hat{C}_3$ & $\hat{C}_4$ \\ \hline
    Normal &-487.71 &  150 & 1202 & 14 &8\\
    Lognormal & -500.04 & 151 & 1085 & 17 & 13 \\
    Gamma &-495.48 & 150 & 1126& 15&15 \\
    \mFFT & -491.83 & 152 & 1140 & 17 &0\\ \hline
  \end{tabular}
\end{center}

\clearpage

\section{Summary and outlook}
In this paper, a framework for modelling single source and multiple
donor forensic DNA samples has been elaborated for STR loci, based on
an idealisation of the steps for forming an \EPG from an initial DNA
sample. It was shown that multivariate probability generating
functions provide a succinct and efficient mathematical representation
of the steps in the process, and that their evaluation can be carried
out efficiently using \FFT. Factorization approximations to make the
mathematical models tractable were shown to introduce some biases for
low template samples, but these biases can be largely removed by a
modification, the \mFFT model, which reintroduces, in a principled
manner, some of the correlation lost by factorization approximations
without affecting computational efficiency.

The examples treated in this paper have assumed that contributors are unrelated individuals. 
The framework may be extended to include relatedness of individuals, which might require that linkage between some 
 loci would need to be taken into account.

 An issue not explored in this paper is the robustness of the
 performance of the models to mis-specifications of model
 parameters. This could be readily explored via simulations in which,
 for example, data is simulated using one set of amplification,
 stutter and selection probabilities, and the simulated data is then
 fitted assuming different probabilities.  Given the reaonably good
 performance of the model on the experimental data from the PROVEDit
 initiative, for which reasonable values were assumed for the
 parameters but without any detailed knowledge of what the true
 experimental values were, one could anticipate that the model will
 prove to be reasonably robust to such parameter mis-specifications.

The current paper has focussed on \EPGs obtained by capillary
electrophoresis, however it may not limited to this. Recently,
\cite{bleka2017open} showed that by substituting read coverages for peak
heights, autosomal SNP mixtures analyzed using massively parallel
sequencing can be interpreted by the open source software Euroformix
\citep{bleka2016euroformix}, originally developed for {\STR}s. By
making the same substitution, that is, by replacing the \RFU scale
factors by `read factors', the framework developed in this paper is
also applicable to such SNP mixtures. Moreover, because SNPs are
biallelic and do not stutter, the factorization approximations are no
longer required.

Finally, other branching processes outside of the forensic applications focussed on in this paper can be treated by the methods
introduced in this paper.

\clearpage
\appendix
\part*{Appendices}
\section{Derivation of moments for amplicon model}
\label{app:ampmoments}
Let $N$ denote the number of target amplicons, and $M$ the number of
stutter amplicons, arising from the amplification of a single target
amplicon. We have that
\begin{align*}
  \E N\cd n &= \frac{\partial F_n(t,s)}{\partial t}\vert_{t=1,s=1}\\
  \E N(N-1)\cd n &= \frac{\partial^2 F_n(t,s)}{\partial t^2}\vert_{t=1,s=1}\\
  \E M\cd n &= \frac{\partial F_n(t,s)}{\partial s}\vert_{t=1,s=1}\\
  \E N(N-1)\cd n &= \frac{\partial^2 F_n(t,s)}{\partial s^2}\vert_{t=1,s=1}\\
  \E NM\cd n &= \frac{\partial^2 F_n(t,s)}{\partial s\partial t}\vert_{t=1,s=1}\\
\end{align*}
from which the variance $\V N$ and $\V M$ may be found, and hence the
correlation $\cor(N,M)$.  We take each in turn.

\subsubsection*{$ \E N\cd n  $:}
Using the chain rule for differentiating, we have
$$
\frac{\partial F_n(t,s)}{\partial t} =
(1-p)\frac{\partial F_{n-1}(t,s)}{\partial t} +
2p(1-\xi)F_{n-1}(t,s)\frac{\partial F_{n-1}(t,s)}{\partial t} + p\xi
\frac{\partial F_{n-1}(t,s)}{\partial t} G_{n-1}(s)
$$
We now substitute $t=1$ and $s=1$, use the property that
$F_n(1,1) = 1$ for all $n$, to obtain a linear difference equation
that can be solved:

\begin{align*}
  \E N\cd n & = (1-p)(\E N\cd n-1) + 2p(1-\xi)(\E N\cd n-1) + p\xi (\E N\cd n-1) \\
            & = (1 + p(1-\xi)) (\E N\cd n-1)\\
            & = (1 + p(1-\xi))^n (\E N\cd 0)\\
            &= (1 + p(1-\xi))^n\\
\end{align*}

This could have been anticipated: a main peak allele is amplified with
probability $p(1-\xi)$, hence starting with one allele and doing $n$
amplifications, the mean number is as shown, $(1 + p(1-\xi))^n$.

\subsubsection*{$ \E N(N-1)\cd n  $ and $\V N\cd n$:}
These could also be written down directly based on the previous
result, however we go through the formal steps.
\begin{align*}
  \E N(N-1)\cd n &= \frac{\partial^2 F_n(t,s)}{\partial t^2}\vert_{t=1,s=1}\\
  \frac{\partial^2 F_n(t,s)}{\partial t^2} &=
                                             (1-p)\frac{\partial^2 F_{n-1}(t,s)}{\partial t^2}  + 
                                             2p(1-\xi)F_{n-1}(t,s)\frac{\partial^2 F_{n-1}(t,s)}{\partial t^2}  +\\& 
                                                                                                                     2p(1-\xi)\left(\frac{\partial F_{n-1}(t,s)}{\partial t}\right)^2
                                                                                                                     + p\xi \frac{\partial^2 F_{n-1}(t,s)}{\partial t^2} G_{n-1}(s)\\
  \E N(N-1)\cd n &= (1 + p(1-\xi)) (\E N(N-1)\cd n-1) + 2p(1-\xi)\left(\E N\cd n-1 \right)^2\\
  \E N^2\cd n &= (1 + p(1-\xi)) (\E N^2\cd n-1) + 2p(1-\xi)\left(\E N\cd n-1 \right)^2\\
                 &= (1 + p(1-\xi)) (\E N^2\cd n-1) + 2(p(1-\xi))(1+p(1-\xi))^{2n-2}\\
                 &= (1 + p(1-\xi))^n (\E N^2\cd 0) + 2(p(1-\xi))\sum_{j=1}^n (1+p(1-\xi))^{2j-2}(1 + p(1-\xi))^{n-j}\\
                 &= (1 + p(1-\xi))^n + 2(p(1-\xi))\sum_{j=0}^{n-1} (1+p(1-\xi))^{2j}(1 + p(1-\xi))^{n-1-j}\\
                 &= (1 + p(1-\xi))^n + 2(p(1-\xi))(1 + p(1-\xi))^{n-1}\sum_{j=0}^{n-1} (1+p(1-\xi))^j\\
                 &= (1 + p(1-\xi))^n + 2(p(1-\xi))(1 + p(1-\xi))^{n-1}\frac{(1+p(1-\xi))^n-1}{(1+p(1-\xi))-1}\\
                 &= (1 + p(1-\xi))^n \left[ 1 + 2\frac{(1+p(1-\xi))^n-1}{1+p(1-\xi)} \right]\\
                 &= (1 + p(1-\xi))^{n-1} \left[ 2(1+p(1-\xi))^n +p(1-\xi)-1 \right]\\
\end{align*} 
 
Hence the variance is
\begin{align*}
  \V N\cd n &= (\E N^2\cd n) - (\E N\cd n)^2\\
            &=(1 + p(1-\xi))^n \left[ 1 + 2\frac{(1+p(1-\xi))^n-1}{1+p(1-\xi)} \right] -(1 + p(1-\xi))^{2n}\\
            &=(1 + p(1-\xi))^n \left[1 + 2\frac{(1+p(1-\xi))^n-1}{1+p(1-\xi)} - (1 + p(1-\xi))^{n}\right]\\
            &= \frac{1-p(1-\xi)}{1+p(1-\xi)}(1 + p(1-\xi))^n \left[ (1+p(1-\xi))^n -1\right]\\
\end{align*}

We also have
\begin{align*}
  \E N(N-1)\cd n &= (\V N\cd n) + (\E N\cd n)^2 - (\E N\cd n) \\
                 &= \frac{1-p(1-\xi)}{1+p(1-\xi)}(1 + p(1-\xi))^n \left[ (1+p(1-\xi))^n -1\right]
                   +(1 + p(1-\xi))^{2n} - (1 + p(1-\xi))^n\\ 
                 &= 2(1 + p(1-\xi))^{n-1}((1 + p(1-\xi))^n-1)\\
\end{align*}

\subsubsection*{$ \E M\cd n  $:}
This follows a similar pattern to the above for $N$:

\begin{align*}
  \E M\cd n &= \frac{\partial F_n(t,s)}{\partial s}\vert_{t=1,s=1}\\
  \frac{\partial F_n(t,s)}{\partial s} &=
                                         (1-p)\frac{\partial F_{n-1}(t,s)}{\partial s}  + 2p(1-\xi)F_{n-1}(t,s)\frac{\partial F_{n-1}(t,s)}{\partial s} 
  \\&
      + p\xi \frac{\partial F_{n-1}(t,s)}{\partial s} G_{n-1}(s)
      + p\xi  F_{n-1}(t,s) \frac{\partial G_{n-1}(s)}{\partial s}
  \\
\end{align*}

Hence

$$ \E M\cd n = (1 +p(1-\xi))(\E M\cd n-1) + p\xi \frac{\partial G_{n-1}(s)}{\partial s}\vert_{s=1}$$

Now $$G_n(s) = (1-p)G_{n-1}(s) + pG_{n-1}^2(s),$$

which is the recursion relation for the generating function of a
single allele amplifying with probability $p$ in the branching
process, hence

$$ \E M\cd n = (1 +p(1-\xi))(\E M\cd n-1) + p\xi (1+p)^{n-1}$$

Now we start with no stutter amplicons, hence $\E M\cd 0 = 0$. Thus

\begin{align*}
  \E M\cd n &= (1 +p(1-\xi))(\E M\cd n-1) + p\xi (1+p)^{n-1}\\
            &=  (1 +p(1-\xi))^2(\E M\cd n-2) + (1 +p(1-\xi))p\xi (1+p)^{n-2}+ p\xi (1+p)^{n-1}\\
            &= \vdots\\
            &=  (1 +p(1-\xi))^n(\E M\cd 0) + p\xi\sum_{j=0}^{n-1}(1 +p(1-\xi))^j(1+p)^{n-j-1}\\
\end{align*}
We now use $\E M\cd 0 = 0$, $1+p -(1+p(1-\xi) = p\xi$, and the
relation
$$a^{n-1} + a^{n-2}b + a^{n-3}b^2 + \cdots ab^{n-2} + b^{n-1}
= \frac{a^n-b^n}{a-b}$$ to deduce that

$$\E M\cd n = p\xi \frac{(1+p)^n - (1+p(1-\xi))^n}{(1+p) - (1+p(1-\xi)}
= (1+p)^n - (1+p(1-\xi))^n$$

When $\xi=0$ this vanishes as it should. Note also that
$\E[N+M\cd n] = (1+p)^{n}$ which is also as expected-- it is as if the
stutter and the main target cannot be distinguished apart. We also
have

$$\frac{\E M\cd n }{\E N+M\cd n }= 1 - \left( 1 -\frac{p\xi}{1+p}\right)^n$$

which $\to 1$ and $n \to \infty$ for $\xi > 0$. Hence ultimately the
stutter allele peak will dominate the original allele peak with enough
cycles, a result found by \cite{weusten2012stochastic}.

\subsubsection*{$ \E M(M-1)\cd n $ and $\V M \cd n$}

Denote $F_{n;s}(t,s) = \partial F_n(t,s)/\partial s$, and $F_{n;ss}$
the second derivative.

Hence
\begin{align*}
  \E M(M-1)\cd n &= F_{n;ss}\vert_{t=1,s=1}\\
  F_{n;ss}&= (1-p) F_{n-1;ss} + 2p(1-\xi)F_{n-1}F_{n-1;ss} + 2p(1-\xi)F_{n-1;s}^2\\
                 &+ p\xi  F_{n-1;ss} G_{n-1} +2 p\xi F_{n-1;s}G_{n-1;s} + p\xi F_{n-1}G_{n-1;ss}\\
  \E M(M-1)\cd n &= (1 + p(1-\xi))(\E M(M-1)\cd n-1) +\\
                 & \left(2p(1-\xi)F_{n-1;s}^2+2 p\xi F_{n-1;s}G_{n-1;s} + p\xi F_{n-1}G_{n-1;ss}\right)\bigg\vert_{s=t=1}\\
                 &= (1 + p(1-\xi))(\E M(M-1)\cd n-1) +\\
                 &
                   2p(1-\xi)\left((1+p)^{n-1} - (1+p(1-\xi))^{n-1}  \right)^2+\\
                 & 
                   2 p\xi \left((1+p)^{n-1} - (1+p(1-\xi))^{n-1}  \right)(1+p)^{n-1} +\\
                 &
                   2p\xi(1+p)^{n-2}[(1+p)^{n-1} -1]\\
\end{align*}

But we have that
$\E M\cd n = (1 +p(1-\xi))(\E M\cd n-1) + p\xi (1+p)^{n-1}$, hence
adding this to both sides we obtain
\begin{align*}
  \E M^2 \cd n &=  (1 +p(1-\xi))(\E M\cd n-1) \\
               &+2p(1-\xi)\left((1+p)^{n-1} - (1+p(1-\xi))^{n-1}  \right)^2+\\
               & 
                 2 p\xi \left((1+p)^{n-1} - (1+p(1-\xi))^{n-1}  \right)(1+p)^{n-1} +\\
               &
                 p\xi(1+p)^{n-2}[2(1+p)^{n-1} +p-1]\\ 
\end{align*}
Expanding this out we have
\begin{align*}
  \E M^2 \cd n &=  (1 +p(1-\xi))(\E M^2\cd n-1) \\
               &+2p(1-\xi)\left((1+p)^{2(n-1)} -2(1+p)^{n-1}(1+p(1-\xi))^{n-1} + (1+p(1-\xi))^{2(n-1)}  \right)\\
               &+
                 2 p\xi \left( (1+p)^{2(n-1)} - (1+p)^{n-1} (1+p(1-\xi))^{n-1} \right)\\
               &+
                 \frac{p\xi}{1+p} \left( 2(1+p)^{2(n-1)} -(1-p)(1+p)^{n-1} \right)\\
\end{align*}
Collecting together similar terms, this reduces to
\begin{align*}
  \E M^2 \cd n &=  (1 +p(1-\xi))(\E M^2\cd n-1) \\
               & +\left(2p + \frac{2p\xi}{1+p}\right)(1+p)^{2(n-1)}\\
               & -
                 \left(4p(1-\xi) +2p\xi)\right)(1+p)^{n-1}(1+p(1-\xi))^{n-1} \\
               &+2p(1-\xi)(1+p(1-\xi))^{2(n-1)}\\
               &-\frac{p(1-p)\xi}{1+p}(1+p)^{n-1} \\
\end{align*}

We may now solve this linear difference equation, noting that the
terms on the right are each of the form $a^{n-1}$ or $a^{2(n-1)}$,
making the particular solution terms of the form $(a^n-b^n)/(a-b)$ or
$(a^{2n}-b^n)/(a^2-b)$, where in each case $b = 1+p(1-\xi)$. We also
use that $(\E M^2\cd 0) =0 $. We obtain
 
 \begin{align*}
   \E M^2 \cd n &= 
                  \left(2p + \frac{2p\xi}{1+p}\right)\frac{(1+p)^{2n} - (1+p(1-\xi))^n}{(1+p)^2-(1+p(1-\xi))}\\
                & -  \left(4p(1-\xi) +2p\xi)\right)(1+p(1-\xi))^{n-1} \frac{(1+p)^n-1}{1+p -1}\\
                & +2p(1-\xi)(1+p(1-\xi))^{n-1}\frac{(1+p(1-\xi))^n -1}{(1+p(1-\xi)) -1} \\
                &-\frac{p(1-p)\xi}{1+p} \frac{(1+p)^{n} - (1+p(1-\xi))^n}{(1+p)-(1+p(1-\xi))}\\
 \end{align*}
 
 Now
 \begin{align*}
   (1+p)-1 = p\\
   (1+p(1-\xi))-1 = p(1-\xi)\\
   (1+p) - (1+p(1-\xi) &= p\xi\\
   (1+p)^2 - (1+p(1-\xi)&= 1 + 2p + p^2-1 - p +p\xi = p(1+p+\xi)\\
 \end{align*}
 
 Hence simplifying further we obtain
 \begin{align*}
   \E M^2 \cd n &=  2\frac{(1+p)^{2n} - (1+p(1-\xi))^n}{1+p}\\
                &-(4 - 2\xi)(1+p(1-\xi))^{n-1} ((1+p)^n-1)\\
                &+ 2(1+p(1-\xi))^{n-1}((1+p(1-\xi))^n -1)\\
                &-\frac{(1-p)}{1+p} ((1+p)^{n} - (1+p(1-\xi))^n) \\
 \end{align*}

 Now we had previously that

$$\E M\cd n =  (1+p)^n - (1 + p(1-\xi))^n$$

Squaring this and subtracting from $E M^2\cd n$ gives the desired
variance $\V M\cd n$. However there does not appear to be a nice
simple reduction in the result, so this is omitted.

For large $n$ we may drop the terms in $(1+p(1-\xi))^n$ and lower, and
we have approximately

$$
\E M^2 \cd \approx 2n (1+p)^{2n-1} -
(4-2\xi)(1+p(1-\xi))^{n-1}(1+p)^{n-1} + 2(1+p(1-\xi))^{2(n-1)}
$$

\subsubsection*{$ \E NM\cd n $:}
\begin{align*}
  \E NM\cd n &= \frac{\partial^2 F_n(t,s)}{\partial s\partial t}\vert_{t=1,s=1}\\
  \frac{\partial^2 F_n(t,s)}{\partial s\partial t}&= 
                                                    (1-p)\frac{\partial^2 F_{n-1}(t,s)}{\partial s\partial t}
                                                    +2p(1-\xi)F_{n-1}(t,s)\frac{\partial^2 F_{n-1}(t,s)}{\partial s\partial t}
                                                    +p\xi\frac{\partial^2 F_{n-1}(t,s)}{\partial s\partial t}G_{n-1}(s)
  \\&+2p(1-\xi)\frac{\partial F_{n-1}(t,s)}{\partial s}\frac{\partial F_{n-1}(t,s)}{\partial t}
  \\&+ p\xi \frac{\partial F_{n-1}(t,s)}{\partial t}\frac{\partial G_{n-1}(s)}{\partial s}\\
  \E NM\cd n&= (1+p(1-\xi))(\E NM\cd n-1 )+2p(1-\xi)(\E N\cd n-1)(\E M\cd n-1)
              +p\xi (\E N\cd n-1)(1+p)^{n-1}\\
             &= (1+p(1-\xi))(\E NM\cd n-1 )
               +2p(1-\xi)(1+p(1-\xi))^{n-1}[ (1+p)^{n-1}-(1+p(1-\xi))^{n-1}]
  \\&+ p\xi (1+p(1-\xi))^{n-1}(1+p)^{n-1}\\
  =&(1+p(1-\xi))(\E NM\cd n-1 ) + p(2-\xi)(1+p(1-\xi))^{n-1}(1+p)^{n-1}
  \\&
      -2p(1-\xi)(1+p(1-\xi))^{n-1}(1+p(1-\xi))^{n-1}\\
\end{align*}

Using $\E NM\cd 0 = 0$, the solution may be written as

\begin{align*}
  \E NM\cd n &=  (2-\xi)p(1 + p(1-\xi))^{n-1}((1+p)^n-1)/(1+p-1)\\
             &
               -2p(1-\xi)(1 + p(1-\xi))^{n-1}\frac{(1 + p(1-\xi))^{n}-1}{(1 + p(1-\xi) -1}\\
             &= (2-\xi)(1 + p(1-\xi))^{n-1}((1+p)^n-1) -2(1 + p(1-\xi))^{n-1}((1 + p(1-\xi))^{n}-1)\\
             &= (1 + p(1-\xi))^{n-1}( (2-\xi)(1+p)^n-1) -2((1 + p(1-\xi))^n-1)\\
             &= (1 + p(1-\xi))^{n-1}(\xi(1-(1+p)^n) + 2( (1+p)^n - (1 + p(1-\xi))^n)
\end{align*}

\subsubsection*{$\cov(M,N)$ }

The covariance $\cov(M,N)$ is given by
\begin{align*}
  \cov(M,N) &= \E NM\cd n  - (\E N\cd n )(\E M\cd n )\\
            &= (1 + p(1-\xi))^{n-1}(\xi(1-(1+p)^n) + 2( (1+p)^n - (1 + p(1-\xi))^n)\\
            &
              - ((1+p)^n - (1 + p(1-\xi))^n))(1 + p(1-\xi))^n\\
            &= (1 + p(1-\xi))^{n-1}\left( (1-p(1-\xi))( (1+p)^n - (1 + p(1-\xi))^n) - \xi((1+p)^n-1)\right)\\
\end{align*}
from which the correlation may be found, by an appropriate scaling
using the variances given earlier.

\subsubsection*{Binomial sampling}
 
The generating function for Binomial sampling with $n$ trials and
success probability $q$ is

\[
  (1-q + qt)^n
\]

Let $F(t,s)$ denote the joint \PGF for main $T$ and stutter $S$ peaks
for $k$ cycles based on an initial single allele. Let $\E N$ denote
the mean number of main alleles $\E S$ the mean number of stutter
allele, and similarly for the variance $\V N$ and $\V S$. Then for an
initial set of $n$ alleles sampled with probability $p$ the joint \PGF
is given by

$$Q(t,s)=(1-q+qF(t,s))^{n}$$

With abuse of notation, let $\E t$ denote the mean number of main
alleles for this \PGF, \etc, so that
$\E t = \partial Q/\partial t \vert_{t=s=1} $, \etc

Thus
 
\[ 
  \frac{{\partial Q(t,s)}}{\partial
    t}=nq(1-q+qF(t,s))^{n-1}\frac{{\partial F(t,s)}}{\partial t}
\]
 
from which it follows that
\[
  \E t = nq \E N
\]
 
Similarly
 
 \[ 
   \frac{{\partial Q(t,s)}}{\partial
     s}=nq(1-q+qF(t,s))^{n-1}\frac{{\partial F(t,s)}}{\partial s}
 \]
 
 from which it follows that
 \[
   \E s = nq \E S
 \]
 
 Taking the second derivatives, we have
 
 \begin{align*}
   \frac{{\partial^{2}Q(t,s)}}{\partial t^{2}}&=n(n-1)q^{2}(1-q+qF(t,s))^{n-2}\left(\frac{{\partial F(t,s)}}{\partial t}\right)^2 + nq(1-q+qF(t,s))^{n-1}\frac{{\partial^2 F(t,s)}}{\partial t^2}\\
   \E t(t-1) &= n(n-1)q^2(\E N)^2 + nq (\E N(N-1)\\
   \V t &=  n(n-1)q^2(\E N)^2 + nq (\E N(N-1) + (nq \E N) - (nq \E N)^2\\
                                              &= nq(\V N + (1-q) (\E N)^2\\
 \end{align*}

 and similarly
 \[
   \V s = nq(\V S + (1-q) (\E S)^2)
 \]

 Finally

\begin{align*}
  \frac{{\partial^{2}Q(t,s)}}{\partial t\partial s}&=nq(1-q+qF(t,s))^{n-1}
                                                     \frac{{\partial F(t,s)}}{\partial t}\frac{{\partial F(t,s)}}{\partial s}
                                                     +
                                                     n(n-1)q^{2}(1-q+qF(t,s))^{n-2}\frac{{\partial^{2}F(t,s)}}{\partial t\partial s}\\
  \E ts &= n(n-1)q^2 (\E N \E S) + nq (\E NS)\\
  \cov(t,s) &=  n(n-1)q^2 (\E N \E S) + nq (\E NS) - (nq \E N)(nq \E S)\\
                                                   &=nq( \E NS  - q(\E N)(\E S))\\
                                                   &= nq(\cov(N,S) + (1-q)(\E N)(\E S))
\end{align*}

Hence

\begin{align*}
  \cor(t,s) &= \frac{nq(\cov(N,S) + (1-q)(\E N)(\E S))}{\sqrt{nq(\V N + (1-q) (\E N)^2}\sqrt{nq(\V S + (1-q) (\E S)^2}}\\
            &= \frac{\cov(N,S) + (1-q)(\E N)(\E S))}{\sqrt{(\V N + (1-q) (\E N)^2}\sqrt{(\V S + (1-q) (\E S)^2}}
\end{align*}

\subsubsection*{Poisson sampling}

Suppose that the number of alleles to be sampled has a Poison
distribution of mean $\lambda$. The using the \PGF for a Poisson
distribution, $\exp(\lambda(t-1))$, we proceed in a similar manner as
for the Binomial sampling, using the joint \PGF

$$Q(t,s)=\exp(\lambda(F(t,s)-1))$$

\begin{align*}
  \frac{{\partial Q(t,s)}}{\partial t} &= \lambda 
                                         \frac{{\partial F(t,s)}}{\partial t} Q(t,s)\\
  \E t &= \lambda \E N\\
  \frac{{\partial Q(t,s)}}{\partial s} &= \lambda 
                                         \frac{{\partial F(t,s)}}{\partial s} Q(t,s)\\
  \E s &= \lambda \E S\\
  \frac{{\partial^2 Q(t,s)}}{\partial t^2} &= \lambda^2 
                                             \left(\frac{{\partial F(t,s)}}{\partial t}\right)^2 Q(t,s)
                                             +
                                             \lambda \frac{{\partial^2 F(t,s)}}{\partial t^2} Q(t,s)\\
  E t(t-1) &= \lambda^2 (\E N)^2 + \lambda (\E N(N-1))\\
  \frac{{\partial^2 Q(t,s)}}{\partial s^2} &= \lambda^2 
                                             \left(\frac{{\partial F(t,s)}}{\partial s}\right)^2 Q(t,s)
                                             +
                                             \lambda \frac{{\partial^2 F(t,s)}}{\partial s^2} Q(t,s)\\
  E s(s-1) &= \lambda^2 (\E S)^2 + \lambda (\E S(S-1))\\
  \frac{{\partial^{2}Q(t,s)}}{\partial t\partial s} &=
                                                      \lambda^2 
                                                      \frac{{\partial F(t,s)}}{\partial t} \frac{{\partial F(t,s)}}{\partial s}Q(t,s)
                                                      +
                                                      \lambda \frac{{\partial^{2}F(t,s)}}{\partial t\partial s} Q(t,s)\\
  \E ts &= \lambda (\E NS) + \lambda^2 (\E N)(\E S)
\end{align*}

from which it follows that
\begin{align*}
  \cov(t,s) &= \lambda (\E NS) + \lambda^2 (\E N)(\E S) - \lambda^2 (\E N)(\E S)\\
            &= \lambda (\E NS) \\
  \V t &= \lambda^2 (\E N)^2 + \lambda (\E N(N-1)) + \lambda(\E N) - (\lambda\E N)^2\\
            &= \lambda (\E N^2)\\
  \V s &= \lambda (\E S^2) \mbox{ similarly},\\
  \cor(t,s) &= \frac{\lambda (\E NS)}{\sqrt{\lambda (\E N^2)}\sqrt{\lambda (\E S^2)}}\\
            &= \frac{\E NS}{\sqrt{(\E N^2)(\E S^2)}}
\end{align*}
which does not depend on $\lambda$. In fact this result is the
limiting result of taking $q\to 0$ in the earlier Binomial sampling.

\section{Derivation of moments for genomic model, no stutters}
\label{app:genomicmoments}

\subsection{Moments of tagged amplicons}
For convenience let $t = t_{a_d}$, and let $D$ denote the derivative
with respect to $t$.  The marginal \PGF for the tagged amplicons is
given by the sum of such amplicons arising from the $g$ and $g_d$
strand: we obtain this by multiplying the {\PGF}s.
$$F(t) = G_n(1,1,1,t)G_{d;n}(1,1, t, 1),$$
from which
$$DF(t) = [DG_n(1,1,1,t)]G_{d;n}(1,1, t, 1) + G_n(1,1,1,t)[DG_{d;n}(1,1, t, 1)]$$
hence the mean number of tagged amplicons is given by the sum of two
terms:
$$E N_{a_d} = [DG_n(1,1,1,t)]_{t=1} + [DG_{d;n}(1,1, t, 1)]_{t=1}$$
where we use $G(1,1,1,) = G_d(1,1,1,1) = 1$.  We may find each term on
the right hand side separately. Using the recurrence relation for the
joint \PGFs of each genomic type, we may derive recurrence relations
for each of these two terms, which turn out to be coupled linear
equations that can be solved either by hand or computer algebra. We
consider each separately:

\subsubsection{$DG$}
We may substitute 1 for every component, except $t_{a_d}$ which we set
to $t$, in the joint \PGF recurrence relations. Assume this is done.
Then differentiating with respect to $t$ and substituting $t=1$ we
obtain the following recurrence relations:
  
\begin{align*}
  DG_{n+1} & = (1-p_g)DG_n + p_g(DG_n + DH_{d;n})\\
           &= DG_n  + p_gDH_{d;n}\\
  DH_{d;n+1} & = (1-p_{h_d})DH_{d;n} + p_{h_d}(DH_{d;n} + DA_n)\\
           &= DH_{d;n}  + p_{h_d} DA_n\\
  DA_{n+1} & = (1-p_{a})DA_n + p_{a}(DA_n +DA_{d;n})\\
           &= DA_n + p_a DA_{d;n}\\
  DA_{d;n+1} & = (1-p_{a_d})DA_{d;n} + p_{a_d}(DA_{d;n}+DA_n)\\
           &= DA_{d;n}  + p_{a_d}DA_n
\end{align*}
with initial conditions $DG_0 = DH_{d;0} = DA_0 = 0$ and
$DA_{d;0} = 1$.

Now consider the last two coupled equations:
\begin{align*}
  DA_{n+1} & =DA_n + p_a DA_{d;n}\\
  DA_{d;n+1} & = DA_{d;n}  + p_{a_d}DA_n
\end{align*}
We may write this in matrix form

\begin{gather*}
  \begin{pmatrix}
    DA_{n+1}\\
    DA_{d;n+1}
  \end{pmatrix}
  \quad= \quad
  \begin{pmatrix}
    1 & p_a\\
    p_{a_d} & 1
  \end{pmatrix}
  \quad
  \begin{pmatrix}
    DA_{n}\\
    DA_{d;n}
  \end{pmatrix}
\end{gather*}
from which we clearly obtain

\begin{gather*}
  \begin{pmatrix}
    DA_{n}\\
    DA_{d;n}
  \end{pmatrix}
  \quad= \quad
  \begin{pmatrix}
    1 & p_a\\
    p_{a_d} & 1
  \end{pmatrix}^n
  \begin{pmatrix}
    0\\
    1
  \end{pmatrix}
\end{gather*}
and hence
\begin{gather*}
  DA_{n} =
  \begin{pmatrix}
    1 & 0
  \end{pmatrix}
  \begin{pmatrix}
    1 & p_a\\
    p_{a_d} & 1
  \end{pmatrix}^n
  \begin{pmatrix}
    0\\
    1
  \end{pmatrix}.
\end{gather*}
Let us denote the square matrix (not raised to the power $n$) by $P$:
\begin{gather*}
  P = \begin{pmatrix}
    1 & p_a\\
    p_{a_d} & 1
  \end{pmatrix}.
\end{gather*}
Then
\begin{gather*}
  DA_{n} =
  \begin{pmatrix}
    1 & 0
  \end{pmatrix}
  P^n
  \begin{pmatrix}
    0\\
    1
  \end{pmatrix}
\end{gather*}
and from the earlier recurrence for $DH_d$;
$$DH_{d;n+1} = DH_{d;n}  + p_{h_d} DA_n$$
we obtain
\begin{gather*}
  DH_{d;n} = p_{h_d}\sum_{j=0}^{n-1}
  \begin{pmatrix}
    1 & 0
  \end{pmatrix}
  P^j
  \begin{pmatrix}
    0\\
    1
  \end{pmatrix}
  = p_{h_d}\begin{pmatrix} 1 & 0
  \end{pmatrix}
  \frac{P^n - I}{P-I}
  \begin{pmatrix}
    0\\
    1
  \end{pmatrix}
\end{gather*}
From the earlier recurrence for $DG$;
$$DG_{n+1} = DG_{n}  + p_{g} DH_{d;n}$$
we therefore obtain
\begin{gather*}
  DG_{n} = p_gp_{h_d}\sum_{j=0}^{n-1}
  \begin{pmatrix}
    1 & 0
  \end{pmatrix}
  \frac{P^j - I}{P-I}
  \begin{pmatrix}
    0\\
    1
  \end{pmatrix}
  = p_gp_{h_d}\begin{pmatrix} 1 & 0
  \end{pmatrix}
  \frac{P^n -I - n(P-I)}{(P-I)^2}
  \begin{pmatrix}
    0\\
    1
  \end{pmatrix}
\end{gather*}
which is our final result for the expected number of amplicons
\textit{arising from the $g$ genomic strand}.

Note that if all the amplification probabilities are equal to 1, we
obtain:
\begin{gather*}
  P^n = \begin{pmatrix}
    1 & 1\\
    1 & 1
  \end{pmatrix}^n =
  \begin{pmatrix}
    2^{n-1} & 2^{n-1}\\
    2^{n-1} & 2^{n-1}
  \end{pmatrix}
\end{gather*}
which leads to $DG_n = 2^{n-1} - n$ obtained earlier.

We now need to do a similar calculation for the $g_d$ strand.

\subsubsection{$DG_d$}
We may substitute 1 for every component, except $t_{a_d}$ which we set
to $t$, in the joint \PGF recurrence relations. Assume this is done.
Then differentiating with respect to $t$ and substituting $t=1$ we
obtain the following recurrence relations:
  
\begin{align*}
  DG_{d;n+1} & = (1-p_{g_d})DG_{d;n} + p_{g_d}(DG_{d;n} + DH_{n})\\
             &= DG_{d;n}  + p_{g_d}DH_{n}\\
  DH_{n+1} & = (1-p_{h})DH_{n} + p_{h}(DH_{n} + DA_{d;n})\\
             &= DH_{n}  + p_{h} DA_{d;n}\\
  DA_{n+1} & = (1-p_{a})DA_n + p_{a}(DA_n +DA_{d;n})\\
             &= DA_n + p_a DA_{d;n}\\
  DA_{d;n+1} & = (1-p_{a_d})DA_{d;n} + p_{a_d}(DA_{d;n}+DA_n)\\
             &= DA_{d;n}  + p_{a_d}DA_n
\end{align*}
with initial conditions $DG_0 = DH_{d;0} = DA_0 = 0$ and
$DA_{d;0} = 1$. Note that the last two coupled equations are as before
for $DG$, and so have the same solution as given above:

\begin{gather*}
  \begin{pmatrix}
    DA_{n}\\
    DA_{d;n}
  \end{pmatrix}
  \quad= \quad
  \begin{pmatrix}
    1 & p_a\\
    p_{a_d} & 1
  \end{pmatrix}^n
  \begin{pmatrix}
    0\\
    1
  \end{pmatrix}
  \quad =P^n
  \begin{pmatrix}
    0\\
    1
  \end{pmatrix}
\end{gather*}
The equation for $H$ reads:
$$ H_{n+1}= DH_{n}  + p_{h} DA_{d;n}$$
but
\begin{gather*}
  DA_{d;n} =\begin{pmatrix} 0 & 1
  \end{pmatrix}
  P^n
  \begin{pmatrix}
    0\\
    1
  \end{pmatrix}
\end{gather*}
Hence
\begin{gather*}
  DH_{n} = p_{h}\sum_{j=0}^{n-1}
  \begin{pmatrix}
    0 & 1
  \end{pmatrix}
  P^j
  \begin{pmatrix}
    0\\
    1
  \end{pmatrix}
  = p_{h}\begin{pmatrix} 0 & 1
  \end{pmatrix}
  \frac{P^n - I}{P-I}
  \begin{pmatrix}
    0\\
    1
  \end{pmatrix}
\end{gather*}
Finally from the recurrence for $DG_d$:

$$DG_{d;n+1} = DG_{d;n}  + p_{g_d} DH_{n}$$
we obtain

\begin{gather*}
  DG_{d;n} = p_{g_d}p_{h}\sum_{j=0}^{n-1}
  \begin{pmatrix}
    0 & 1
  \end{pmatrix}
  \frac{P^j - I}{P-I}
  \begin{pmatrix}
    0\\
    1
  \end{pmatrix}
  = p_{g_d}p_{h}\begin{pmatrix} 0 & 1
  \end{pmatrix}
  \frac{P^n -I - n(P-I)}{(P-I)^2}
  \begin{pmatrix}
    0\\
    1
  \end{pmatrix}
\end{gather*}
which is our final result for the expected number of amplicons
\textit{arising from the $g_d$ genomic strand}.

Note that is all amplification efficiencies are equal to 1, then we
obtain $DG_{d;n} = 2^{n-1}-1$ obtained earlier.

\subsubsection{$DG$ and $DG_d$ combined}

The mean number of tagged amplicons is given by the sum of the two
results obtained above, which may be written in the form:

\begin{gather*}
  \begin{pmatrix}
    p_{g}p_{h_d},& p_{g_d}p_{h}
  \end{pmatrix}
  \frac{P^n -I - n(P-I)}{(P-I)^2}
  \begin{pmatrix}
    0\\
    1
  \end{pmatrix}
\end{gather*}

If $p_{g}p_{h_d}= p_{g_d}p_{h}$, which will happen if $p_{g}= p_{g_d}$
and $p_{h_d}= p_{h}$, this simplifies to:
\begin{gather*}
  p_{g}p_{h}
  \begin{pmatrix}
    1 & 1
  \end{pmatrix}
  \frac{P^n -I - n(P-I)}{(P-I)^2}
  \begin{pmatrix}
    0\\
    1
  \end{pmatrix}
\end{gather*}
with
\begin{gather*}
  P = \begin{pmatrix}
    1 & p_a\\
    p_{a_d} & 1
  \end{pmatrix}
\end{gather*}
If $p_a = p_{a_d} = p$ then
\begin{gather*}
  P = \begin{pmatrix}
    1 & p\\
    p & 1
  \end{pmatrix}
\end{gather*}
then
\begin{gather*}
  P^n = \frac{1}{2}
  \begin{pmatrix}
    (1+p)^n+(1-p)^n & (1+p)^n - (1-p)^n\\
    (1+p)^n - (1-p)^n & (1+p)^n + (1-p)^n
  \end{pmatrix}
\end{gather*}
and
$$(P-I)^{-2} = 
\begin{pmatrix} p^{-2} & 0\\
  0 & p^{-2}
\end{pmatrix}
= \frac{1}{p^2}I
 $$
 
 After a little matrix algebra
 we obtain that the mean number of tagged amplicons is given by

$$E A_d = \frac{p_gp_h}{p^2}( (1+p)^n - np -1)$$
 
Note that if $p_g = p_h = p = 1$ this reduces to $2^n-n-1$ obtained
earlier for the Eulerian numbers.  If we have instead $p_g = p_h = p $
we obtain
$$E A_d = (1+p)^n - np -1,$$
which can be compared to the value $(1+p)^n$ of the simple \PCR model
that starts with a single amplicon.

Note that, if starting with $m$ amplicons these mean values are simply
multiplied by $m$.

Returning to the more general result:

\begin{gather*}
  \begin{pmatrix}
    p_{g}p_{h_d},& p_{g_d}p_{h}
  \end{pmatrix}
  \frac{P^n -I - n(P-I)}{(P-I)^2}
  \begin{pmatrix}
    0\\
    1
  \end{pmatrix}
\end{gather*}
with
\begin{gather*}
  P = \begin{pmatrix}
    1 & p_a\\
    p_{a_d} & 1
  \end{pmatrix}
\end{gather*}
The eigenvalues of $P$ are $1 + \sqrt{p_ap_{a_d}}$ and
$1 - \sqrt{p_ap_{a_d}}$, and it can be shown that:

\begin{gather*}
  P^n = \frac{1}{2}
  \begin{pmatrix}
    (1+\sqrt{p_ap_{a_d}})^n+(1-\sqrt{p_ap_{a_d}})^n &,& \sqrt{\frac{p_a}{p_{a_d}}}((1+\sqrt{p_ap_{a_d}})^n - (1-\sqrt{p_ap_{a_d}})^n)\\
    \sqrt{\frac{p_{a_d}}{p_{a}}}((1+\sqrt{p_ap_{a_d}})^n -
    (1-\sqrt{p_ap_{a_d}})^n)&, & (1+\sqrt{p_ap_{a_d}})^n +
    (1-\sqrt{p_ap_{a_d}})^n
  \end{pmatrix}
\end{gather*}
and
$$(P-I)^{-2} =\frac{1}{p_ap_{a_d}} I
 $$
 
 More matrix algebra
 shows that

\begin{gather*}
  \left(P^n -I - n(P-I)\right)
  \begin{pmatrix}
    0\\
    1
  \end{pmatrix}
  =
  \begin{pmatrix}
    \sqrt{\frac{p_a}{p_{a_d}}}[(1+\sqrt{p_ap_{a_d}})^n-(1-\sqrt{p_ap_{a_d}})^n])/{2}-np_a\\
    [(1+\sqrt{p_ap_{a_d}})^n+(1-\sqrt{p_ap_{a_d}})^n]/{2}-1
  \end{pmatrix}
\end{gather*}

Hence the mean number of amplicons for the general case is given by
the row and column matrix inner product:

\begin{gather*}
  \begin{pmatrix}
    \frac{p_{g}p_{h_d}}{p_ap_{a_d}},& \frac{p_{g_d}p_{h}}{p_ap_{a_d}}
  \end{pmatrix}
  \begin{pmatrix}
    \sqrt{\frac{p_a}{p_{a_d}}}[(1+\sqrt{p_ap_{a_d}})^n-(1-\sqrt{p_ap_{a_d}})^n]/{2}-np_a\\
    [(1+\sqrt{p_ap_{a_d}})^n+(1-\sqrt{p_ap_{a_d}})^n]/{2}-1
  \end{pmatrix}
\end{gather*}
that is:
$$
\frac{p_{g}p_{h_d}}{p_ap_{a_d}}\left(
  \sqrt{\frac{p_a}{p_{a_d}}}\frac{(1+\sqrt{p_ap_{a_d}})^n-(1-\sqrt{p_ap_{a_d}})^n}{2}-np_a
\right) + \frac{p_{g_d}p_{h}}{p_ap_{a_d}}\left(
  \frac{(1+\sqrt{p_ap_{a_d}})^n+(1-\sqrt{p_ap_{a_d}})^n}{2}-1 \right)
$$

If $p_a = p_{a_d} = p$, this simplifies to:

$$
\frac{p_{g}p_{h_d}}{p^2}\left( \frac{(1+p)^n-(1-p)^n}{2}-np \right) +
\frac{p_{g_d}p_{h}}{p^2}\left( \frac{(1+p)^n+(1-p)^n}{2}-1 \right)
$$

\subsubsection{Variance of the number of amplicons}

Recall that if $F(t)$ is a \PGF for a random variable $X$ then

\begin{align*}
  \E X &= \frac{dF(t)}{dt} \vert_{t=1} = F'(1)\\
  \E X(X-1) &= \frac{dF(t)}{dt}\vert_{t=1} = F''(1)\\
  \V X &= F''(1) + F'(1) - (F'(1))^2
\end{align*}

We can apply this to
$$F(t) = G_n(1,1,1,t)G_{d;n}(1,1, t, 1),$$
however is is simpler to consider the variances of the number of
tagged amplicons arising separately from the individual $g$ and $g_d$
strands and add them: because they amplify independently their
variances will add to give the total variance of interest. This we now
do.

Recall that our basic vectorial \PGF has the form
  
\begin{align*}
  G_{n+1} & \to G_n(1-p_g) + p_gG_nH_{d;n}\\
  H_{d;n+1} & \to H_{d;n}(1-p_{h_d}) + p_{h_d}H_{d;n}A_n\\
  A_{n+1} & \to A_n(1-p_{a}) + p_{a}A_nA_{d;n}\\
  A_{d;n+1} & \to A_{d;n}(1-p_{a_d}) + p_{a_d}A_{d;n}A_n
\end{align*}
with
\begin{align*}
  G_0 &= t_g\\
  H_{d,0} &= t_{h_d}\\
  A_{d} &= t_{a}\\
  A_{d,0} &= t_{a_d}
\end{align*}

Differentiating each twice with respect to $ t_{a_d}$ and setting all
the $t$'s equal to 1 gives this set of linear recurrence relations

\begin{align*}
  D^2G_{n+1} &= D^2 G_n + 2p_gDG_nDH_{d;n} + p_g D^2H_{d;n}\\
  D^2H_{d;n+1} & =D^2H_{d;n} +2 p_{h_d}DH_{d;n}DA_n + p_{h_d}D^2A_n\\
  D^2A_{n+1} & = D^2A_n + 2p_{a}DA_nDA_{d;n} + p_{a}D^2A_{d;n}\\
  D^2A_{d;n+1} & = D^2 A_{d;n} +2 p_{a_d}DA_{d;n}DA_n + p_{a_d}D^2A_n
\end{align*}

As before, we consider the last two coupled set of equations. These
may be written in matrix form as follows:

\begin{gather}
  \begin{pmatrix}
    D^2A_{n+1}\\
    D^2A_{d;n+1}
  \end{pmatrix}
  \quad= \quad
  \begin{pmatrix}
    1 & p_a\\
    p_{a_d} & 1
  \end{pmatrix}
  \begin{pmatrix}
    D^2A_{n}\\
    D^2A_{d;n}
  \end{pmatrix}
  \quad + \quad 2\begin{pmatrix}
    p_{a}DA_nDA_{d;n}\\
    p_{a_d}DA_{d;n}DA_n
  \end{pmatrix} \label{eq:dtwoa}
\end{gather}
Previously we had the first order solution
\begin{gather*}
  \begin{pmatrix}
    DA_{n}\\
    DA_{d;n}
  \end{pmatrix}
  \quad= \quad
  \begin{pmatrix}
    1 & p_a\\
    p_{a_d} & 1
  \end{pmatrix}^n
  \begin{pmatrix}
    0\\
    1
  \end{pmatrix}
  \quad =P^n
  \begin{pmatrix}
    0\\
    1
  \end{pmatrix}
\end{gather*}
where
\begin{gather*}
  P^n = \frac{1}{2}
  \begin{pmatrix}
    (1+\sqrt{p_ap_{a_d}})^n+(1-\sqrt{p_ap_{a_d}})^n &,& \sqrt{\frac{p_a}{p_{a_d}}}((1+\sqrt{p_ap_{a_d}})^n - (1-\sqrt{p_ap_{a_d}})^n)\\
    \sqrt{\frac{p_{a_d}}{p_{a}}}((1+\sqrt{p_ap_{a_d}})^n -
    (1-\sqrt{p_ap_{a_d}})^n)&, & (1+\sqrt{p_ap_{a_d}})^n +
    (1-\sqrt{p_ap_{a_d}})^n
  \end{pmatrix}
\end{gather*}
hence
\begin{gather*}
  \begin{pmatrix}
    DA_{n}\\
    DA_{d;n}
  \end{pmatrix}
  \quad= \quad \frac{1}{2}\begin{pmatrix}
    \sqrt{\frac{p_a}{p_{a_d}}}((1+\sqrt{p_ap_{a_d}})^n - (1-\sqrt{p_ap_{a_d}})^n)\\
    (1+\sqrt{p_ap_{a_d}})^n + (1-\sqrt{p_ap_{a_d}})^n
  \end{pmatrix}
\end{gather*}
and it follows that
$$
DA_nDA_{d;n} = \frac{1}{2}\sqrt{\frac{p_{a_d}}{p_{a}}} \left(
  (1+\sqrt{p_ap_{a_d}})^{2n} - (1-\sqrt{p_ap_{a_d}})^{2n} \right)
$$
This could be substituted into \eqref{eq:dtwoa} and the resulting
in-homogeneous difference equation may be solved (initial conditions
are that the second derivatives $D^2A_0 = D^2A_{d;0} = 0$). It gets a
bit complicated, so it is simpler to work from the following scalar
equation and the substitute matrices for scalar.

So consider the recurrence relation
$$y_{n+1} = p y_{n} + c(a^{n} - b^{n})$$
where $y_0 = 0$. Iteration gives the solution sequence:
\begin{align*}
  y_1 &= 0\\
  y_2 &= (a-b)c\\
  y_3 &= p(a-b)c + (a^2-b^2)c\\
  y_4 &= p^2(a-b)c + p(a^2-b^2)c + (a^3-b^3)c\\
      &\vdots \\
  y_n &= \sum_{i=0}^{n-2} p^{n-i}(a^{i+1} - b^{i+1})c\\
      &= \left(a\frac{p^{n-1}-a^{n-1}}{p-a} - b\frac{p^{n-1}-b^{n-1}}{p-b} \right)c\\
\end{align*}

Referring back , we may substitute
$c = \sqrt{p_{a_d}/p_a}\binom{p_a}{p_{a_d}}$,
$a = (1+\sqrt{p_{a_d}p_a})^2 I$, $b = (1-\sqrt{p_{a_d}p_a})^2I$, where
$I$ is the $2\times 2$ identity matrix, and $p=P$, to recover a matrix
formula for the solution of $y_n \equiv \binom{D^2A_n}{D^2A_{d;n}}$.

The result is a fairly complex matrix expression that has to be picked
apart to substitute various terms into the equations to solve for the
$h$ and $h_d$, which themselves in turn have to be substituted into
equations to solve for $g$ and $g_d$. This could possibly be done
using a computer algebra system, the results can be expected to be
messy - even more so when we introduce stutter - so we stop the
algebraic analysis here.

It seems appropriate therefore to generate matrix expressions at the
outset which would be more amenable to numerical evaluation. We now
follow this more direct approach. We could proceed by considering the
matrix formulation for $g$ and $g_d$ separately, or by using a
combined approach. We do the latter, they lead to the same results.

\subsection{Moments from matrix analysis}
\subsubsection{First moment: Mean as matrix expression}

The recurrence relation for the mean may be expressed as

\begin{gather*}
  \begin{pmatrix}
    DG_{n+1}\\
    DG_{d;n+1}\\
    DH_{n+1}\\
    DH_{d;n+1}\\
    DA_{n+1}\\
    DA_{d;n+1}
  \end{pmatrix}
  \quad= \quad
  \begin{pmatrix}
    1 & 0 & 0 & p_g & 0 & 0\\
    0 & 1 & p_{g_d} & 0 & 0 & 0\\
    0 & 0 & 1 & 0 & 0 & p_h\\
    0 & 0 & 0 & 1 & p_{h_d} & 0\\
    0 & 0 & 0 & 0 & 1 & p_a\\
    0 & 0 & 0 & 0 & p_{a_d} & 1\\
  \end{pmatrix}
  \quad
  \begin{pmatrix}
    DG_{n}\\
    DG_{d;n}\\
    DH_{n}\\
    DH_{d;n}\\
    DA_{n}\\
    DA_{d;n}
  \end{pmatrix}
  \quad = P \begin{pmatrix}
    DG_{n}\\
    DG_{d;n}\\
    DH_{n}\\
    DH_{d;n}\\
    DA_{n}\\
    DA_{d;n}
  \end{pmatrix}
\end{gather*}

where $P$ is now the $6 \times 6$ matrix. If we denote the column
matrix on the right by $Y_n$, then we have
$$Y_{n+1} = PY_n \to  Y_n = P^n Y_0$$
where $Y_0$ is the transpose of the row vector $(0,0,0,0,0,1)$, and
the mean number of tagged amplicons is given by the sum of the
elements in the first two rows of $Y_n$.

Numerically this is straightforward to extract given the values of the
various amplification probabilities.

\subsubsection{Second moment: as matrix expression}

The matrix equation for the second moment follows the pattern of the
first order moment matrix equation.  Let $Z_n$ denote the column
vector of second derivatives, that is

\begin{gather*}
  Z_n =
  \begin{pmatrix}
    D^2G_{n}\\
    D^2G_{d;n}\\
    D^2H_{n}\\
    D^2H_{d;n}\\
    D^2A_{n}\\
    D^2A_{d;n}
  \end{pmatrix}
\end{gather*}

Then the recurrence equations for the second moments may be written as

\begin{gather*}
  Z_{n+1} = P Z_n + 2
  \begin{pmatrix}
    p_g DG_{n}DH_{d;n}\\
    p_{g_d} DG_{d;n}DH_{n}\\
    p_{h} DH_{n}DA_{d;n}\\
    p_{h_d} DH_{d;n}DA_{n}\\
    p_{a} DA_{n}DA_{d;n}\\
    p_{a_d} DA_{d;n}DA_{n}
  \end{pmatrix}
  = PZ_n + 2
  \begin{pmatrix}
    p_gY_n(1)Y_n(4)\\
    p_{g_d} Y_n(2)Y_n(3)\\
    p_{h} Y_n(3)Y_n(6) \\
    p_{h_d}Y_n(4)Y_n(5) \\
    p_{a} Y_n(5)Y_n(6) \\
    p_{a_d} Y_n(6)Y_n(5)
  \end{pmatrix}
\end{gather*}

where $Y_n(i)$ is the element in the $i^{th}$ row of the column vector
$Y_n$. We have the initial condition $Z_0 = (0,0,0,0,0,0)^T$.

Thus we can iterate to solve sequentially for the pairs of column
vectors $Y_1, Z_1$; $Y_2, Z_2$; \ldots; $Y_n, Z_n$.

Now $Y_n(1)$ will denote the mean number of tagged amplicons arising
from the $g$ strand, and $Z_n(1)$ will denote the corresponding
expectation $E M(M-1)$ for the number of tagged amplicons arising for
the $g$ strand. Hence the variance of tagged amplicons arising form
the $g$ strand will be
$$ Z_n(1) + Y_n(1) - Y^2_n(1)$$
Similarly the variance of tagged amplicons from the $g_d$ strand will
be
$$ Z_n(2) + Y_n(2) - Y^2_n(2)$$
Hence the total variance of tagged amplicons will be
$$ Z_n(1) + Y_n(1) - Y^2_n(1) + Z_n(2) + Y_n(2) - Y^2_n(2)$$
If there are $m$ genomic strands to begin with, we simply multiply
this variance expression by $m$ to get the required total variance.

\section{Derivation of moments for genomic model, single stutters}
\label{app:gmom2}

\subsection{Moments}

The combined set of iterative equations of the branching process is:

\begin{align*}
  t_g & \to t_g(1-p_g) + p_g(1-\xi)t_gt_{h_d} + p_g\xi t_gt_{h_{sd}}\\
  t_{g_d} & \to t_{g_d}(1-p_{g_d}) + p_{g_d}(1-\xi)t_{g_d}t_{h} + p_{g_d}\xi t_{g_d}t_{h_{s}}\\
  t_{h_d} & \to t_{h_d}(1-p_{h_d}) + p_{h_d}(1-\xi)t_{h_d}t_{a}+ p_{h_d}\xi t_{h_d}t_{{a_s}}\\
  t_{h_{sd}} & \to t_{h_{sd}}(1-p_{h_{sd}}) + p_{h_{sd}} t_{h_{sd}}  t_{a_s} \\
  t_{h} & \to t_{h}(1-p_{h}) + p_{h}(1-\xi)t_{h}t_{a_d}+ p_{h}\xi t_{h}t_{{a_{sd}}}\\
  t_{h_{s}} & \to t_{h_{s}}(1-p_{h_{s}}) + p_{h_{s}} t_{h_{s}} t_{a_s} \\
  t_{a} & \to t_{a}(1-p_{a}) + p_{a}(1-\xi)t_at_{a_d} + p_{a}\xi t_{a}t_{a_{sd}} \\
  t_{a_d} & \to t_{a_d}(1-p_{a_d}) + p_{a_d}(1-\xi) t_{a_d}t_{a} + p_{a_d}\xi t_{a_d}t_{a_s}\\
  t_{a_s} & \to t_{a_s}(1-p_{a_s}) + p_{a_s}t_{a_s}t_{a_{sd}}\\
  t_{a_{sd}} & \to t_{a_{sd}}(1-p_{a_{sd}}) + p_{a_{sd}}t_{a_{sd}}t_{a_s}
\end{align*}
which lead directly to the coupled set of equations for the vectorial
generating function:
\begin{align*}
  G_{n+1} & = G_n[(1-p_g)+ p_g(1-\xi) H_{n;d} + p_g\xi  H_{n;sd}]\\
  G_{n+1;d} &= G_{n;d} [ (1-p_{g_d}) + p_{g_d}(1-\xi)H_n + p_{g_d}\xi H_{n;s}] \\
  H_{n+1}& = H_{n}[ (1-p_{h}) + p_{h}(1-\xi)A_{n;d}+ p_{h}\xi A_{n;sd}]\\
  H_{n+1;d}& = H_{n;d}[ (1-p_{h_d}) + p_{h_d}(1-\xi)A_{n}+ p_{h_d}\xi A_{n;s}]\\
  H_{n+1;s} &=  H_{n+1;s} [(1-p_{h_{s}}) + p_{h_{s}}A_{n;sd}] \\
  H_{n+1;sd}& = H_{n;sd}[(1-p_{h_{sd}}) + p_{h_{sd}} A_{n;s}] \\
  A_{n+1} & = A_n[ (1-p_{a}) + p_{a}(1-\xi)A_{n;d} + p_{a}\xi A_{n;sd}] \\
  A_{n+1;d} & = A_{n;d} [ (1-p_{a_d}) + p_{a_d}(1-\xi) A_{n} + p_{a_d}\xi A_{n;s}] \\
  A_{n+1;s} & = A_{n;s}[(1-p_{a_s}) + p_{a_s}A_{n;sd} ]\\
  A_{n+1;sd} & = A_{n;sd}[(1-p_{a_{sd}}) + p_{a_{sd}}A_{n;s}]
\end{align*}

Differentiating once for the mean, we obtain the matrix equation
\begin{gather*}
  \begin{pmatrix}
    DG_{n+1}\\
    DG_{n+1;d}\\
    DH_{n+1}\\
    DH_{n+1;d}\\
    DH_{n+1;s}\\
    DH_{n+1;sd}\\
    DA_{n+1}\\
    DA_{n+1;d} \\
    DA_{n+1;s}\\
    DA_{n+1;sd}
  \end{pmatrix}
  =
  \begin{pmatrix}
    1& 0 & 0 & p_g(1-\xi) & 0 &p_g\xi &0&0&0&0\\
    0 & 1 & p_{g_d}(1-\xi) & 0 & p_{g_d}\xi & 0 & 0 &0 &0 &0\\
    0 & 0 & 1 & 0 & 0 & 0 & 0 & p_{h}(1-\xi) & 0 & p_{h}\xi  \\
    0 & 0 & 0 & 1& 0 & 0 & p_{h_d}(1-\xi) & 0 & p_{h_d}\xi & 0\\
    0 & 0 & 0 & 0 & 1 &  0 & 0 & 0 & 0 & p_{h_s}\\
    0 & 0 & 0 & 0 & 0 &  1 & 0 & 0  & p_{h_{sd}}&0 \\
    0 & 0 & 0 & 0 & 0 &   0& 1 & p_a(1-\xi) & 0 & p_a\xi \\
    0 & 0 & 0 & 0 & 0 &   0& p_{a_d}(1-\xi) & 1 & p_{a_d}\xi & 0\\
    0 & 0 & 0 & 0 & 0 &   0& 0 & 0 & 1 & p_{a_s} \\
    0 & 0 & 0 & 0 & 0 &   0& 0 & 0 &  p_{a_{sd}} & 1 \\
  \end{pmatrix}
  \begin{pmatrix}
    DG_{n}\\
    DG_{n;d}\\
    DH_{n}\\
    DH_{n;d}\\
    DH_{n;s}\\
    DH_{n;sd}\\
    DA_{n}\\
    DA_{n;d} \\
    DA_{n;s}\\
    DA_{n;sd}
  \end{pmatrix}
\end{gather*}

Denoting the $10 \times 10$ square matrix by $P$, and the column
matrix on the right by $Y_n$, then we have
$$Y_{n+1} = PY_n \to  Y_n = P^n Y_0$$
where $Y_0$ is the transpose of the row vector
$(0,0,0,0,0,0,0,0,0,1)$, and the mean number of tagged amplicons is
given by the sum of the elements in the first two rows of $Y_n$. This
will be equal to $P^n[1,10] + P^n[2,10]$.


Let $Z$ denote the column of second derivatives, then we have the
following in-homogeneous matrix recurrence relation that is readily
solved numerically by iteration:

\begin{gather*}
  Z_{n+1} = P Z_n + 2
  \begin{pmatrix}
    p_gDG_{n}[(1-\xi) DH_{n:d} + \xi DH_{n;sd}] \\
    p_{g_d}DG_{n;d}[(1-\xi) DH_{n} + \xi DH_{n;s}]\\
    p_hDH_{n}[ (1-\xi)DA_{n;d} + \xi DA_{n;sd}]\\
    p_{h_d}DH_{n;d}[ (1-\xi)DA_{n} + \xi DA_{n;s}]\\\
    p_{h_s}DH_{n;s}DA_{n;sd}\\
    p_{h_{sd}}DH_{n;sd}DA_{n;s}\\
    p_aDA_{n}[ (1-\xi)D A_{n;d} + \xi DA_{n;sd}]\\
    p_{a_d}DA_{n;d}[ (1-\xi) DA_n + \xi DA_{n;s}] \\
    p_{a_s}DA_{n;s} DA_{n;sd}\\
    p_{a_{sd}}DA_{n;sd} DA_{n;s}
  \end{pmatrix}
\end{gather*}

and similarly to before, the total variance of tagged amplicons in
stutter position is given by
$$ Z_n(1) + Y_n(1) - Y^2_n(1) + Z_n(2) + Y_n(2) - Y^2_n(2)$$

\subsection{Covariance with main peak}

The vectorial generating function may be used to find the mean and
variance of the main peak, and also the covariance between main peak
and stutter peak.  For the main peak, the mean height is given by
multiplying $P^n$ into the column matrix that has zero everywhere
except for a 1 on the seventh row, and then adding the values in the
first two rows of the resulting column matrix.  The variance is
similarly found.

The second derivatives obey the following recurrence relation.
{\small
  \begin{gather*}
    \begin{pmatrix}
      DDsG_{n+1}\\
      DDsG_{n+1;d}\\
      DDsH_{n+1}\\
      DDsH_{n+1;d}\\
      DDsH_{n+1;s}\\
      DDsH_{n+1;sd}\\
      DDsA_{n+1}\\
      DDsA_{n+1;d} \\
      DDsA_{n+1;s}\\
      DDsA_{n+1;sd}
    \end{pmatrix}
    = P
    \begin{pmatrix}
      DDsG_{n}\\
      DDsG_{n;d}\\
      DDsH_{n}\\
      DDsH_{n;d}\\
      DDsH_{n;s}\\
      DDsH_{n;sd}\\
      DDsA_{n}\\
      DDsA_{n;d} \\
      DDsA_{n;s}\\
      DDsA_{n;sd}
    \end{pmatrix}
    +
    \begin{pmatrix}
      p_gDG_{n}[(1-\xi) DsH_{n:d} + \xi DsH_{n;sd}] \\
      p_{g_d}DG_{n;d}[(1-\xi) DsH_{n} + \xi DsH_{n;s}]\\
      p_hDH_{n}[ (1-\xi)DsA_{n;d} + \xi DsA_{n;sd}]\\
      p_{h_d}DH_{n;d}[ (1-\xi)DsA_{n} + \xi DsA_{n;s}]\\\
      p_{h_s}DH_{n;s}DsA_{n;sd}\\
      p_{h_{sd}}DH_{n;sd}DsA_{n;s}\\
      p_aDA_{n}[ (1-\xi)Ds A_{n;d} + \xi DsA_{n;sd}]\\
      p_{a_d}DA_{n;d}[ (1-\xi) DsA_n + \xi DsA_{n;s}] \\
      p_{a_s}DA_{n;s} DsA_{n;sd}\\
      p_{a_{sd}}DA_{n;sd} DsA_{n;s}
    \end{pmatrix}
    +
    \begin{pmatrix}
      p_gDsG_{n}[(1-\xi) DH_{n:d} + \xi DH_{n;sd}] \\
      p_{g_d}DsG_{n;d}[(1-\xi) DH_{n} + \xi DH_{n;s}]\\
      p_hDsH_{n}[ (1-\xi)DA_{n;d} + \xi DA_{n;sd}]\\
      p_{h_d}DsH_{n;d}[ (1-\xi)DA_{n} + \xi DA_{n;s}]\\\
      p_{h_s}DsH_{n;s}DA_{n;sd}\\
      p_{h_{sd}}DsH_{n;sd}DA_{n;s}\\
      p_aDsA_{n}[ (1-\xi)D A_{n;d} + \xi DA_{n;sd}]\\
      p_{a_d}DsA_{n;d}[ (1-\xi) DA_n + \xi DA_{n;s}] \\
      p_{a_s}DsA_{n;s} DA_{n;sd}\\
      p_{a_{sd}}DsA_{n;sd} DA_{n;s}
    \end{pmatrix}
  \end{gather*}
}

\section{A further look at target and stutter peak height correlations}
\label{sec:brightstutter}


In \cite{bright2013developing} the authors looked at developing a
model for the stutter ratio, defined as the ratio of the stutter peak
to main peak, such that the mean stutter ratio is linearly dependent
on the \textit{longest uninterrupted sequence} (LUS) of repeats. They used
controlled experimental single-source samples, 289 in all, with a
target of 1ng of DNA and amplified with a 25 RFU detection
limit. After discarding loci which had heterozygous loci one repeat
apart, 2323 heterozygous loci were left for analysis.

It is important to note that they define stutter ratio as the ratio of
the stutter peak height to the main peak height:
$$SR_a = \frac{O_{a-1}}{O_a}$$
They define the total allelic product to be
$$T_a = O_{a-1}+ O_a$$

They propose a lognormal distribution for the ratio of observed to
expected peak height for main and stutter peaks (they comment that a
gamma model is not suitable because of the heavy tails). From their
data they conclude that there is little evidence of correlation
between the main and stutter peaks given the target amount of
DNA. They write: ``Unexpectedly the scatter plots
  in Fig. 8a and b indicate that there is no detectable correlation
  between stutter and allele in this biological model''. They found a
low correlation of around 0.11.

Now clearly having little to no correlation is quite at variance with
the branching \PCR model.  We suggest that this is an artefact of
grouping the data together within and across loci. First we recast
their results. In their Appendix 2 they give the following table for
the linear regression of their stutter ratio model (using the NGM
SELect system), in which the stutter ratio of an allele $i$, having  LUS $L_i$,  from a locus is equal to
$ SR_i = a + b\times L_i$:

\begin{center}
  \begin{tabular}{rlrr}
    \hline
    & Locus & Intercept $a$ & Slope $b$ \\ 
    \hline
    1 & D10S1248 & -0.0576 & 0.0089 \\ 
    2 & D12S391 & -0.0571 & 0.0107 \\ 
    3 & D16S539 & -0.0502 & 0.0088 \\ 
    4 & D18S51 & -0.0297 & 0.0066 \\ 
    5 & D19S433 & -0.0302 & 0.0074 \\ 
    6 & D1S1656 & -0.0699 & 0.0106 \\ 
    7 & D21S11 & -0.0079 & 0.0059 \\ 
    8 & D22S1045 & -0.0881 & 0.0139 \\ 
    9 & D2S1338 & -0.0073 & 0.0062 \\ 
    10 & D2S441 & 0.0004 & 0.0031 \\ 
    11 & D3S1358 & -0.0455 & 0.0092 \\ 
    12 & D8S1179 & -0.0148 & 0.0062 \\ 
    13 & FGA & -0.0344 & 0.0066 \\ 
    14 & SE33 & 0.0129 & 0.0041 \\ 
    15 & TH01 & -0.0208 & 0.0052 \\ 
    16 & vWA & -0.0354 & 0.0078 \\ 
    \hline
  \end{tabular}
\end{center}

Now from the simple amplicon model, with amplification probability $p$
and conditional stutter probability $\xi$, the expected number of
amplicons heights for main and stutter peaks given $k$ cycles and
starting from $n_0$ amplicons, are given by

\begin{align*}
  ET &= n_0(1 +p(1-\xi))^k\\
  ES &= n_0[(1 +p)^k -  (1 +p(1-\xi))^k] \\
\end{align*}

Thus we can approximate, with $\xi$ dependent on $L$:

$$SR = a + b L \approx \frac{ES}{ET} =  \left(\frac{1+p}{(1+p(1-\xi))}\right)^k - 1
\approx \frac{kp}{1+p} \xi + \frac{k(k-1)p^2}{2(1+p)^2}\xi^2 +
0(\xi^3)$$

From this we obtain to a good approximation, by taking the linear term
only, the incremental change in mean stutter ratio by a unit increase
in LUS:
$$\Delta\xi = b\frac{1+p}{kp}$$

For $p \in [0.8,1]$, $(1+p)/p$ varies between 2 and
2.25, hence to a good approximation $\Delta\xi = 2b/k$.

So now we can collect results. Given the nature of the experimental
setup, with a target DNA amount of 1ng, (corresponding to around 152
cells) we may take the number of alleles from a heterozygous marker to
be $Pois(152)$. In addition the LUS in their plots varies from 10-15
for most markers, some having a larger range, some smaller.  So for
simplicity we shall simulate LUS values from the same range for each
of the 16 markers, (also corresponding therefore to a uniform
distribution of alleles).  The following table collects individual
marker correlations, based on around 2000 simulations on each marker,
under various scenarios, but using the $\xi$ values for each marker
using the SR formula above; we take $k=28$ and $p=0.85$ in all
simulations.

\begin{center}
  \begin{tabular}{l|cccc}
    Locus& 
           LUS=12&
                   LUS=12&
                           LUS$\sim$Unif[8,15]&
                                                LUS$\sim$Unif[8,15] \\ 
         &$n_0=152$&
                     $n_0\sim Pois(152)$&
                                          $n_0=152$&
                                                     $n_0\sim Pois(152)$ \\ \hline
    D10S1248 &0.2005406 &0.7255809 &-0.5541554 &-0.07269729 \\
    D12S391 &0.2290437 &0.773266 &-0.6050955 &-0.04926867 \\
    D16S539 &0.2054748 &0.7379974 &-0.5366395 &-0.03364616 \\
    D18S51 &0.1665038 &0.7108246 &-0.4260814 &0.06089547 \\
    D19S433 &0.248041 &0.7182632 &-0.4844209 &0.06972643 \\
    D1S1656 &0.1848627 &0.7401926 &-0.6481533 &-0.0976838 \\
    D21S11 &0.1952987 &0.7507546 &-0.3596321 &0.2097556 \\
    D22S1045 &0.217241 &0.7738013 &-0.7170921 &-0.1489542 \\
    D2S1338 &0.2145392 &0.7708068 &-0.3787377 &0.1656755 \\
    D2S441 &0.1509088 &0.6713596 &-0.1463518 &0.2585785 \\
    D3S1358 &0.2333797 &0.7477239 &-0.5480802 &-0.03041745 \\
    D8S1179 &0.2451524 &0.731677 &-0.3884186 &0.1381175 \\
    FGA &0.1308529 &0.6779035 &-0.4587745 &0.06434562 \\
    SE33 &0.1765904 &0.6558634 &-0.2304545 &0.1803573 \\
    TH01 &0.1666116 &0.6661396 &-0.3677087 &0.1420518 \\
    vWA &0.1850046 &0.7455873 &-0.5083684 &0.0339069 \\ \hline
  \end{tabular}
\end{center}

We see that keeping LUS and $n_0$ fixed we get moderate correlations
as in earlier chapter. Making the starting number of alleles Poisson
distributed and keeping LUS fixed leads to high correlation between
the stutter and peak amplicon numbers, also as found earlier. Now
making LUS random (to simulate a range of genotypes) but keeping $n_0$
fixed the correlations are negative. Combining the randomization of
LUS and making $n_0\sim Pois(132)$ yields the correlations in the
final column, which are largely quite small with some positive and
some negative.

Finally we do a simulation in which we aggregate the various loci
simulations: here are 16 runs of such a simulation which mirrors the
experimental data of \cite{bright2013developing}. We see that all the
correlations are negative, but very close to zero.

\begin{center}
  \begin{tabular}{cccc} \hline
    -0.003916072 &-0.02691701 &-0.0257117 &-0.03113958 \\
    -0.02115133 &-0.02109356 &-0.02308784 &-0.05802165 \\
    -0.02720094 &-0.0288306 &-0.0345473 &-0.07301637 \\
    -0.01966967 &-0.031646 &-0.0649325 &-0.02202189 \\ \hline
  \end{tabular}
\end{center}

The following scatterplot was obtained from one simulation, in which
we plot the log of the sumulated number of amplicons divided by their
expected number (itself estimated from simulation) for the target and
stutter alleles, so that it corresponds to the quantities plotted in
Figure~8b of \cite{bright2013developing}. The plot below has a smaller
range than their Figure~8b, which could be because we have simulated
from the amplicon model rather than the genomic model.  The sample
correlation of the points in the plot is 0.157, close to the figure of
0.11 they found from their experimental data.

\begin{center}
  \includegraphics[scale=0.8]{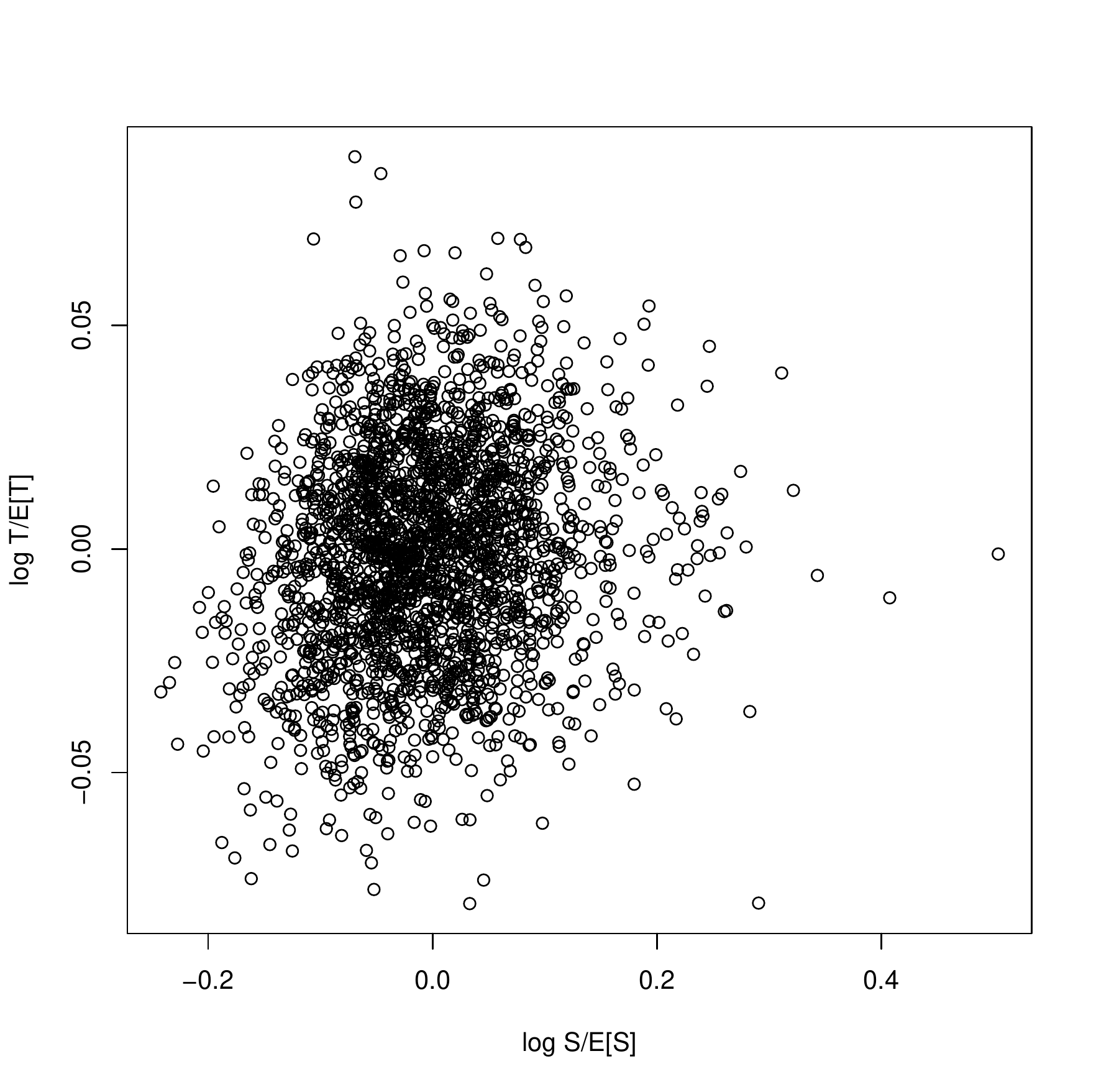}
\end{center}

Hence, on the basis of this
simulation,  we deduce that the bivariate  \PCR branching process model showing high main allele 
and stutter-peak correlations on individual alleles is
compatible with the experimental results of 
\cite{bright2013developing} which aggregates observations of  alleles within and between loci.

Support for this may be found in a dataset of experimental single
source and mixed profiles released by Boston University in
2012\footnote{see
  \url{http://www.bu.edu/dnamixtures/pages/help/introduction/}} These
were samples prepared from extracted DNA from 4 anonymous individuals
(labelled A,B,C and D) diluted to prepare controlled DNA amounts, and
amplified for four kits. In the single source samples there many
independent amplifications of each individual's DNA, allowing the
target-stutter peak height data to be found for individual alleles for
a given amount of initial DNA, making sure that neither are
"contaminated" by neighbouring alleles.  Looking at the genotypes , we
see that for person A allele 25 on the locus D2D1338 is a possible
target allele. Taking all the single source samples on 1ng and 10
seconds injection time for person A amplified with the
Identifiler\texttrademark kit (chosen here because it yields the most
number of data points), the author extracted target (25) and stutter
(24) peak height values to produce the following scatterplot , having
a sample correlation of 0.929.

\begin{center}
\includegraphics[scale=0.5]{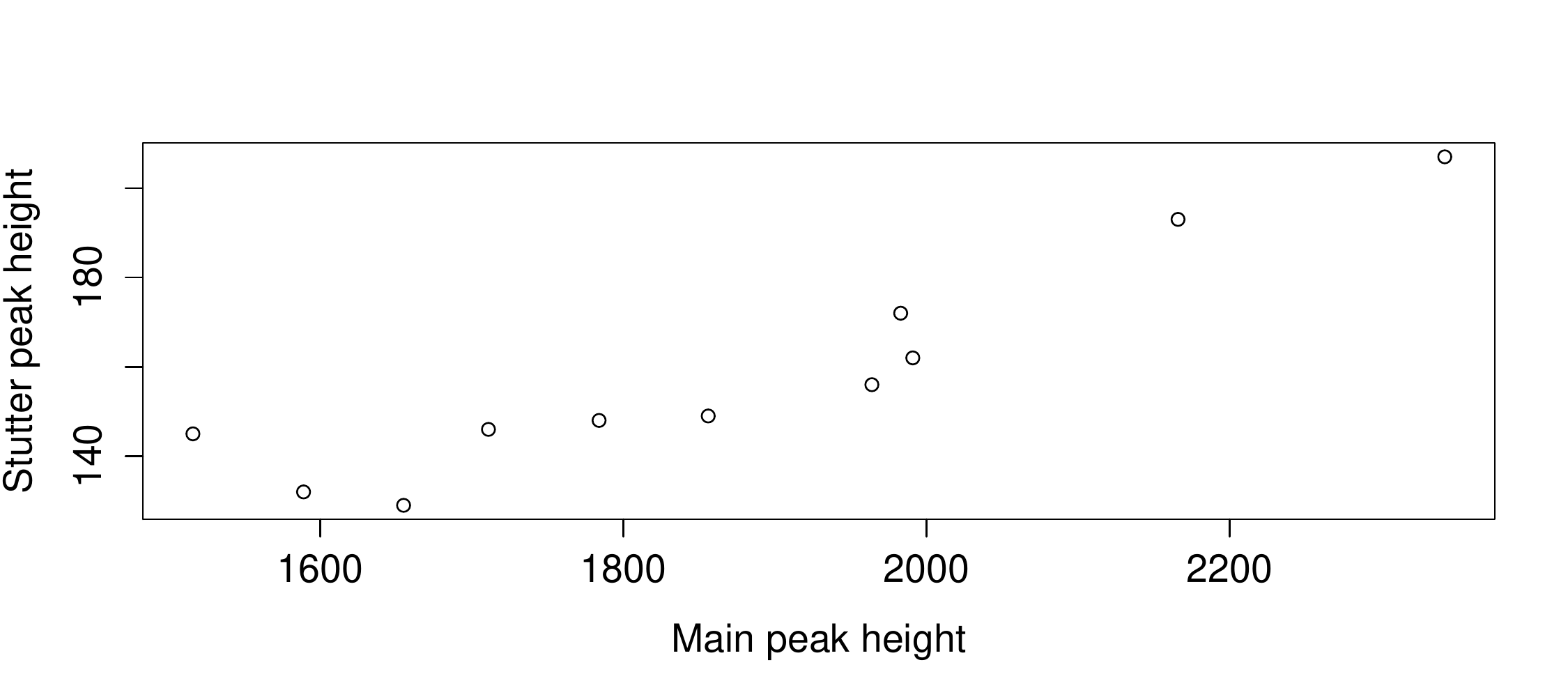}
\end{center}

Bearing in mind the way the samples were prepared, we should expect
Poisson sampling for the initial number of amplicons prior to \PCR,
and hence expect a high correlation to result as found. Looking at the
stutter proportions (stutter/total) which average around 0.08 for the
11 data points, we see that the results are very consistent with \PCR
branching modelling approach for stutter.

In conclusion, we suggest that the combined effects of aggregating
across loci and alleles might be giving a misleading impression that
there is no significant correlation.  Dedicated experiments
concentrating on specific alleles and their stutter product, that do
not aggregate data over different alleles either within or between
loci, could be carried out to confirm or refute this.

\newpage
\section{Sample code for generating distributions and plots}
\subsection{Python scripts}

\subsubsection{code for \figref{fig:ampkd}.}
\label{sec:py:ampkd}
\begin{verbatim}
import numpy as np
import matplotlib.pyplot as plt
from scipy.stats.kde import gaussian_kde

K = 28
M=1
phi = 2.0/9
p = 0.8

N = np.ones(100000).astype(int)
N = N*M
n = np.random.binomial(N,phi)

for k in range(K):
    m = np.random.binomial(n, p)
    n = n+m
 
# find total dropout probability, 
pdrop = float(len(n[n==0]))/len(n)
# remove zeros for the kernel density estimate
n = n[n != 0]

# find the kernel density estimate
pcr_dist = gaussian_kde(n)

# plot it out
x = np.linspace(1,3.0e7,1000)

fig = plt.figure()
plt.plot(x, pcr_dist(x),'r')
\end{verbatim}

\subsubsection{code for \figref{fig:ampfft28}}
\label{sec:py:ampfft}

\begin{verbatim}
import numpy as np
import matplotlib.pyplot as plt

def  singleAmpliconPCR(p,K, M, phi):
   N = M*2**K
   a = np.zeros(N)
   a[1] = 1.0
   afft = np.fft.fft(a)
   for k in xrange(K):
       afft = (1-p)*afft + p*afft*afft
   afft = (1-phi + phi*afft)**M
   pn  = np.fft.ifft(afft)
   pn = pn.real
   return pn

pn = singleAmpliconPCR(p=0.8,K=28,M=1, phi=2.0/11)
n = range(len(pn))
pdrop = pn[0]
pn[0] = 0
plt.plot(n,pn,'-')
\end{verbatim}

\subsubsection{code for
  \algref{alg-ampliconjointfft}} \label{py:alg-ampliconjointfft}
\label{sec-ampliconjointfftPy}

\begin{alg}[{\sc Joint distribution for target and stutter amplicons:
    Python code}]
  \label{alg-ampliconjointfftPy}
\begin{verbatim}
import numpy as np
from scipy.fftpack import fftn, ifftn

def  ampliconStutterPCR(p, xi, K, M, phi):
   N = M*2**K
   NT = N
   NS = N
   F = np.zeros((NT,NS))
   F[1,0] = 1
   F = fftn(F)
   
   G = np.zeros(NS)
   G[1] = 1
   G = fftn(G)
   for k in xrange(K):
       for g in xrange(NS):
          for  f in xrange(NT) :
              F[f,g] = F[f,g]*(1-p + p*(1-xi)*F[f,g]+ p*xi*G[g])
          G[g] = G[g]*(1-p + p*G[g])
       F[f,g] = (1-phi + phi*F[f,g])**M
   F = ifftn(F)
   F = F.real
   return F
\end{verbatim}
  \sbackup
\end{alg}
Note that even for $M=1$ and with a typical $K=28$ cycles in \PCR, the
size of the 2-dimensional array $F$ to hold the joint distribution
will be $N=2^{56} \approx 7.2\times 10^{16}$, which is far in excess
of the memory of any computer.  To fill the array with double
precision values would require approximation 4.2 billion Gigabytes of
Ram. Filling the array would also take an excessive amount of
time. Hence this should be used only for small $K$ and $M$ values.

\subsubsection{code for
  \algref{alg-ampliconstuttermargfft}} \label{py:alg-ampliconstuttermargfft}

\begin{alg}[{\sc Marginal distribution for stutter amplicons: Python
    code}]
  \label{alg-ampliconstuttermargfftPy}
\begin{verbatim}
import numpy as np
from scipy.fftpack import fftn, ifftn

def  ampliconStutterMarginalPCR(p, xi, K, M, phi):
   N = M*2**K
   F = np.zeros(N)
   F[0] = 1
   F = fftn(F)
   G = np.zeros(N)
   G[1] = 1
   G = fftn(G)
   for k in xrange(K):
      F *= 1-p + p*(1-xi)*F+ p*xi*G
      G *= 1-p + p*G
   F = (1-phi + phi*F)**M
   F = ifftn(F)
   F = F.real
   return F
\end{verbatim}
  \sbackup
\end{alg}

\newpage
\subsection{R scripts}

\subsubsection{code for generating target marginal distribution, no
  stutters} \label{r-alg-gillampliconfRcode}

R code for the distribution of the total number of amplicons arising
from $K $ amplification cycles, amplification probability $p$ on each
cycle, starting with $M$ amplicons binomially sampled prior to
amplification with probability $\phi$.  Note that $R$ vector indices
start at 1 instead of 0, hence on line 3 we set $F[2] = 1$ (rather
than $F[1]=1$).

\begin{alg}[{\sc Amplicon probability distribution using R}]
  \label{alg-gillampliconfRcode}
\begin{verbatim}

N = M*2**K
F = rep(0,N)
F[2] = 1														
F = fft(F,inverse=FALSE)
for (k in 1:K) F = (1 - p)*F + p*F*F    # K amplifications cycles
F = (1-phi + phi*F)**M                  # binomial sampling
F = Re(fft(F,inverse=TRUE)) /N            # real part of inverse
\end{verbatim}
\end{alg}

\subsubsection{code to generate
  \figref{fig:dropoutamp}}\label{r-fig:dropoutamp}
\begin{verbatim}
K=15
Th= 40000
M = 200
phi = 0.1
p = 0.85
N = M*2**K
F = rep(0,N)
F[2] = 1  													
F = fft(F,inverse=FALSE)
for (k in 1:K) F = (1 - p)*F + p*F*F    # K amplifications cycles
Fphi = 1-phi + phi*F
pdhet = NULL
pdhom = NULL
for (m in 5:M){
# heterozygous
G = Fphi**m                 # binomial sampling
G = Re(fft(G,inverse=TRUE)) /N            # real part of inverse
G = cumsum(G)
pdhet = c(pdhet, G[Th])
# use double m for homozygous
G = Fphi**(2*m)     
G = Re(fft(G,inverse=TRUE))  /N
G = cumsum(G)
pdhom = c(pdhom, sum(G[Th]))
cat ("m = ", m, "\n")
}
m = 5:M
plot(pdhet~m,log="x", ylim = c(0,1), typ="l", xlab = "Number of cells", 
			ylab = "Dropout probability P(D)")
par(new=T)
plot(pdhom~m, col="red", log="x", ylim=c(0,1), typ="l", 
			xlab = "Number of cells", 	ylab = "Dropout probability P(D)")
\end{verbatim}

\subsubsection{code for generating marginal distributions for the
  genomic model, implementing \algref{alg:gemomemarg}
} \label{sec:r-fullgill09}

\begin{verbatim}
N = M*2^K
  g = rep(0,N)
  g[1]=1
  g = FFT(g)
  
  gd = h = hd = hs = hsd  = a = ad = as = g
  asd = rep(0,N)
  asd[2]=1 			# note use of index 2.
  asd = FFT(asd)

for (k in 1:K){
  g = g*(1-q + q*hd)
  gd = gd*(1-q + q*h)
  hd = hd*(1-p + p*a)
  h = h*(1-p + p*ad)
  at  = a   	# temporary copy to break the cyclic dependency of a and ad
  atd = ad    # temporary copy to break the cyclic dependency of a and ad
  ad = atd*(1-p + p*at)
  a = at*(1-p + p*atd)
}

# the final probability distribution of interest.
pn = Re(IFFT( (1 -phi + phi*g*gd)^M ))  

\end{verbatim}

\newpage
\subsection{Julia scripts}

\subsubsection{code for generating \figref{fig:fulltraunc}}\label{jl:full trunc}
Julia, like R, uses arrays with indices starting from 1.  You will
need to install the PyPlot package to run this script, and will
require approximately 10Gb of free ram to do the full \FFT evaluation
with $K=28$.

\begin{verbatim}
using PyPlot

# function for full fft analysis
function maindist(p::Float64, K::Int64)
    N = 2^K
    y= zeros(N)
    y[2] = 1
    y=fft(y)
    for i in 1:N
        for k in 1:K
            y[i] = y[i]*(1-p+p*y[i])
        end
    end
    y=ifft(y)
    return y
end

# truncated analysis
function maintrunc1(p::Float64, K::Int64, npos::Int64,  L::Int64)
    N = 2^K
    pn = 1.0
    for j in 2:L 
        x = exp(-2pi*im*(j-1)/N)
        for k in 1:K
            x = x*(1-p + p*x)
        end
        x = x*exp(2pi*im*(npos-1)*(j-1)/N)
        pn = pn + 2*real(x)
    end
    return pn /N  
end

p = 0.85
K = 28
cy = maindist(p, 28)

# pick a subset of points for the plot, to speed up the plotting
step = 60000
xv = [ i*step for i in 1:1000]
pnfull = [real(cy[i*step]) for i in 1:1000]
# generate the truncated values
pntruncated =  [maintrunc1(p, K, i*step,1024)   for i in 1:1000]

xlabel("Number of amplicons, n") 
ylabel("P(n)")
fig, ax = subplots()
ax[:plot](xv,pnfull, label = "FFT")
ax[:plot](xv,pntruncated , label = "truncated")
ax[:legend]()
title("p = 0.85, K = 28")
\end{verbatim}

\cleardoublepage \addcontentsline{toc}{section}{\bibname}

\let\myindtmp\indexspace 
\renewcommand{\indexspace}{\myindtmp\vspace*{-2pt}}
 
\cleardoublepage \markboth{\indexname}{\indexname} \printindex

\end{document}